\documentclass[12pt,preprint]{emulateapj}

\shorttitle{ALMA HCN/HCO$^{+}$ J=3--2 observations of Seyferts and LIRGs} 
\shortauthors{Imanishi et al.}

\begin{document}


\title{ALMA HCN and HCO$^{+}$ J=3--2 observations of optical Seyfert and 
luminous infrared galaxies  
-- Confirmation of elevated HCN-to-HCO$^{+}$ flux ratios in AGNs --}


\author{Masatoshi Imanishi \altaffilmark{1,2}}
\affil{National Astronomical Observatory of Japan, 
National Institutes of Natural Sciences (NINS), 
2-21-1 Osawa, Mitaka, Tokyo 181-8588, Japan}
\email{masa.imanishi@nao.ac.jp}

\author{Kouichiro Nakanishi \altaffilmark{2}}
\affil{National Astronomical Observatory of Japan, 
National Institutes of Natural Sciences (NINS), 
2-21-1 Osawa, Mitaka, Tokyo 181-8588, Japan}

\and

\author{Takuma Izumi}
\affil{Institute of Astronomy, School of Science, The University of Tokyo,
2-21-1 Osawa, Mitaka, Tokyo 181-0015, Japan}

\altaffiltext{1}{Subaru Telescope, National Astronomical Observatory of
Japan, National Institutes of Natural Sciences (NINS), 
650 North A'ohoku Place, Hilo, Hawaii, 96720, U.S.A.} 
\altaffiltext{2}{Department of Astronomical Science,
The Graduate University for Advanced Studies (SOKENDAI), 
Mitaka, Tokyo 181-8588, Japan}

\begin{abstract}
We present the results of our ALMA observations of three AGN-dominated
nuclei in optical Seyfert 1 galaxies (NGC 7469, I Zw 1, and IC
4329 A) and eleven luminous infrared galaxies (LIRGs) with various
levels of infrared estimated energetic contributions by AGNs 
at the HCN and HCO$^{+}$ J=3--2 emission lines.  
The HCN and HCO$^{+}$ J=3--2 emission lines are clearly detected at the
main nuclei of all sources, except for IC 4329 A. 
The vibrationally excited (v$_{2}$=1f) HCN J=3--2 and HCO$^{+}$ J=3--2
emission lines are simultaneously covered, and HCN v$_{2}$=1f J=3--2
emission line signatures are seen in the main nuclei of two LIRGs,
IRAS 12112$+$0305 and IRAS 22491$-$1808, neither of which
show clear buried AGN signatures in the infrared. 
If the vibrational excitation is dominated by infrared radiative
pumping, through the absorption of infrared 14 $\mu$m photons, primarily
originating from AGN-heated hot dust emission, then these two LIRGs may
contain  infrared-elusive, but (sub)millimeter-detectable, extremely
deeply buried AGNs.
These vibrationally excited emission lines are not detected in the three
AGN-dominated optical Seyfert 1 nuclei. 
However, the observed HCN v$_{2}$=1f to v=0 flux ratios in these optical
Seyferts are still consistent with the intrinsic flux ratios in LIRGs
with detectable HCN v$_{2}$=1f emission lines.
The observed HCN-to-HCO$^{+}$ J=3--2 flux ratios tend to be higher in
galactic nuclei with luminous AGN signatures compared with
starburst-dominated regions, as previously seen at J=1--0 and J=4--3.
\end{abstract}


\section{Introduction}

According to widely accepted cold dark matter-based galaxy formation
scenarios, small gas-rich galaxies collide, merge, and then
evolve into massive galaxies, as seen in the present day universe
\citep{whi78}.
Recent observations have shown that supermassive black holes 
(SMBHs) are ubiquitously present in the spheroidal components of 
present-day galaxies, and that there is a correlation between the mass
of SMBHs and the spheroidal stellar  
components \citep{mag98,fer00,gul09,mcc13}.
When gas-rich galaxies containing SMBHs at their center collide and
merge, both active star formation and mass accretion onto SMBHs 
(= active galactic nucleus (AGN) activity) are predicted to occur, but
while deeply embedded in dust and gas \citep{hop05,hop06}. 
Observations at low dust extinction wavelengths are necessary to
investigate these types of obscured activity in gas-rich galaxy mergers.  

Molecular rotational J-transition line flux ratios in the
(sub)millimeter wavelength range can be a powerful tool for this
purpose because dust extinction effects are usually negligible, unless
obscuration is extremely high with a hydrogen column density N$_{\rm H}$ 
$>>$ 10$^{25}$ cm$^{-2}$ \citep{hil83}.  
In particular, interferometric observations can probe the properties of
nuclear molecular gas in the close vicinity of an AGN, by minimizing
contamination from spatially extended (kpc-scale) starburst emission in the
host galaxy, and have provided an indication, based on the
observations of nearby bright starburst and Seyfert (= modestly luminous
AGNs) galaxies, that enhanced HCN J=1--0 emission could be a good AGN
signature \citep{koh05,kri08}.  
Based on this, interferometric HCN and HCO$^{+}$ J=1--0 observations
have been extensively performed for nearby gas-rich merging luminous
infrared galaxies (LIRGs; infrared 8--1000 $\mu$m luminosity 
L$_{\rm IR}$ $>$ 10$^{11}$L$_{\odot}$), and it has been confirmed that LIRGs
with infrared-identified energetically important obscured AGN signatures 
tend to display higher HCN-to-HCO$^{+}$ flux ratios than
starburst-dominated LIRGs \citep{ima04,ima06b,in06,ima07a,ima09}.
HCN and HCO$^{+}$ have similar dipole moments ($\mu$ = 3.0 debye
and 3.9 debye, respectively), so that they also have similar critical
densities at the same J-transitions.
It is expected that HCN and HCO$^{+}$ emission arises from similar regions
inside galaxies.  
Interpretations of the observed HCN-to-HCO$^{+}$ flux ratios are less
ambiguous than the flux comparison among molecules with largely different 
dipole moments (e.g., HCN vs. CO).
Using ALMA, a similar enhancement of observed HCN-to-HCO$^{+}$ flux
ratios was also found at J=4--3 for LIRGs that are infrared-diagnosed
to be AGN-important 
\citep{ima13a,ima13b,ima14b,ion13,gar14,izu15,izu16}.  
These observations suggest that elevated HCN-to-HCO$^{+}$ flux
ratios could be used to identify AGNs, including deeply buried (= obscured
in virtually all directions) ones. 

The physical origin of HCN flux enhancement in an AGN is not yet
completely understood.   
Compared with a starburst (nuclear fusion), an AGN (mass accretion onto a
SMBH) shows stronger X-ray emission when normalized to the ultraviolet
luminosity \citep{sha11,ran03}.
This strong X-ray emission may enhance the HCN abundance, relative to
HCO$^{+}$ \citep{mei05,lin06}, and may be responsible for 
the enhanced HCN emission in AGNs.
Next, because the radiative energy generation efficiency of a
mass-accreting SMBH in an AGN (6--42\% of Mc$^{2}$) is much higher than
the nuclear fusion reaction inside stars in a starburst ($\sim$0.7\% of
Mc$^{2}$), an AGN can produce much higher surface brightness emission,
and thereby a larger amount of hot dust ($>$ 100 K) in its close
vicinity, than a starburst. 
HCN abundance enhancement due to high gas/dust temperature chemistry
\citep{har10} may also be the cause of the HCN flux enhancement in
AGNs.
It is also argued that the HCO$^{+}$ abundance can decrease in highly
turbulent molecular gas in the close vicinity of a strongly
X-ray-emitting AGN \citep{pap07}, which may result in an elevated  
HCN-to-HCO$^{+}$ abundance ratio.   
In fact, \citet{yam07} and \citet{izu16} made non-LTE calculations of
HCN and HCO$^{+}$ emission over a wide parameter range and found that
an enhanced HCN abundance is required to account for the high
HCN-to-HCO$^{+}$ flux ratios observed in AGNs.  
HCN flux enhancement by infrared radiative pumping
\citep{car81,ziu86,aal95,gar06,sak10} is also suggested. 
Namely, HCN can be vibrationally excited by infrared 14 $\mu$m
photons that are 
strongly emitted from AGN-heated hot dust, and through its decay back to
the vibrational ground level (v=0) the HCN rotational J-transition flux at
v=0 could be higher than that of collisional excitation alone \citep{ran11}. 
However, chemical models also predict that the HCN and HCO$^{+}$
abundances around an AGN can vary strongly, depending on the surrounding
molecular gas parameters \citep{mei05,har13}. 
The proposed decrease in HCO$^{+}$ abundance could also occur in turbulent
molecular gas in a starburst with strong cosmic rays \citep{pap07}. 
In addition, HCO$^{+}$, as well as HCN, can be vibrationally excited by
absorbing infrared 12 $\mu$m photons \citep{dav84,kaw85}, and 
the HCO$^{+}$ J-transition flux at v=0 could also be increased through
the action of infrared radiative pumping.  
Given these remaining theoretical ambiguities, further detailed
interferometric observations of galactic nuclei where the energetic roles
of AGNs are reasonably well estimated are an important step toward better
clarifying whether elevated HCN-to-HCO$^{+}$ flux ratios
can indeed operate as good indicators of AGNs. 

In light of this, we conducted ALMA Cycle 1 HCN J=3--2 (with a
rest-frame frequency of 
$\nu_{\rm rest}$ = 265.89 GHz) and HCO$^{+}$ J=3--2 
($\nu_{\rm rest}$ = 267.56 GHz) observations of AGN-dominated nuclei in
band 6 (211--275 GHz).
These high-spatial-resolution ALMA observations enable us to minimize the 
contamination from spatially extended starburst activity in host
galaxies, and thus provide a clearer view on whether AGNs 
indeed show enhanced HCN-to-HCO$^{+}$ flux ratios, compared to starburst
galaxies. 
As spatially resolved pre-ALMA HCN and HCO$^{+}$ J=3--2 data for
galaxies are still very limited in the literature
\citep{sak10,hsi12,aal15a}, our ALMA Cycle 2 and 3 LIRG data, with
various AGN and starburst contributions, are also included so that we
can compare the HCN-to-HCO$^{+}$ J=3--2 flux ratios of AGN-dominated
nuclei to regions of strong starburst contributions.

Observations of HCN and HCO$^{+}$ at J=3--2 have several important
advantages compared with other J-transition lines. 
First, our main targets are nearby LIRGs, AGNs, and starbursts, whose
redshifts are as large as $z \sim$ 0.3.  
HCN and HCO$^{+}$ J=3--2 lines can be observed
simultaneously for these targets using ALMA.
We note that (1) HCN and HCO$^{+}$ J=1--0 lines cannot be covered by
ALMA if the target redshifts exceed $z \sim$ 0.06, because these lines
are shifted beyond the frequency coverage of ALMA band 3 (84--116 GHz), and 
(2) HCN and HCO$^{+}$ J=2--1 line observations of nearby galaxies are 
impossible in ALMA Cycles 1, 2, and 3, because ALMA band 5 (163--211 GHz)
is not yet open. 
Second, compared with HCN/HCO$^{+}$ J=4--3 observations in ALMA band 7
(275--373 GHz) for
nearby galaxies, HCN/HCO$^{+}$ J=3--2 observations in band 6 are less
affected by the Earth's atmospheric background noise and thus enable us to
obtain higher quality, higher signal-to-noise (S/N) ratio data.
Third, the effects of the precipitable water vapor value at the ALMA
observing site are smaller in band 6 than in band 7, so that the
probability of observation execution is expected to be higher for
HCN/HCO$^{+}$ J=3--2 in band 6 than HCN/HCO$^{+}$ J=4--3 in band 7. 
Finally, in ALMA band 6, in addition to the HCN J=3--2 and HCO$^{+}$ J=3--2
lines, vibrationally excited v$_{2}$=1, l=1f (hereafter v$_{2}$=1f) 
emission lines of HCN ($\nu_{\rm rest}$ = 267.20 GHz) and HCO$^{+}$
($\nu_{\rm rest}$ = 268.69 GHz) can also be simultaneously covered in
one shot, with the 5 GHz-wide correlator unit. 
These v$_{2}$=1f emission lines can be used to investigate how
infrared radiative pumping works in observed galaxies and affects
rotational excitation at v=0 \citep{sak10,ima13b,aal15a,aal15b,ima16a,mar16}. 
The simultaneous observations of all of these lines make their flux
ratios reliable, with the effect of possible systematic
uncertainties being minimized.

In this paper, we report the results of our ALMA Cycle 1, 2, and 3
observations of 
AGN-dominated galactic nuclei, starburst-dominated regions, and
the nuclei of AGN-starburst composite LIRGs. 
Throughout this paper, we adopt H$_{0}$ $=$ 71 km s$^{-1}$ Mpc$^{-1}$, 
$\Omega_{\rm M}$ = 0.27, and $\Omega_{\rm \Lambda}$ = 0.73
\citep{kom09}.
Molecular lines without the notation of v (the vibrational level)
refer to v=0 (i.e., the vibrational ground level). 
HCN refers to H$^{12}$C$^{14}$N.

\section{Targets}

Even though our high-spatial-resolution ALMA interferometric
observations can eliminate the effects of spatially extended (kpc-scale)
{\it circum-nuclear} starburst activity in the host galaxies, 
emission from {\it nuclear} starbursts at the central $<$200--300 pc of
galaxies cannot be clearly separated.
To purely probe the emission properties of AGN-affected molecular 
gas, we have to observe AGNs with the least contamination from {\it nuclear}
starburst activity. 
Infrared slit spectroscopy at $>$3 $\mu$m using ground-based telescopes
is one of the most powerful tools for selecting such almost pure AGNs,
because the presence of nuclear starburst activity can be investigated
by polycyclic aromatic hydrocarbon (PAH) emission features, which are
not emitted from AGNs but can be produced from nuclear starbursts
occurring at a location sufficiently shielded from the AGN's X-ray
radiation \citep{voi92,moo86,gen98,roc91,imd00,esq14}.
We have performed extensive ground-based infrared 3--4 $\mu$m
($L$-band) spectroscopy of optical Seyfert nuclei using narrow
($<$1.6$''$) slits \citep{ima02,ima03,rod03,iw04,ima06c,ima11a}, and
identified a large number of AGNs with undetectable nuclear starburst
signatures. 

Our next step is to further select nuclear-starburst-free AGNs which,
with ALMA, are expected to emit a detectable amount of molecular
emission lines.  
As our interest lies in AGN-affected molecular gas in the close vicinity
of the central AGN engine, we have to estimate the amount of such
nuclear AGN-affected molecular gas, rather than the entire amount of
molecular gas in the host galaxy. Assuming that dense molecular gas
and dust spatially co-exist in the dusty molecular tori around AGNs,
AGNs with strong {\it nuclear} AGN-heated hot dust radiation in the
infrared 3--10 $\mu$m are expected to accompany bright AGN-affected
nuclear molecular emission lines.  

Finally, to clearly detect the vibrationally excited (v$_{2}$=1f) HCN
emission line without veiling by the much brighter HCO$^{+}$ emission
line at v=0 (only $\sim$400 km s$^{-1}$ separation) \citep{sak10},
AGN-dominated nuclei with {\it small molecular line widths} are
appropriate targets.  
AGNs whose molecular line full widths at 10\% intensity are $<$500 km
s$^{-1}$ are desirable.    
We target Type 1 unobscured AGNs, because their dusty molecular tori
in the close vicinity of AGNs are expected to be preferentially viewed
from a {\it face-on} direction from our line-of-sight, making
rotation-originated nuclear molecular line widths small.
As long as turbulence-originated line widths are not very large, 
small observed nuclear molecular line widths are expected. 

In summary, Type 1 unobscured AGNs without nuclear PAH emission features
and with bright nuclear infrared 3--10 $\mu$m continuum fluxes are
suitable targets for our purpose.
NGC 7469, I Zw 1, and IC 4329 A meet these criteria.

NGC 7469 (z=0.0164) is included in both the Center for Astrophysics (CfA) 
\citep{huc92} and 12 $\mu$m \citep{rus93} Seyfert galaxy samples, and is
classified optically as a Seyfert 1 \citep{vei95,yua10}. 
It displays circum-nuclear starburst activity at 1--2 arcsec 
($\sim$0.3--0.7 kpc) in radius around the central AGN
\citep{soi03,gal05,dia07,reu10}. 
Although the 3.3 $\mu$m PAH emission feature, a signature of starbursts,
is detected in large aperture ($>$4'') infrared spectroscopy
\citep{ima10}, ground-based infrared $<$1''-wide slit spectroscopy
targeting the nuclear region shows no detectable 3.3 $\mu$m PAH
emission feature \citep{iw04}, suggesting that nuclear starburst
activity is very weak and undetectable. 
The nuclear infrared 3--4 $\mu$m flux of NGC 7469 is one of the highest
among observed Seyfert 1 galaxies with no detectable 3.3 $\mu$m PAH
emission features \citep{iw04}.
Recent sensitive ground-based 8--13 $\mu$m spectroscopy has shown some
signatures of an 11.3 $\mu$m PAH emission feature from the NGC 7469
nucleus \citep{hon10,esq14}, but the estimated nuclear starburst
luminosity is only $\sim$20\% of the AGN luminosity \citep{esq14,gen95}. 
Interferometric CO J=2--1 and HCN/HCO$^{+}$ J=4--3 observations of the 
NGC 7469 nucleus revealed that molecular emission lines at the nuclear
position are narrow with a full width at 10\% intensity $<$400 km s$^{-1}$
\citep{dav04,izu15}. 
Detailed results of ALMA Cycle 2 sub-arcsec spatial-resolution
observations of NGC 7469 at HCN J=4--3 and HCO$^{+}$ J=4--3 were
reported by \citet{izu15}.   

I Zw 1 (z=0.0611) is included in the Palomar-Green quasi-stellar objects
(PG QSO) catalog and is classified
optically as a Seyfert 1 \citep{sch83}.   
\citet{ima11a} performed infrared 3--4 $\mu$m ($L$-band) 1$\farcs$6-wide
slit spectroscopy and did not detect a 3.3 $\mu$m PAH emission feature
in the nuclear region of I Zw 1. 
The observed nuclear infrared 3--4 $\mu$m flux is high among
PAH non-detected AGNs. 
No detectable 11.3 $\mu$m PAH emission feature is found at the nucleus
\citep{bur13}.  
Single-dish CO J=1--0 and HCN J=1--0 observations of I Zw 1 with the IRAM
30-m telescope show narrow molecular line emission with full width at 10\%
intensity of $<$400 km s$^{-1}$ \citep{eva06}.

IC 4329 A (z=0.0160) is an optical Seyfert 1 galaxy in the 12 $\mu$m
Seyfert galaxy sample \citep{rus93}.
The nuclear 3--4 $\mu$m ($L$-band) spectrum taken with a 1$\farcs$6-wide
slit is dominated by a PAH-free featureless continuum, with one of
the highest observed fluxes for this class of object \citep{iw04}.  
The 11.3 $\mu$m PAH emission is also not detected in the nuclear region 
\citep{hon10}. 
The detection of molecular rotational J-transition emission lines in the
(sub)millimeter wavelength range toward the IC 4329 A nucleus has not been
reported in the published literature. 

As a comparison of these AGN-dominated optical Seyfert 1 galactic nuclei,
we observed nearby merging LIRGs, because they often display
starburst activity, in addition to deeply obscured AGNs. 
The relative energetic contributions from starbursts and obscured AGNs
vary greatly among different LIRGs, but are reasonably estimated based
on infrared spectroscopy, as long as emission from obscured AGNs
contributes to the observed infrared flux.

IRAS 08572$+$3915 ($z =$ 0.0580) is an ultraluminous infrared galaxy
(ULIRG; L$_{\rm IR}$ $\gtrsim$10$^{12}$L$_{\odot}$) with L$_{\rm IR}$
$\sim$ 10$^{12.1}$L$_{\odot}$ (Table 1),  
classified optically as a low-ionization nuclear emission-line region
(LINER) galaxy \citep{vei99} or a Seyfert 2 
\citep{yua10}. It shows a merging double nuclear morphology
(northwestern [NW] and southeastern [SE]) with a separation of $\sim$5
arcsec \citep{sco00,kim02,ima14a}.
The NW nucleus (IRAS 08572+3915 NW) is argued to contain 
an energetically dominant buried AGN, based on infrared spectroscopic and
imaging energy diagnostic methods
\citep{dud97,imd00,soi00,spo06,ima06a,arm07,ima07b,vei09,nar10}, and so 
has been selected as a target to scrutinize elusive buried AGNs
through various methods \citep{ima07a,ten09,ten15}.
We conducted ALMA Cycle 0 observations of IRAS 08572$+$3915 at
HCN/HCO$^{+}$/HNC J=4--3 lines and reported their molecular line flux
ratios \citep{ima14b}; 
detailed descriptions about IRAS 08572$+$3915 can be found in the
paper cited.  
The CO J=1--0 emission line was detected in only the NW nucleus, based on
pre-ALMA interferometric observations \citep{eva02}.

The Superantennae (IRAS 19254$-$7245; $z =$ 0.0617) is a ULIRG with 
L$_{\rm IR}$ $\sim$ 10$^{12.1}$L$_{\odot}$ (Table 1), classified
optically as a Seyfert 2 \citep{mir91}.
Infrared \citep{ris03,ima10} and X-ray \citep{bra03,bra09}
observations support the presence of an energetically important obscured 
AGN.
Due to its very low declination ($-$72$^{\circ}$), no interferometric
molecular gas observations were reported in the published literature
prior to ALMA. 

IRAS 12112$+$0305 ($z =$ 0.0730) is a ULIRG with L$_{\rm IR}$ $\sim$
10$^{12.3}$L$_{\odot}$ (Table 1), classified optically as a LINER
\citep{vei99} or a Seyfert 2 \citep{yua10}. 
It consists of two merging nuclei (northeastern [NE] and southwestern
[SW]), with a separation of $\sim$3 arcsec \citep{kim02}.  
Infrared spectra at 3--35 $\mu$m from nuclear emission, both combined
and separately at individual nuclei, are dominated by strong PAH emission
features \citep{ima06a,ima07b,nar09,vei09,ima10}. 
Such infrared spectral shapes are usually seen when the observed
infrared spectra are dominated by starburst emission
\citep{moo86,gen98,imd00}. 
No clear observational AGN signatures are recognizable in infrared
energy diagnostic methods.   
If a luminous AGN is present in either of the nuclei, the AGN must be
deeply buried in gas and dust so as to be infrared elusive.  
The PAH-to-infrared luminosity ratio is smaller than that expected from
modestly obscured starburst-dominated galaxies by a factor of 3--6,
depending on the 3.3 $\mu$m, 6.2 $\mu$m, and 11.3 $\mu$m PAH features  
\citep{ima07b,ima10}, suggesting that PAH emission is more
flux-attenuated than in the comparison starburst galaxies, or that an elusive
AGN contributes to the infrared luminosity without producing PAH
emission.  
Interferometric observations detected CO J=1--0 emission from both
nuclei, with a flux approximately three times higher at the NE nucleus
than at the SW nucleus \citep{eva02}. 

IRAS 22491$-$1808 ($z =$ 0.0776) is a ULIRG with L$_{\rm IR}$ $\sim$
10$^{12.2}$L$_{\odot}$ (Table 1). 
We reported the results of our ALMA Cycle 0 observations of IRAS
22491$-$1808 at HCN/HCO$^{+}$/HNC J=4--3 lines in \citet{ima14b}, and
detailed information relating to this ULIRG is presented in that paper.  
In summary, similar to IRAS 12112$+$0305, the infrared spectra of IRAS
22491$-$1808 at 3--35 $\mu$m are characterized by large equivalent width
PAH emission features \citep{ima07b,vei09,ima10}. 
The observed infrared emission at 3--35 $\mu$m is interpreted as originating
from starburst activity, and any putative AGN in this ULIRG must be
particularly deeply embedded to elude infrared spectroscopic observations.
Similar to IRAS 12112$+$0305, the PAH-to-infrared luminosity ratio is
smaller than that expected from modestly obscured
starburst-dominated galaxies by a factor of 3--8, depending on the 3.3
$\mu$m, 6.2 $\mu$m, and 11.3 $\mu$m PAH features \citep{ima07b,ima10}.  
The observed HCN-to-HCO$^{+}$ J=4--3 flux ratios are substantially
higher than those from starburst regions \citep{ima14b}. 
Obtaining further molecular line data is desirable for a better
understanding of the
origin of the observed molecular line emission in this ULIRG.

NGC 1614 (IRAS 04315$-$0840) at $z =$ 0.0160 is a well-studied nearby
LIRG with L$_{\rm IR}$ $\sim$ 10$^{11.7}$L$_{\odot}$ (Table 1). 
We observed NGC 1614 at HCN/HCO$^{+}$/HNC J=4--3 lines in ALMA Cycle 0,
as representative of starburst-dominated galaxies \citep{ima13a};
detailed properties of NGC 1614 can be found in the paper cited.  
In summary, NGC 1614 shows optical and infrared 2--35 $\mu$m spectra
whose emission properties are explained solely by nuclear starburst
activity, with a spatial extension of a few arcsec
\citep{alo01,mil96,soi01,dia08,ima11b,per15}.  
In general, dust extinction is smaller in LIRGs than ULIRGs
\citep{soi00,soi01}, so that a quantitative discussion of starbursts and
any putative AGNs is easier in LIRGs.   
The detected starburst-origin emission line luminosities in the infrared
are high enough to account for the observed total infrared luminosity
\citep{ima10}, leaving little room for the presence of an energetically
significant AGN in NGC 1614, unlike the ULIRGs, IRAS 12112$+$0305, and
IRAS 22491$-$1808.
Low HCN-to-HCO$^{+}$ J=4--3 flux ratios are found in our ALMA Cycle 0
data at multiple starburst regions in NGC 1614 \citep{ima13a}

The above five LIRGs were observed in ALMA Cycle 2. 
In addition to them, six new LIRGs have been observed in ALMA Cycle 3
at HCN/HCO$^{+}$ J=3--2, and their data have recently been sent to us. 
We include these six LIRGs (IRAS 12127$-$1412, IRAS 15250$+$3609, PKS
1345$+$12, IRAS 06035$-$7102, IRAS 13509$+$0442, and IRAS 20414$-$1651)
in this paper. 

IRAS 12127$-$1412 ($z =$ 0.1332) is a ULIRG with L$_{\rm IR}$ $\sim$
10$^{12.2}$L$_{\odot}$ (Table 1).
It shows two nuclei (NE and SW) with a separation of $\sim$10 arcsec
\citep{kim02,ima14a}. 
The optical classification of this ULIRG is a LINER \citep{vei99} or an 
HII-region \citep{yua10}. 
The NE nucleus is thought to contain a luminous buried AGN, based on
various infrared energy diagnostic methods
\citep{ima06a,ima07a,vei09,ima10,nar08,nar09,nar10}.  
The results of our ALMA Cycle 0 observations of IRAS 12127$-$1412 at
HCN/HCO$^{+}$/HNC J=4--3 lines were reported in \citet{ima14b}.
These molecular emission lines were detected at the NE nucleus, but not
at the SW nucleus. 

IRAS 15250$+$3609 ($z =$ 0.0552) is a ULIRG with L$_{\rm IR}$ $\sim$
10$^{12.0}$L$_{\odot}$ (Table 1).  
It is classified optically as a LINER \citep{vei95} or an
AGN-starburst composite \citep{yua10}.
The presence of a luminous buried AGN is suggested from 
(1) a steeply rising continuum at $>$3.5 $\mu$m, most likely due to 
AGN-heated hot dust emission \citep{ima10}, 
(2) low equivalent widths of PAH emission features at $>$5 $\mu$m
\citep{spo02,nar10,ima11b,sti13}, and (3) a high emission surface
brightness \citep{ima11b}. 
\citet{nar10} estimated that $\sim$50\% of the bolometric luminosity 
of IRAS 15250$+$3609 is explained by the buried AGN.

PKS 1345$+$12 (IRAS 13451$+$1232; $z =$ 0.1215) is a ULIRG with 
L$_{\rm IR}$ $\sim$ 10$^{12.3}$L$_{\odot}$ (Table 1). 
It is classified optically as a Seyfert 2 \citep{vei99,yua10}.
It has two nuclei with a separation of $\sim$2 arcsec along the
east-west (EW) direction \citep{sco00,kim02,ima14a}, and CO J=1--0
emission is detected in the western nucleus \citep{eva99}.  
Infrared $\sim$2 $\mu$m spectroscopy detected an AGN-origin broad (full
width at half maximum [FWHM] $>$ 2000 km s$^{-1}$) Pa$\alpha$ (1.875
$\mu$m) emission line, due to smaller dust extinction effects compared
with the optical \citep{vei97}, 
suggesting that the hidden AGN is only modestly obscured.
Based on the luminosity of the detected broad Pa$\alpha$ emission line, 
\citet{vei97} argued that the AGN could be responsible for the bulk of the
bolometric luminosity of PKS 1345$+$12.
A rising infrared 2.5--5 $\mu$m spectrum with a non-detectable 3.3 $\mu$m
PAH emission feature also supports the AGN-dominated nature of this
ULIRG \citep{ima10}.
PKS 1345$+$12 shows an excess of the radio-to-infrared luminosity ratio,
compared with the majority of LIRGs \citep{dra04}, suggesting that
the AGN is radio-loud. 

IRAS 06035$-$7102 ($z =$ 0.0795) is a ULIRG with L$_{\rm IR}$ $\sim$
10$^{12.2}$L$_{\odot}$ (Table 1). 
It is classified optically as a LINER \citep{duc97}, although there are
several indications that an optically elusive luminous buried 
AGN exists due to (1) low PAH emission equivalent widths and strong
absorption features by dust and ice at $>$3 $\mu$m
\citep{spo02,dar07,far09,ima10} and (2) a steeply rising continuum at
$>$3.5 $\mu$m \citep{ima10}. 
\citet{nar10} argued that the putative buried AGN accounts for
$\sim$20\% of the bolometric luminosity of IRAS 06035$-$7102.

IRAS 13509$+$0442 ($z =$ 0.1364) is a ULIRG with L$_{\rm IR}$ $\sim$
10$^{12.3}$L$_{\odot}$ (Table 1). 
It is classified optically as an HII region \citep{vei99} 
or an AGN-starburst composite \citep{yua10}. 
Its infrared 2.5--35 $\mu$m spectrum is characterized by large 
equivalent width PAH emission features \citep{ima07b,ima10}, 
suggesting that the observed 2.5--35 $\mu$m infrared flux is dominated
by starburst emission, with no obvious AGN signatures.
However, the observed PAH-to-infrared luminosity ratio is a factor of
2--4 lower than that expected from modestly obscured starburst-dominated
galaxies \citep{ima07b,ima10}.   

IRAS 20414$-$1651 ($z=$0.0870) is a ULIRG with L$_{\rm IR}$ $\sim$
10$^{12.3}$L$_{\odot}$ (Table 1), classified optically as an HII 
region \citep{vei99} or an AGN-starburst composite \citep{yua10}.  
Similar to IRAS 13509$+$0442, the PAH-emission-dominated infrared
2.5--35 $\mu$m spectrum of IRAS 20414$-$1651 \citep{ima07b,nar09,vei09,ima10} 
suggests that the observed infrared 2.5--35 $\mu$m flux is dominated
by starburst activity, without any discernible AGN signatures in the
infrared. 
Yet, the observed PAH-to-infrared luminosity ratio is low by a factor of
5--12, compared to modestly obscured starburst-dominated galaxies
\citep{ima07b,ima10}.

Table 1 presents the basic properties of these three optical Seyfert 1
galaxies and eleven LIRGs.

\section{Observations and Data Analysis}

Our observations of the three AGN-dominated Seyfert 1 galactic nuclei
(NGC 7469, I Zw 1, IC 4329 A) in band 6 were made within
the scope of our ALMA Cycle 1 program 2012.1.00034.S (PI = M. Imanishi).
Observations of the eleven LIRGs were performed within our ALMA
Cycle 2 program 2013.1.00032.S (PI = M. Imanishi) and Cycle 3 program
2015.1.00027.S (PI = M. Imanishi).  
Our Cycle 1 program was transferred to Cycle 2, and actual
observations were conducted during the ALMA Cycle 2 observing period.   
Observation details are summarized in Table 2.
Although HCN/HCO$^{+}$ J=4--3 and HNC J=3--2/J=4--3 data were also taken 
for LIRGs during our Cycle 2 and 3 programs, we focus on the
HCN/HCO$^{+}$ 
J=3--2 emission line properties in this paper, because only HCN/HCO$^{+}$
J=3--2 emission line data were obtained for the AGN-dominated optical
Seyfert 1 nuclei. 
The remaining emission lines of LIRGs will be discussed by \citet{ima16c}. 

We adopted the widest 1.875-GHz-width mode in each spectral window 
to cover as wide a frequency as possible.
To simultaneously observe our target lines, 
HCN J=3--2 ($\nu_{\rm rest}$ = 265.89 GHz), 
HCO$^{+}$ J=3--2 ($\nu_{\rm rest}$ = 267.56 GHz), 
HCN v$_{2}$=1f J=3--2 ($\nu_{\rm rest}$ = 267.20 GHz), and 
HCO$^{+}$ v$_{2}$=1f J=3--2 ($\nu_{\rm rest}$ = 268.69 GHz), 
we used three spectral windows, each of which covered 
(1) HCN J=3--2, (2) HCN v$_{2}$=1f J=3--2 and HCO$^{+}$ J=3--2, and (3)
HCO$^{+}$ v$_{2}$=1f J=3--2, respectively.
Because the interferometer in ALMA band 6 can cover 5 GHz at the upper
sideband or lower sideband,  
HCN and HCO$^{+}$ at both v=0 (the vibrational ground level) and v$_{2}$=1
(the vibrationally excited level) could be observed simultaneously at J=3--2,
which is not possible at J=4--3 in ALMA band 7. 
 
We began our data analysis from calibrated data provided by the Joint ALMA 
Observatory, using CASA (https://casa.nrao.edu).
We first checked the visibility plots to see if the signatures of 
the targeted bright emission lines were visible.
The presence of HCN J=3--2 and HCO$^{+}$ J=3--2 emission lines was
evident in the visibility plots of all observed galaxies, except for IC
4329 A. 
We estimated the continuum flux level using channels that were
unaffected by strong emission lines.    
The estimated continuum levels were subtracted using the task
``uvcontsub''; the task ``clean'' was then applied to create final 
continuum-subtracted molecular line data.
The clean task was also applied to the extracted continuum data
themselves.  
For these clean procedures, 40-channel binning of velocity
channel with $\sim$0.5 km s$^{-1}$ spacing for each (resulting
resolution is $\sim$20 km s$^{-1}$) and a pixel scale of 0$\farcs$1
pixel$^{-1}$ were basically utilized.   
For LIRGs observed in ALMA Cycle 3, data were taken and delivered 
with four channels binning (Spec Avg. = 4 in the ALMA Observing Tool)
to reduce the ALMA data rate, for which we used 20-channel
spectral binning. The net velocity resolution was $\sim$40 km s$^{-1}$. 
Since these LIRGs observed in ALMA Cycle 3 showed large line widths with
$>$200 km s$^{-1}$ in FWHM for bright molecular emission lines (see
$\S$3), we can properly trace their emission line profiles and estimate
their fluxes based on Gaussian fitting.  
As some of our targets displayed spatially extended structures, we
applied the primary beam correction to all data, although its effects are
relatively limited as long as we discuss emission from the central
few arcsec regions.  
When we created the $\sim$20 km s$^{-1}$ resolution spectrum of The
Superantennae (the broadest molecular line source) from the cleaned
data, it was found that continuum levels in the spectral windows that
contained the HCN J=3--2 and HCO$^{+}$ J=3--2 emission lines were negative in
our first continuum determination due to continuum
over-subtraction. We thus re-determined the continuum levels in these
spectral windows using only a small number of data points at the edge
of each spectral window, sufficiently separated from the broad
HCN/HCO$^{+}$ J=3--2 emission line peaks.
In the revised determination, signals outside the emission lines in the
continuum-subtracted spectra became almost zero.
However, it was uncertain whether data at the edge of these spectral
windows were completely line-free. 
For The Superantennae, continuum determination ambiguities could be larger
than for other sources.  
According to the ALMA Cycle 1, 2, and 3 Proposer's Guides, the
absolute calibration uncertainty of our ALMA band 6 data should be $<$10\%. 
The position reference frames are FK5 for objects observed in ALMA
Cycles 1 and 2, and ICRS for those observed in ALMA Cycle 3.

\section{Results}

Figure 1 displays continuum maps of the three optical Seyfert 1
galaxies, NGC 7469, I Zw 1, and IC 4329 A. 
Continuum emission at $\sim$250--260 GHz (observed frame) is clearly
detected at the nuclear regions in all sources. 
In NGC 7469, continuum emission is also detected in the three
starburst regions, which we denote as SB1, SB2, and SB3. 
The observed continuum emission properties are summarized in Table 3. 

Continuum maps of the LIRGs observed in ALMA Cycles 2 and 3 are 
shown in Figures 2 and 3, respectively.  
Continuum emission is detected in the nuclei of all galaxies. 
In IRAS 12112$+$0305, the emission is seen both at the 
NE and SW nuclei, with the NE nucleus brighter than
that of the SW. 
In NGC 1614, continuum emission is spatially extended, and 
arises primarily from two distinct regions. 
We investigate the possible continuum emission signal for the fainter
IRAS 08572$+$3915 SE and IRAS 12127$-$1412 SW nuclei, but detection
is $<$3$\sigma$ within 1'' around the near-infrared position
defined by \citet{kim02}.  
The fainter northern nucleus of The Superantennae, located at
$\sim$1$\farcs$5 west and $\sim$8'' north of the main
southern nucleus \citep{reu07,jia12}, is also undetected with
$>$3$\sigma$ in the continuum map within 1'' around the expected position.
For IRAS 22491$-$1808, we attribute the detected nuclear position in the
continuum to the eastern nucleus \citep{kim02,haa11}. The
fainter secondary western nucleus, $\sim$2'' separated from the main
eastern nucleus \citep{kim02}, is not clearly detected. 
For PKS 1345$+$12, the coordinate of the ALMA continuum emission peak
agrees with that of a VLBI radio emission peak \citep{ma98}, and we
regard that the detected continuum emission originates from the CO
J=1--0 detected western nucleus \citep{eva99}. 

For IRAS 13509$+$0442 in Figure 3, continuum emission is detected not only
at the nucleus but also at (13 53 31.66, $+$04 28 13.1)J2000,
$\sim$1$''$ east and 8--9$''$ north of the IRAS 13509$+$0442 nucleus. 
This source is detected also in our band 7 continuum map at $\sim$313
GHz at the same position with $\sim$3.6 mJy \citep{ima16c}. 
In the Sloan Digital Sky Survey (SDSS) optical image, two sources were
detected at $\sim$10$''$ 
north of IRAS 13509$+$0442, identified as SDSS
J135331.89$+$042816.4 (z=0.186 galaxy) and J135331.44$+$042813.8, but
their positions were significantly ($>$3$''$) offset from our ALMA
continuum-detected source.  
Further high-spatial-resolution multiple-frequency data will help to
unravel the nature of this optically faint, millimeter-bright source. 

In Figure 4, we show the full frequency coverage spectra, within the beam
size, at the nuclear positions of three optical Seyfert 1 galaxies and
three starburst regions (SB1, SB2, and SB3) of NGC 7469.
To investigate the emission properties from the entire starburst ring of
NGC 7469, we also present an area-integrated spectrum in Figure 4e, by
integrating signals at the annular region with 0$\farcs$8--2$\farcs$5
radius, centered at the nuclear position of NGC 7469.  

The full frequency coverage spectra for the eleven LIRGs are shown in 
Figure 5. 
Spectra at the nuclear positions are shown, within the beam size, with
the exception of NGC 1614. 
In the spectrum of IRAS 15250$+$3609 in Figure 5(j), a broad dip
is visible at the higher frequency side of the HCO$^{+}$ J=3--2
emission.  
A similar profile was observed for HCO$^{+}$ J=4--3 and J=3--2 emission
lines in the ULIRG Arp 220 \citep{sak09}, and was interpreted as the P
Cygni profile due to outflow. 
It is likely that IRAS 15250$+$3609 displays similar 
HCO$^{+}$ outflow activity.
A similar dip, however, is not clearly evident at the higher frequency
side of the HCN J=3--2 emission line, suggesting that a larger 
fraction of HCO$^{+}$ is in the outflow component than HCN. 
A narrow dip at the lower frequency side of the HCO$^{+}$ J=3--2
emission is interpreted to be of inflow origin, as seen in some fraction
LIRGs \citep{vei13}.
In Figure 5(n), no significant molecular emission line was observed at
the position of the NE source detected in the continuum map of IRAS
13509$+$0442. 

Integrated intensity (moment 0) maps were created for the HCN J=3--2 and
HCO$^{+}$ J=3--2 emission lines, by summing velocity channels with
discernible signal signs, with no cut-off in signal-to-noise ratios. 
Because NGC 7469 and NGC 1614 show morphologies with multiple
emission components, their moment 0 maps are presented in Figures 6  
and 7, respectively. 
Figure 8 shows the moment 0 maps of the remaining galaxies whose
molecular emission predominantly arises from nuclear regions only.
The HCN J=3--2 and HCO$^{+}$ J=3--2 emission line properties in these
moment 0 maps are summarized in Tables 4 and 5, respectively.  
The peak positions of the HCN J=3--2 and HCO$^{+}$ J=3--2 emission lines 
spatially agree with the continuum emission peaks within one pixel
(0$\farcs$1) in both the RA and DEC directions, with the exception
of NGC 1614.  
The secondary nuclei of IRAS 08572+3915, The Superantennae, IRAS
22491$-$1808, IRAS 12127$-$1412 and PKS 1345$+$12 are not detected
with $>$3$\sigma$ in the moment 0 maps of the HCN J=3--2 and HCO$^{+}$
J=3--2 emission lines. 

For NGC 1614, the peak positions of the HCN J=3--2 and HCO$^{+}$ J=3--2
moment 0 maps in Figure 7 are significantly offset from the continuum
peak positions in Figure 2. 
The molecular gas emission peaks in the moment 0 maps of NGC 1614 agree
within one pixel between HCN J=3--2 and HCO$^{+}$ J=3--2. 
We denote the northern peak at (04 34 00.00, $-$08 34 44.4)J2000 and 
southern peak at (04 33 59.98, $-$08 34 45.2)J2000 as SB1 and SB2,
respectively, based on the HCN J=3--2 moment 0 map of NGC 1614. 
Full frequency coverage spectra of NGC 1614 at the SB1 and SB2 peaks,
within the beam size, are shown in Figure 5(f) and 5(g), respectively.   
An area-integrated spectrum in a circular region with a radius of
2$\farcs$5 around (04 34 00.03, $-$08 34 44.6)J2000 is also shown
in Figure 5(h).   

Gaussian fits for the HCN J=3--2 and HCO$^{+}$ J=3--2 emission lines 
in the spectra, within the beam size, at the individual multiple
positions of NGC 7469 and NGC 1614 are shown in Figures 6 and 7,
respectively.  
Figure 9 displays Gaussian fits for the remaining sources.
Estimates of the HCN J=3--2 and HCO$^{+}$ J=3--2 emission line fluxes are 
based on the peak pixel values in the moment 0 maps and Gaussian fits
within the beam size, and are summarized in Tables 4 and 5, respectively.
Both estimates generally agree within $\sim$20\%. 
We will adopt the fluxes estimated from Gaussian fits for these detected
emission lines. 
For the area-integrated spectra of NGC 7469 and NGC 1614, only flux
estimates based on Gaussian fits are available, and will be used for our
discussion.  

Zoom-in spectra of the AGN-dominated nuclei of the two optical Seyfert 1
galaxies with clearly detectable HCN and HCO$^{+}$ J=3--2 emission lines
(NGC 7469 and I Zw 1) are displayed in Figure 10, to better investigate
the HCN and HCO$^{+}$ v$_{2}$=1f J=3--2 emission lines. Their presence
is, however, unclear. 
We created their moment 0 maps by summing ten velocity channels at the
expected frequencies of the HCN and HCO$^{+}$ v$_{2}$=1f J=3--2 lines, but
there is no signature of these emission lines at all.
The 3$\sigma$ upper limit, relative to the rms noise, is shown
in Table 6.   

In the spectra of IRAS 08572$+$3915, IRAS 12112$+$0305 NE, IRAS
22491$-$1808, and IRAS 20414$-$1651 in Figure 5, an emission tail
is recognizable at the lower frequency side of the HCO$^{+}$ J=3--2
emission line. 
Figure 10 displays zoomed-in spectra around the HCO$^{+}$ J=3--2
emission lines for these four sources, to better distinguish the
emission tails.  
As these tails are not visible at the higher frequency side of the
HCO$^{+}$ J=3--2 emission lines, the most natural interpretation for
the lower frequency side of the tails is the contribution from the HCN
v$_{2}$=1f J=3--2 emission line \citep{aal15a,aal15b}.

IRAS 15250$+$3609 shows a weak emission sub-peak at the expected
frequency of the HCN v$_{2}$=1f J=3--2 line, at the lower frequency side
of the strong HCO$^{+}$ J=3--2 emission (Figure 5).  
However, a similar weak emission sub-peak is also observed at the lower
frequency side of the HCN J=3--2 emission line (Figure 5). 
Figure 11 compares the velocity profile of the HCN J=3--2 and
 HCO$^{+}$ J=3--2 emission lines, after normalizing the Gaussian-fit
peak flux for the main bright emission component.  
The emission line flux of the sub-peak component, relative to the main
component, is higher for HCO$^{+}$ J=3--2 than HCN J=3--2. 
Although this could be due partly to the contribution from the HCN
v$_{2}$=1f J=3--2 emission to the sub-peak component of HCO$^{+}$
J=3--2, the expected HCN v$_{2}$=1f J=3--2 emission peak for $z=$0.0552
is slightly offset from the observed sub-peak component of HCO$^{+}$ in
velocity (Figure 11). 
It is more likely that the outflow-origin redshifted and blueshifted
components, relative to the main molecular emission component at the
nucleus, are stronger for HCO$^{+}$ than HCN, which produces stronger P
Cygni profile for HCO$^{+}$.
A similar trend is seen also at the J=4--3 data of HCN and HCO$^{+}$ for
IRAS 15250$+$3690 \citep{ima16c}.
We made moment 0 maps of the sub-peak component of the HCN
J=3--2 (10 velocity channels) and HCO$^{+}$ J=3--2 (7 velocity channels)
and detected these emission lines with 0.70 [Jy beam$^{-1}$ km s$^{-1}$] 
(5.8$\sigma$) and 0.72 [Jy beam$^{-1}$ km s$^{-1}$] (9.2$\sigma$)
for HCN J=3--2 and HCO$^{+}$ J=3--2, respectively. 
The peak position of the HCO$^{+}$ J=3--2 sub-peak component spatially
coincides with the continuum peak, but that of HCN J=3--2 is shifted to
the north with 2 pix (0$\farcs$2). However, the position determination
accuracy of the HCN J=3--2 sub-peak component is on the order of 
(beam-size)/(signal-to-noise-ratio) $\sim$ 1.21/5.8 $\sim$ 0$\farcs$21. 
We see no clear evidence that the sub-peak component is spatially
offset from the continuum peak at the nucleus.
If the sub-peak component is of outflow origin, the outflow is compact
and is located close to the nucleus in the moment 0 maps.
Since the continuum flux at $\sim$250 GHz is $\sim$11 mJy beam$^{-1}$
(Table 3), the flux attenuation by outflow of HCO$^{+}$ gas at J=3--2 is 
$\sim$20\%, or an optical depth with $\tau$ $\sim$ 0.2, for the broad
absorption component at v$_{\rm opt}$ $=$ 15800--16400 km s$^{-1}$
(Figure 11). 
For the narrow absorption component at v$_{\rm opt}$ $\sim$ 16800 km
s$^{-1}$ (Figure 11), which is likely to be of inflow origin, 
the flux attenuation is $>$40\%. 

For IRAS 08572$+$3915, IRAS 12112$+$0305 NE, IRAS 22491$-$1808, and
IRAS 20414$-$1651, we created the integrated intensity (moment 0) maps
by integrating signals marked with the horizontal solid straight lines
(bracketed by the short vertical solid lines) in Figure 10, where the
bulk of the possible HCN v$_{2}$=1f J=3--2 emission lines are covered,
while contamination from HCO$^{+}$ J=3--2 emission lines is minimized, based
on our Gaussian fits of the HCO$^{+}$ J=3--2 emission lines.  
In the HCN v$_{2}$=1f J=3--2 emission line moment 0 maps of IRAS
12112$+$0305 NE and IRAS 22491$-$1808, the HCN v$_{2}$=1f J=3--2 emission
line was detected with 4.5$\sigma$ and 4$\sigma$, respectively (Table 6),
although a possible contribution from the bright HCO$^{+}$ v=0 J=3--2
emission line cannot completely be ruled out.
These moment 0 maps are shown in Figure 12.  
For IRAS 08572$+$3915 and IRAS 20414$-$1651, we barely see possible
sign of HCN v$_{2}$=1f J=3--2 emission at the position close to the
continuum peak, but the detection significance is $\sim$2$\sigma$ in the
moment 0 maps, partly because not all the HCN v$_{2}$=1f J=3--2 emission
components can be used to create the moment 0 maps, due to the
contamination from the nearby much brighter HCO$^{+}$ v=0 J=3--2
emission line. 
We see no discernible signature of the HCN v$_{2}$=1f J=3--2
emission line in the remaining LIRGs.
There is no clear sign of the HCO$^{+}$ v$_{2}$=1f J=3--2
emission line in any of the observed galaxies.

The deconvolved, intrinsic emission sizes, estimated using the CASA task
``imfit'', for spatially not-clearly resolved sources are shown in
Table 7. 
The luminosities of the vibrational ground (v=0) HCN/HCO$^{+}$ J=3--2
emission lines are provided in Table 8.
Table 9 tabulates the luminosities of vibrationally excited (v$_{2}$=1f)
HCN J=3--2 emission lines for sources with $>$3$\sigma$ detection (IRAS
12112$+$0305 NE and IRAS 22491$-$1808).

Intensity-weighted mean velocity (moment 1) and intensity-weighted
velocity dispersion (moment 2) maps for HCN/HCO$^{+}$ J=3--2 (v=0)
emission lines are shown in Figure 13.
The similar emission morphology and dynamics in moment 0, 1, and 2 maps
between HCN J=3--2 and HCO$^{+}$ J=3--2 support our previous view
($\S$1) that these emission lines originate in spatially similar regions
within individual galaxies. 

\section{Discussion}

\subsection{Vibrationally excited HCN/HCO$^{+}$ J=3--2 emission lines} 

\subsubsection{Luminous infrared galaxies}

Because the energy levels of the vibrationally excited (v$_{2}$=1) state
for HCN ($\sim$1030 K) and HCO$^{+}$ ($\sim$1200 K) are too high to be 
excited by collisions, infrared radiative pumping is thought to be
necessary for vibrational excitation \citep{sak10}. 
Due to a large amount of AGN-heated hot ($>$ few 100 K) dust emission, 
the 14 $\mu$m luminosity in an AGN is significantly higher than that in
a starburst for the same bolometric luminosity \citep{mar07,veg08}.  
Thus, HCN vibrational excitation, through the absorption of infrared
14 $\mu$m photons, is expected to occur more efficiently in an AGN than
in a starburst.   
In fact, the HCN v$_{2}$=1f emission lines at J=3--2 and/or J=4--3 have
recently been detected in gas/dust-rich LIRGs, which most likely, or
plausibly, contain luminous AGNs, i.e., NGC 4418 \citep{sak10,cos15}, IRAS
20551$-$4250 \citep{ima13b,ima16a}, Mrk 231 \citep{aal15a}, and a few
further LIRGs \citep{aal15b,mar16}, demonstrating that this infrared
radiative pumping mechanism actually works in some AGNs.  

As the frequencies of HCN v$_{2}$=1f and HCO$^{+}$ v=0 are very close 
to each other at J=3--2 and J=4--3, we can only clearly separate these
lines for galaxies with small molecular line widths, such as NGC 4418 
\citep{sak10}, IRAS 20551$-$4250 \citep{ima13b,ima16a}, and 
IC 860 \citep{aal15b}.
For the majority of the other galaxies, these lines are blended. 
Even if the HCN v$_{2}$=1f J=3--2 or J=4--3 emission line is detected, 
it is recognized as a tail at the lower frequency side of the much
brighter HCO$^{+}$ v=0 J=3--2 or J=4--3 emission line 
\citep{aal15a,aal15b,mar16}. 
The four ULIRGs, IRAS 08572$+$3915, IRAS 12112$+$0305 NE, IRAS
22491$-$1808, and IRAS 20414$-$1651, show these profiles. 
In particular, IRAS 12112$+$0305 NE and IRAS 22491$-$1808 can be
categorized as sources that display detectable HCN v$_{2}$=1f J=3--2
emission lines, given the $>$4$\sigma$ detection in the moment 0 maps
(Figure 12).
For IRAS 08572$+$3915 and IRAS 22491$-$1808, similar HCN v$_{2}$=1f
J=4--3 emission tails were not clearly seen in our ALMA Cycle 0 band 7
data \citep{ima14b}; however, this is not surprising due to 
the improved performance of ALMA Cycle 2 data and intrinsically lower
noise in band 6 with lower background emission, than in band 7.
For IRAS 12112$+$0305 NE, a similar signature of the HCN v$_{2}$=1f
J=4--3 emission line at the lower frequency side of HCO$^{+}$ v=0 J=4--3
was observed in our ALMA Cycle 2 data, despite a lower detection
significance than J=3--2, while no significant emission tail is
recognizable at the lower frequency side of HCN v=0 J=4--3
\citep{ima16c}.

While IRAS 08572$+$3915 is classified as a ULIRG possessing a luminous
buried AGN in the infrared spectrum, IRAS 12112$+$0305 NE, IRAS
22491$-$1808, and IRAS 20414$-$1651 display no clear infrared
buried AGN signatures. 
The signatures of the HCN v$_{2}$=1f J=3--2 emission lines in
our data suggest the presence of strong mid-infrared 14 $\mu$m
continuum-emitting sources at the nuclei of these two LIRGs. 
The HCN v$_{2}$=1f J=3--2 to infrared luminosity ratios are $>$7
$\times$ 10$^{-9}$ and $\sim$9 $\times$ 10$^{-9}$ for IRAS 12112$+$0305
NE and IRAS 22491$-$1808, respectively.
These ratios are several factors higher than that in the Galactic
active ($>$10 L$_{\odot}$/M$_{\odot}$) and luminous
($>$10$^{7}$L$_{\odot}$) star-forming region, W49A ($<$1.2 $\times$
10$^{-9}$) \citep{nag15,ima16a}.
A luminous buried AGN is a plausible origin, although the possibility
of a very compact extreme starburst cannot be completely ruled out 
\citep{aal15b}.  
IRAS 12112$+$0305 NE and IRAS 22491$-$1808 are candidates that 
contain extremely deeply buried AGNs whose signatures are not seen in
infrared 5--35 $\mu$m spectroscopic energy diagnostic methods due to
dust extinction, but are revealed by our (sub)millimeter method because
of the reduced effects of dust extinction \citep{dra84}. 
If this is the case, (sub)millimeter observations could be an even more
powerful method for detecting extremely deeply buried AGNs in LIRGs. 
IRAS 20414$-$1651 may also belong to this class, but higher quality
data are needed to quantitatively better estimate the HCN v$_{2}$=1f
J=3--2 emission line luminosity.

The HCO$^{+}$ v$_{2}$=1f J=3--2 emission line is not clearly detected in
any of the observed three optical Seyfert 1s and eleven LIRGs.
Like HCN, infrared radiative pumping should also work for HCO$^{+}$,
because HCO$^{+}$ can be excited to the v$_{2}$=1 level by absorbing
infrared 12 $\mu$m photons \citep{dav84,kaw85}. 
The infrared radiative pumping rate (P$_{\rm IR}$) is 
\begin{eqnarray}
P_{IR} & \propto & B_{v2=0-1,vib} \times F_{\nu (IR)} \times N_{v=0}, 
\end{eqnarray}
where B$_{v2=0-1,vib}$ is the Einstein B coefficient from v=0 to
v$_{2}$=1, F$_{\nu (IR)}$ is the infrared flux in [Jy] used for the
infrared radiative pumping of HCN and HCO$^{+}$, and N$_{v=0}$ is the
column density at the v=0 level. 
Here, the possible difference in population between HCN and HCO$^{+}$ at
rotational J-levels within v=0 and v$_{2}$=1 is not considered. 
As discussed in \citet{ima16a}, the B$_{v2=0-1,vib}$ values are
comparable within 10\% between HCN and HCO$^{+}$.
The F$_{\rm \nu (IR)}$ values at 12 $\mu$m and 14 $\mu$m for the observed
galaxies are derived from their Spitzer IRS low-resolution spectra
\citep{bra06,ima07b,wu09}. 
For sources with strong 9.7 $\mu$m silicate dust absorption features, 
power law continua determined from data points outside the broad 9.7
$\mu$m absorption features are utilized to estimate the intrinsic
infrared flux at 14 $\mu$m and 12 $\mu$m, which are used for the
vibrational excitation to v$_{2}$=1 of HCN and HCO$^{+}$, respectively.  
In none of the galaxies was the intrinsic 14 $\mu$m flux $>$30\% larger
than the intrinsic 12 $\mu$m flux.  
Therefore, at least for the observed galaxies in this paper, the term 
B$_{v2=0-1,vib}$ $\times$ F$_{\nu (IR)}$ does not differ greatly between
HCN and HCO$^{+}$ and thereby, under similar HCN and HCO$^{+}$
abundance, the infrared radiative pumping rate is comparable
between HCN and HCO$^{+}$.
If the flux of the HCN v$_{2}$=1f J=3--2 emission line is significantly
higher than that of HCO$^{+}$ v$_{2}$=1f J=3--2, HCN should then have
a significantly higher column density, and thereby a higher abundance,
than HCO$^{+}$, unless excitation conditions at J=3 significantly 
differ between HCN and HCO$^{+}$ \citep{ima16a}.

The upper limit of the HCO$^{+}$ v$_{2}$=1f J=3--2 flux (Table 6) is
only 10--20\% lower than the flux of the HCN v$_{2}$=1f J=3--2 emission
lines, even for IRAS 12112$+$0305 NE and IRAS 22491$-$1808. 
Thus, our only constraint is that the HCN abundance is at least comparable to
HCO$^{+}$ or possibly higher.

\subsubsection{Optical Seyfert galaxies}

The original science goal of our ALMA Cycle 1 program was to detect 
the HCN v$_{2}$=1f J=3--2 emission lines from the AGN-dominated nuclear
regions of the three Seyfert 1 galaxies, NGC 7469, I Zw 1, and IC 4329 A.
In NGC 7469 and I Zw 1, the clear detection of the HCN and HCO$^{+}$
J=3--2 emission lines at v=0 suggests that at least a modest amount of
dense molecular gas is present at the nuclei. 
If the infrared radiative pumping mechanism is commonly working in AGNs,
it is expected that a number of HCN and HCO$^{+}$ v$_{2}$=1f emission
lines are produced. 
We did not detect HCN/HCO$^{+}$ v$_{2}$=1f J=3--2 emission
lines in these AGN-dominated Seyfert 1 nuclei, and this requires some
quantitative consideration. 
For the NGC 7469 nucleus, the HCN v$_{2}$=1f J=4--3 emission line was
also undetected by ALMA observations \citep{izu15}. 

For the non-detected v$_{2}$=1f J=3--2 emission lines of HCN and
HCO$^{+}$, we use the 3$\sigma$ upper limits from the moment 0 maps
tabulated in Table 6 for our discussion.  
As there is no existing report for the detection of the HCO$^{+}$ v$_{2}$=1f
J=3--2 emission line in external galaxies, we focus here on the HCN
v$_{2}$=1f J=3--2 line. 
The observed v$_{2}$=1f to v=0 flux ratios at J=3--2 for HCN are $<$0.02
and $<$0.04 for NGC 7469 and I Zw 1 nuclei, respectively.  
For LIRGs with detected HCN v$_{2}$=1f emission lines, the observed 
HCN v$_{2}$=1f to v=0 flux ratios at J=3--2 or J=4--3 are $\sim$0.04 in
IRAS 20551$-$4250 and Mrk 231 \citep{ima13b,aal15a,ima16a}, 
and 0.1--0.2 for the other sources \citep{sak10,aal15b,mar16}. 
The observed ratios in NGC 7469 ($<$0.02) and I Zw 1 ($<$0.04) are
lower than these ratios.

We consider that a plausible scenario for the non-detection of the HCN
v$_{2}$=1f J=3--2 emission line in NGC 7469 and I Zw 1 is the small line 
opacity of the HCN v=0 J=3--2 emission.
Thus far, the HCN v$_{2}$=1f J=3--2 or J=4--3 emission lines have been
detected in LIRGs with buried AGNs whose signatures are unclear in
the optical spectroscopic classification, except for Mrk 231
\citep{sak10,ima13b,aal15a,aal15b,mar16}.   
Although Mrk 231 is classified optically as a Seyfert 1 galaxy due to the 
detection of broad optical emission lines \citep{vei99,yua10}, the AGN
emission in Mrk 231 is estimated to be highly obscured in infrared and
X-ray data \citep{arm07,ten14}.
It is likely that Mrk 231 is not a bona fide unobscured Seyfert 1
galaxy, but rather an obscured AGN.

For the obscured AGN-hosting LIRGs with detectable HCN v$_{2}$=1f
emission lines, significant flux attenuation by line
opacity of the HCN v=0 emission 
is indicated \citep{sak10,aal15a,aal15b,ima16a,mar16}.
For NGC 4418, IRAS 20551$-$4250, and Mrk 231, the line-opacity-corrected
intrinsic HCN v$_{2}$=1f to v=0 flux ratios are quantitatively estimated
to be $\sim$0.01 \citep{sak10,aal15a,ima16a}.  
These ratios are smaller than the upper limits at the NGC 7469 and I Zw
1 nuclei.  
NGC 7469 and I Zw 1 are classified optically as Seyfert 1 (=
unobscured AGNs); thus, the direction along our line of sight in
front of the AGN is at least clear of gas and dust.
It is likely that molecular gas and dust are present in the close
vicinity of the AGNs in the direction perpendicular to our sightline.
If the ratio of rotational to random velocity of molecular gas does not
differ greatly between unobscured optical Seyfert 1 AGNs and buried AGNs
in LIRGs, the column density ratio along the maximum and minimum column 
density directions is not dissimilar.
The presence of a transparent direction suggests that the total amount of
nuclear molecular gas in unobscured AGNs is smaller than buried AGNs in
LIRGs \citep{ima07b,ima10}. 
Thus, the flux attenuation of the HCN and HCO$^{+}$ v=0 emission by line
opacity is also expected to be smaller in unobscured AGNs.
\citet{izu15} estimated the line
opacity for the HCN v=0 J=4--3 emission to be $<$3.5 for the NGC 7469
nucleus.   
Even if unobscured AGNs and buried AGNs show intrinsically similar HCN
v$_{2}$=1f to v=0 flux ratios, the {\it observed} HCN v$_{2}$=1f to v=0
flux ratios in buried AGNs can become larger due to higher HCN v=0 flux
attenuation. 
The upper limits of the observed HCN v$_{2}$=1f to v=0 flux ratios at
J=3--2 at the NGC 7469 and I Zw 1 nuclei are still consistent with the
scenario that the efficiency of infrared radiative pumping in these
unobscured-AGN-dominant nuclei is as high as that of buried AGNs with
detected HCN v$_{2}$=1f emission lines.
If this scenario is indeed the case, then HCN v$_{2}$=1f J=3--2 emission
lines should be detected from the nuclei of NGC 7469 and I Zw 1 in
data with a factor of 5--10 better sensitivity, even in the case
that the line opacity correction of HCN v=0 J=3--2 emission is negligible. 
Future higher sensitivity observations and line opacity estimates for the
HCN v=0 J=3--2 emission line will help to quantify how infrared
radiative pumping works in various types of 
AGNs, including unobscured AGNs in optical Seyfert 1 galaxies and buried
AGNs in LIRGs. 

\subsection{HCN to HCO$^{+}$ J=3--2 flux ratios}

\subsubsection{Observed ratios}

The HCN-to-HCO$^{+}$ J=3--2 flux ratios at individual positions in
individual galaxies are displayed in Figure 14.  
The NGC 7469 nucleus and I Zw 1 are classified as Seyfert 1s. 
In NGC 7469, SB1, SB2, SB3, and the SB ring (0$\farcs$8--2$\farcs$5
annular region) are taken to be starburst-dominated.
For NGC 1614, all regions are regarded as starburst-dominated. 
Among the other LIRGs, IRAS 08572$+$3915, The Superantennae, 
IRAS 12127$-$1412, IRAS 15250$+$3609,  PKS 1345$+$12, and IRAS
06035$-$7102 are categorized as obscured AGNs, based on infrared
spectroscopic energy diagnostic methods ($\S$2). 
IRAS 12112$+$0305 NE and IRAS 22491$-$1808 are now classified as
infrared-elusive, but (sub)millimeter-detectable, extremely deeply
buried AGN candidates ($\S$5.1.1).  
We tentatively include IRAS 20414$-$1651 in this category as well,
because its spectrum in Figure 10 shows a more clearly discernible HCN
v$_{2}$=1f J=3--2 emission signature at the lower frequency part of the
bright HCO$^{+}$ v=0 J=3--2 emission than other ULIRGs (IRAS
12112$+$0305 SW, IRAS 
12127$-$1412, PKS 1345$+$12, IRAS 06035$-$7102, IRAS 13509$+$0442).
IRAS 12112$+$0305 SW and IRAS 13509$+$0442 show no AGN signature in
either infrared or our new ALMA (sub)millimeter data.
Our ALMA Cycle 2 results of the buried-AGN-hosting ULIRG IRAS
20551$-$4250 \citep{ima16a}, and multiple AGN-dominated nuclear regions
of the optical Seyfert 2 galaxy, NGC 1068 \citep{ima16b}, are also
plotted. In addition to these ALMA data, HCN J=3--2 and HCO 
$^{+}$ J=3--2 simultaneous observational data for NGC 4418
\citep{sak10} and NGC 1097 \citep{hsi12} taken with Submillimeter Array
(SMA), and those for Mrk 231 \citep{aal15a} obtained with IRAM 
Plateau de Bure Intermerometer (PdBI), are added, by classifying NGC
4418, the NGC 1097 nucleus, the NGC 1097 starburst ring, 
and Mrk 231 as a buried AGN, Seyfert 1, starburst, 
and an obscured AGN, respectively.  

In Figure 14, we see a clear trend for AGNs, including infrared-elusive
buried AGN candidates, to show elevated 
HCN-to-HCO$^{+}$ J=3--2 flux ratios, compared with starburst regions. 
Multiple starburst regions in NGC 7469 and NGC 1614, and other starburst
galaxies (IRAS 12112$+$0305 SW, IRAS 13509$+$0442, and NGC 1097
off-nuclear starburst) consistently show low HCN-to-HCO$^{+}$ J=3--2
flux ratios. Hence, the low ratios are interpreted to be a general
property of starbursts, rather than a specific property of a particular
starburst region. The excess of the flux ratios in AGN-dominant nuclei
in optical Seyferts and LIRGs with luminous obscured AGN signatures,
compared with starbursts, is taken to be a robust result. 

A similar HCN-to-HCO$^{+}$ flux enhancement in AGNs at J=1--0 has been
proposed \citep{koh05,kri08,cos11,pri15,izu16}, and was the basis for our
ALMA observations. 
The enhanced HCN-to-HCO$^{+}$ flux ratios in AGNs appear to be common at
different J-transition lines.
As mentioned in $\S$1, the HCN-to-HCO$^{+}$ flux comparison at J=3--2 is 
applicable to many interesting nearby LIRGs at z=0.06--0.3 
\citep{kim98}, whose HCN and HCO$^{+}$ observations at J=1--0 are not
possible with ALMA.

\subsubsection{Interpretation}

The observed HCN-to-HCO$^{+}$ flux enhancement in AGNs,
compared with starbursts, is naturally explained if (1) the HCN
abundance is enhanced and/or (2) HCN excitation to J=3 is higher, as
discussed by \citet{ima16a}.
Regarding scenario (1), it is clear that enhanced molecular
abundance generally produces a higher flux of that molecule in the
optically thin regime.  
When line opacity becomes significant, the emission line flux does not
increase proportionally to the increased abundance.
However, adopting the widely accepted clumpy molecular gas model, 
where molecular clouds consist of randomly moving clumps with a small
volume filling factor and the line opacity is primarily inside each
clump, rather than different clumps in the foreground with different
velocities \citep{sol87}, an increasing HCN abundance will result in an
increased HCN flux even in the optically thick regime, if each clump has
a decreasing radial density profile \citep{ima07a}.  
This behavior differs from molecular clouds with a smooth gas
distribution, where the observed molecular line flux saturates at some
point when the opacity exceeds a certain threshold.  

Regarding scenario (2), the critical density of HCN is a factor of
$\sim$5 higher than HCO$^{+}$ at the same J-transition
\citep{mei07,gre09}, under the same line opacity.   
For molecular gas with the same temperature and density, HCO$^{+}$
J=3--2 is more easily excited (to close to the thermalized condition)
than HCN J=3--2.  
As the temperature of molecular gas in the close vicinity of an
AGN can be higher than that in a starburst due to the AGN's higher
emission surface brightness ($\S$1), if HCO$^{+}$ J=3--2 is thermally
excited and 
HCN J=3--2 is only sub-thermally excited in some starbursts, and if AGNs
can excite HCN J=3--2 closer to the thermal condition, then the observed
HCN-to-HCO$^{+}$ J=3--2 flux ratios in AGNs can show some excess,
relative to those in starbursts, even under similar abundance.  
Among the galaxies observed in our ALMA programs, HCN J=1--0 flux data
are available for IRAS 08572$+$3916 by pre-ALMA interferometric
observations \citep{ima07a} and for I Zw 1 by single-dish telescope
observations \citep{eva06}.   
The HCN J=3--2 to J=1--0 flux ratios are $\sim$1.1 and $\sim$2.2 for 
IRAS 08572$+$3915 and I Zw 1, respectively. 
These values are substantially lower than the ratio of nine, which is
expected from thermally excited optically thick molecular gas.
The deviation is larger for IRAS 08572$+$3915 than I Zw 1.  
For IRAS 08572$+$3915, the J=4--3 to J=3--2 flux ratios are 1.3$\pm$0.1
for HCN and 1.4$\pm$0.1 for HCO$^{+}$ \citep{ima14b}, both of which are
lower than the 1.8 (=16/9) expected for thermally excited optically thick
gas. The significant sub-thermal excitation of IRAS 08572$+$3915 may be
partly responsible for the relatively low observed HCN-to-HCO$^{+}$
J=3--2 flux ratio, compared with other AGNs (Figure 14). 

It has been argued that an infrared radiative pumping mechanism can
enhance the observed HCN v=0 J=3--2 flux \citep{car81,aal95,ran11}.  
Although this is true, such infrared radiative pumping should also work
for HCO$^{+}$ ($\S$1). 
As described in $\S$5.1.1 and \citet{ima16a}, under the same abundance,
the rate of infrared radiative pumping to the v$_{2}$=1 level does not
differ a great deal between HCN and HCO$^{+}$ for the observed galaxies.
Even though this infrared radiative pumping may work more efficiently in
AGNs than in starbursts and increase the {\it absolute fluxes} of HCN
and HCO$^{+}$ v=0 J=3--2 emission lines compared with collisional 
excitation alone, it is not clear whether this may be largely
responsible for the elevated HCN-to-HCO$^{+}$ v=0 J=3--2 
{\it flux ratios} in AGNs.  

\subsubsection{Line opacity}

If HCN abundance enhancement is (at least partly) responsible for the
elevated HCN-to-HCO$^{+}$ J=3--2 flux ratios in AGNs, the HCN line
opacity could be higher than HCO$^{+}$. 
Even though some AGNs show high observed HCN-to-HCO$^{+}$ flux ratios, 
other AGNs may not, due to higher HCN flux attenuation than HCO$^{+}$ by
line opacity.
Thus, the selection of AGN-important galaxies, based on the observed high
HCN-to-HCO$^{+}$ flux ratios, may be incomplete and miss some fraction
of AGNs unless HCN line opacity is properly corrected for.

In the widely accepted clumpy molecular gas model \citep{sol87}
mentioned in $\S$5.2.2, the opacity in a molecular cloud mostly comes
from each clump.
If the properties of each clump inside a molecular cloud are assumed to
be uniform \citep{sol87}, the observed molecular line flux from a
molecular cloud is attenuated without significantly changing the
observed velocity profiles.   
An effective way to estimate the HCN line opacity is the comparison of
molecular isotopologues such HCN and H$^{13}$CN, assuming a certain 
intrinsic $^{12}$C-to-$^{13}$C abundance ratio. 
The H$^{13}$CN J=3--2 emission line was detected for the AGN-hosting
LIRG IRAS 20551$-$4250, and it was estimated that line opacity
correction causes the {\it intrinsic} HCN-to-HCO$^{+}$ J=3--2 flux ratio to be
substantially larger than the {\it observed} flux ratio \citep{ima16a}.
Among obscured-AGN-classified LIRGs, PKS 1345$+$12 and IRAS 06035$-$7102
also show not-as-high observed HCN-to-HCO$^{+}$ J=3--2 flux ratios in
Figure 14. It is not clear at this stage whether the {\it intrinsic}
HCN-to-HCO$^{+}$ J=3--2 flux ratios are similarly non-high or higher
than the observed ratios for these two sources.
Line opacity correction is definitely required to refine the AGN
selection based on the HCN-to-HCO$^{+}$ J=3--2 flux ratios. 

In the nuclei of some LIRGs, the concentration of molecular gas could be
extreme, and so the volume filling factor of molecular gas clumps in
molecular clouds could be large.  
The molecular gas geometry may be better approximated by a spatially
smooth distribution \citep{dow93,sco15} rather than the clumpy
structure. In this case, double-peaked molecular emission line profiles
could be produced through self-absorption by foreground molecular gas,  
which works preferentially for the most abundant central velocity
component \citep{aal15b}.   
IRAS 12112$+$035 NE and IRAS 20414$-$1651 display such
double-peaked emission line 
profiles for HCN J=3--2 and HCO$^{+}$ J=3--2 (Figures 5 and 9).
For IRAS 12112$+$0305 NE, a similar profile is also evident at J=4--3 of
HCN and HCO$^{+}$, albeit at lower S/N ratios \citep{ima16c}. 
The origin of this double-peaked line profile could be (1)
self-absorption and/or (2) emission being dominated by molecular 
gas in a prominent rotating disk.  

For case (1), we tried single Gaussian fits using data 
not significantly affected by the central dips. 
These fits are shown as dotted curved lines in the spectra of IRAS
12112$+$0305 NE and IRAS 20414$-$1651 in Figure 9. 
The estimated HCN J=3--2 and HCO$^{+}$ J=3--2 emission line fluxes,
based on the single Gaussian component fits, are included in Tables 4
and 5, respectively. 
For IRAS 12112$+$0305 NE, the fluxes based on the two Gaussian
component fits of the double-peaked profiles are smaller than those of
the single Gaussian fits by $\sim$25\% and $\sim$50\% for HCN J=3--2 and
HCO$^{+}$ J=3--2, respectively. If these flux differences are due to
self-absorption by foreground molecular gas inside IRAS 12112$+$0305 NE,
then it is estimated that the HCN J=3--2 and HCO$^{+}$ J=3--2 fluxes are
attenuated by a factor of $\sim$1.3 and $\sim$2, respectively.
For IRAS 20414$-$1651, the fluxes based on the two Gaussian fits are
smaller than those of the single Gaussian fits by a factor of 3--5 for
both HCN J=3--2 and HCO$^{+}$ J=3--2.

For IRAS 12112$+$0305 NE and IRAS 20414$-$1651, H$^{13}$CN J=3--2
data have been taken in ALMA Cycle 2 and been marginally detected 
in the moment 0 maps with 0.28 [Jy beam$^{-1}$ km s$^{-1}$]
(3.2$\sigma$) and 0.20 [Jy beam$^{-1}$ km s$^{-1}$] (3.3$\sigma$) 
\citep{ima16c}, respectively, which correspond to the observed
HCN-to-H$^{13}$CN J=3--2 flux ratios with $\sim$30 and $\sim$20. 
Assuming that the intrinsic $^{12}$C/$^{13}$C abundance ratios in these
ULIRGs are 50--100 \citep{hen93a,hen93b,mar10,hen14} and that 
H$^{13}$CN J=3--2 emission is optically thin, it is suggested that 
HCN J=3--2 emission is flux-attenuated with a factor of 1.5--3 and
2.5--5 for IRAS 12112$+$0305 NE and IRAS 20414$-$1651, respectively.
Hence, if the central dips detected in the HCN J=3--2 and HCO$^{+}$
J=3--2 emission in IRAS 12112$+$0305 NE and IRAS 20414$-$1651 are 
due to self-absorption, the flux attenuation estimated from the
comparison of Gaussian fittings is smaller than that derived from the
HCN-to-H$^{13}$CN J=3--2 flux comparison by a factor of 1--2 for IRAS
12112$+$0305 NE, while these two estimates look comparable within
uncertainty for IRAS 20414$-$1651. 
However, despite limited signal-to-noise ratios, double-peaked emission
line profiles with similar velocity peaks to the bright HCN J=3--2 and
HCO$^{+}$ J=3--2 emission lines or top-flat type line profiles, rather
than a centrally-peaked single Gaussian profile, are seen for the
H$^{13}$CN J=3--2 and CS J=7--6 emission lines in IRAS 12112$+$0305 NE
and for H$^{13}$CN J=3--2 in IRAS 20414$-$1651 \citep{ima16c}.
Since the self-absorption effect is expected to be much smaller for 
the fainter H$^{13}$CN J=3--2 emission line than HCN J=3--2 and
HCO$^{+}$ J=3--2, it is not clear whether the observed double-peaked
emission line profiles detected in IRAS 12112$+$0305 NE and IRAS
20414$-$1651 are explained solely by the self-absorption. 

We next consider the second rotating disk scenario.  
In the intensity-weighted mean velocity (moment 1) maps of IRAS
12112$+$0305 NE and IRAS 22491$-$1651 in Figure 13, the signature of a
rotational motion is marginally seen along the north-east to south-west
direction, with a velocity difference of $\sim$200--300 km s$^{-1}$ and
$\sim$400 km s$^{-1}$, respectively. 
They are comparable to the observed velocity difference of the double
peaks in Figure 9. 
Figure 15 shows the spectra within the beam size, at 3 pix
(0$\farcs$3) north and 3 pix (0$\farcs$3) east (i.e., $\sim$0$\farcs$4
north-east), and at 3 pix (0$\farcs$3) south and 3 pix (0$\farcs$3) west
(i.e., $\sim$0$\farcs$4 south-west), relative to the continuum peak
positions, for IRAS 12112$+$0305 NE and IRAS 22491$-$1651. 
It is shown that the red (blue) component is relatively strong at the  
0$\farcs$4 north-east (south-west) position, as is expected from the
moment 1 maps of both objects.
We interpret that compact rotating disks which are not clearly
resolved with our ALMA beam size (0$\farcs$5--0$\farcs$8) can also
contribute significantly to the observed double-peaked emission line
profiles in IRAS 12112$+$0305 NE and IRAS 20414$-$1651.

In summary, AGNs tend to show elevated HCN-to-HCO$^{+}$ J=3--2 flux
ratios, but some AGNs have non-high observed HCN-to-HCO$^{+}$ J=3--2
flux ratios. This could be explained by a larger flux attenuation caused
by line opacity for HCN than HCO$^{+}$, if the HCN abundance is higher
than HCO$^{+}$.  
In this respect, although we may be able to say that the elevated
observed HCN-to-HCO$^{+}$ J=3--2 flux ratios are good AGN signatures, 
not all AGNs are selected based on the observed high HCN-to-HCO$^{+}$
J=3--2 flux ratios.  
Line opacity correction will make our method even more powerful and
complete by reducing the number of missing AGNs.
Clear double-peaked HCN J=3--2 and HCO$^{+}$ J=3--2 emission line 
profiles are seen in IRAS 12112$+$0305 NE and IRAS 20414$-$1651, which
we interpret that rotating disks contribute significantly, in addition
to a possible self-absorption effect.

\subsection{Non-detection of molecular gas in IC 4329 A}

The non-detection of the HCN J=3--2 and HCO$^{+}$ J=3--2 emission lines
in IC 4329 A was unexpected.   
In this subsection, we briefly consider its possible causes.
The three Seyfert 1 galaxies, NGC 7469, I Zw 1, and IC 4329 A, were
selected because their nuclear infrared $L$-band (3--4 $\mu$m) emission
is thought to be dominated by AGN-heated hot dust emission and their
observed fluxes are high ($\S$2). 
If dust and molecular gas spatially coexist in the nuclear region, 
strong collisionally excited molecular gas emission is also expected
there. The observed nuclear $L$-band (3--4 $\mu$m) flux of IC 4329 A is
about $\sim$5 times higher than those of NGC 7469 and I Zw 1
\citep{iw04,ima11a}.  
At a first-order approximation, the HCN J=3--2 peak flux in IC 4329 A is
expected to be higher than NGC 7469 and I Zw 1.
However, the observed HCN J=3--2 emission peak is more than a factor of
ten and five smaller than those of NGC 7469 and I Zw 1, respectively
(Figure 4).

The nuclear HCN J=3--2 emission peak is roughly predicted from the
nuclear infrared emission, based on their correlation \citep{ima14b}. 
If we assume that the observed infrared luminosity in Table 1 originates
from the nuclear region, then the expected HCN J=1--0 emission peaks are  
$\sim$18 mJy, $\sim$3 mJy, and $\sim$4 mJy for NGC 7469, I Zw 1, and IC
4329 A, respectively. 
If the HCN J=3--2 emission peak is nine times larger than that of HCN
J=1--0, which is expected in thermally excited optically thick gas, then  
the expected HCN J=3--2 flux peaks are $\sim$165 mJy, $\sim$25 mJy, and
$\sim$35 mJy, for NGC 7469, I Zw 1, and IC 4329 A, respectively. 
For NGC 7469, since it is estimated that about one third of the infrared 
luminosity originates from the nuclear region \citep{gen95}, 
the expected HCN J=3--2 flux peak from the NGC 7469 nucleus is $\sim$55
mJy.
For the NGC 7469 nucleus and I Zw 1, the observed HCN J=3--2 peak
fluxes ($\sim$25 mJy and $\sim$10 mJy, respectively) agree with the
above expected values within a factor of 2--3. 
However, the observed HCN J=3--2 peak flux of IC 4329 A is more than an
order of magnitude smaller than the above expectation.
Possible explanations include (1) infrared emission is spatially 
extended, rendering the fraction of the nuclear component small, and 
(2) HCN J=3--2 is only sub-thermally excited, and the HCN J=3--2 to 
J=1--0 flux ratio is thus considerably (more than an order of 
magnitude) smaller than nine. 
With regard to (1), as no clear spatially extended off-nuclear emission
is detected at infrared 10 $\mu$m \citep{asm14}, this seems
unlikely. Regarding (2), because IC 4329 A contains a luminous
X-ray-emitting AGN \citep{bri11,bre14}, this also seems unlikely for
nuclear molecular gas.   

One scenario that could explain the weak HCN and HCO$^{+}$ J=3--2
emission in IC 4329 A is that the observed featureless nuclear infrared
$L$-band (3--4 $\mu$m) continuum is not dominated by AGN-heated hot dust
emission, but by other emission mechanisms such as synchrotron emission.  
Figure 16 displays the spectral energy distributions of NGC 7469, I Zw
1, and IC 4329 A in the infrared and radio wavelength ranges.  
The radio emission at $<$20 GHz is usually dominated by synchrotron
emission. The q-value, defined as the decimal logarithm of the
far-infrared (40--500 $\mu$m) to radio flux ratio \citep{con91}, is
often used to detect radio-loud AGNs, which show stronger 
synchrotron emission than the majority of radio-quiet AGNs. 
While the q-values of NGC 7469 and I Zw 1 (Table 10) are within the
range of starburst-dominated galaxies and many radio-quiet AGNs (q
$\sim$ 2.3--2.4) \citep{con91,bar96,cra96,roy98}, that of IC 4329 A (q
$\sim$ 1.5) (Table 10) is substantially lower, suggesting that IC 4329 A
contains a radio-loud AGN. 
However, the observed infrared $L$-band (3--4 $\mu$m or $\sim$10$^{5}$
GHz) flux is well above the extrapolation from the synchrotron emission
component at $<$20 GHz. 
The strong infrared excess at 200--10$^{5}$ GHz in IC 4329 A, as well as
NGC 7469 and I Zw 1, suggests that their infrared $L$-band (3--4 $\mu$m)
emission is dominated by AGN-heated hot dust emission
\citep{alo11,ich15}, rather than synchrotron emission. 
Thus, this possibility also seems unlikely.

A second scenario is that the column density of the obscuring gas and
dust surrounding the central AGN of IC 4329 A is very small. 
The infrared $L$-band (3--4 $\mu$m) continuum emission primarily arises
from hot dust with $>$a few 100 K, located at the innermost region of the
obscuring material, with limited contribution from outer cooler dust.
On the other hand, collisionally excited HCN and HCO$^{+}$ J=3--2
emission lines can still be produced at the outer regions, whose gas/dust
temperature is several 10 K to a few 100 K.  
If the obscuring gas/dust column density around an AGN is
substantially smaller in IC 4329 A than in NGC 7469 and I Zw 1, then the
smaller than expected HCN and HCO$^{+}$ J=3--2 emission line fluxes 
from the infrared $L$-band (3--4 $\mu$m) continuum flux could thus be
explained. From the infrared spectral energy distribution, \citet{ich15}
estimated that the outer to innermost radius ratio of nuclear-obscuring
dust in IC 4329 A is a factor of $\sim$3 smaller than NGC 7469,
suggesting that this scenario is a possibility.

As a third possibility, if the molecular line widths of IC 4329 A are
much larger than those of NGC 7469 and I Zw 1, the observed HCN/HCO$^{+}$
J=3--2 emission peak could be small, even if their fluxes are large. 
If IC 4329 A follows the similar HCN J=3--2 to nuclear infrared
luminosity correlation and has similar molecular line widths to NGC 7469
and I Zw 1, the HCN J=3--2 flux peak is then expected to be $>$15 mJy.
To explain the actual observed HCN J=3--2 flux peak with $<$1.5 mJy,
molecular line widths $>$10 times larger than those of NGC 7469 and I Zw
1, i.e., at least a FWHM $\sim$ 2000--3000 km s$^{-1}$, are required.
This is more than a factor of 2 larger than the highly turbulent
ongoing major merger ULIRG, The Superantennae \citep{mir91}. 
Because IC 4329 A is classified as a fairly settled spiral or S0 galaxy
with a nuclear dust lane, but with no obvious highly disturbed
morphology \citep{mal98}, such an extremely large molecular line width
seems unlikely.
For IC 4329 A, there has been no molecular gas detection reported in the
published literature, even for CO J=1--0 and J=2--1.
Future ALMA high-sensitivity observations of bright CO emission at the
IC 4329 A nucleus may help to test this scenario, if detection is
realized. 

In Figure 16, the observed flux increases with decreasing frequency from  
10$^{5}$ GHz (3 $\mu$m) to 10$^{4}$ GHz (30 $\mu$m). 
However, the spectral energy distribution is flatter in IC 4329 A than
NGC 7469 and I Zw 1, which means that the temperature of the dust
thermal emission in this frequency range is higher in IC 4329 A.
For given dust thermal radiation luminosity, if the dust temperature is
higher, then the required dust mass can be smaller; consequently, the
molecular mass becomes smaller, if dust and molecular gas spatially
coexist in a similar manner. 
This could contribute to the observed weaker-than-expected molecular gas
emission in IC 4329 A.
The 2--10 keV X-ray luminosities of NGC 7469, I Zw 1, and IC
4329 A are $\sim$2 $\times$ 10$^{43}$ ergs s$^{-1}$, 
$\sim$8 $\times$ 10$^{43}$ ergs s$^{-1}$, and 
$\sim$6 $\times$ 10$^{43}$ ergs s$^{-1}$, respectively
\citep{pin05,bri11,bre14}
\footnote{
For these Seyfert 1 galaxies, absorption-corrected and -uncorrected 
2--10 keV X-ray luminosities are comparable, due to the estimated low
X-ray-absorbing hydrogen column density (N$_{\rm H}$).
}.
The high 2--10 keV X-ray luminosity of IC 4329 A indicates the presence
of a luminous AGN.
This luminous AGN, together with the estimated small
outer-to-inner-radius ratio for nuclear dust, in IC 4329 A may be
related to its higher dust effective temperature derived
from the 10$^{4}$--10$^{5}$ GHz (3--30 $\mu$m) data.

In summary, the considerably smaller-than-expected HCN and HCO$^{+}$
J=3--2 emission line flux peak in IC 4329 A could be due to some 
combination of (1) a low column density of obscuring gas/dust around an
AGN and/or (2) a small dust mass to infrared $L$-band (3--4 $\mu$m)
luminosity ratio, due to a high dust effective temperature.
In either scenario, our (sub)millimeter energy diagnostic
method is not sensitive to almost bare AGNs with a very limited amount
of surrounding molecular gas, such as IC 4329A, because at least a
detectable amount of molecular line emission is required for our method
to be effective.

\section{Summary}

We conducted HCN and HCO$^{+}$ J=3--2 observations in our ALMA
Cycle 1, 2, and 3 programs, at both vibrational ground (v=0) and 
vibrationally excited (v$_{2}$=1f) levels, of three optical Seyfert 1 
galactic nuclei, diagnosed to be energetically dominated by
unobscured AGNs  
(NGC 7469, I Zw 1, and IC 4329 A) and eleven LIRGs with different
levels of estimated energetic contribution by an AGN (IRAS 08572$+$3915,
The Superantennae, IRAS 12112$+$0305, IRAS 22491$-$1808, NGC 1614, 
IRAS 12127$-$1412, IRAS 15250$+$3609, PKS 1345$+$12, IRAS
06035$-$7102, IRAS 13509$+$0442, and IRAS 20414$-$1651).  
Among the LIRGs, NGC 1614 is dominated by starburst activity; 
IRAS 08572$+$3915, The Superantennae, IRAS 12127$-$1412, IRAS
15250$+$3609, PKS 1345$+$12, and IRAS 06035$-$7102 show infrared
signatures of luminous obscured AGNs; and IRAS 12112$+$0305, IRAS
22491$-$1808, IRAS 13509$+$0442, and IRAS 20414$-$1651 
display no obvious AGN signatures in the infrared.  
Our primary scientific goals were to investigate 
(1) if enhanced HCN-to-HCO$^{+}$ flux ratios at J=3--2 are found in
AGNs, compared to starburst-dominated regions, as previously seen at
J=1--0 and J=4--3, and  
(2) if vibrationally excited v$_{2}$=1f J=3--2 emission lines of HCN and
HCO$^{+}$ are detected in AGNs, as HCN v$_{2}$=1f emission lines
had been detected in several obscured AGNs. 
We found the following main results:

\begin{enumerate}

\item 
HCN and HCO$^{+}$ J=3--2 emission lines at v=0 were clearly
detected in all observed primary galactic nuclei, except for IC 4329 A. 
In NGC 7469, these emission lines were also clearly detected at three
distinct starburst ring regions in the host galaxy.
The spatially extended HCN J=3--2 and HCO$^{+}$ J=3--2 emission was 
also detected in the starburst-dominated LIRG, NGC 1614. 
In IRAS 12112$+$0305, the emission lines were detected in both the
brighter main NE nucleus and fainter secondary SW nucleus.

\item 
The HCO$^{+}$ J=3--2 emission lines of IRAS 12112$+$0305 NE and IRAS
22491$-$1808 showed significant lower frequency tails, which we interpret
to be due to the contribution from vibrationally excited (v$_{2}$=1f) HCN
J=3--2 emission lines. 
In the moment 0 maps, the HCN v$_{2}$=1f J=3--2 emission lines were
detected with $>$4$\sigma$ in both objects.
The vibrational excitation of HCN is believed to originate in infrared
radiative pumping, by absorbing infrared 14 $\mu$m photons.
Because for given bolometric luminosity, an AGN can produce much
stronger infrared 14 $\mu$m photons than a starburst, an AGN is a
plausible origin for the infrared 14 $\mu$m continuum-emitting source,
even though a very compact extreme starburst cannot be completely ruled out.
If the AGN scenario is the case, IRAS 12112$+$0305 NE and IRAS
22491$-$1808 may contain extremely deeply buried AGNs whose signatures
were not seen in previous infrared spectroscopic energy diagnostic
methods, but were first detected in our ALMA (sub)millimeter data.

\item 
The signature of a similar lower frequency tail was recognizable
for the HCO$^{+}$ J=3--2 emission line of IRAS 08572$+$3915 and 
IRAS 20414$-$1651, but its detection significance in the moment 0
maps is lower than the above two LIRGs. 
IRAS 15250$+$3609 displays an emission sub-peak at close to the
expected frequency of the HCN v$_{2}$=1f J=3--2 emission line, but
it seems more likely that outflow origin HCO$^{+}$ v=0 J=3--2 emission
is largely contributing to this sub-peak.
The remaining LIRGs with infrared-obscured AGN signatures, 
The Superantennae, IRAS 12127$-$1412, PKS 1345$+$12 and IRAS
06035$-$7102, and the starburst-classified LIRGs, NGC 1614, IRAS
12112$+$0305 SW, and IRAS 13509$-$0442, show no 
discernible signature of the vibrationally excited HCN v$_{2}$=1f
J=3--2 emission line.  

\item 
Vibrationally excited HCN v$_{2}$=1f J=3--2 emission lines were not
clearly seen in the two unobscured AGN-dominated Seyfert 1 nuclei, NGC
7469 and I Zw 1, despite our original expectation for the detection.  
The upper limits of the HCN v$_{2}$=1f to v=0 J=3--2 flux ratios were
lower than the observed ratios in several LIRGs with detected HCN
v$_{2}$=1f J=3--2 emission lines.
However, when the intrinsic ratios after line opacity correction for HCN
v=0 J=3--2 emission were derived in these LIRGs, the upper limits in NGC
7469 and I Zw 1 were still larger than the intrinsic ratios. 
We interpret that the line opacity and flux attenuation of HCN v=0
emission lines are relatively small in optical Seyfert 1 AGNs, compared
with deeply buried AGNs in LIRGs.   
Currently, we do not see any evidence that the infrared radiative
pumping efficiency is different between these two types of AGNs.

\item
The vibrationally excited HCO$^{+}$ v$_{2}$=1f J=3--2 emission lines
were not recognizable in any of the observed galactic nuclei.  

\item 
We identified that galaxies that do (or may) contain AGNs tend to
display higher HCN-to-HCO$^{+}$ J=3--2 flux ratios than starburst
regions, as previously argued at J=1--0 and J=4--3.
However, a small fraction of AGN-classified sources do not
necessarily show very high observed HCN-to-HCO$^{+}$ J=3--2 flux
ratios. Line-opacity-corrected intrinsic flux ratios are needed for all
sources to refine the AGN selection in this method.

\item
IRAS 15250$+$3609 showed a significant dip at the higher frequency 
(blueshifted) side of the HCO$^{+}$ J=3--2 emission line and a
sub-emission-peak at the lower frequency (redshifted) side of the
HCO$^{+}$ J=3--2 emission line. 
These features are interpreted to be
due to the P Cygni profile by outflow activity of HCO$^{+}$.  

\item In the continuum map of IRAS 13509$+$0442 at $\sim$235 GHz, a
source even brighter than IRAS 13509$+$0442 was detected at $\sim$1$''$
eastern and 8--9$''$ northern side. 
No obvious optical counterpart was identified in the SDSS. 
No emission lines were seen in our ALMA spectra.   
The nature of this source is still enigmatic. 

\end{enumerate}  

We thank Dr. K. Saigo, Y. Ao, and A. Kawamura for their support
regarding analysis of the ALMA data, and K. Sakamoto for useful discussions. 
We are grateful to the anonymous referee for his/her constructive 
comment which has helped a lot to improve the clarity of the
descriptions of this manuscript.  
M.I. was supported by JSPS KAKENHI Grant Number 23540273, 15K05030 and
the ALMA Japan Research Grant of the NAOJ Chile Observatory,
NAOJ-ALMA-0001, 0023, 0072.
T.I. is thankful for the fellowship received from the Japan Society for
the Promotion of Science (JSPS). 
This paper makes use of the following ALMA data:
ADS/JAO.ALMA\#2012.1.00034.S, 2013.1.00032.S, 2013.1.00188.S, and
2015.1.00027.S.  
ALMA is a partnership of ESO (representing its member states), NSF (USA) 
and NINS (Japan), together with NRC (Canada), NSC and ASIAA
(Taiwan), and KASI (Republic of Korea), in cooperation with the Republic
of Chile. The Joint ALMA Observatory is operated by ESO, AUI/NRAO, and
NAOJ. 
This research has made use of NASA's Astrophysics Data System and the
NASA/IPAC Extragalactic Database (NED) which is operated by the Jet
Propulsion Laboratory, California Institute of Technology, under
contract with the National Aeronautics and Space Administration. 
This research has also made use of Sloan Digital Sky Survey (SDSS)
data. 
Funding for the Sloan Digital Sky Survey (SDSS) has been provided by the 
Alfred P. Sloan Foundation, the Participating Institutions, the National
Aeronautics and Space Administration, the National Science Foundation,
the U.S. Department of Energy, the Japanese Monbukagakusho, and the Max
Planck Society. The SDSS Web site is http://www.sdss.org/. 
The SDSS is managed by the Astrophysical Research Consortium (ARC) for
the Participating Institutions. The Participating Institutions are The
University of Chicago, Fermilab, the Institute for Advanced Study, the
Japan Participation Group, The Johns Hopkins University, Los Alamos
National Laboratory, the Max-Planck-Institute for Astronomy (MPIA), the
Max-Planck-Institute for Astrophysics (MPA), New Mexico State
University, University of Pittsburgh, Princeton University, the United
States Naval Observatory, and the University of Washington.

\clearpage

\begin{deluxetable}{lccrrrrccc}
\tabletypesize{\scriptsize}
\tablecaption{Basic properties of observed galaxies \label{tbl-1}}
\tablewidth{0pt}
\tablehead{
\colhead{Object} & \colhead{Redshift} & \colhead{Physical scale} & 
\colhead{f$_{\rm 12}$}   & 
\colhead{f$_{\rm 25}$}   & 
\colhead{f$_{\rm 60}$}   & 
\colhead{f$_{\rm 100}$}  & 
\colhead{log L$_{\rm IR}$} & 
\colhead{Optical} &
\\
\colhead{} & \colhead{} & \colhead{[kpc/$''$]}  & \colhead{[Jy]}
& \colhead{[Jy]} & \colhead{[Jy]} & \colhead{[Jy]}  &
\colhead{[L$_{\odot}$]} & \colhead{Class} \\  
\colhead{(1)} & \colhead{(2)} & \colhead{(3)} & \colhead{(4)} & 
\colhead{(5)} & \colhead{(6)} & \colhead{(7)} & \colhead{(8)} & 
\colhead{(9)} 
}
\startdata
NGC 7469 & 0.0164 & 0.33 & 1.35 & 5.79 & 25.87 & 34.90 & 11.6 & Sy1 $^{a}$ \\
I Zw 1   & 0.0611 & 1.2 & 0.51 & 1.21 & 2.24  & 2.63  & 11.9 & Sy1 $^{b}$ \\
IC 4329 A  & 0.0160 & 0.32 & 1.08 & 2.21 & 2.03 & 1.66  & 10.9 & Sy1 $^{c}$ \\ \hline
IRAS 08572$+$3915 & 0.0580 & 1.1 & 0.32 & 1.70 & 7.43  & 4.59  & 12.1 & LI$^{d}$(Sy2$^{e}$) \\  
Superantennae (IRAS 19254$-$7245) & 0.0617 & 1.2 & 0.22 & 1.24 & 5.48 &
5.79 & 12.1 & Sy2 $^{f}$ \\    
IRAS 12112$+$0305 & 0.0730 & 1.4 & 0.12 & 0.51 & 8.50 & 9.98 & 12.3 & LI
$^{d}$ (Sy2 $^{e}$)\\     
IRAS 22491$-$1808 & 0.0776 \tablenotemark{A} (0.076) & 1.5 & 0.05 & 0.55 & 5.44
& 4.45 & 12.2 & HII$^{d,e}$ \\  
NGC 1614 & 0.0160 & 0.32 & 1.38 & 7.50 & 32.12 & 34.32 & 11.7 & HII
$^{a,g}$ (Cp$^{e}$)\\ \hline
IRAS 12127$-$1412 & 0.1332 \tablenotemark{A} (0.133) & 2.3 & $<$0.13 & 0.24 &
1.54 & 1.13 & 12.2 & LI $^{d}$ (HII $^{e}$) \\
IRAS 15250$+$3609 & 0.0552 \tablenotemark{A} (0.054) & 1.1 & 0.16 & 1.31 & 7.10
& 5.93 & 12.0 & LI $^{a}$ (Cp$^{e}$) \\
PKS 1345$+$12 (IRAS 13451$+$1232) & 0.1215 \tablenotemark{A} (0.122) & 2.2 &
0.14 & 0.67 & 1.92 & 2.06 & 12.3 & Sy2 $^{d,e}$ \\ 
IRAS 06035$-$7102 & 0.0795 & 1.5 & 0.12 & 0.57 & 5.13
& 5.65 & 12.2 & LI $^{h}$ \\
IRAS 13509$+$0442 & 0.1364 \tablenotemark{A} (0.136) & 2.4 & 0.10 & $<$0.23 &
1.56 & 2.53 & 12.3 & HII $^{d}$ (Cp$^{e}$) \\
IRAS 20414$-$1651 & 0.0870 \tablenotemark{A} (0.086) & 1.6 & $<$0.65 & 0.35 &
4.36 & 5.25 & 12.3 & HII $^{d}$ (Cp$^{e}$) \\
\hline 
\enddata

\tablenotetext{A}{
The observed peak frequencies of the detected HCN J=3--2 and
HCO$^{+}$ J=3--2 emission lines in our ALMA spectra are significantly
offset from those expected from optically-derived redshifts
\citep{soi87,str92,kim95,kim98,vei99}.  
Our high quality ALMA molecular line data provide accurate redshifts
with four effective digits, which are adopted in this paper.
Optically-derived redshifts \citep{kim95,kim98} are shown in parentheses.
}

\tablecomments{
Col.(1): Object name. 
The top three sources are optical Seyfert 1 galaxies. 
The next five and bottom six sources are LIRGs observed in ALMA Cycle 2
and Cycle 3, respectively.
Col.(2): Redshift. 
Col.(3): Physical scale in [kpc arcsec$^{-1}$]. 
Col.(4)--(7): f$_{12}$, f$_{25}$, f$_{60}$, and f$_{100}$ are 
{\it IRAS} fluxes at 12 $\mu$m, 25 $\mu$m, 60 $\mu$m, and 100 $\mu$m,
respectively, taken from the IRAS Faint Source Catalog (FSC). 
Col.(8): Decimal logarithm of infrared (8$-$1000 $\mu$m) luminosity
in units of solar luminosity (L$_{\odot}$), calculated with
$L_{\rm IR} = 2.1 \times 10^{39} \times$ D(Mpc)$^{2}$
$\times$ (13.48 $\times$ $f_{12}$ + 5.16 $\times$ $f_{25}$ +
$2.58 \times f_{60} + f_{100}$) [ergs s$^{-1}$] \citep{sam96}.
Col.(9): Optical spectroscopic classification.
``Sy1'', ``Sy2'', ``LI'', ``HII'', and ``Cp'' refer to Seyfert 1, Seyfert
2, LINER, HII-region, and starburst$+$AGN composite, respectively.
$^{a}$: \citet{vei95}.
$^{b}$: \citet{sch83}.
$^{c}$: \citet{rus93}.
$^{d}$: \citet{vei99}.
$^{e}$: \citet{yua10}.
$^{f}$: \citet{mir91}.
$^{g}$: \citet{kew01}.
$^{h}$: \citet{duc97}.
}

\end{deluxetable}

\begin{deluxetable}{llccc|ccc}
\tabletypesize{\scriptsize}
\tablecaption{Log of our ALMA observations \label{tbl-2}} 
\tablewidth{0pt}
\tablehead{
\colhead{Object} & \colhead{Date} & \colhead{Antenna} & 
\colhead{Baseline} & \colhead{Integration} & \multicolumn{3}{c}{Calibrator} \\ 
\colhead{} & \colhead{[UT]} & \colhead{Number} & \colhead{[m]} &
\colhead{[min]} & \colhead{Bandpass} & \colhead{Flux} & \colhead{Phase}  \\
\colhead{(1)} & \colhead{(2)} & \colhead{(3)} & \colhead{(4)} &
\colhead{(5)} & \colhead{(6)} & \colhead{(7)}  & \colhead{(8)} 
}
\startdata 
NGC 7469 & 2014 April 29 & 35 & 21--558 & 25 & J2148$+$0657 & Neptune & J2257$+$0743 \\   
& 2014 May 3 & 31 & 17--532 & 25 & J2148$+$0657 & Neptune & J2257$+$0743 \\   
& 2014 July 22 & 34 & 18--784 & 24 & J2253$+$1608 & J2232$+$117 & J2253$+$1608 \\   
I Zw 1 & 2014 May 4 & 39 & 17--558 & 22 & J2148$+$0657  & Neptune & J0121$+$1149 \\   
& 2014 May 4 & 39 & 17--558 & 22 & J0006$-$0623  & Neptune & J0121$+$1149 \\   
& 2014 July 16 & 34 & 20--650 & 14 & J0237$+$2848  & J0238$+$166 & J0121$+$1149 \\   
IC 4329 A & 2014 April 10 & 32 & 21--484 & 27 & J1107$-$4449 & Pallas &
J1342$-$2900 \\ \hline 
IRAS 08572$+$3915 & 2015 June 9 & 36 & 21--784 & 11 & J0854$+$2006 & Ganymede & J0927$+$3902 \\ 
Superantennae & 2015 May 18 & 37 &  21--555 & 10 & J2056$-$4714 & Ceres & J1933$-$6942 \\
IRAS 12112$+$0305 & 2015 May 14 & 37 & 23--558 & 5 & J1058$+$0133 & Ganymede & J1229$+$0203 \\
IRAS 22491$-$1808 & 2015 May 24 & 34 & 21--539 & 5 & J2258$-$2758 & Titan & J2056$-$4714 \\
NGC 1614 & 2015 May 24 & 36 & 21--539 & 5& J0423$-$0120 & Callisto & J0423$-$0120  \\ \hline
IRAS 12127$-$1412 & 2016 March 3 & 41 & 15--396 & 29 & J1256$-$0547 & J1256$-$0547 & J1215$-$1731 \\
IRAS 15250$+$3609 & 2016 March 13 & 38 & 15--460 & 11 & J1751$+$0939 & Titan & J1453$+$2648 \\
PKS 1345$+$12 & 2016 March 3 & 41 & 15--396 & 18 & J1229$+$0203 & Titan & J1415$+$1320 \\
IRAS 06035$-$7102 & 2016 March 14 & 43 & 15--460 & 12 & J0519$-$4546 & J0519$-$4546 & J0601$-$7036 \\ 
IRAS 13509$+$0442 & 2016 March 3 & 39 & 15--396 & 24 & J1256$-$0547 & J1256$-$0547 & J1359$+$0159 \\ 
IRAS 20414$-$1651 & 2016 March 26 & 38 & 15--460 & 11 & J1924$-$2914 & Pallas & J2011$-$1546 \\ \hline
\enddata

\tablecomments{ 
Col.(1): Object name. 
Col.(2): Observing date in UT. 
The bottom six sources were observed in ALMA Cycle 3.
Col.(3): Number of antennas used for observations. 
Col.(4): Baseline length in meters. Minimum and maximum baseline lengths are shown.  
Col.(5): Net on-source integration time in min.
Cols.(6), (7), and (8): Bandpass, flux, and phase calibrator for the 
target source, respectively.}

\end{deluxetable}

\begin{deluxetable}{cccrccl}
\tabletypesize{\scriptsize}
\tablecaption{Continuum emission properties \label{tbl-2}}
\tablewidth{0pt}
\tablehead{
\colhead{Object} & \colhead{Position} & \colhead{Frequency} & \colhead{Flux} & 
\colhead{Peak Coordinate} & \colhead{rms} & \colhead{Beam} \\
\colhead{} & \colhead{} & \colhead{[GHz]} & \colhead{[mJy beam$^{-1}$]} & 
\colhead{(RA,DEC)J2000} & \colhead{[mJy beam$^{-1}$]} & 
\colhead{[$''$ $\times$ $''$] ($^{\circ}$)} \\  
\colhead{(1)} & \colhead{(2)} & \colhead{(3)}  & \colhead{(4)}  &
\colhead{(5)} & \colhead{(6)} & \colhead{(7)}  
}
\startdata 
NGC 7469 & Nucleus & 260.8--265.4 & 3.4 (33$\sigma$) & (23 03 15.62, +08 52 26.1)
& 0.10 & 0.57$\times$0.52 (72$^{\circ}$) \\
 & SB1 & 260.8--265.4 & 1.2 (12$\sigma$) & (23 03 15.52, +08 52 25.1) &
0.10 & 0.57$\times$0.52 (72$^{\circ}$) \\ 
 & SB2 & 260.8--265.4 & 0.71 (7$\sigma$) & (23 03 15.69, +08 52 27.1)
& 0.10 & 0.57$\times$0.52 (72$^{\circ}$) \\ 
 & SB3 & 260.8--265.4 & 0.86 (8$\sigma$)& (23 03 15.58, +08 52 27.4) & 0.10 & 0.57$\times$0.52 (72$^{\circ}$) \\
I Zw 1 & Nucleus  & 249.7--254.2 &  1.2 (21$\sigma$) & (00 53 34.94, +12 41 36.0) &
0.057 & 0.67$\times$0.60 (60$^{\circ}$) \\  
IC 4329 A & Nucleus & 260.7--265.4 & 13.0 (69$\sigma$) & 
(13 49 19.26, $-$30 18 34.2) & 0.19   & 0.95$\times$0.53 (83$^{\circ}$) \\ \hline
IRAS 08572$+$3915 & Nucleus & 250.4--254.9 & 2.5 (23$\sigma)$ & (09 00 25.36, $+$39
03 53.9) & 0.11 & 0.89$\times$0.47 ($-$19$^{\circ}$) \\
Superantennae & Nucleus & 249.5--254.4 & 5.0 (29$\sigma)$ & (19 31 21.41, $-$72
39 21.6) & 0.17 & 0.87$\times$0.61 (21$^{\circ}$) \\
IRAS 12112$+$0305 & NE  & 246.8--251.3 & 8.3 (46$\sigma)$ & (12 13 46.06, $+$02
48 41.5) & 0.18 & 0.73$\times$0.55 (66$^{\circ}$) \\
 & SW & 246.8--251.3 & 0.80 (4$\sigma)$ & (12 13 45.94, $+$02 48 39.1) & 0.19 &
0.73$\times$0.55 (66$^{\circ}$) \\ 
IRAS 22491$-$1808 & Nucleus & 246.1--250.7 & 4.7 (19$\sigma)$ & (22 51 49.35, $-$17 
52 24.2) & 0.25 & 0.92$\times$0.59 (67$^{\circ}$) \\
NGC 1614 & S-peak & 260.8--265.5 & 2.6 (8$\sigma)$ & (04 34 00.01, $-$08 34 45.7) & 0.31 &
1.06$\times$0.58 ($-$75$^{\circ}$) \\
         & N-peak & 260.8--265.5 & 2.5 (8$\sigma)$ & (04 33 59.99, $-$08 34 44.7) & 0.31 &
1.06$\times$0.58 ($-$75$^{\circ}$) \\ \hline
IRAS 12127$-$1412 & Nucleus & 233.7--238.4 & 1.4 (25$\sigma)$ & (12 15
19.13, $-$14 29 41.7) & 0.057 & 0.94$\times$0.74 (71$^{\circ}$) \\
IRAS 15250$+$3609 & Nucleus & 251.0--255.8 & 11 (41$\sigma$) & (15 26 59.42, $+$35 58
37.4) & 0.27 & 1.21$\times$0.72 ($-$0$^{\circ}$) \\
PKS 1345+12 & Nucleus & 236.0--240.7 & 364 (70$\sigma$) & (13 47 33.37, $+$12 17 24.2)
& 5.2 & 0.92$\times$0.85 ($-$24$^{\circ}$) \\ 
IRAS 06035$-$7102 & Nucleus & 245.4--250.2 & 2.4 (19$\sigma$) & (06 02 53.95, $-$71 03
10.2) & 0.13 & 1.12$\times$0.78 (49$^{\circ}$) \\
IRAS 13509$+$0442 & Nucleus & 233.1--237.7 & 1.5 (20$\sigma$) & (13 53 31.57, $+$04 28
04.8) & 0.076 & 0.99$\times$0.81 (46$^{\circ}$) \\
 & NE source & 233.1--237.7 & 1.9 (26$\sigma$) & (13 53 31.66, $+$04 28 13.1) &
0.076 & 0.99$\times$0.81 (46$^{\circ}$) \\ 
IRAS 20414$-$1651 & Nucleus & 243.9--248.6 & 5.2 (38$\sigma)$ & (20 44
18.17, $-$16 40 16.8) & 0.14 & 0.76$\times$0.66 ($-$89$^{\circ}$) \\ 
\enddata

\tablecomments{
Col.(1): Object name.
Col.(2): Position inside galaxy. 
Col.(3): Observed frequency used for continuum extraction in
[GHz]. Frequencies at obvious emission lines are removed.
Col.(4): Flux in [mJy beam$^{-1}$] at the emission peak. 
The value at the highest flux pixel (0$\farcs$1 pixel$^{-1}$) in the continuum map is adopted. The detection significance relative to the rms noise is shown in parentheses, where possible systematic uncertainty is not included.  
The possible systematic ambiguity in The Superantennae could be larger 
than the remaining sources due to the very broad HCN/HCO$^{+}$ J=3--2
emission line profiles, extending almost all frequency coverage of each of the corresponding spectral windows.  
Col.(5): The coordinate of the continuum emission peak in J2000.
Col.(6): The rms noise level (1$\sigma$) in [mJy beam$^{-1}$], 
derived from the standard deviation of sky signals in each 
continuum map. 
Col.(7): Synthesized beam in [arcsec $\times$ arcsec] and position angle
in [degree]. 
The position angle is 0$^{\circ}$ along the north--south direction
and increases in the counterclockwise direction.}

\end{deluxetable}

\begin{deluxetable}{ll|llcl|cccc}
\tabletypesize{\scriptsize}
\tablecaption{Flux of the HCN J=3--2 emission line \label{tbl-4}} 
\tablewidth{0pt}
\tablehead{
\colhead{Object} & \colhead{Pos} & 
\multicolumn{4}{c}{Integrated intensity (moment 0) map} & 
\multicolumn{4}{c}{Gaussian line fit} \\  
\colhead{} & \colhead{} & \colhead{Peak} &
\colhead{rms} & \colhead{Ch} & \colhead{Beam} & \colhead{Velocity}
& \colhead{Peak} & \colhead{FWHM} & \colhead{Flux} \\ 
\colhead{} & \colhead{} &
\multicolumn{2}{c}{[Jy beam$^{-1}$ km s$^{-1}$]} & 
\colhead{Sum} & \colhead{[$''$ $\times$ $''$]} &
\colhead{[km s$^{-1}$]} & \colhead{[mJy]} & \colhead{[km s$^{-1}$]} &
\colhead{[Jy km s$^{-1}$]} \\  
\colhead{(1)} & \colhead{(2)} & \colhead{(3)} & \colhead{(4)} & 
\colhead{(5)} & \colhead{(6)} & \colhead{(7)} & \colhead{(8)} &
\colhead{(9)} & \colhead{(10)}
}
\startdata 
NGC 7469 & Nuc & 5.4 (59$\sigma$) & 0.092 & 21 & 
0.58$\times$0.52 & 4909$\pm$2 & 24$\pm$1 & 220$\pm$3 & 
5.5$\pm$0.1 \\     
 & SB1 & 0.96 (10$\sigma$) & 0.092 & 21 &
0.58$\times$0.52 & 4966$\pm$1 & 15$\pm$1 & 63$\pm$1 & 
0.96$\pm$0.03 \\     
 & SB2 & 0.70 (8$\sigma$) & 0.092 & 21 &
0.58$\times$0.52 & 4916$\pm$1 & 11$\pm$1 & 65$\pm$2 & 
0.72$\pm$0.03 \\     
 & SB3 & 0.67 (7$\sigma$) & 0.092 & 21 &
0.58$\times$0.52 & 5028$\pm$1 & 12$\pm$1 & 51$\pm$2 & 
0.65$\pm$0.03 \\     
 & ring & --- & --- & --- & --- & 4952$\pm$5 & 45$\pm$2 & 214$\pm$12 & 10.1$\pm$1.0 \\ 
I Zw 1 & Nuc & 2.6 (31$\sigma$) & 0.086 & 21 &
0.69$\times$0.63 & 18335$\pm$3 & 10$\pm$1 & 296$\pm$7 & 
2.9$\pm$0.1 \\ \hline    
08572$+$3915 & Nuc & 2.5 (14$\sigma$) & 0.18 & 27 &
0.91$\times$0.47 & 17501$\pm$7 & 6.6$\pm$0.2 & 382$\pm$17 & 
2.6$\pm$0.1 \\     
Superant & Nuc & 9.6 (20$\sigma$) & 0.47 & 58 &
0.88$\times$0.66 & 18540$\pm$7 & 12.8$\pm$0.2 & 821$\pm$17 & 
10.6$\pm$0.3 \tablenotemark{E} \\     
12112$+$0305 & NE & 8.3 (28$\sigma$) & 0.30 & 34 &
0.73$\times$0.54 & 21661$\pm$18,21958$\pm$20 &
15.0$\pm$1.4,14.6$\pm$1.2 & 283$\pm$34,273$\pm$46 & 8.1$\pm$1.2 \\     
 & & & & & & 21807$\pm$4 \tablenotemark{C} & 24$\pm$1 \tablenotemark{C}
& 431$\pm$14 \tablenotemark{C} & 10.4$\pm$0.7 \tablenotemark{C} \\     
 & SW & 0.72 (4$\sigma$) & 0.17 \tablenotemark{B} & 16 \tablenotemark{B} &
0.73$\times$0.54 & 21970$\pm$34 & 2.0$\pm$0.5 &
323$\pm$75 & 0.64$\pm$0.22 \\      
22491$-$1808 & Nuc & 7.3 (26$\sigma$) & 0.28 & 29 &
0.95$\times$0.60 & 23309$\pm$6 & 18.0$\pm$0.5 & 458$\pm$13 & 
8.1$\pm$0.3 \\     
NGC 1614 & SB1 & 0.86 (4$\sigma$) & 0.21 & 16 &
1.07$\times$0.58 & 4847$\pm$7 & 8.1$\pm$1.2 & 103$\pm$18 & 
0.87$\pm$0.20 \\     
 & SB2 & 1.0 (5$\sigma$) & 0.21 & 16
& 1.07$\times$0.58 & 4763$\pm$8 & 7.2$\pm$1.0 &
128$\pm$23 & 0.85$\pm$0.22 \\      
 & 2$\farcs$5 & --- & --- & --- & --- & 4812$\pm$23 &
19.1$\pm$3.8 & 258$\pm$66 & 5.1$\pm$1.7 \\ \hline
12127$-$1412 & Nuc & 1.1 (12$\sigma$) & 0.087 & 14 &
0.94$\times$0.74 & 39961$\pm$14 & 2.6$\pm$0.1 & 524$\pm$33
& 1.3$\pm$0.1 \\
15250$+$3609 & Nuc & 4.7(32$\sigma$) \tablenotemark{A} & 0.15
\tablenotemark{A} & 9 \tablenotemark{A} & 1.21$\times$0.73
& 16575$\pm$5 \tablenotemark{A} & 
19$\pm$1 \tablenotemark{A} & 274$\pm$10 \tablenotemark{A} & 5.2$\pm$0.3
\tablenotemark{A} \\
PKS 1345 & Nuc & 2.3(20$\sigma$) & 0.11 & 12 & 0.92$\times$0.86
& 36456$\pm$15 & 6.0$\pm$0.4 & 480$\pm$34 & 2.7$\pm$0.3 \\
06035$-$7102 & Nuc & 4.4(25$\sigma$) & 0.17 & 13 &
1.12$\times$0.78 & 23853$\pm$5 & 14$\pm$1 & 367$\pm$12 &
4.9$\pm$0.2 \\
13509$+$0442 & Nuc & 0.96(13$\sigma$) & 0.074 & 9 & 0.99$\times$0.81
& 40937$\pm$6 & 4.5$\pm$0.2 & 268$\pm$15 & 1.1$\pm$0.2 \\
20414$-$1651 & Nuc & 3.9 (26$\sigma)$ & 0.15 & 17 &
0.76$\times$0.66 & 25829$\pm$11,26216$\pm$12 & 7.4,8.2
& 250$\pm$24,309$\pm$28 & 4.3$\pm$0.3 \\
 & & & & & & 26050$\pm$4 \tablenotemark{D} & 53$\pm$15 \tablenotemark{D}
& 336$\pm$20 \tablenotemark{D} & 17$\pm$5 \tablenotemark{D} \\
\hline

\enddata

\tablenotetext{A}{
Only the bright main emission component at 251.8--252.2 GHz in
Figure 5(j) is considered.  
The fainter sub-peak component at 251.3--251.8 GHz, which is likely
to be of outflow origin, is not included.
}

\tablenotetext{B}{The frequency range of significant signal detection is
slightly different between the NE and SW nuclei. 
We created the moment 0 map of the SW nucleus by summing channels with
significant signal detection at the SW nucleus.}

\tablenotetext{C}{Single Gaussian fit after removing data points
affected by the central dips, based on the assumption of
self-absorption.  
Data at 21360--21650 km s$^{-1}$ and 21970--22260 km s$^{-1}$ in Figure
9 are used for the fit.}

\tablenotetext{D}{
Single Gaussian fit after removing data points affected by the 
central dips, based on the assumption of self-absorption.  
Data at 25200--25760 km s$^{-1}$ and 26340--27000 km s$^{-1}$ in Figure
9 are used for the fit.
}

\tablenotetext{E}{
Because the HCN J=3--2 emission line is extremely broad, extending
almost all frequency coverage of the ALMA spectral window, the ambiguity
of the continuum determination is large. The possible systematic
uncertainty is expected to be larger than for other objects.}

\tablecomments{ 
Col.(1): Object.
Col.(2): Position. 
For NGC 7469, ``ring'' refers to an annular region with a radius of 
0$\farcs$8--2$\farcs$5. 
For NGC 1614, ``2$\farcs$5'' refers to a 2$\farcs$5 radius circular region.
Col.(3): Integrated intensity of the HCN J=3--2 emission line 
($\nu_{\rm rest}$=265.89 GHz) in [Jy beam$^{-1}$ km s$^{-1}$] at the 
emission peak. 
The detection significance relative to the rms noise (1$\sigma$) in the 
moment 0 map is shown in parentheses. 
Possible systematic uncertainty is not included. 
Col.(4): rms noise (1$\sigma$) level in the moment 0 map in 
[Jy beam$^{-1}$ km s$^{-1}$], derived from the standard deviation 
of sky signals in each moment 0 map. 
Col.(5): The number of velocity channels summed to create moment 0 maps.
Each velocity channel has a width of $\sim$20 MHz ($\sim$20 km s$^{-1}$)
for objects observed in ALMA Cycles 1 and 2, or $\sim$40 MHz
($\sim$40 km s$^{-1}$) for those observed in ALMA Cycle 3 (bottom six 
sources).
Col.(6): Beam size in [arcsec $\times$ arcsec].
Cols.(7)--(10): Gaussian fits of emission lines in the spectra at the 
continuum peak position (except for NGC 1614), within the beam size, or
integrated starburst regions for NGC 7469 and NGC 1614. 
For NGC 1614, spectra are taken at the HCN J=3--2 emission peaks.  
Col.(7): Optical local standard of rest (LSR) velocity (v$_{\rm opt}$) of the emission peak in [km s$^{-1}$]. 
Col.(8): Peak flux in [mJy]. 
Col.(9): Observed FWHM in [km s$^{-1}$] in the left panels of Figure 9.  
Col.(10): Flux in [Jy km s$^{-1}$]. The observed FWHM in 
[km s$^{-1}$] in column 9 is divided by ($1+z$) to obtain the
intrinsic FWHM in [km s$^{-1}$].}

\end{deluxetable}

\clearpage

\begin{deluxetable}{ll|llcl|cccc}
\tabletypesize{\scriptsize}
\tablecaption{Flux of the HCO$^{+}$ J=3--2 emission line \label{tbl-4}} 
\tablewidth{0pt}
\tablehead{
\colhead{Object} & \colhead{Pos} &  
\multicolumn{4}{c}{Integrated intensity (moment 0) map} & 
\multicolumn{4}{c}{Gaussian line fit} \\  
\colhead{} & \colhead{} & \colhead{Peak} &
\colhead{rms} & \colhead{Ch} & \colhead{Beam} & \colhead{Velocity}
& \colhead{Peak} & \colhead{FWHM} & \colhead{Flux} \\ 
\colhead{} & \colhead{} &
\multicolumn{2}{c}{[Jy beam$^{-1}$ km s$^{-1}$]} & 
\colhead{Sum} & \colhead{[$''$ $\times$ $''$]} &
\colhead{[km s$^{-1}$]} & \colhead{[mJy]} & \colhead{[km s$^{-1}$]} &
\colhead{[Jy km s$^{-1}$]} \\  
\colhead{(1)} & \colhead{(2)} & \colhead{(3)} & \colhead{(4)} & 
\colhead{(5)} & \colhead{(6)} & \colhead{(7)} & \colhead{(8)} &
\colhead{(9)} & \colhead{(10)} 
}
\startdata 
NGC 7469 & Nuc & 4.6 (49$\sigma$) & 0.094 & 19 & 
0.57$\times$0.52 & 4914$\pm$2 & 21$\pm$1 & 217$\pm$3 & 
4.8$\pm$0.1 \\     
 & SB1 & 1.2 (13$\sigma$) & 0.094 & 19 & 
0.57$\times$0.52 & 4966$\pm$1 &  22$\pm$1 & 57$\pm$1 & 
1.3$\pm$0.03 \\     
 & SB2 & 0.97 (10$\sigma$) &  0.094 & 19 &
0.57$\times$0.52 & 4914$\pm$1 & 15$\pm$1 & 63$\pm$1 & 
0.96$\pm$0.03 \\     
 & SB3 & 0.86 (9$\sigma$) & 0.094 & 19 &
0.57$\times$0.52 & 5028$\pm$1 & 19$\pm$1 & 46$\pm$1 & 
0.93$\pm$0.03 \\  
 & ring & --- & --- & --- & --- & 4948$\pm$6 & 63$\pm$3 & 217$\pm$12
& 14.4$\pm$1.0 \\  
I Zw 1 & Nuc & 1.2 (23$\sigma$) & 0.053 & 18 &  
0.66$\times$0.59 & 18328$\pm$5 & 4.5$\pm$0.1 & 314$\pm$13  
& 1.4$\pm$0.1 \\ \hline    
08572$+$3915 & Nuc & 2.8 (13$\sigma$) & 0.21 & 19 & 
0.89$\times$0.47 & 17482$\pm$7 & 9.7$\pm$0.5 & 300$\pm$16  
& 2.9$\pm$0.2 \\     
Superant & Nuc & 6.6 (19$\sigma$) & 0.35 & 48 & 
0.88$\times$0.62 & 18555$\pm$10 & 9.9$\pm$0.3 & 767$\pm$23  
& 7.6$\pm$0.3 \tablenotemark{E} \\     
12112$+$0305 & NE & 4.5 (17$\sigma$) & 0.26 & 25 & 
0.76$\times$0.62 & 21665$\pm$7,21979$\pm$6 &
11.1$\pm$0.7,12.4$\pm$0.8 & 185$\pm$18,208$\pm$20 & 4.6$\pm$0.4 \\     
 & & & & & & 21829$\pm$5 \tablenotemark{C} & 27$\pm$3 \tablenotemark{C}
& 331$\pm$19 \tablenotemark{C} & 9.0$\pm$0.1 \tablenotemark{C} \\     
 & SW & 1.4 (7$\sigma$) & 0.21 \tablenotemark{B} & 18 \tablenotemark{B}
& 0.76$\times$0.62 & 
21888$\pm$17, 22071$\pm$16 & 4.5$\pm$0.5, 4.3$\pm$0.8 & 186$\pm$37,
125$\pm$33 & 1.4$\pm$0.3 \\     
22491$-$1808 & Nuc & 3.5 (14$\sigma$) & 0.25 & 26 & 
0.95$\times$0.60 & 23255$\pm$11 & 9.7$\pm$0.5 & 371$\pm$30  
& 3.6$\pm$0.3 \\     
NGC 1614 & SB1 & 2.0 (7$\sigma$) & 0.28 & 18 & 
1.06$\times$0.57 & 4846$\pm$1 & 27.3$\pm$1.2 & 65$\pm$4  
& 1.9$\pm$0.1 \\     
 & SB2 & 2.0 (7$\sigma$) & 0.28 & 18 & 1.06$\times$0.57
& 4769$\pm$4 & 15.4$\pm$1.2 & 112$\pm$11 & 1.8$\pm$0.2  \\
 & 2$\farcs$5 & --- & --- & --- & --- & 4766$\pm$13 & 57$\pm$4 &
263$\pm$18 & 15.7$\pm$1.5 \\ \hline
12127$-$1412 & Nuc & 0.80 (11$\sigma)$ & 0.075 & 11 &
0.94$\times$0.74 & 39980$\pm$16 & 2.3$\pm$0.2 &
457$\pm$32 & 0.97$\pm$0.09 \\
15250$+$3609 & Nuc & 2.0(21$\sigma$) \tablenotemark{A} & 0.094 
\tablenotemark{A} & 7 \tablenotemark{A} & 1.21$\times$0.72
& 16568$\pm$9 
\tablenotemark{A} & 11$\pm$1 \tablenotemark{A} & 186$\pm$19
\tablenotemark{A} & 2.0$\pm$0.3 \tablenotemark{A} \\
PKS 1345 & Nuc & 3.3(29$\sigma$) & 0.12 & 12 & 0.92$\times$0.85
& 36454$\pm$13 & 8.7$\pm$0.5 & 486$\pm$27 &
4.0$\pm$0.3 \\
06035$-$7102 & Nuc & 5.6(32$\sigma$) & 0.18 & 13 &
1.12$\times$0.78 & 23868$\pm$4 & 17$\pm$1 & 378$\pm$9 &
6.3$\pm$0.2 \\
13509$+$0442 & Nuc & 1.2(18$\sigma$) & 0.063 & 7 & 0.97$\times$0.80
& 40923$\pm$6 & 5.9$\pm$0.3 & 251$\pm$13 & 1.4$\pm$0.1 \\
20414$-$1651 & Nuc & 2.6 (13$\sigma$) & 0.21 & 15 &
0.76$\times$0.67 & 25834$\pm$12,26259$\pm$12 &
5.8$\pm$0.5,6.3$\pm$0.5 & 235$\pm$31,273$\pm$32 & 3.0$\pm$0.3 \\
 & & & & & & 26055$\pm$3 \tablenotemark{D} & 37$\pm$9 \tablenotemark{D}
& 348$\pm$17 \tablenotemark{D} & 13$\pm$3 \tablenotemark{D} \\
\enddata

\tablenotetext{A}{
Only the bright main emission component at 253.4--253.7 GHz in
Figure 5(j) is considered.  
The fainter sub-peak component at 253.0--253.3 GHz, which is likely
to be of outflow origin, is not included.
}

\tablenotetext{B}{The frequency range for significant signal detection is
slightly different between the NE and SW nuclei. 
We created the moment 0 map of the SW nucleus by summing channels with
significant signal detection at the SW nucleus.}

\tablenotetext{C}{
Single Gaussian fit after removing data points
affected by the central dips, based on the assumption of 
self-absorption.  
Data at 21500--21660 km s$^{-1}$ and 21970--22130 km s$^{-1}$ in Figure
9 are used for the fit.}

\tablenotetext{D}{
Single Gaussian fit after removing data points affected by the central
dips, based on the assumption of self-absorption.  
Data at 25200--25760 km s$^{-1}$ and 26340--27000 km s$^{-1}$ in Figure
9 are used for the fit.
}

\tablenotetext{E}{
Because the HCO$^{+}$ J=3--2 emission line is extremely broad, extending
across almost all frequency coverage of the ALMA spectral window, the
ambiguity in continuum determination is large. The possible systematic
uncertainty is expected to be larger than that of other objects.}

\tablecomments{ 
Col.(1): Object.
Col.(2): Position. 
Definitions are the same as those in Table 4.
Col.(3): Integrated intensity of the HCO$^{+}$ J=3--2 emission 
($\nu_{\rm rest}$ = 267.56 GHz) in [Jy beam$^{-1}$ km s$^{-1}$] at the 
emission peak. 
The detection significance relative to the rms noise (1$\sigma$) in the 
moment 0 map is shown in parentheses. 
Possible systematic uncertainty is not included. 
Col.(4): rms noise (1$\sigma$) level in the moment 0 map in 
[Jy beam$^{-1}$ km s$^{-1}$], derived from the standard deviation 
of sky signals in each moment 0 map. 
Col.(5): The number of velocity channels summed to create moment 0 maps.
Each velocity channel has a width of $\sim$20 MHz ($\sim$20 km s$^{-1}$)
for objects observed in ALMA Cycles 1 and 2, or $\sim$40 MHz
($\sim$40 km s$^{-1}$) for those observed in ALMA Cycle 3 (bottom six 
sources).  
Col.(6): Beam size in [arcsec $\times$ arcsec].
Cols.(7)--(10): Gaussian fits of emission lines in the spectra, made in
the same way as Table 4.
Col.(7): Optical LSR velocity (v$_{\rm opt}$) of the emission peak in [km s$^{-1}$]. 
Col.(8): Peak flux in [mJy]. 
Col.(9): Observed FWHM in [km s$^{-1}$] in the right panels of Figure 9.  
Col.(10): Flux in [Jy km s$^{-1}$].}

\end{deluxetable}

\clearpage

\begin{deluxetable}{lll|llcl}
\tabletypesize{\scriptsize}
\tablecaption{Flux of vibrationally excited (v$_{2}$=1f) HCN and 
HCO$^{+}$ J=3--2 emission lines for selected sources \label{tbl-6}}  
\tablewidth{0pt}
\tablehead{
\colhead{Object} & \colhead{Line} & \colhead{$\nu_{\rm rest}$} & 
\multicolumn{4}{c}{Integrated intensity (moment 0) map} \\  
\colhead{} & \colhead{} & \colhead{} & \colhead{Peak} &
\colhead{rms} & \colhead{Channels} & \colhead{Beam} \\ 
\colhead{} & \colhead{} & \colhead{[GHz]} &
\multicolumn{2}{c}{[Jy beam$^{-1}$ km s$^{-1}$]} & 
\colhead{Summed} & \colhead{[$''$ $\times$ $''$] ($^{\circ}$)} \\
\colhead{(1)} & \colhead{(2)} & \colhead{(3)} & \colhead{(4)} & 
\colhead{(5)} & \colhead{(6)} & \colhead{(7)} 
}
\startdata 
NGC 7469 (nucleus) & HCN v$_{2}$=1f J=3--2 & 267.20 & $<$0.10 ($<$3$\sigma$) & 0.032 & 10 & 
0.57$\times$0.52 (78$^{\circ}$) \\
 & HCO$^{+}$ v$_{2}$=1f J=3--2 & 268.69 & $<$0.11 ($<$3$\sigma$) &
0.034 & 10 & 0.56$\times$0.50 (54$^{\circ}$) \\
I Zw 1 & HCN v$_{2}$=1f J=3--2 & 267.20 & $<$0.10 ($<$3$\sigma$) & 0.032 & 10
& 0.66$\times$0.59 (61$^{\circ}$) \\
 & HCO$^{+}$ v$_{2}$=1f J=3--2 & 268.69 & $<$0.11 ($<$3$\sigma$) &
0.035 & 10 & 0.62$\times$0.60 ($-$31$^{\circ}$) \\ 
IRAS 08572$+$3915 & HCN v$_{2}$=1f J=3--2 & 267.20 & 0.28 (2.2$\sigma$)
\{$<$0.38 ($<$3$\sigma$)\} & 0.13 & 15
& 0.89$\times$0.47 ($-$19$^{\circ}$) \\
 & HCO$^{+}$ v$_{2}$=1f J=3--2 & 268.69 & $<$0.21 ($<$3$\sigma$)
& 0.067 & 10 & 0.88$\times$0.46 ($-$19$^{\circ}$) \\
IRAS 12112$+$0305 NE & HCN v$_{2}$=1f J=3--2 & 267.20 & 0.52 (4.5$\sigma$)
& 0.11 & 13 & 0.76$\times$0.62 (68$^{\circ}$) \\
 & HCO$^{+}$ v$_{2}$=1f J=3--2 & 268.69 & $<$0.41 ($<$3$\sigma$)
& 0.13 & 10 & 0.72$\times$0.54 (68$^{\circ}$) \\
IRAS 22491$-$1808 & HCN v$_{2}$=1f J=3--2 & 267.20 & 0.45 (4.0$\sigma$)
& 0.11 & 9 & 0.91$\times$0.60 (67$^{\circ}$) \\
 & HCO$^{+}$ v$_{2}$=1f J=3--2 & 268.69 & $<$0.40 ($<$3$\sigma$)
& 0.13 & 10 & 0.89$\times$0.59 (67$^{\circ}$) \\
IRAS 20414$-$1651 & HCN v$_{2}$=1f J=3--2 & 267.20 & 0.13 (2.0$\sigma$)
\{$<$0.20 ($<$3$\sigma$)\}
& 0.064 & 4 & 0.76$\times$0.67 ($-$88$^{\circ}$) \\
 & HCO$^{+}$ v$_{2}$=1f J=3--2 & 268.69 & $<$0.19 ($<$3$\sigma$)
& 0.063 & 6 & 0.75$\times$0.66 (90$^{\circ}$) \\
\hline 

\enddata

\tablecomments{ 
Col.(1): Object. 
Col.(2): Observed molecular line. 
Col.(3): Rest-frame frequency of each molecular line in [GHz]. 
Col.(4): Integrated intensity in [Jy beam$^{-1}$ km s$^{-1}$] at the 
emission peak. 
Detection significance relative to the rms noise (1$\sigma$) in the 
moment 0 map is shown in parentheses. 
Possible systematic uncertainty is not included. 
Col.(5): rms noise (1$\sigma$) level in the moment 0 map in 
[Jy beam$^{-1}$ km s$^{-1}$], derived from the standard deviation 
of sky signals in each moment 0 map. 
Col.(6): The number of velocity channels summed to create moment 0 maps.
Each velocity channel has a width of $\sim$20 MHz ($\sim$20 km s$^{-1}$)
except for IRAS 20414$-$1651, which has a width of $\sim$40 MHz
($\sim$40 km s$^{-1}$).
Col.(7): Beam size in [arcsec $\times$ arcsec] and position angle in
[degree]. The position angle is 0$^{\circ}$ along the north-south direction, 
and increases counterclockwise.}

\end{deluxetable}

\begin{deluxetable}{lccc}
\tabletypesize{\scriptsize}
\tablecaption{Intrinsic emission size after deconvolution \label{tbl-7}}
\tablewidth{0pt}
\tablehead{
\colhead{Object} & \colhead{HCN J=3--2} & \colhead{HCO$^{+}$ J=3--2} &
\colhead{continuum}  
\\
\colhead{} & \colhead{[mas $\times$ mas] ($^{\circ}$)} & \colhead{[mas
    $\times$ mas] ($^{\circ}$)} & \colhead{[mas $\times$ mas]
  ($^{\circ}$)} \\   
\colhead{(1)} & \colhead{(2)} & \colhead{(3)} & \colhead{(4)} 
}
\startdata
NGC 7469 nucleus & 680$\pm$60, 510$\pm$50 (48$\pm$17) & 810$\pm$100,
570$\pm$80 (46$\pm$18) & 2480$\pm$440, 710$\pm$150 (53$\pm$4) \\
I Zw 1   & 340$\pm$50, 190$\pm$90 (37$\pm$22) & 370$\pm$50, 290$\pm$60
(40$\pm$55) & 650$\pm$60, 340$\pm$50 (53$\pm$7) \\
IC 4329 A & --- & --- & 190$\pm$30, 110$\pm$60 (42$\pm$21) \\ 
IRAS 08572$+$3915 & 490$\pm$150, 330$\pm$110 (145$\pm$67) & 540$\pm$160,
300$\pm$160 (140$\pm$31) & 500$\pm$80, 360$\pm$90 (125$\pm$28) \\
Superantennae & 310$\pm$70, 100$\pm$160 (130$\pm$39) & 520$\pm$80,
130$\pm$60 (143$\pm$13) & 590$\pm$60, 340$\pm$80 (163$\pm$12) \\
IRAS 12112$+$0305 NE & 360$\pm$30, 140$\pm$60 (155$\pm$13) & 370$\pm$90,
150$\pm$140 (128$\pm$27) & 250$\pm$40, 230$\pm$40 (85$\pm$84) \\
IRAS 22491$-$1808 & 210$\pm$100, 150$\pm$80 (10$\pm$64) & could not
deconvolve & $<$430, $<$110 (---) \\
IRAS 12127$-$1412 & $<$370, $<$240 (---) & could not deconvolve &
360$\pm$90, 230$\pm$130 (66$\pm$62) \\
IRAS 15250$+$3609 & 380$\pm$100, 270$\pm$130 (149$\pm$50) & 280$\pm$130,
140$\pm$150 (117$\pm$70) & 270$\pm$100, 240$\pm$170 (179$\pm$78) \\
PKS 1345$+$12 & $<$430, $<$200 (---) & 340$\pm$60, 90$\pm$130
(88$\pm$16) & 150$\pm$30, 100$\pm$70 (81$\pm$83) \\
IRAS 06035$-$7102 &  $<$380, $<$240 (---) & 540$\pm$50, 480$\pm$90
(148$\pm$73) & 460$\pm$110, 380$\pm$200 (103$\pm$66) \\
IRAS 13509$+$0442 & $<$670, $<$270 (---) & 550$\pm$100, 320$\pm$200
(174$\pm$23) & 430$\pm$80, 400$\pm$100 (78$\pm$85) \\
IRAS 13509$+$0442 NE & --- & --- & $<$390, $<$230 (---) \\
IRAS 20414$-$1651 & 470$\pm$60, 300$\pm$110 (178$\pm$59) & 520$\pm$120,
410$\pm$160 (105$\pm$82) & 400$\pm$40, 320$\pm$50 (168$\pm$25) \\
\enddata

\tablecomments{
Col.(1): Object name. 
Cols.(2), (3), and (4): Intrinsic emission size of HCN J=3--2, HCO$^{+}$
J=3--2, and continuum in [mas], respectively, after deconvolution using
the CASA task ``imfit''. 
The position angle in [degree] is shown in parentheses.}

\end{deluxetable}

\begin{deluxetable}{lcccc}
\tabletypesize{\scriptsize}
\tablecaption{Luminosity of the HCN and HCO$^{+}$ v=0 J=3--2 emission lines
  \label{tbl-8}} 
\tablewidth{0pt}
\tablehead{
\colhead{Object} & \colhead{HCN J=3--2} & \colhead{HCO$^{+}$ J=3--2} &
\colhead{HCN J=3--2} & \colhead{HCO$^{+}$ J=3--2} \\
\colhead{} & \colhead{10$^{4}$ [L$_{\odot}$]} & \colhead{10$^{4}$ [L$_{\odot}$]} & 
\colhead{10$^{7}$ [K km s$^{-1}$ pc$^{2}$]} & \colhead{10$^{7}$ [K km s$^{-1}$ pc$^{2}$]} \\       
\colhead{(1)} & \colhead{(2)} & \colhead{(3)} & \colhead{(4)} & \colhead{(5)} 
}
\startdata
NGC 7469 nucleus & 0.74$\pm$0.01 & 0.65$\pm$0.01 & 1.2$\pm$0.1 &
1.1$\pm$0.1 \\
NGC 7469 SB ring & 1.33$\pm$0.01 & 1.92$\pm$0.01 & 2.2$\pm$0.1 &
3.1$\pm$0.1 \\
I Zw 1   & 5.5$\pm$0.2 & 2.7$\pm$0.2 & 9.2$\pm$0.3 & 4.4$\pm$0.3 \\
IRAS 08572$+$3915 & 4.5$\pm$0.2 & 5.0$\pm$0.3 & 7.4$\pm$0.3 & 8.2$\pm$0.6 \\
Superantennae & 17.9$\pm$0.8 & 10.8$\pm$0.6 & 29.7$\pm$1.3 & 17.5$\pm$1.0 \\
IRAS 12112$+$0305 NE & 16.7$\pm$0.2 & 9.5$\pm$0.8 & 27.8$\pm$4.1 &
15.6$\pm$1.4 \\
IRAS 12112$+$0305 SW & 1.8$\pm$0.6 & 3.9$\pm$0.8 & 2.9$\pm$1.0 & 6.3$\pm$1.3 \\
IRAS 22491$-$1808 & 21.1$\pm$0.9 & 11.2$\pm$1.9 & 35.0$\pm$1.5 &
18.3$\pm$3.1 \\
NGC 1614 (2$\farcs$5 radius) & 0.65$\pm$0.22 & 2.0$\pm$0.2 & 1.1$\pm$0.4 & 3.3$\pm$0.3 \\
IRAS 12127$-$1412 & 12.1$\pm$0.9 & 9.1$\pm$0.8 & 20.2$\pm$1.6 &
14.9$\pm$1.4 \\
IRAS 15250$+$3609 & 7.8$\pm$0.5 & 3.1$\pm$0.5 & 13.0$\pm$0.8 & 5.0$\pm$0.8 \\
PKS 1345$+$12 & 21.1$\pm$2.4 & 31.5$\pm$2.4 & 35.1$\pm$3.9 &
51.4$\pm$3.9 \\
IRAS 06035$-$7102 & 15.8$\pm$0.6 & 20.4$\pm$0.6 & 26.2$\pm$1.1 &
33.2$\pm$1.1 \\
IRAS 13509$+$0442 & 10.8$\pm$2.0 & 13.8$\pm$1.0 & 17.9$\pm$3.3 &
22.5$\pm$1.6 \\
IRAS 20414$-$1651 & 16.8$\pm$1.2 & 11.8$\pm$1.2 & 28.0$\pm$2.0 &
19.3$\pm$1.9 \\
\enddata

\tablecomments{
Col.(1): Object name. 
Cols.(2) and (3): Luminosity of the HCN J=3--2 and HCO$^{+}$ J=3--2 emission
lines in units of [L$_{\odot}$], respectively. 
Cols.(4) and (5): Luminosity of the HCN J=3--2 and HCO$^{+}$ J=3--2 emission
lines in units of [K km s$^{-1}$ pc$^{2}$], respectively.}

\end{deluxetable}

\begin{deluxetable}{lcc}
\tabletypesize{\scriptsize}
\tablecaption{Luminosity of the HCN v$_{2}$=1f J=3--2 emission
  line\label{tbl-9}} 
\tablewidth{0pt}
\tablehead{
\colhead{Object} & \colhead{L(HCN v$_{2}$=1f J=3--2)} & \colhead{L'(HCN v$_{2}$=1f J=3--2)} \\
\colhead{} & \colhead{10$^{4}$ [L$_{\odot}$]} & 
\colhead{10$^{7}$ [K km s$^{-1}$ pc$^{2}$]} \\
\colhead{(1)} & \colhead{(2)} & \colhead{(3)} 
}
\startdata
IRAS 12112$+$0305 NE & 1.4 & 2.3 \\
IRAS 22491$-$1808 & 1.4 & 2.3 \\
\enddata

\tablecomments{
Col.(1): Object name. 
Col.(2): Luminosity of the HCN J=3--2 v$_{2}$=1f emission line in units of
[L$_{\odot}$]. 
Col.(3): Luminosity of the HCN J=3--2 v$_{2}$=1f emission line in units of
[K km s$^{-1}$ pc$^{2}$].}

\end{deluxetable}

\begin{deluxetable}{lcc}
\tabletypesize{\scriptsize}
\tablecaption{Radio properties of NGC 7469, I Zw 1, and IC 4329 A
  \label{tbl-1}} 
\tablewidth{0pt}
\tablehead{
\colhead{Object} & \colhead{S$_{\rm 1.4 GHz}$} & \colhead{q} 
\\
\colhead{} & \colhead{[mJy]} & \colhead{} \\  
\colhead{(1)} & \colhead{(2)} & \colhead{(3)} 
}
\startdata
NGC 7469 & 181 & 2.3 \\
I Zw 1   & 8.8 & 2.5 \\
IC 4329 A & 66.8 & 1.5 \\ \hline
\enddata

\tablecomments{
Col.(1): Object name. 
Col.(2): Radio 1.4 GHz flux in [mJy] \citep{con98}. 
Col.(3): Far-infrared-to-radio flux ratio, parameterized as the q-value
\citep{con91}.}

\end{deluxetable}

\clearpage

\begin{figure}
\begin{center}
\includegraphics[angle=0,scale=.41]{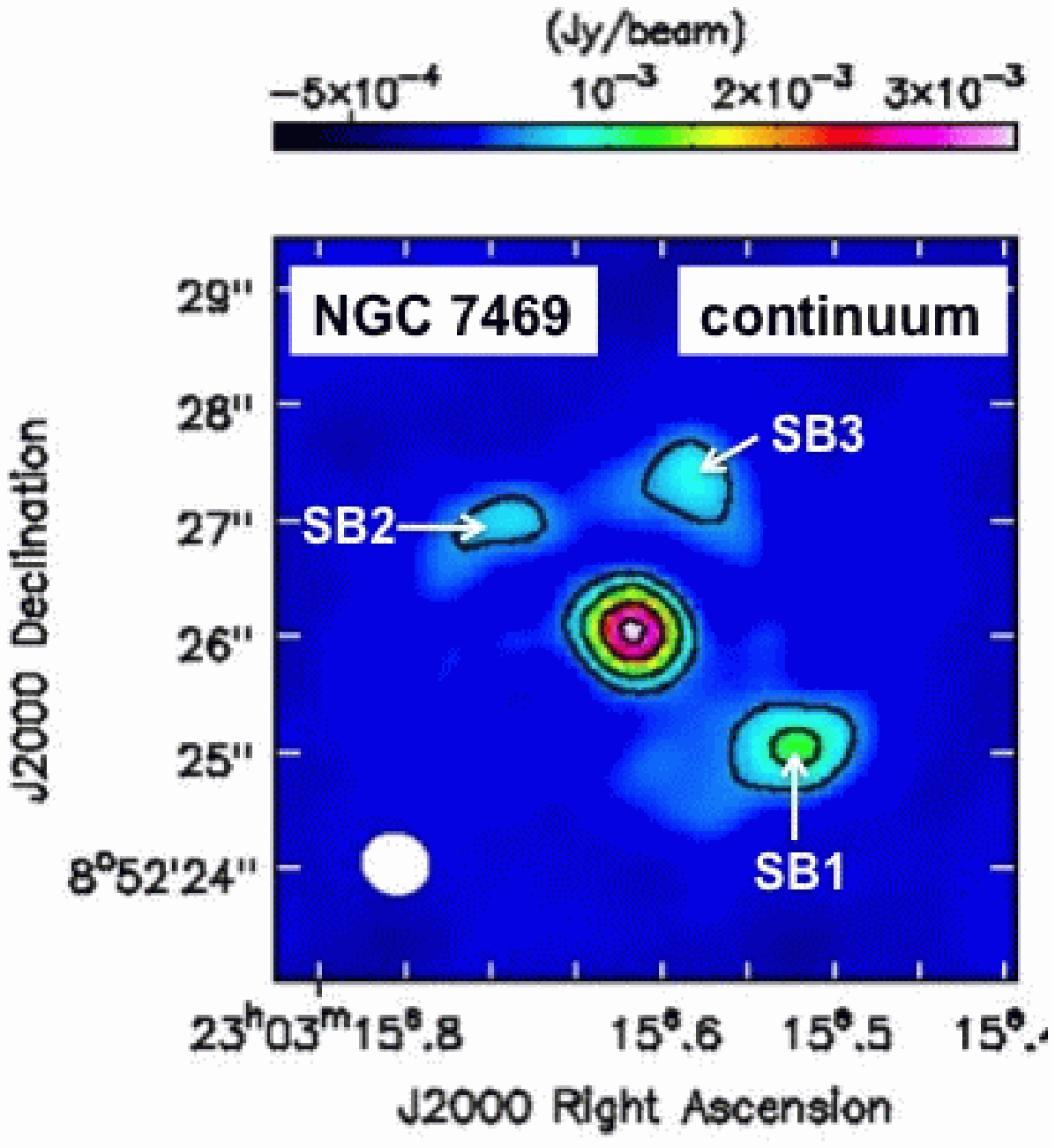} 
\includegraphics[angle=0,scale=.41]{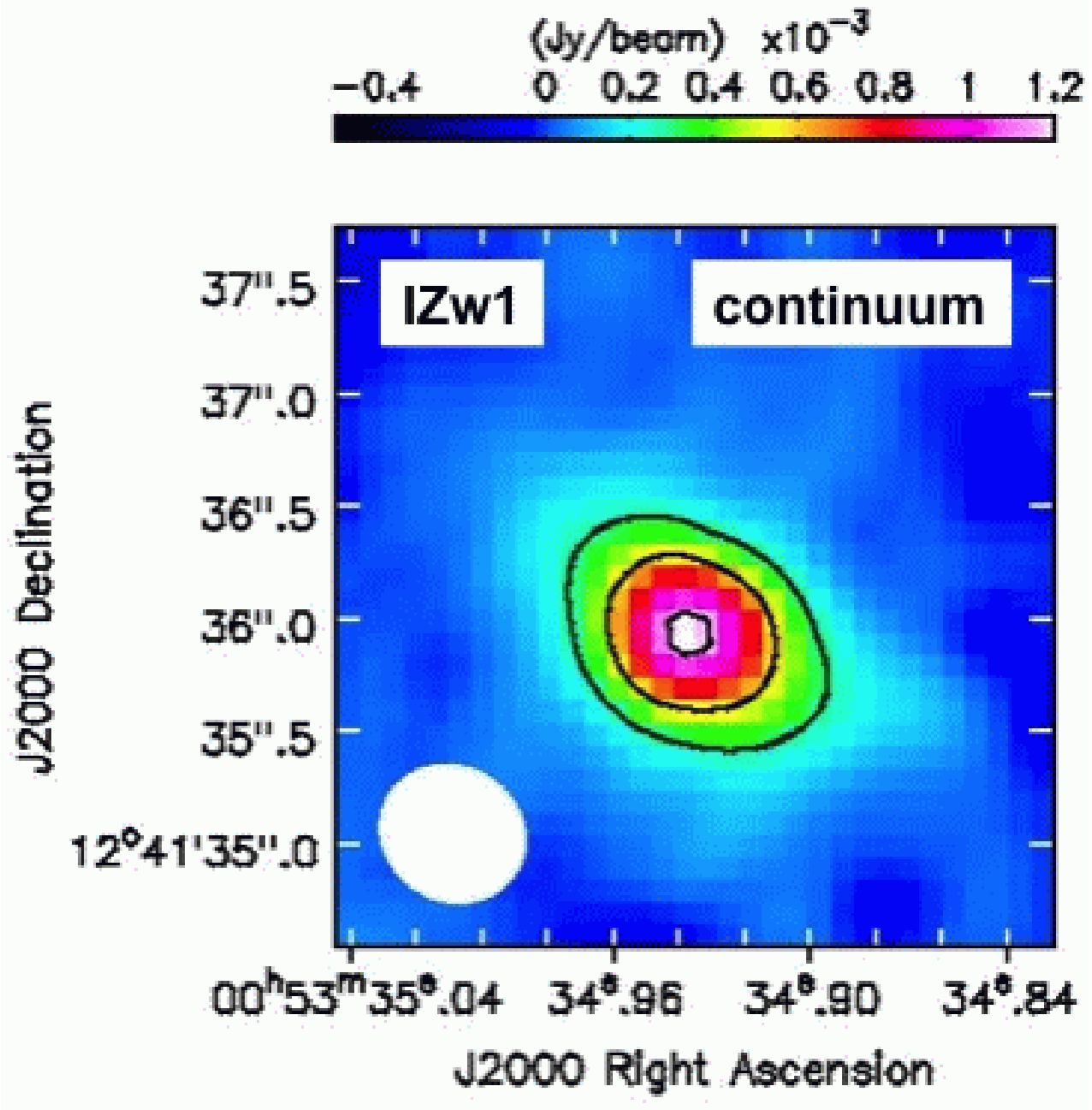} 
\includegraphics[angle=0,scale=.41]{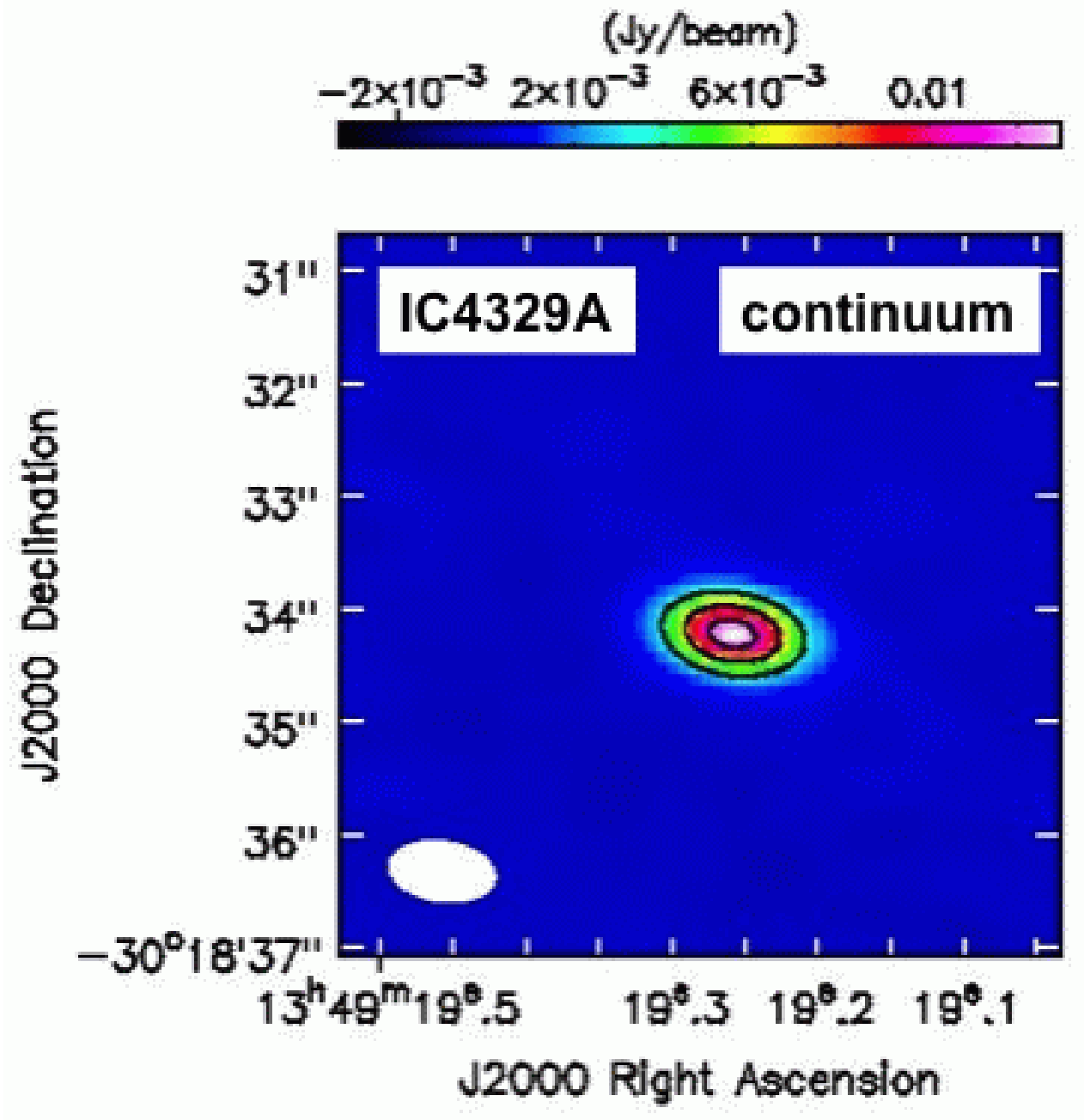} \\
\end{center}
\caption{Continuum maps of NGC 7469, I Zw 1, and IC 4329 A.
For NGC 7469, 5$\sigma$, 10$\sigma$, 20$\sigma$, and 30$\sigma$
contours, as well as the locations of the starburst knots SB1, SB2, and
SB3, are shown. 
The plotted contours are 5$\sigma$, 10$\sigma$, 20$\sigma$ 
for I Zw 1, and 20$\sigma$, 40$\sigma$, 60$\sigma$ for IC 4329 A.}
\end{figure}

\begin{figure}
\begin{center}
\includegraphics[angle=0,scale=.41]{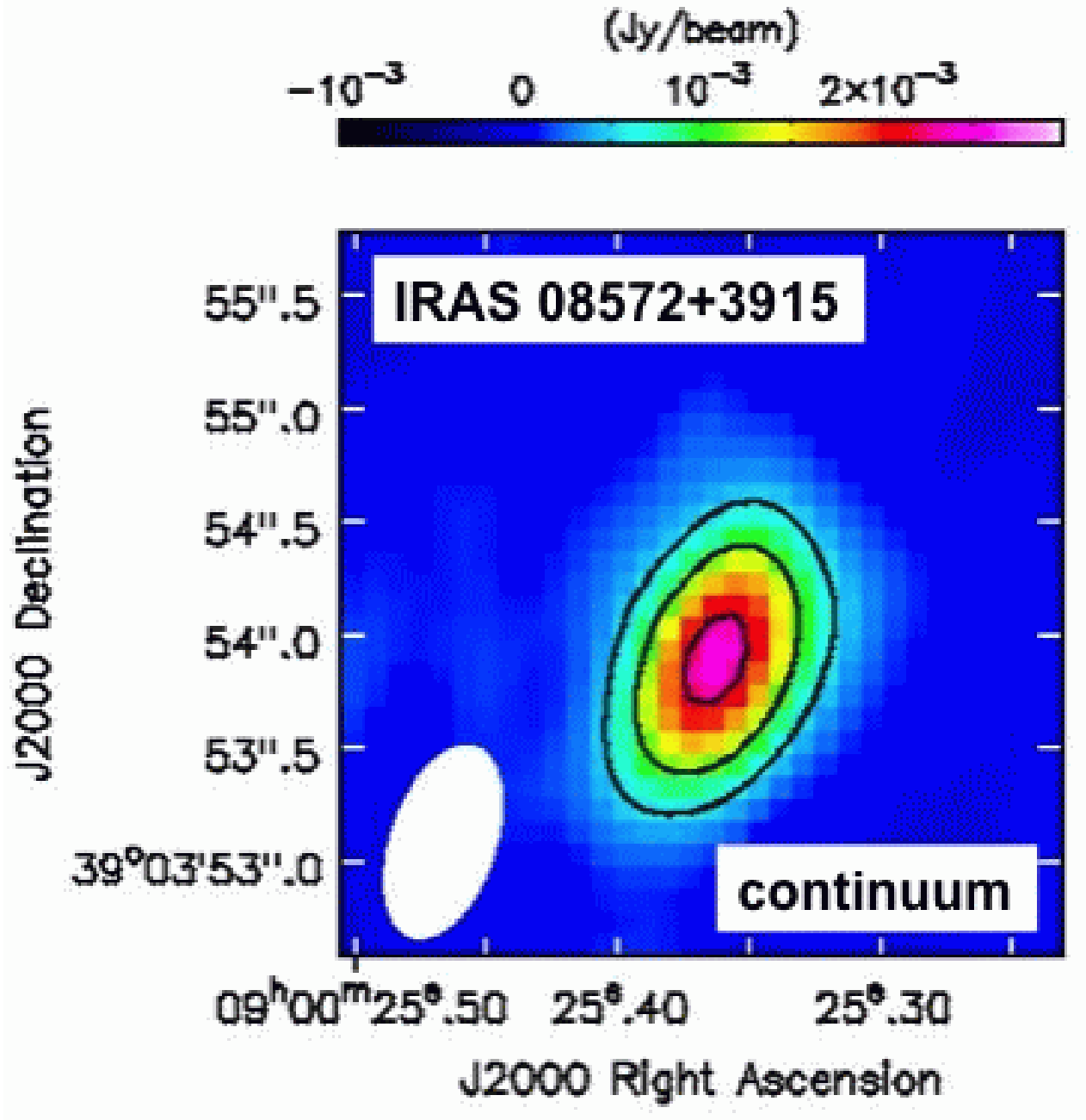} 
\includegraphics[angle=0,scale=.41]{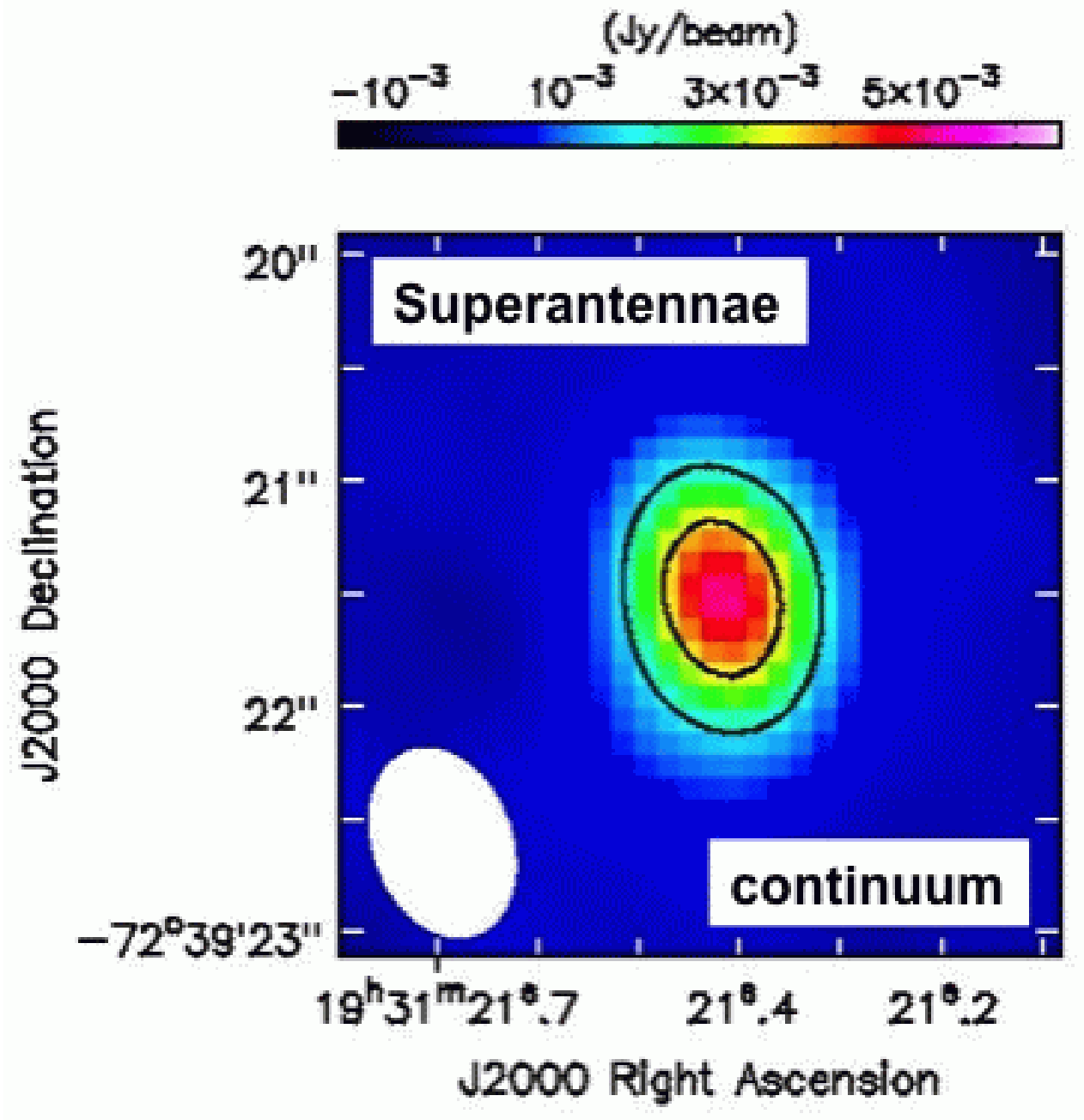} 
\includegraphics[angle=0,scale=.41]{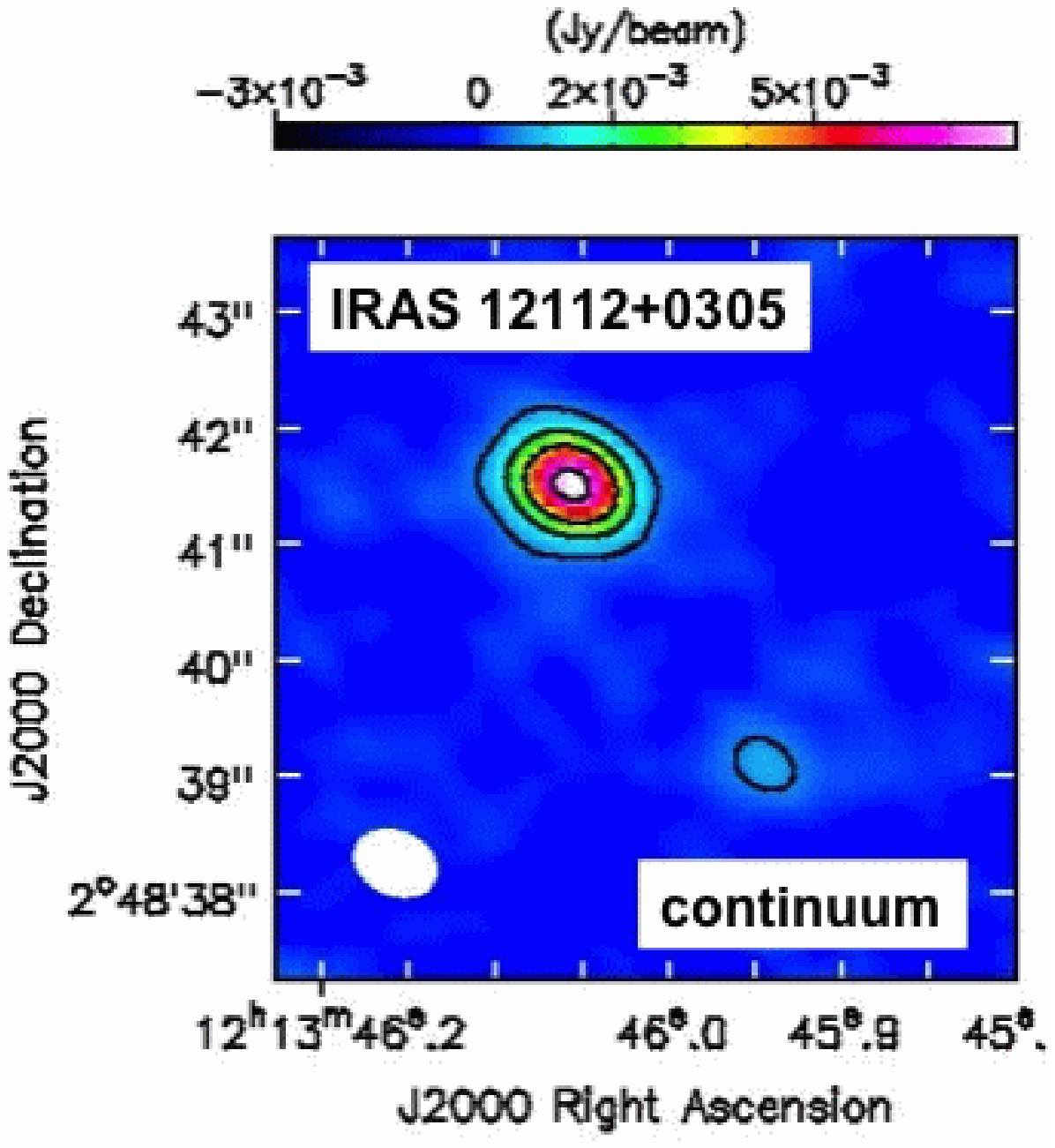}\\
\includegraphics[angle=0,scale=.41]{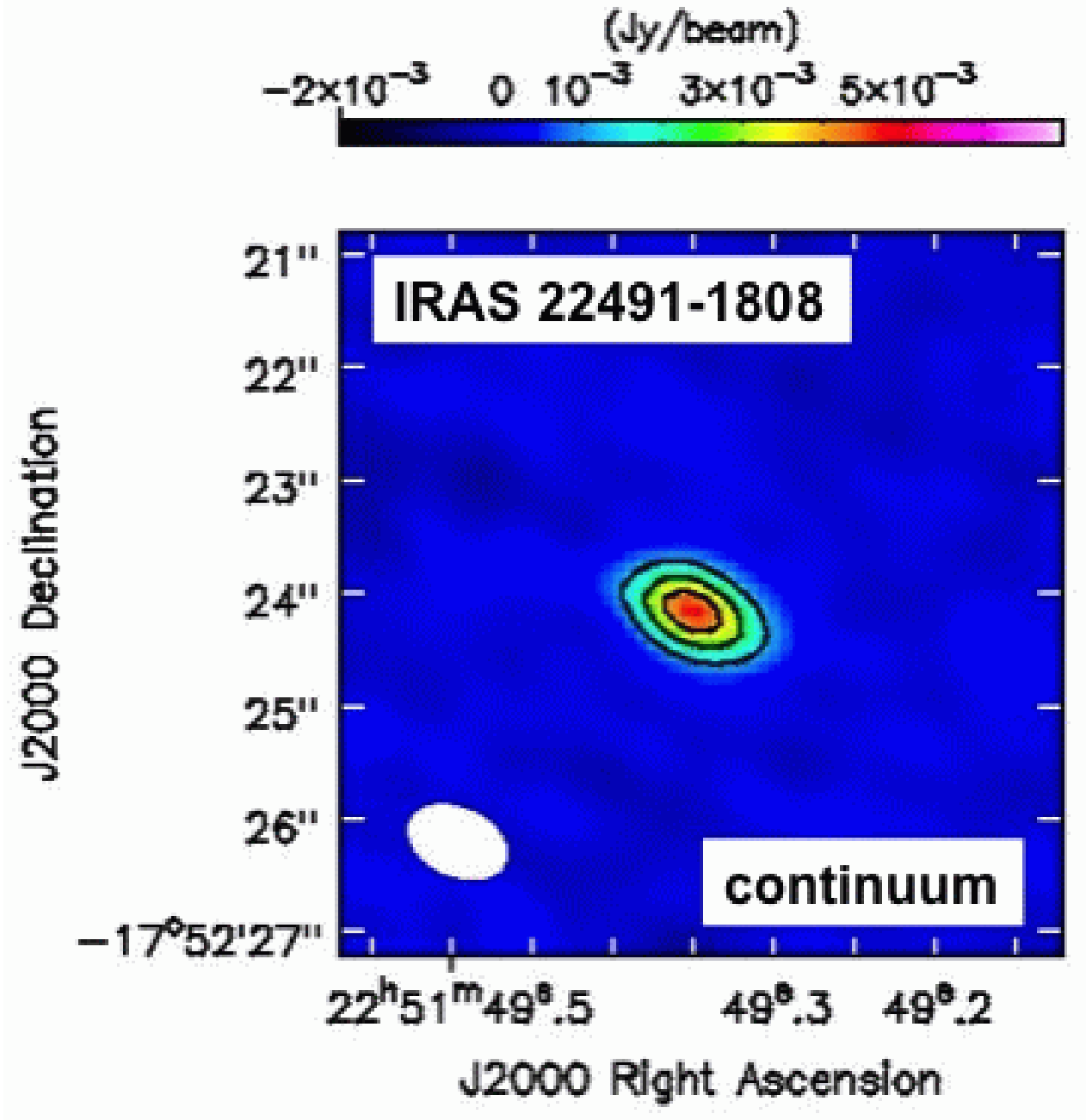} 
\includegraphics[angle=0,scale=.41]{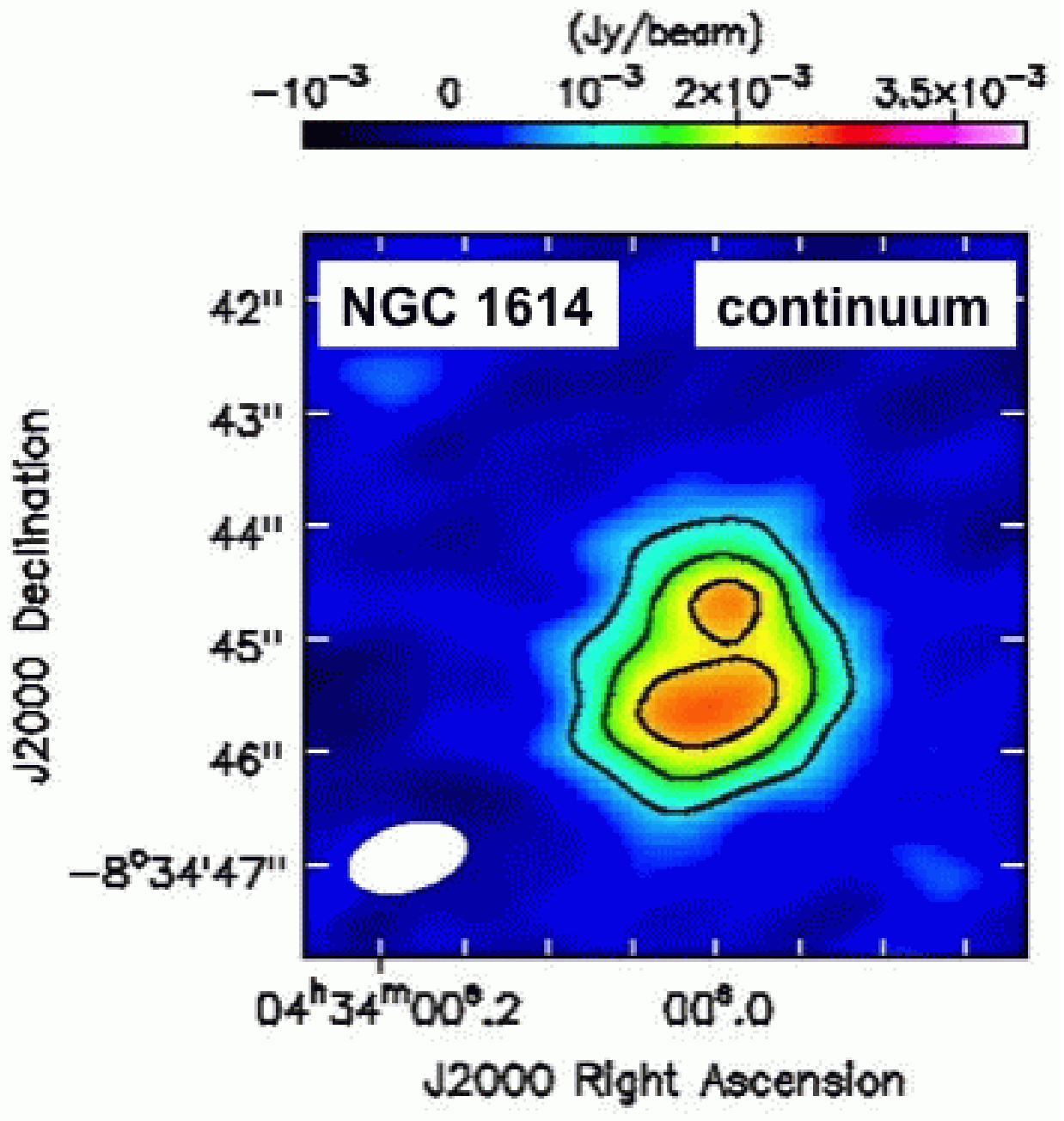} \\
\end{center}
\caption{Continuum maps of LIRGs observed in ALMA Cycle 2. 
The displayed apparent scale depends on the structure of the individual
objects and the beam size of the individual data. 
The contours are 5$\sigma$, 10$\sigma$, 20$\sigma$ for IRAS 08572$+$3915, 
10$\sigma$, 20$\sigma$ for The Superantennae, 
3$\sigma$, 10$\sigma$, 20$\sigma$, 40$\sigma$ for IRAS 12112$+$0305, 
5$\sigma$, 10$\sigma$, 15$\sigma$ for IRAS 22491$-$1808, and
3$\sigma$, 5$\sigma$, 7$\sigma$ for NGC 1614.}
\end{figure}

\begin{figure}
\begin{center}
\includegraphics[angle=0,scale=.39]{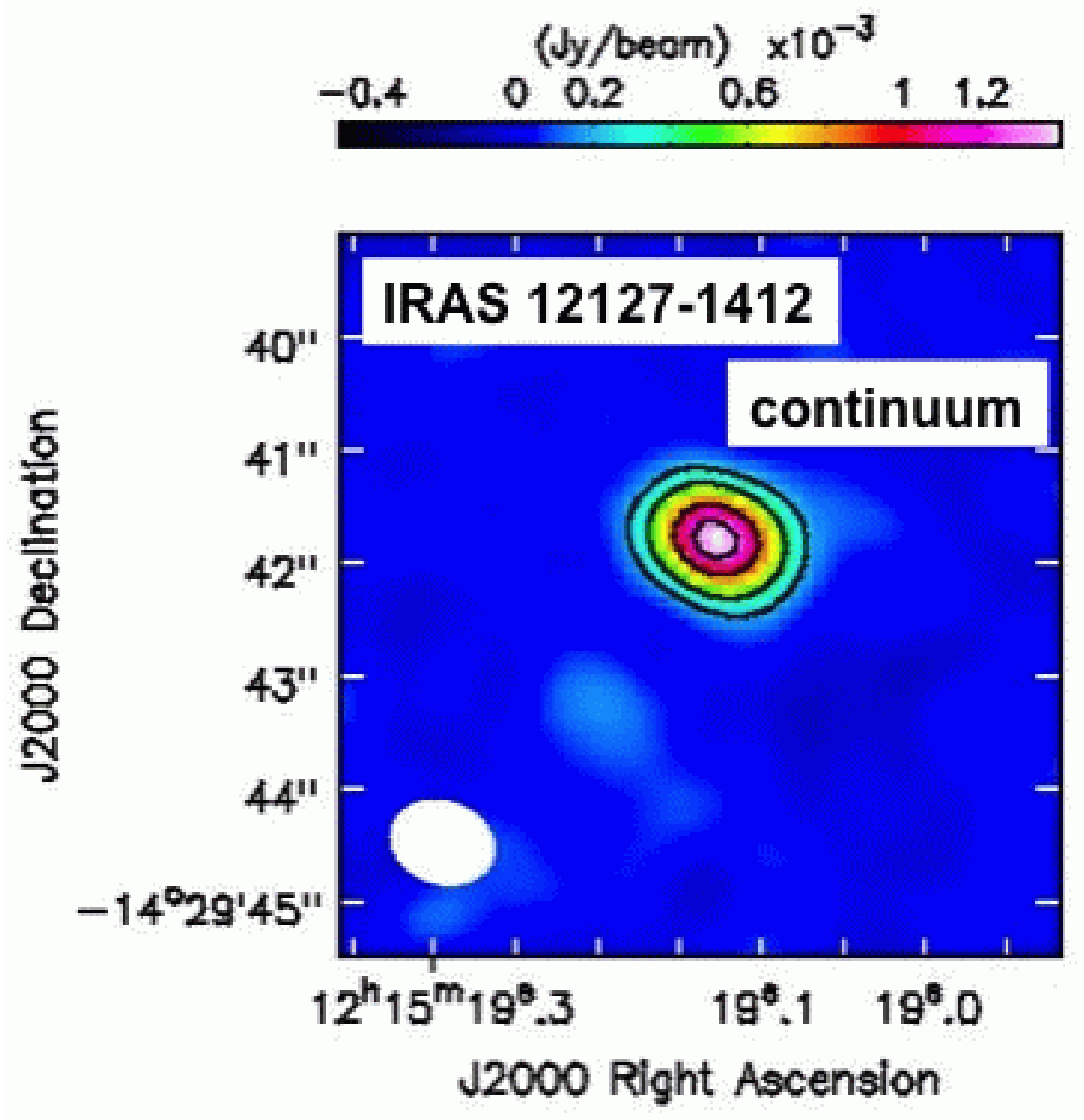} 
\includegraphics[angle=0,scale=.39]{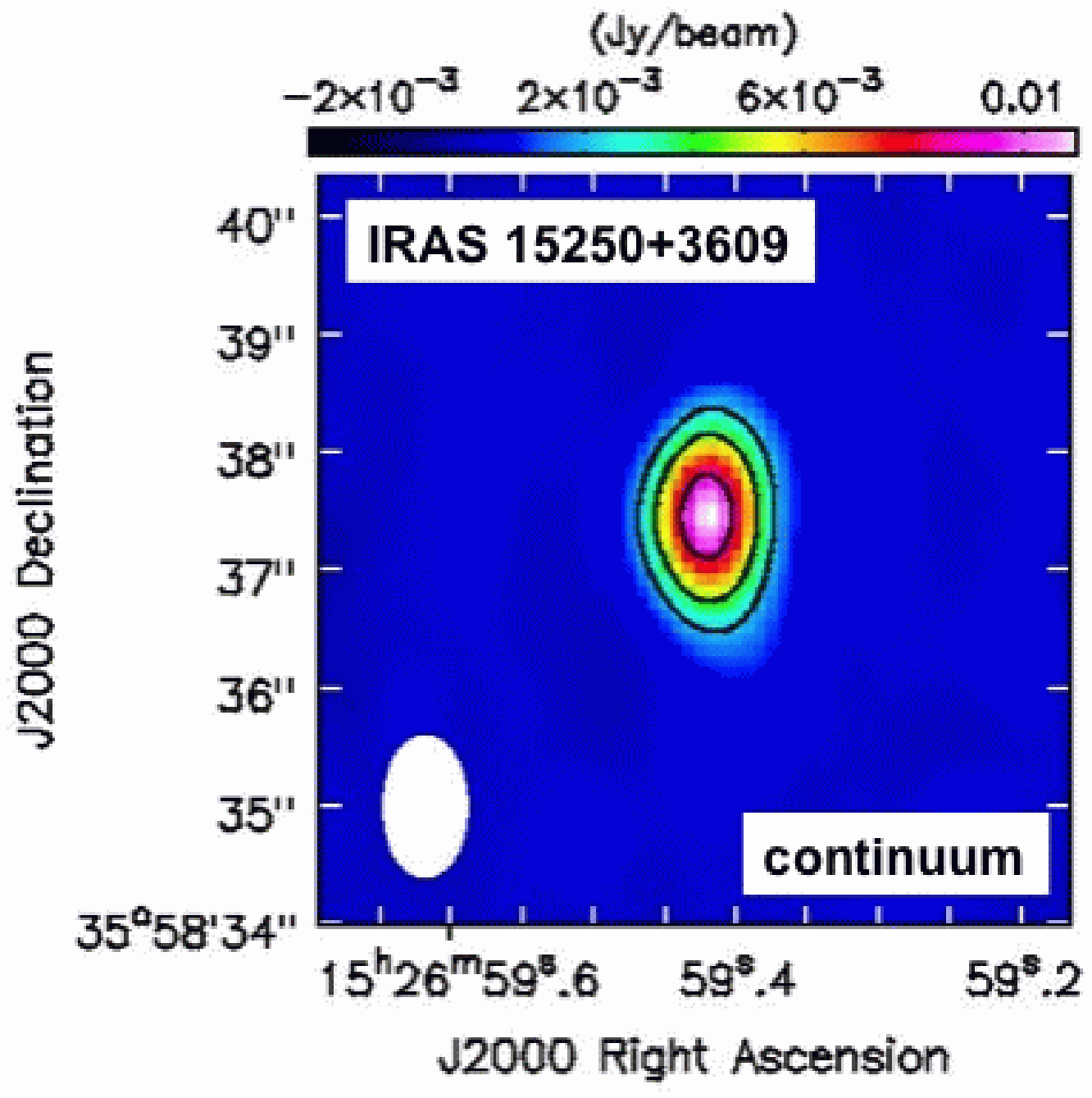} 
\includegraphics[angle=0,scale=.39]{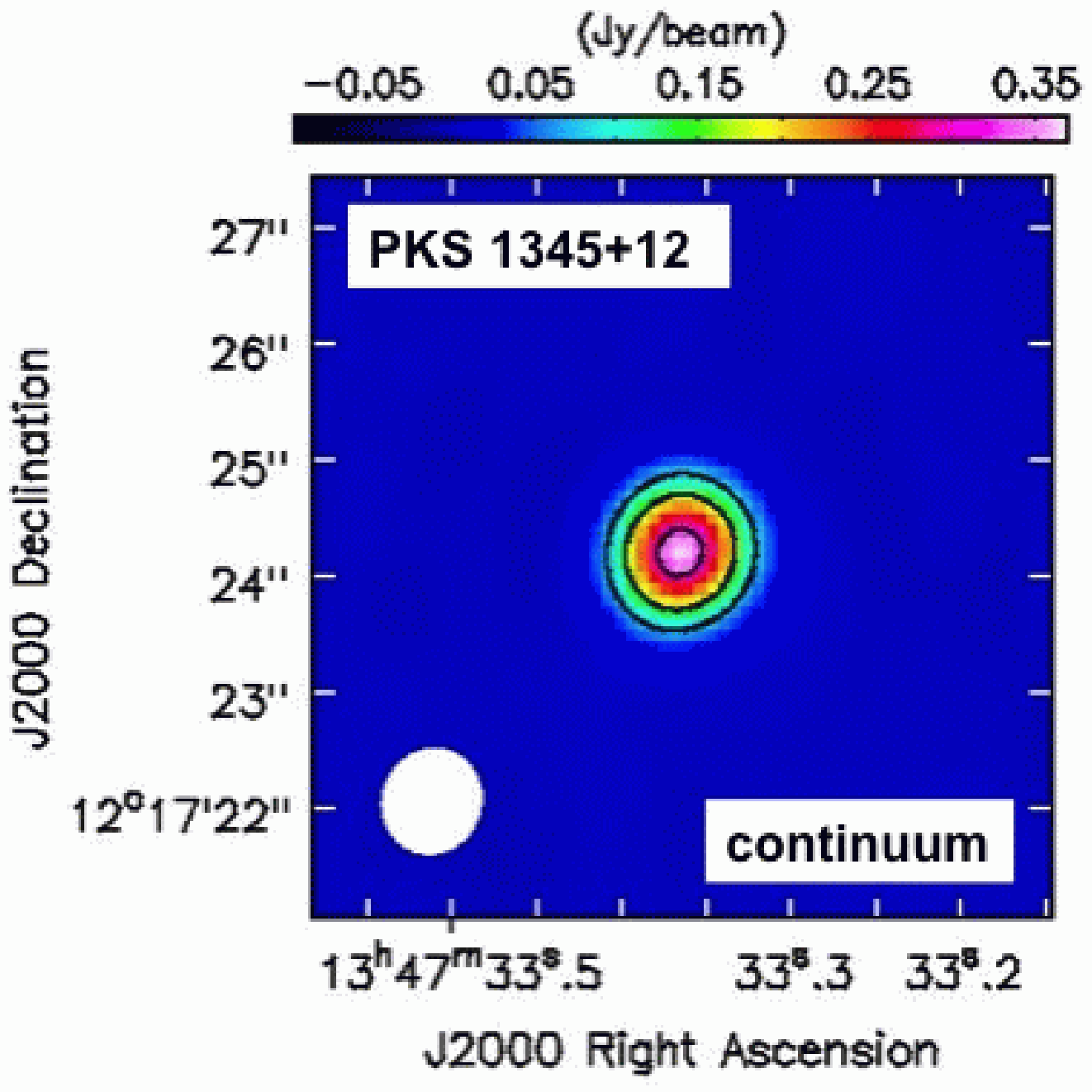} \\
\includegraphics[angle=0,scale=.39]{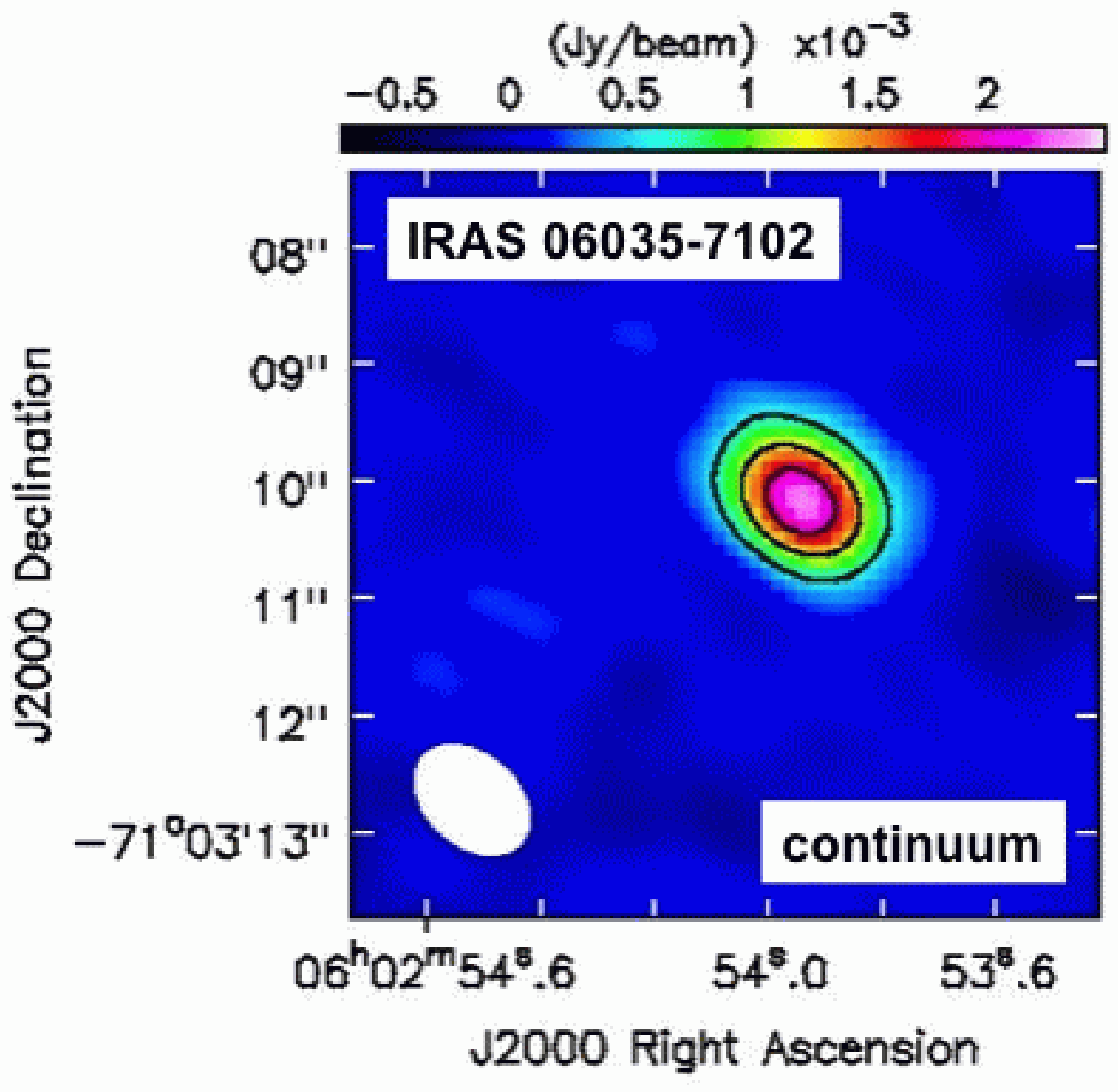} 
\includegraphics[angle=0,scale=.39]{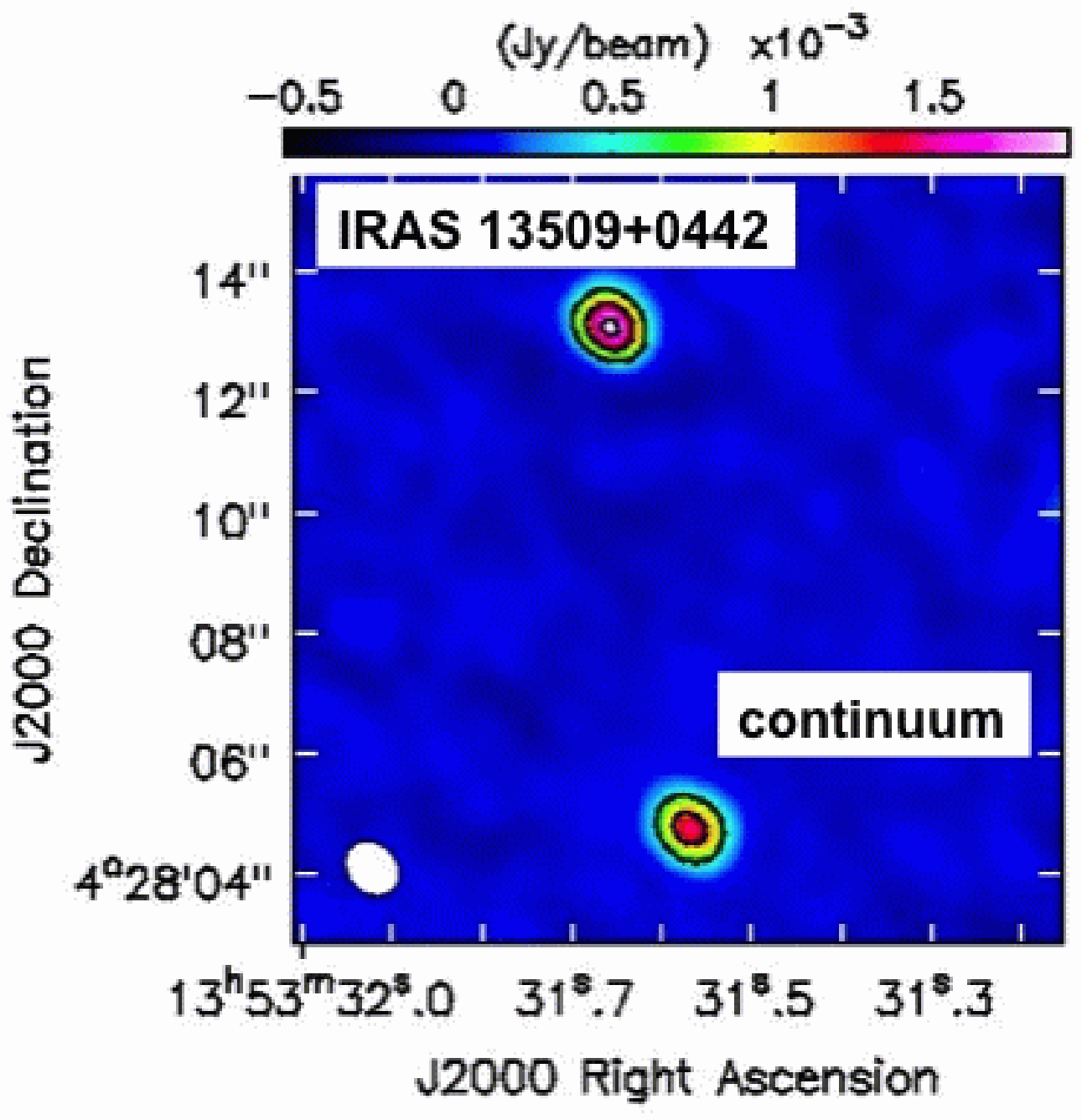} 
\includegraphics[angle=0,scale=.39]{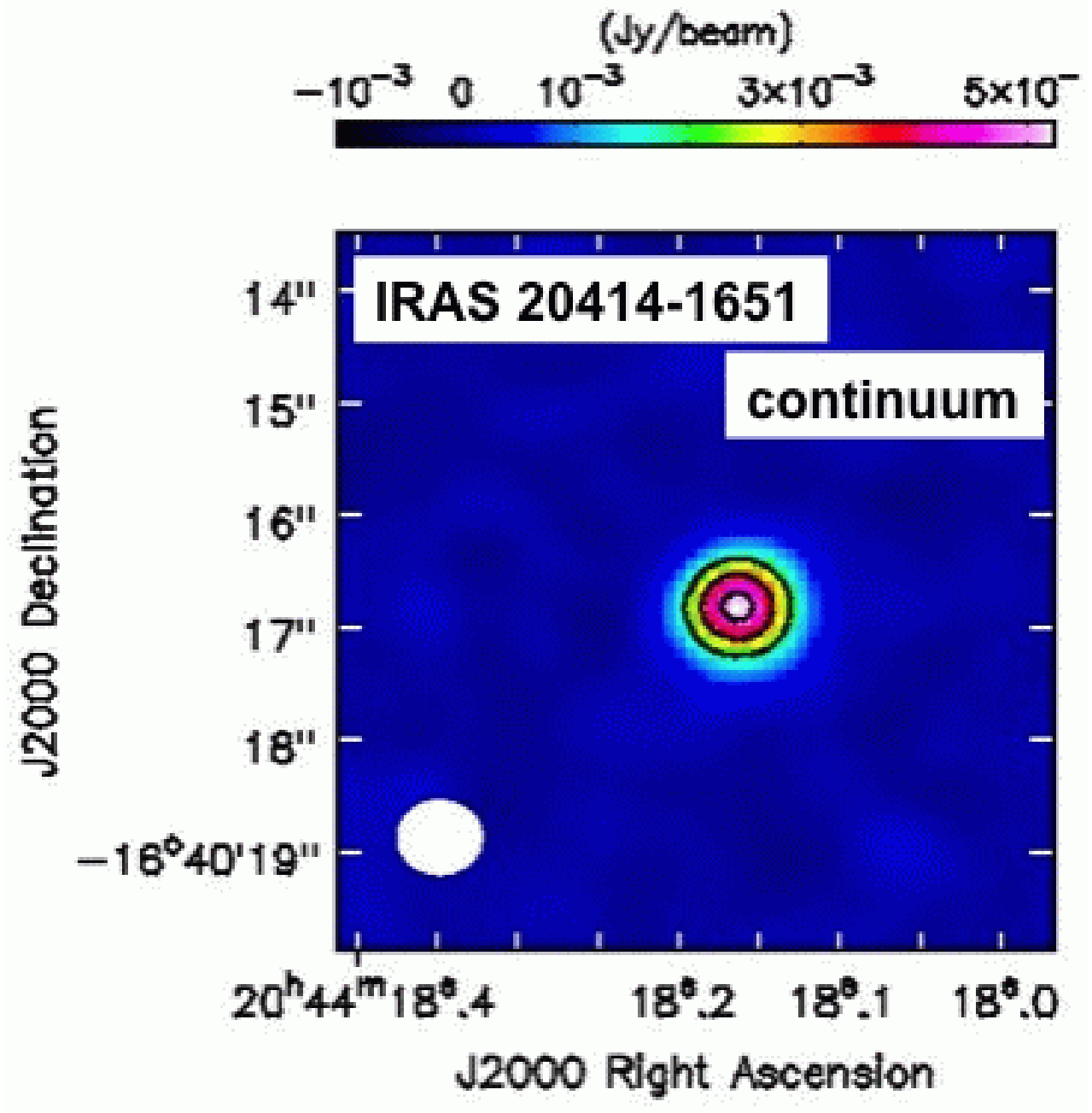} \\
\end{center}
\caption{Continuum maps of LIRGs observed in ALMA Cycle 3. 
The displayed apparent scale depends on the structure of the individual
objects and the beam size of the individual data. 
The contours are 
4$\sigma$, 8$\sigma$, 16$\sigma$, 22$\sigma$ for IRAS 12127$-$1412,
8$\sigma$, 16$\sigma$, 32$\sigma$ for IRAS 15250$+$3609,
15$\sigma$, 30$\sigma$, 60$\sigma$ for PKS 1345$+$12,
5$\sigma$, 10$\sigma$, 15$\sigma$ for IRAS 06035$-$7102, 
8$\sigma$, 16$\sigma$, 24$\sigma$ for IRAS 13509$+$0442, and 
15$\sigma$, 25$\sigma$, 35$\sigma$ for IRAS 20414$-$1651. 
IRAS 13509$+$0442 is the galaxy seen in the lower part of the image, and
the continuum emission in the upper part is from another optically faint
source.}
\end{figure}

\begin{figure}
\begin{center}
\includegraphics[angle=0,scale=.35]{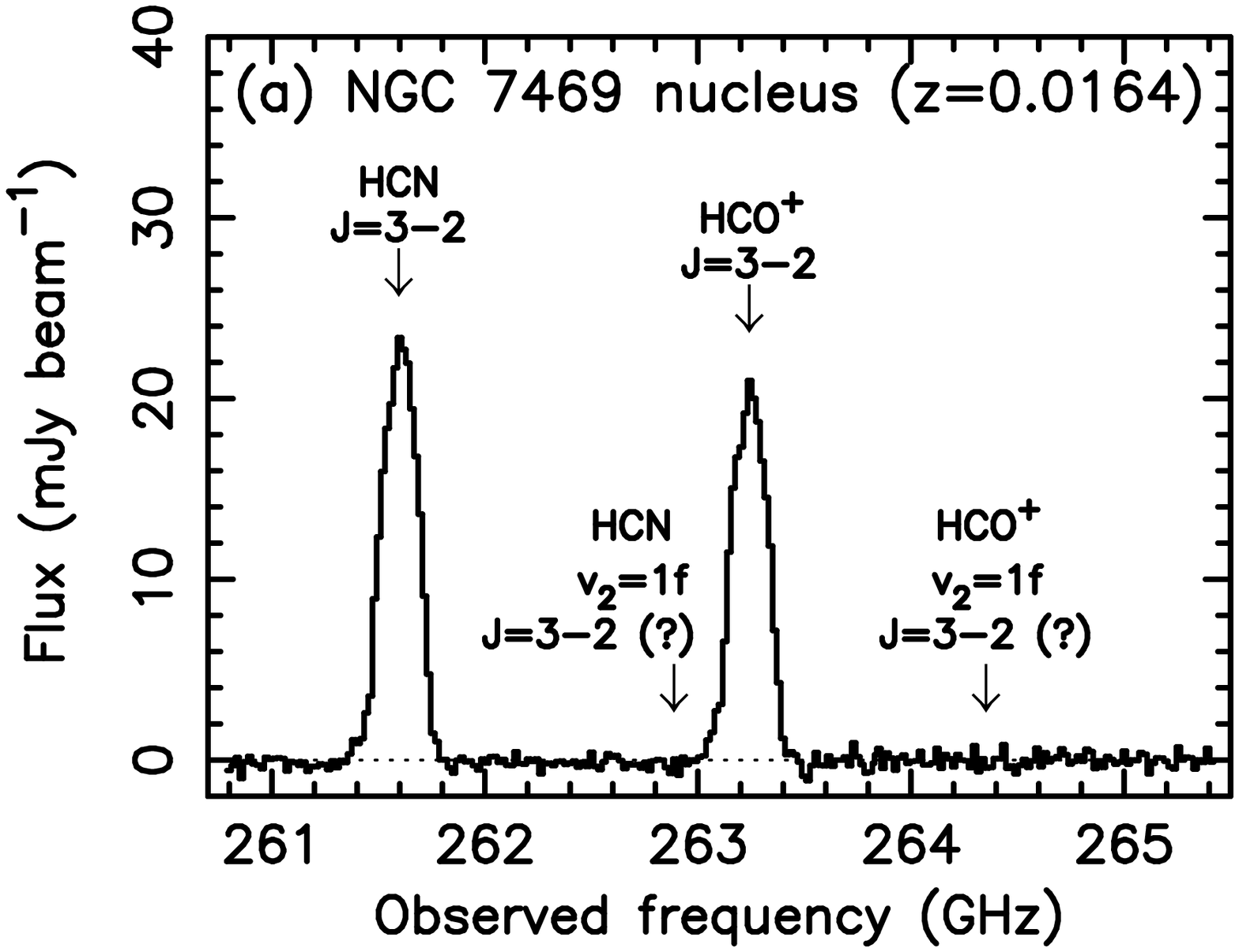} 
\includegraphics[angle=0,scale=.35]{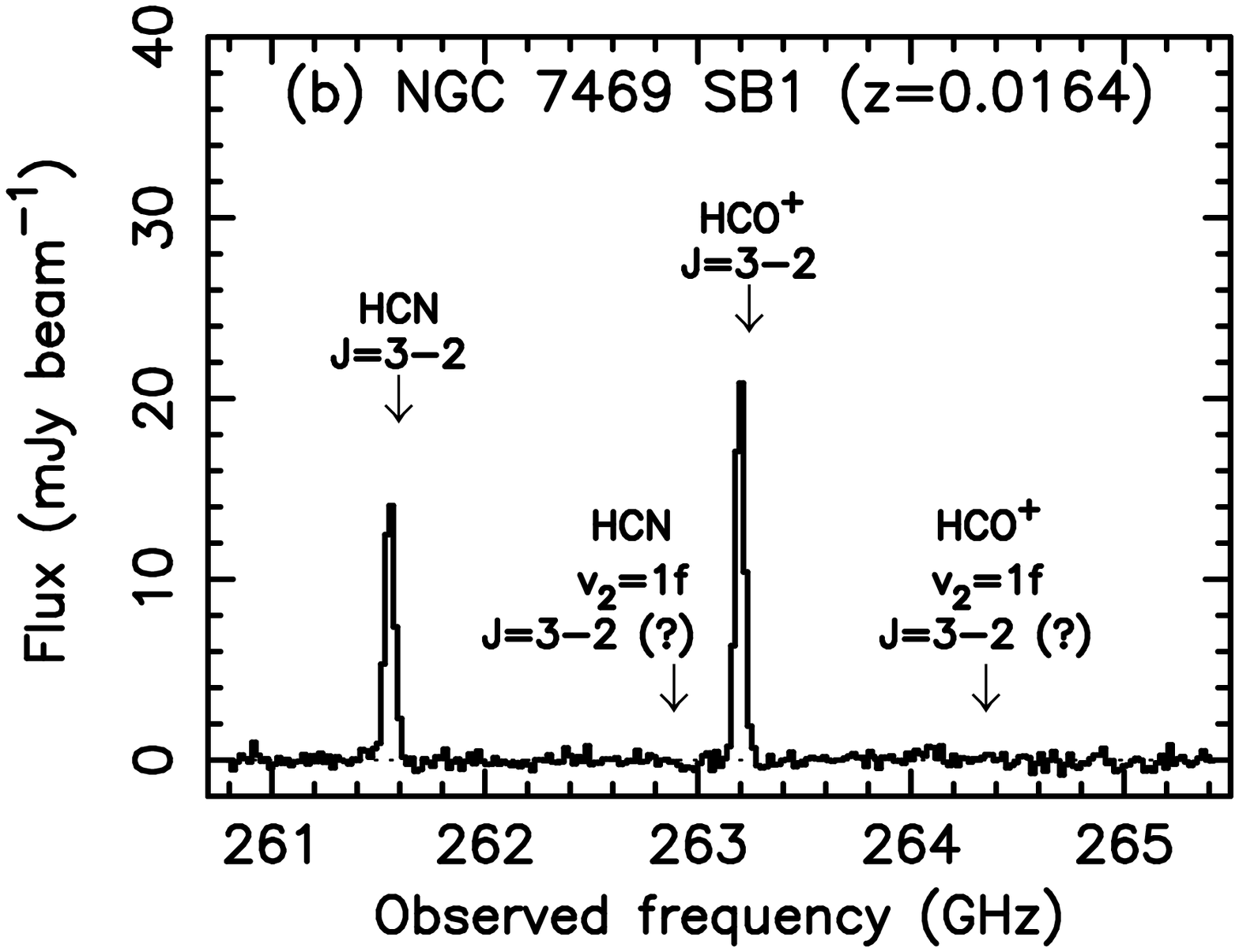} \\
\vspace{-0.4cm}
\includegraphics[angle=0,scale=.35]{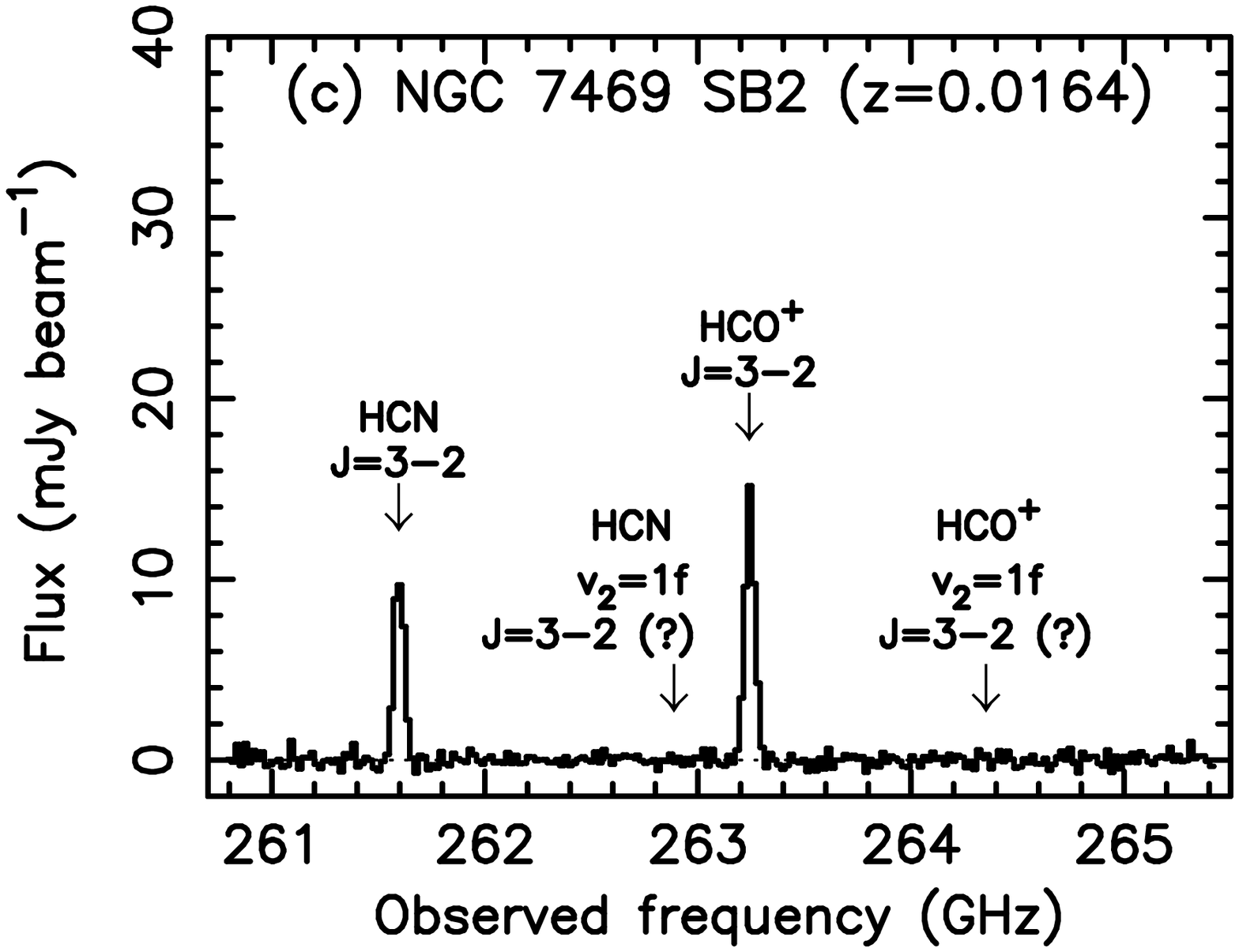} 
\includegraphics[angle=0,scale=.35]{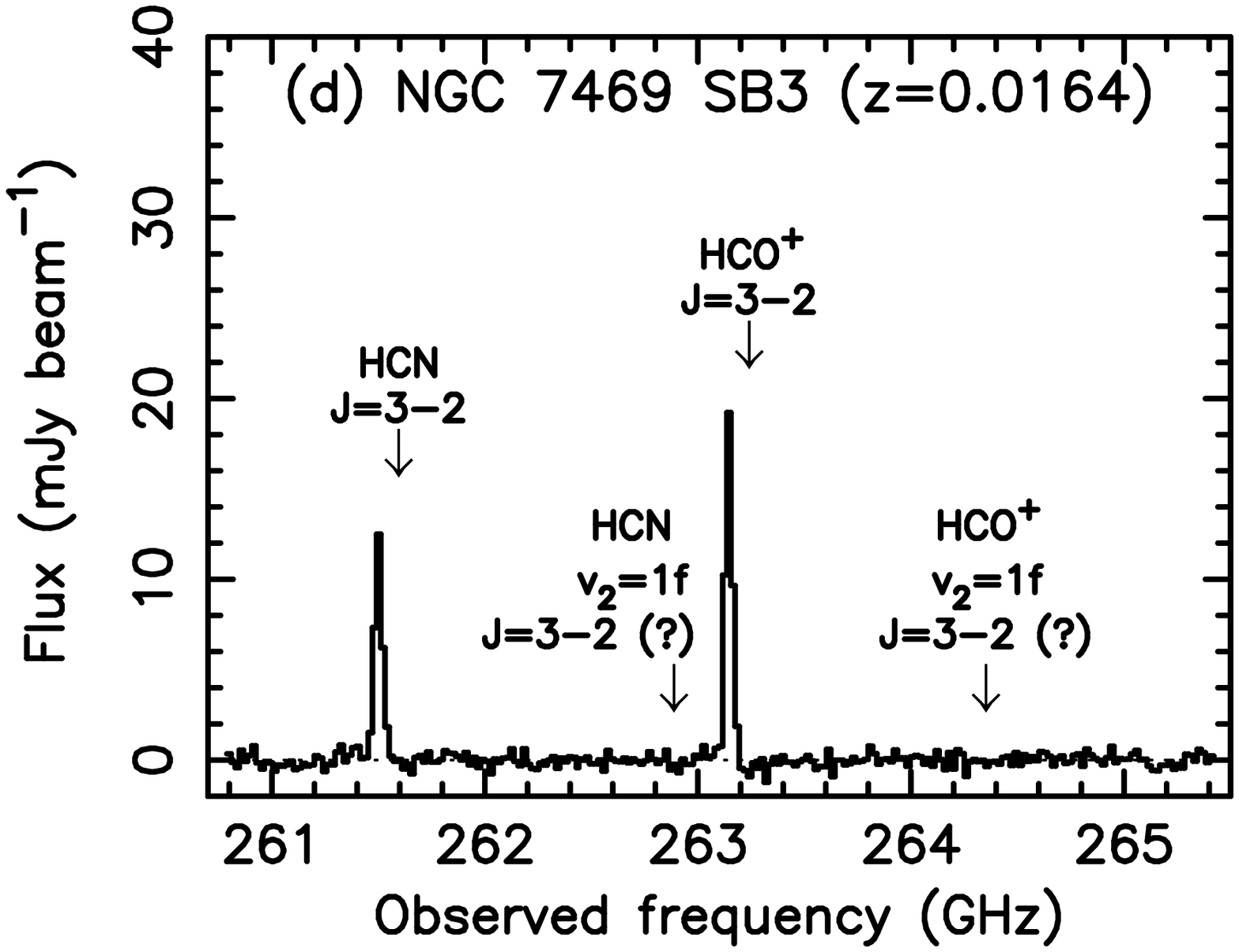} \\
\vspace{-0.4cm}
\includegraphics[angle=0,scale=.35]{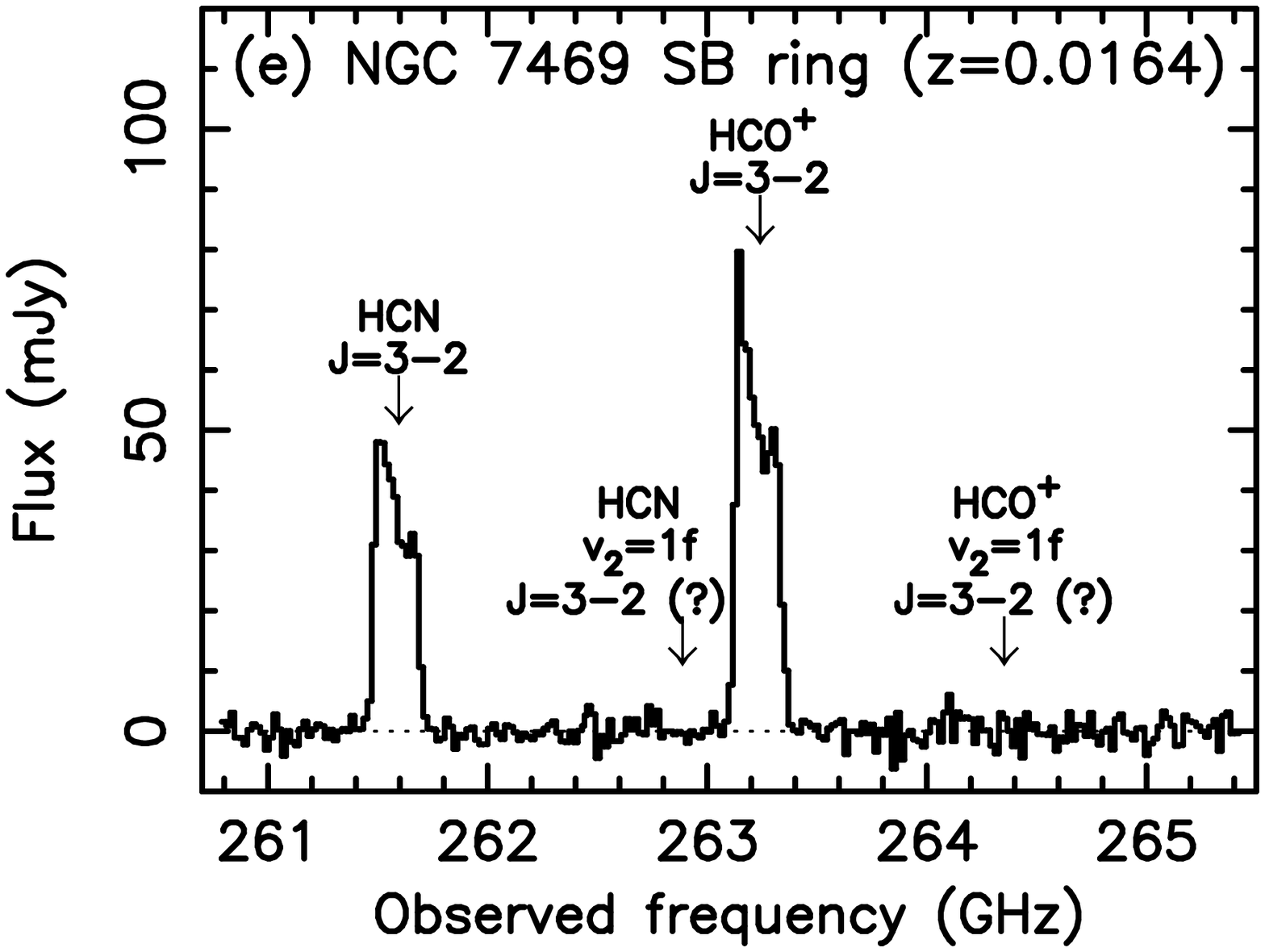}
\includegraphics[angle=0,scale=.35]{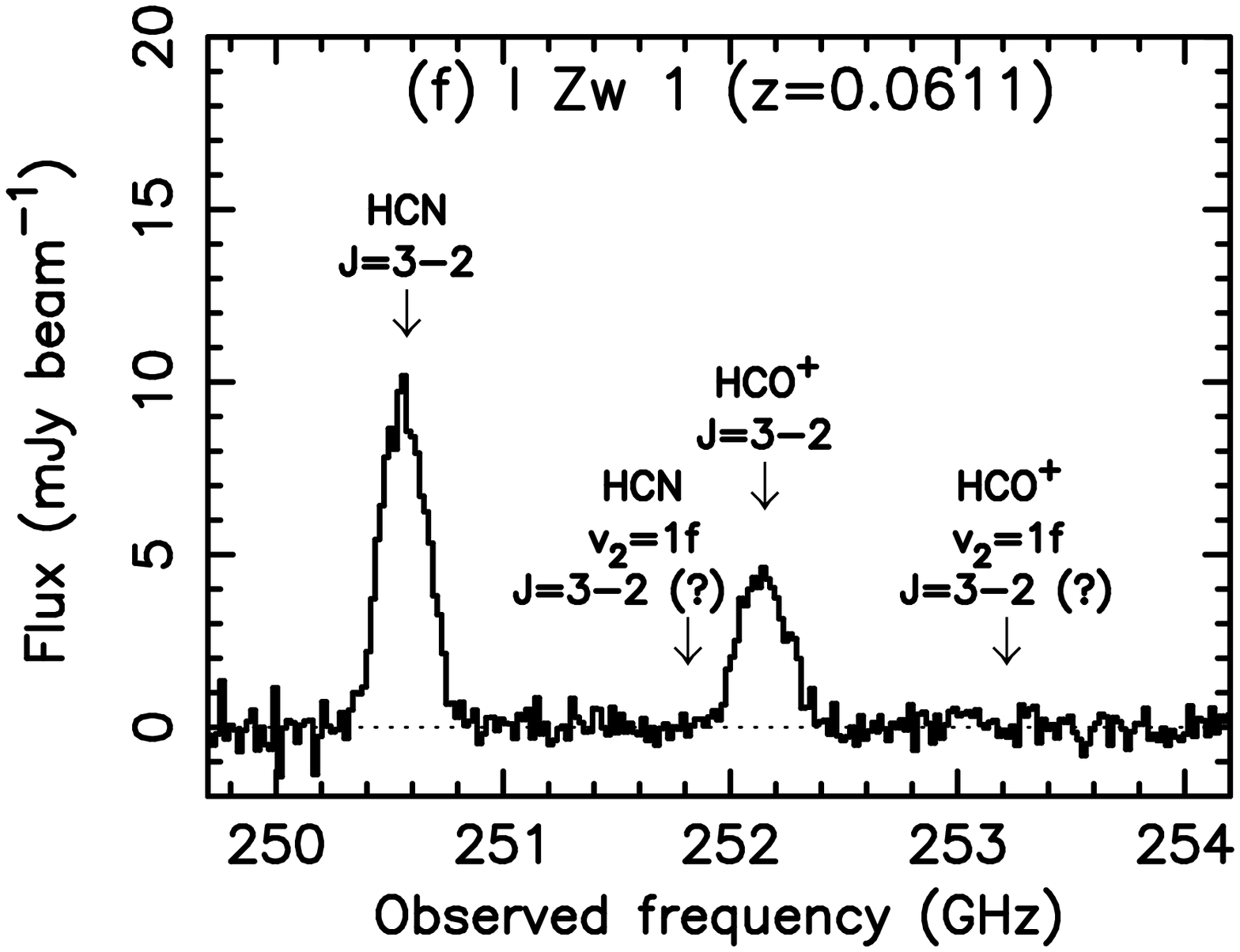} \\
\vspace{-0.4cm}
%
\includegraphics[angle=0,scale=.35]{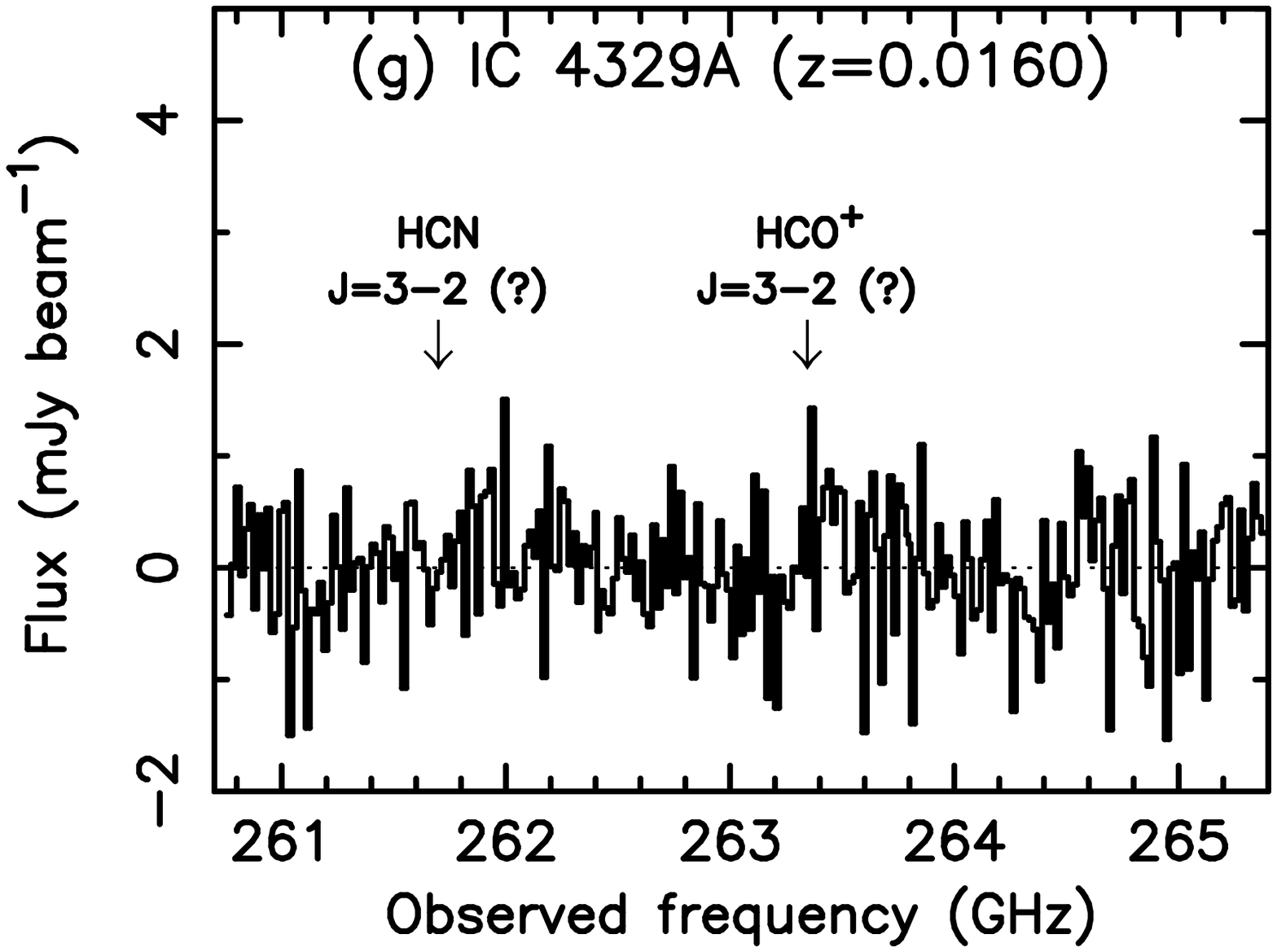} \\
\end{center}
\vspace{-0.4cm}
\caption{Full frequency coverage spectra at interesting regions 
of optical Seyfert 1 galaxies.
Spectra are taken within the beam size, except for (e).
(a), (b), (c), and (d) are spectra at the NGC 7469 nucleus defined as the
continuum emission peak, SB1, SB2, and SB3, respectively.
(e) is the area-integrated spectrum of the annular region at
0$\farcs$8--2$\farcs$5 radius from the nucleus of NGC 7469.
(f) and (g) are spectra of the nuclei of I Zw 1 and IC 4329 A (defined
as the continuum emission peaks), respectively.
Down arrows are shown at the expected observed frequency of some emission
lines for the adopted redshifts (Table 1).}
\end{figure}

\begin{figure}
\begin{center}
\includegraphics[angle=0,scale=.35]{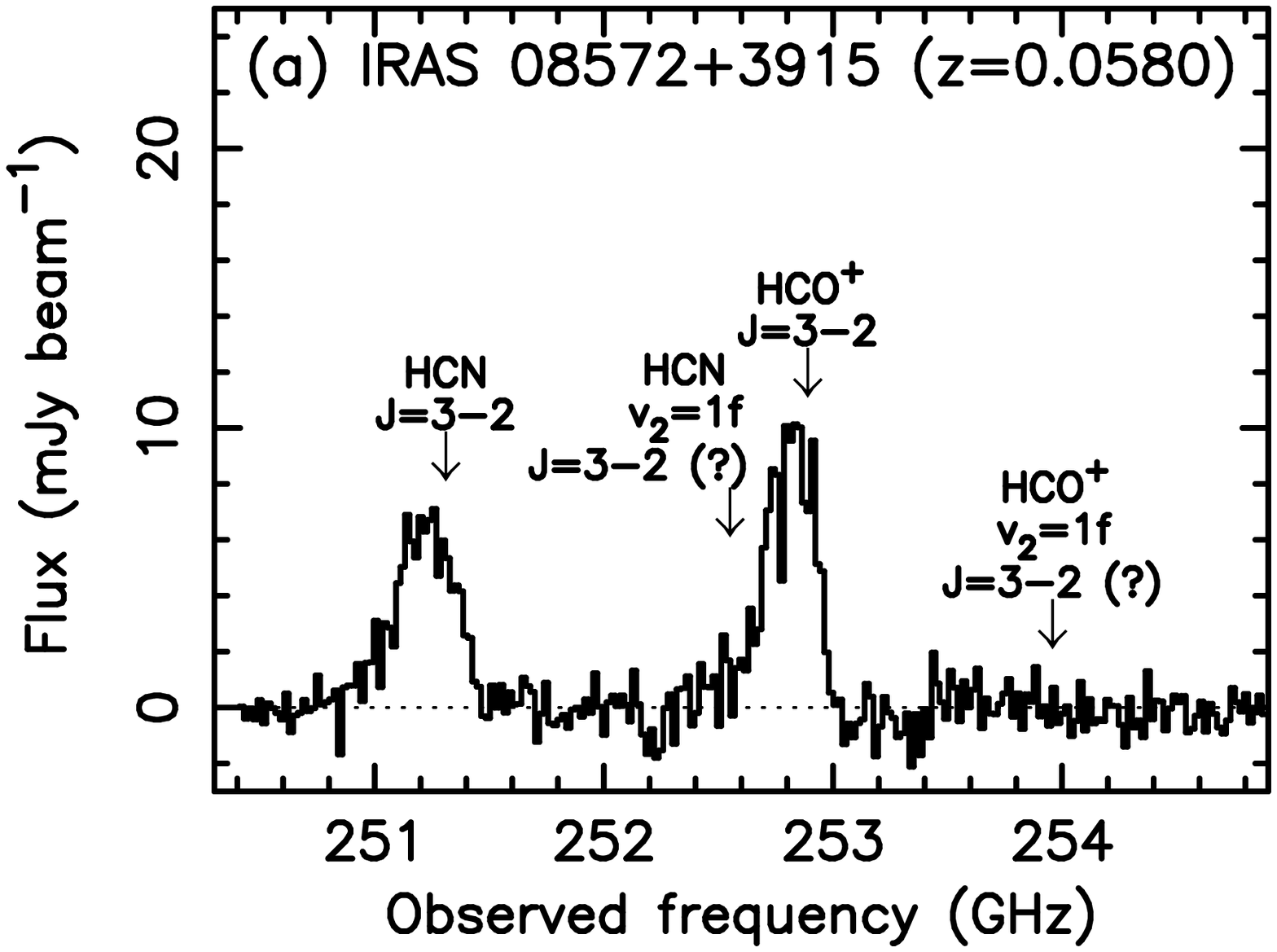} 
\includegraphics[angle=0,scale=.35]{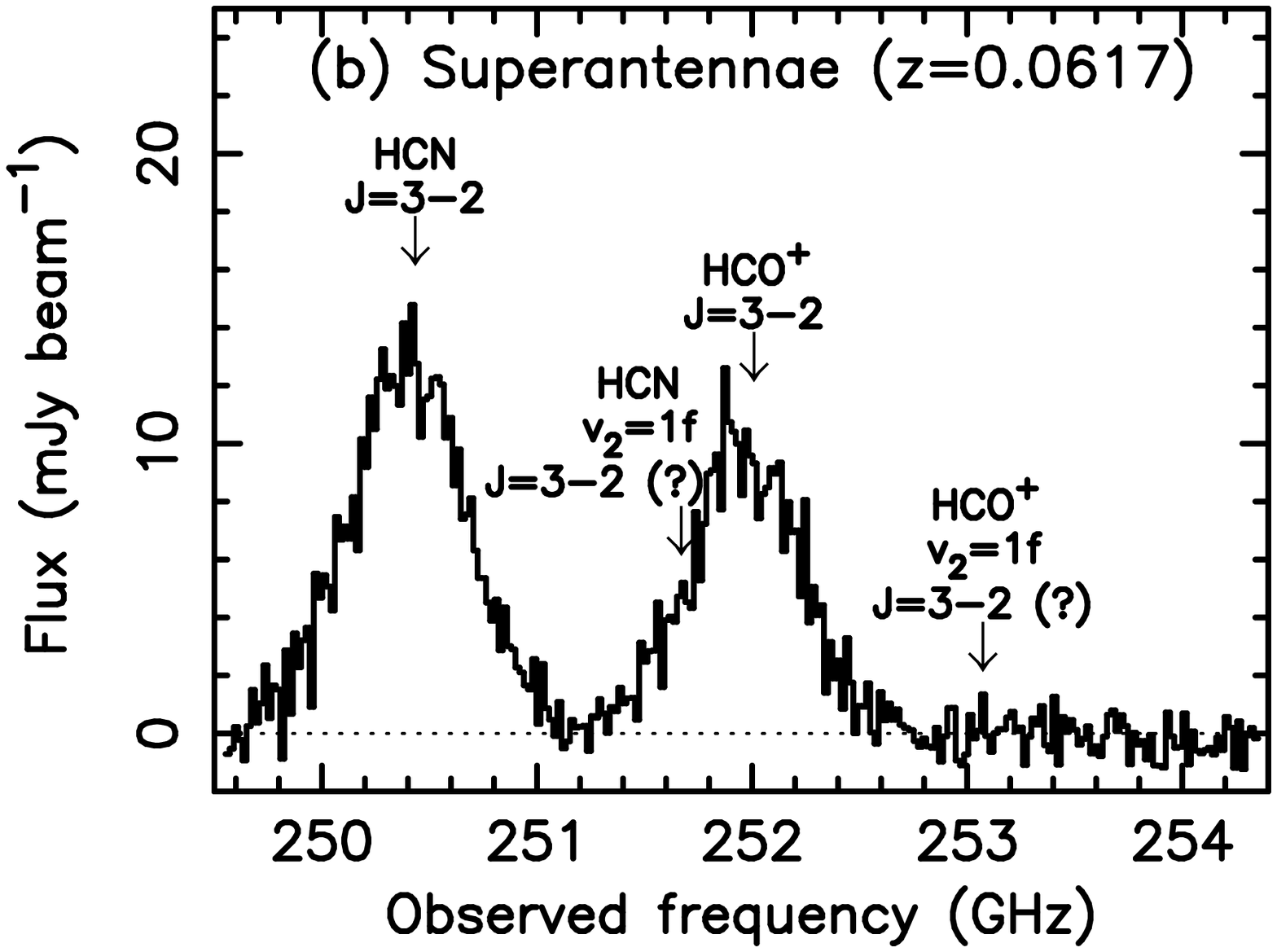} \\
\vspace{-0.4cm}
\includegraphics[angle=0,scale=.35]{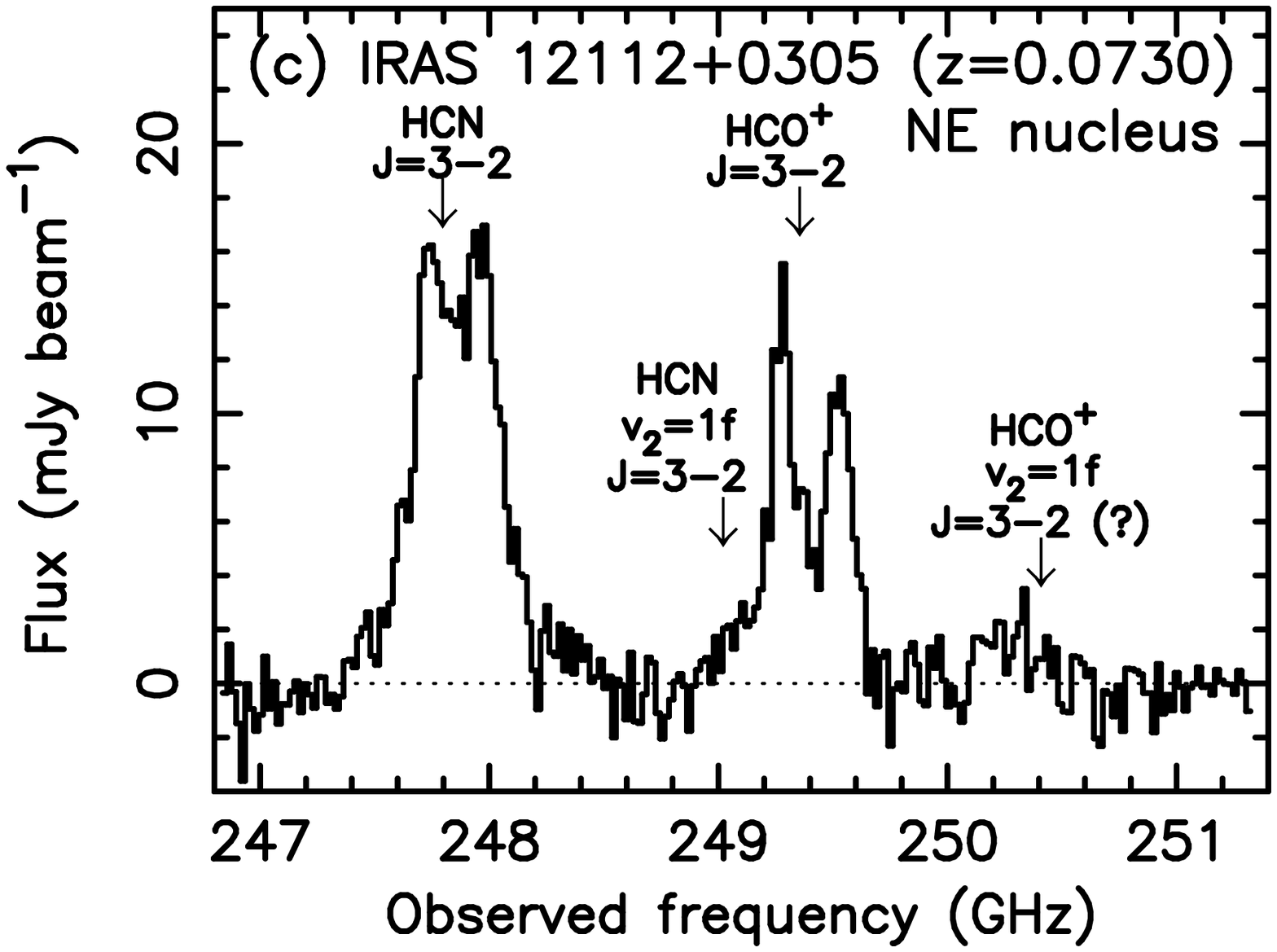} 
\includegraphics[angle=0,scale=.35]{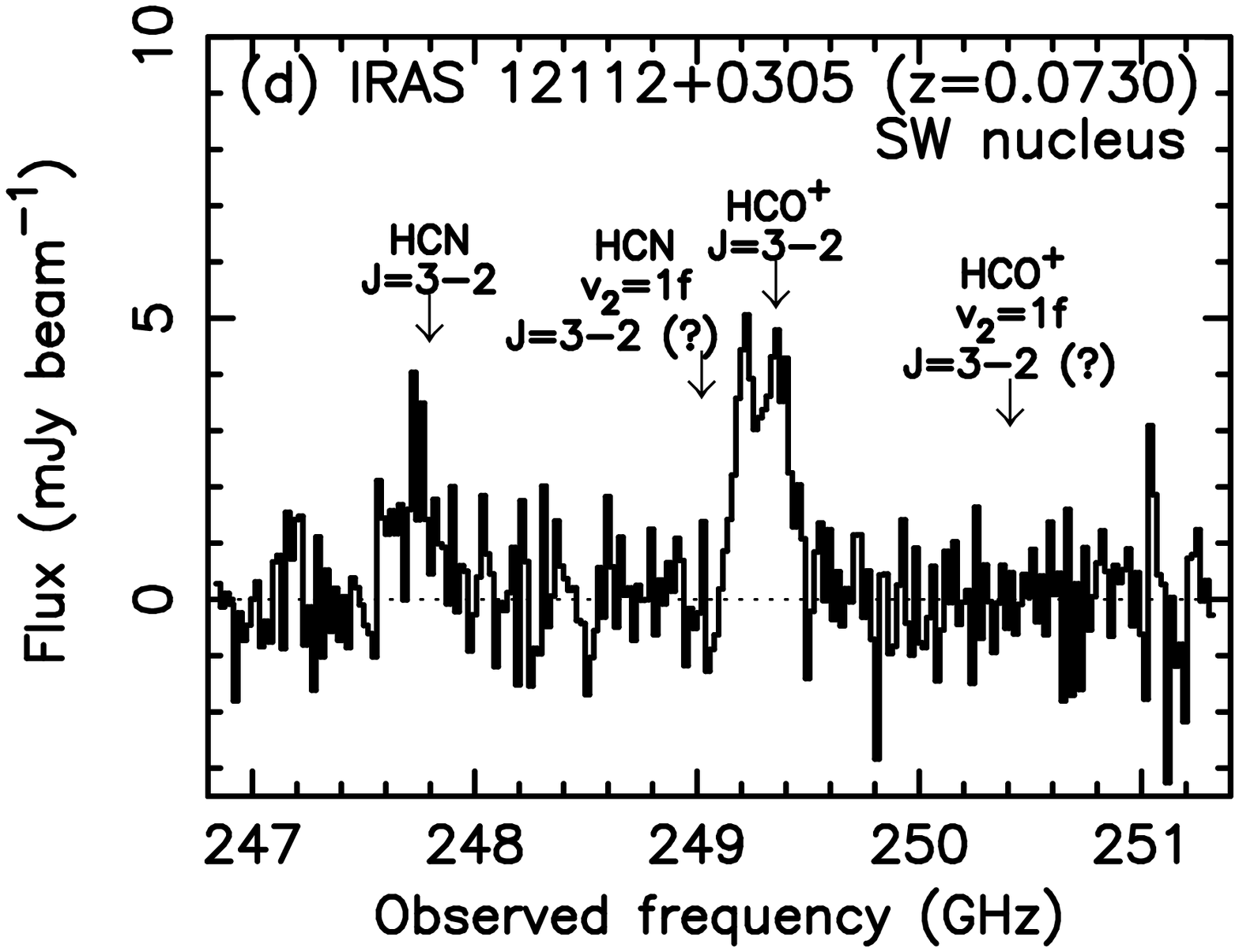} \\
\vspace{-0.4cm}
\includegraphics[angle=0,scale=.35]{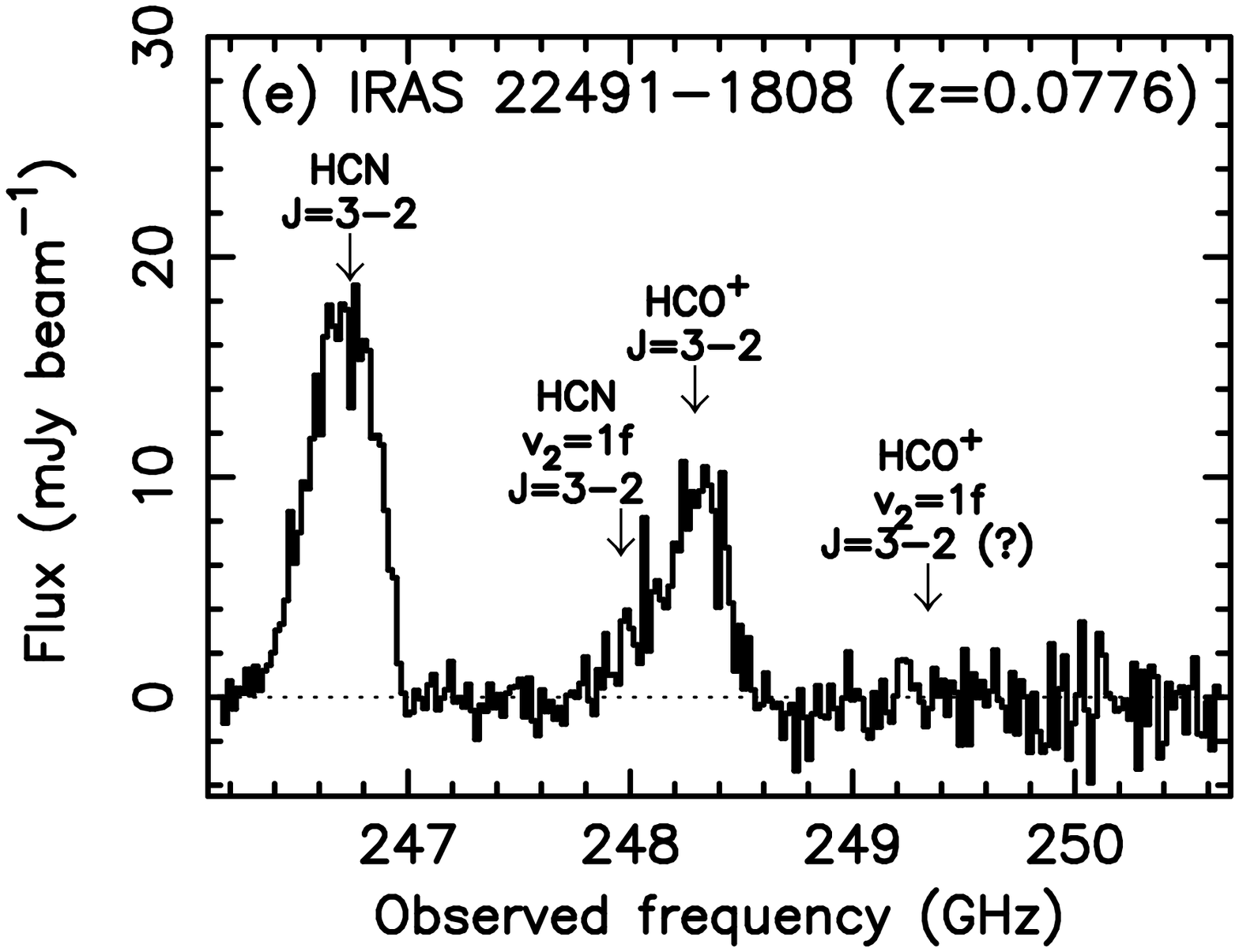} 
\includegraphics[angle=0,scale=.35]{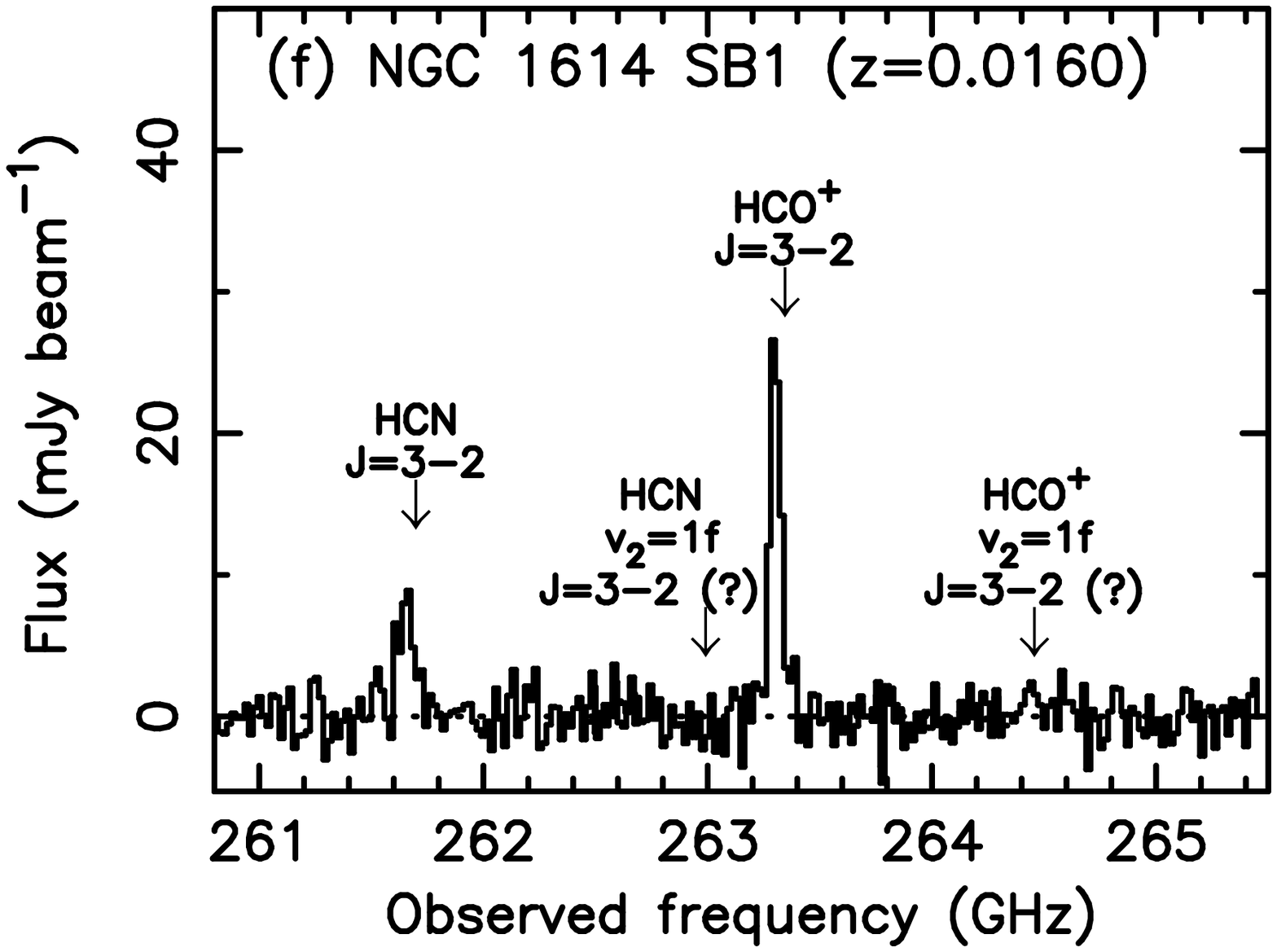} \\
\vspace{-0.4cm}
\includegraphics[angle=0,scale=.35]{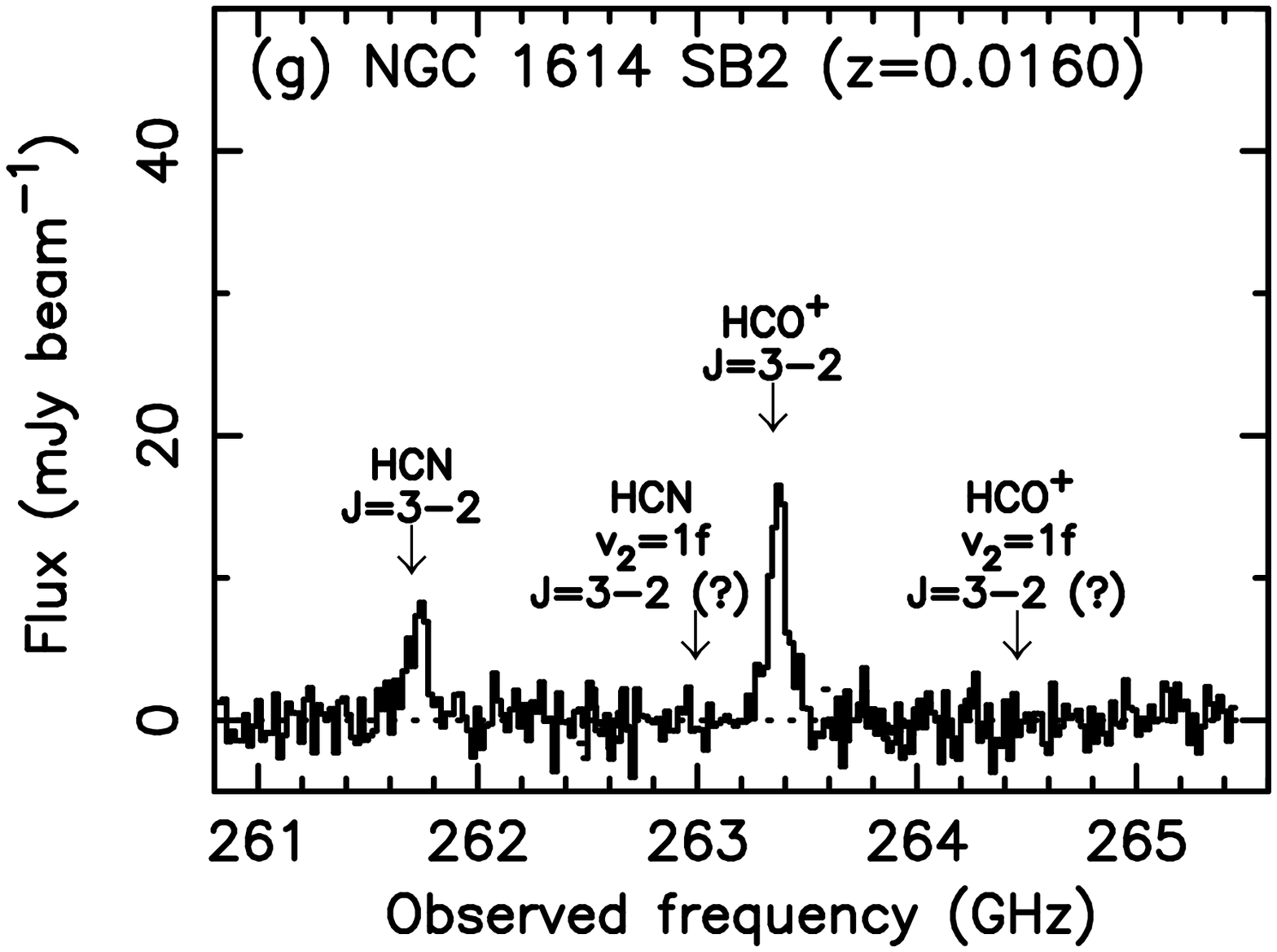} 
\includegraphics[angle=0,scale=.35]{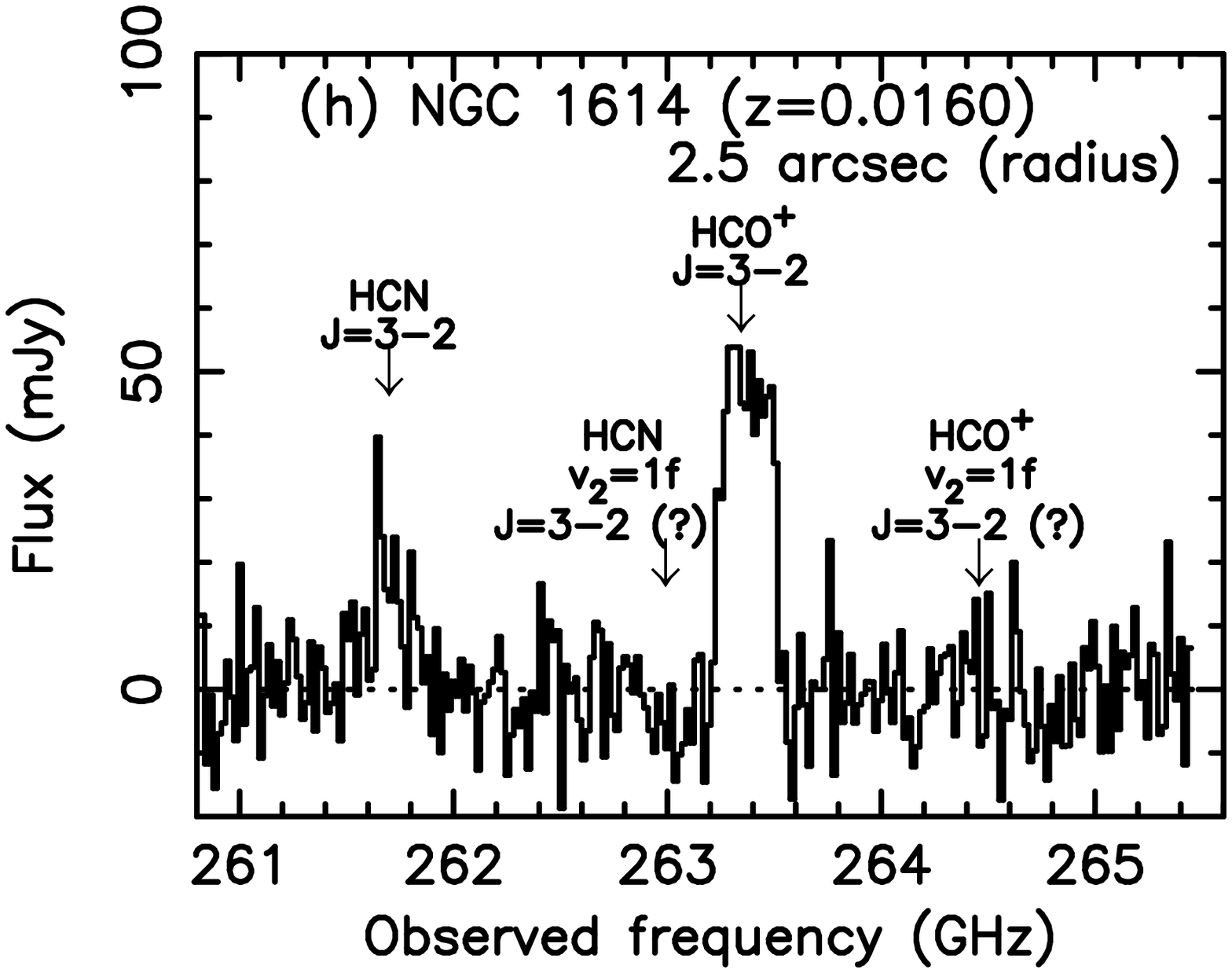} \\ 
\end{center}
\end{figure}

\clearpage

\begin{figure}
\begin{center}
\includegraphics[angle=0,scale=.35]{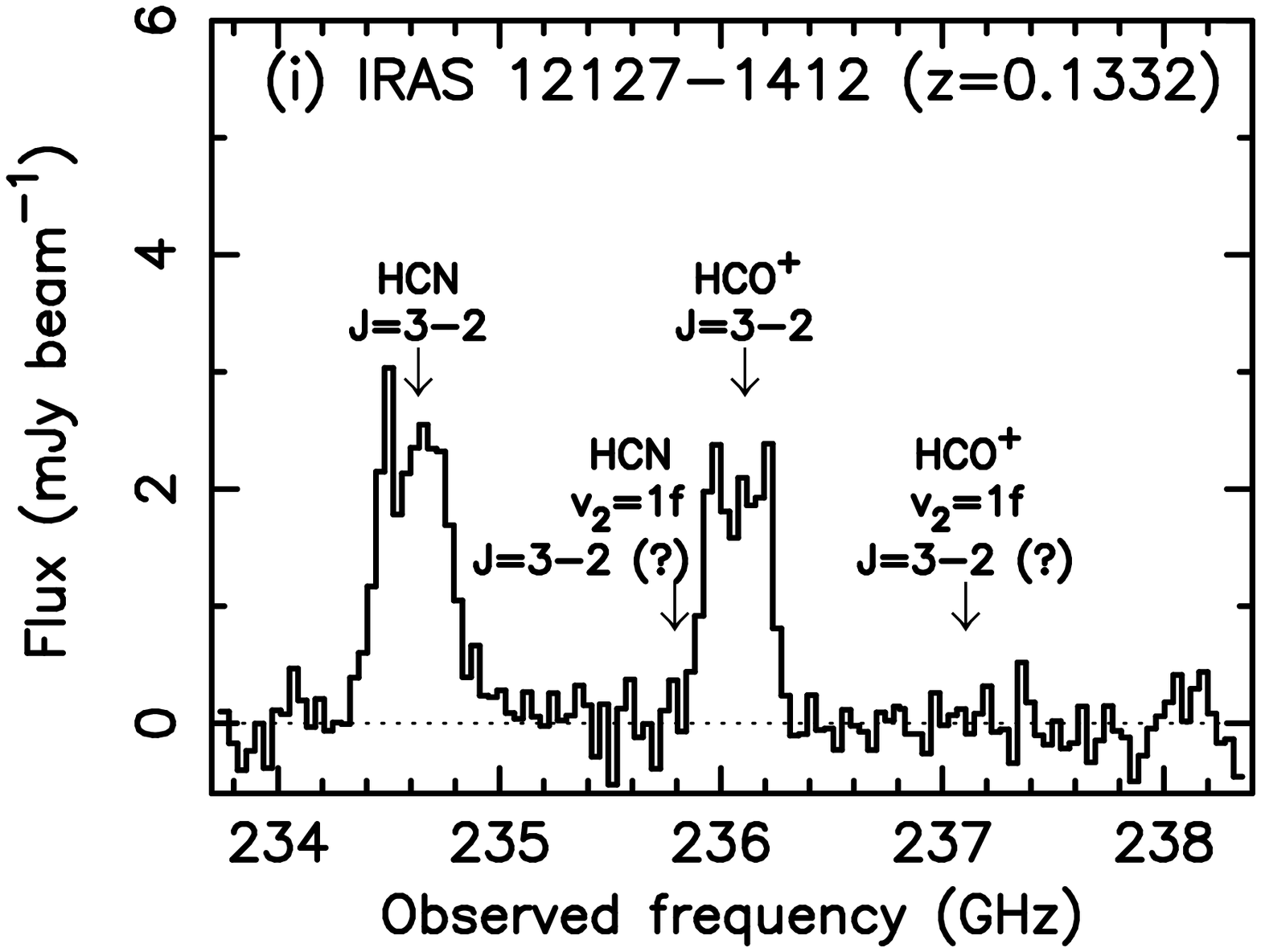} 
\includegraphics[angle=0,scale=.35]{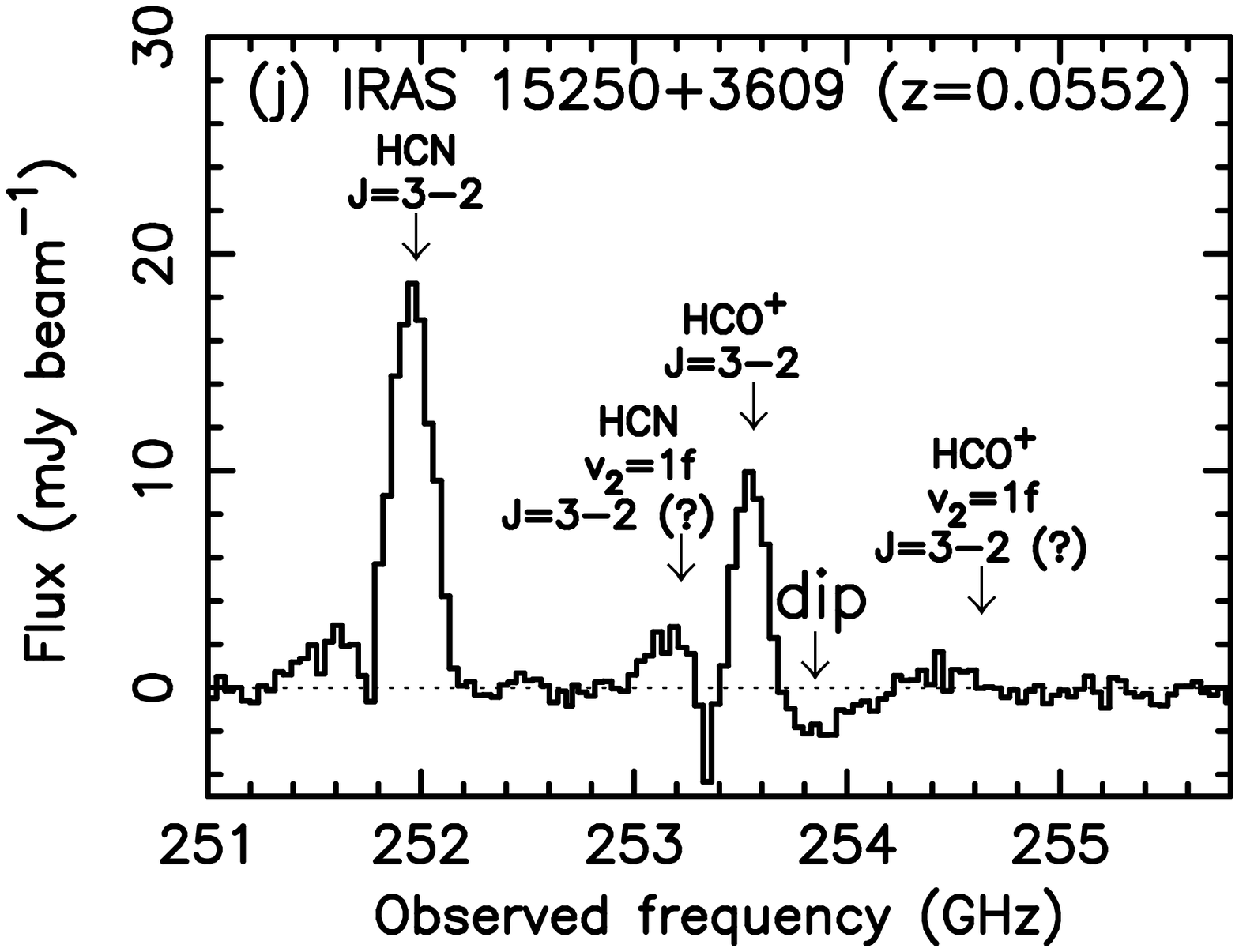} \\
\vspace{-0.4cm}
\includegraphics[angle=0,scale=.35]{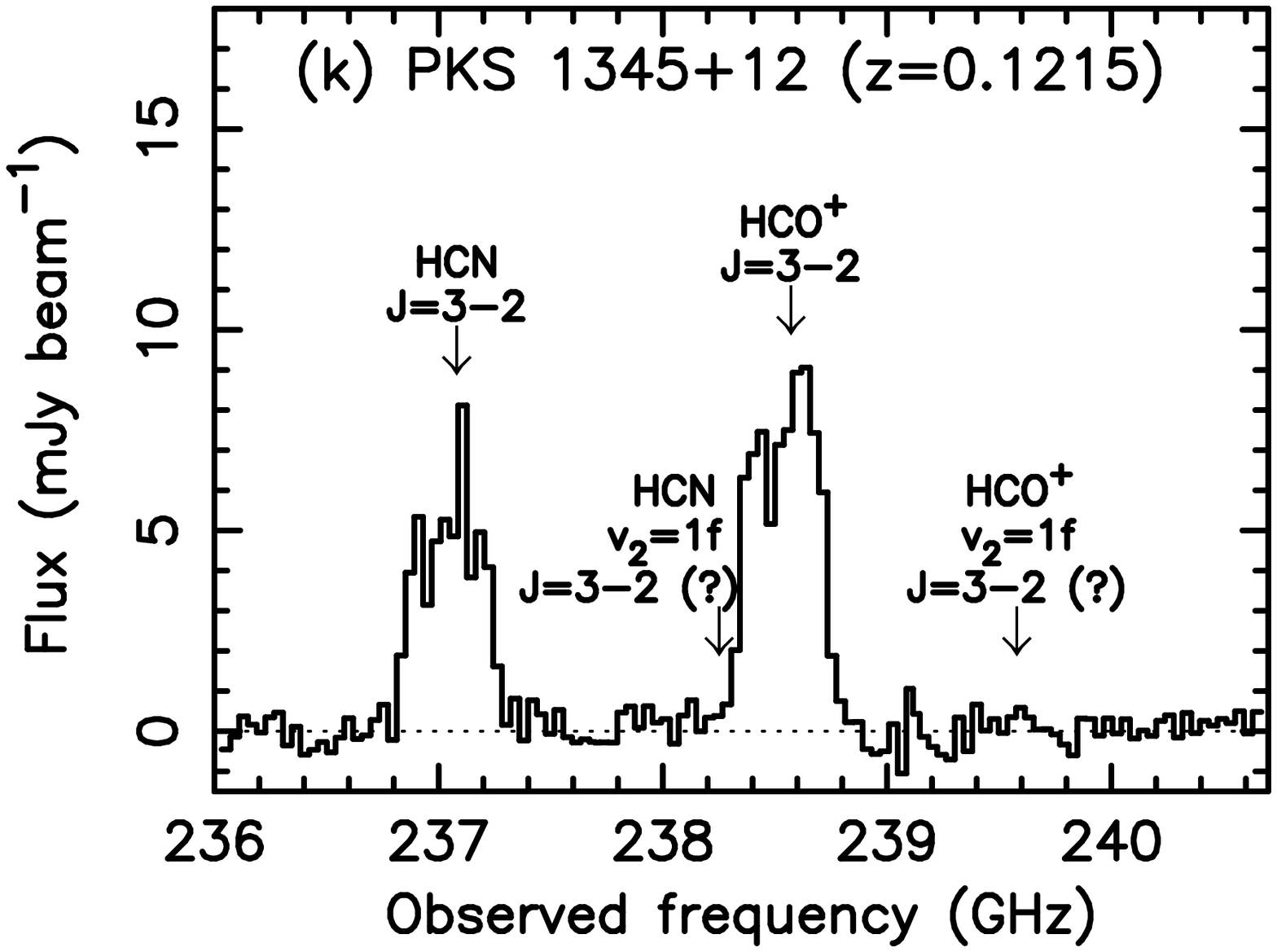} 
\includegraphics[angle=0,scale=.35]{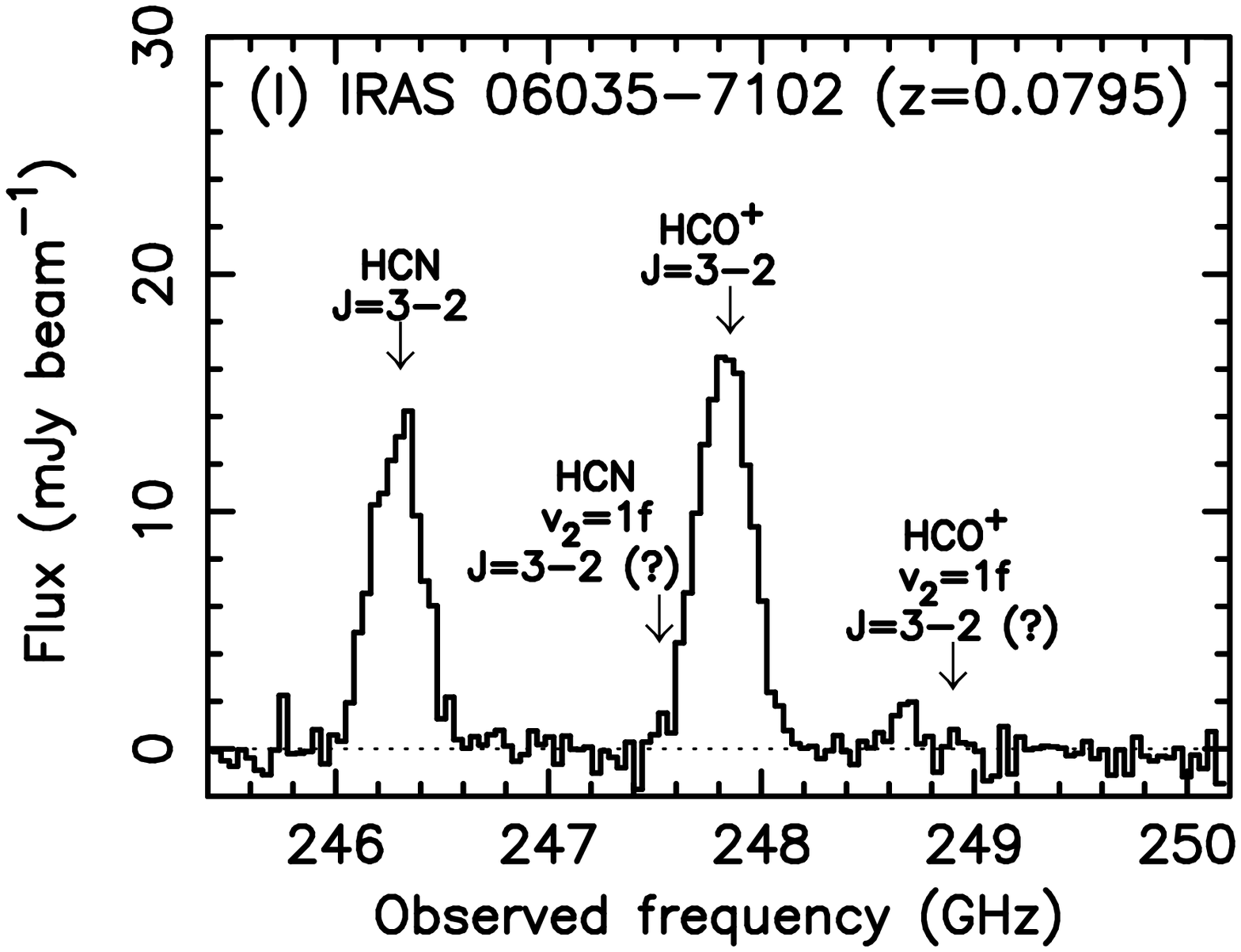} \\ 
\vspace{-0.4cm}
\includegraphics[angle=0,scale=.35]{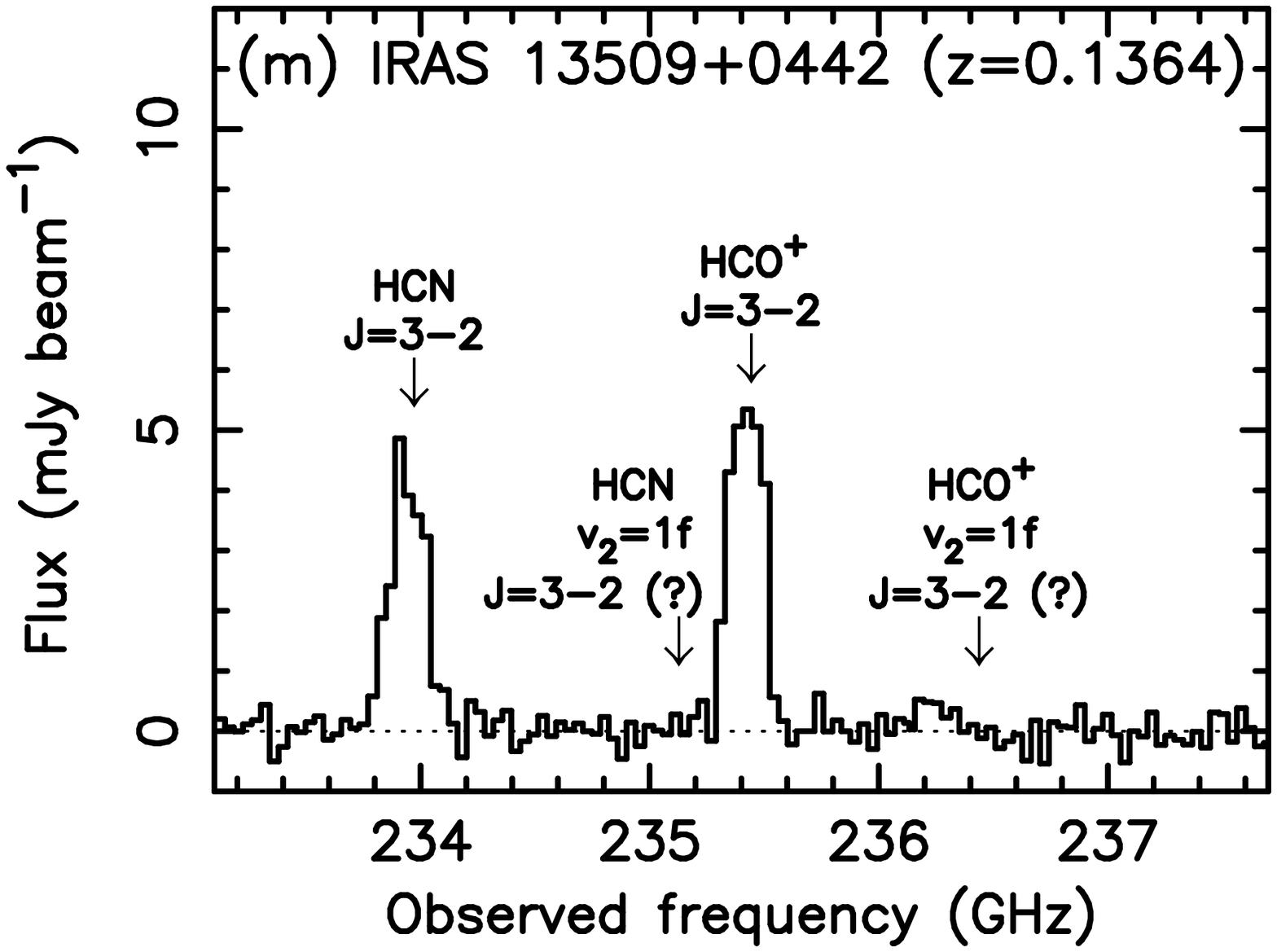} 
\includegraphics[angle=0,scale=.35]{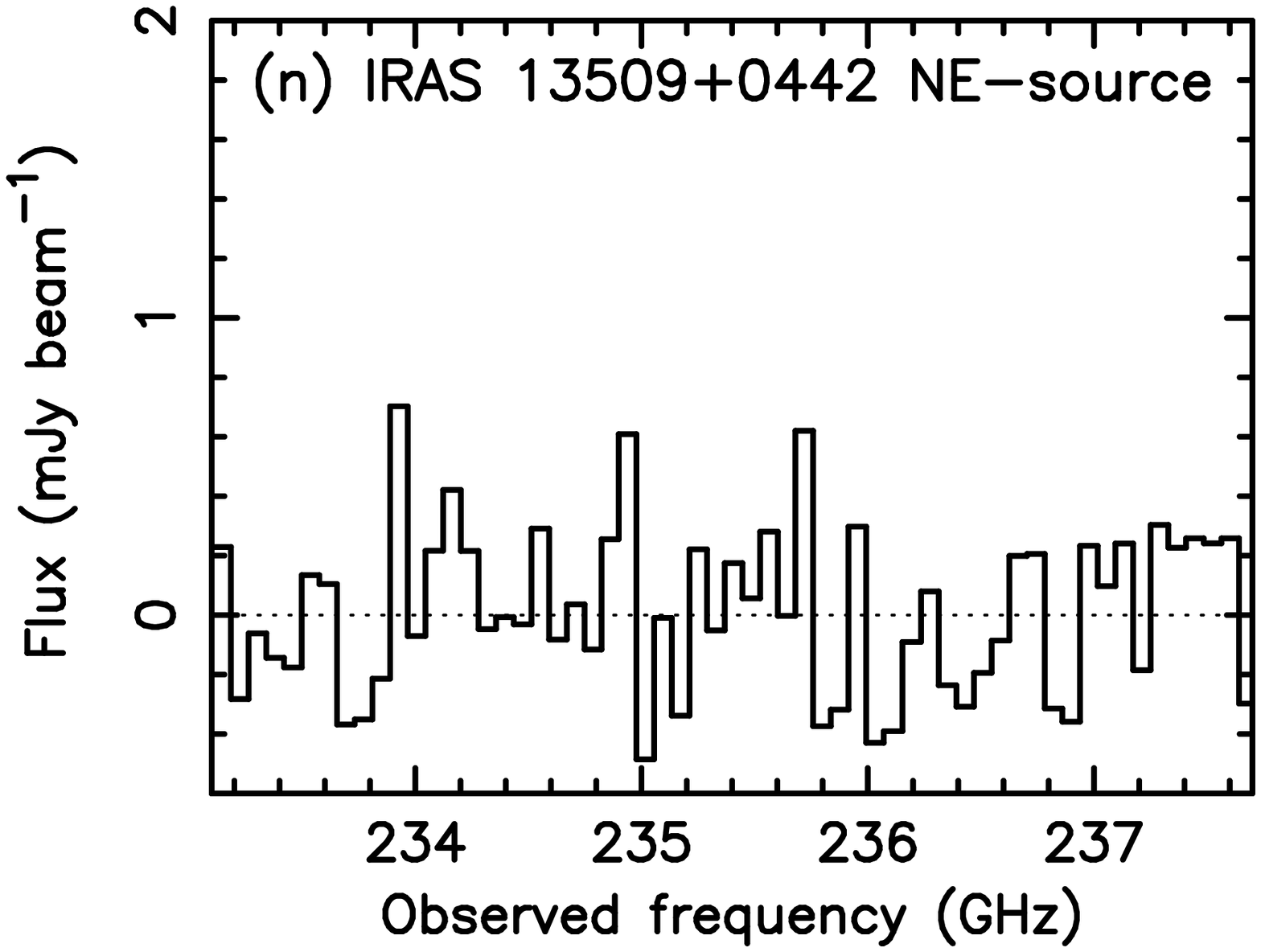} \\ 
\vspace{-0.4cm}
\includegraphics[angle=0,scale=.35]{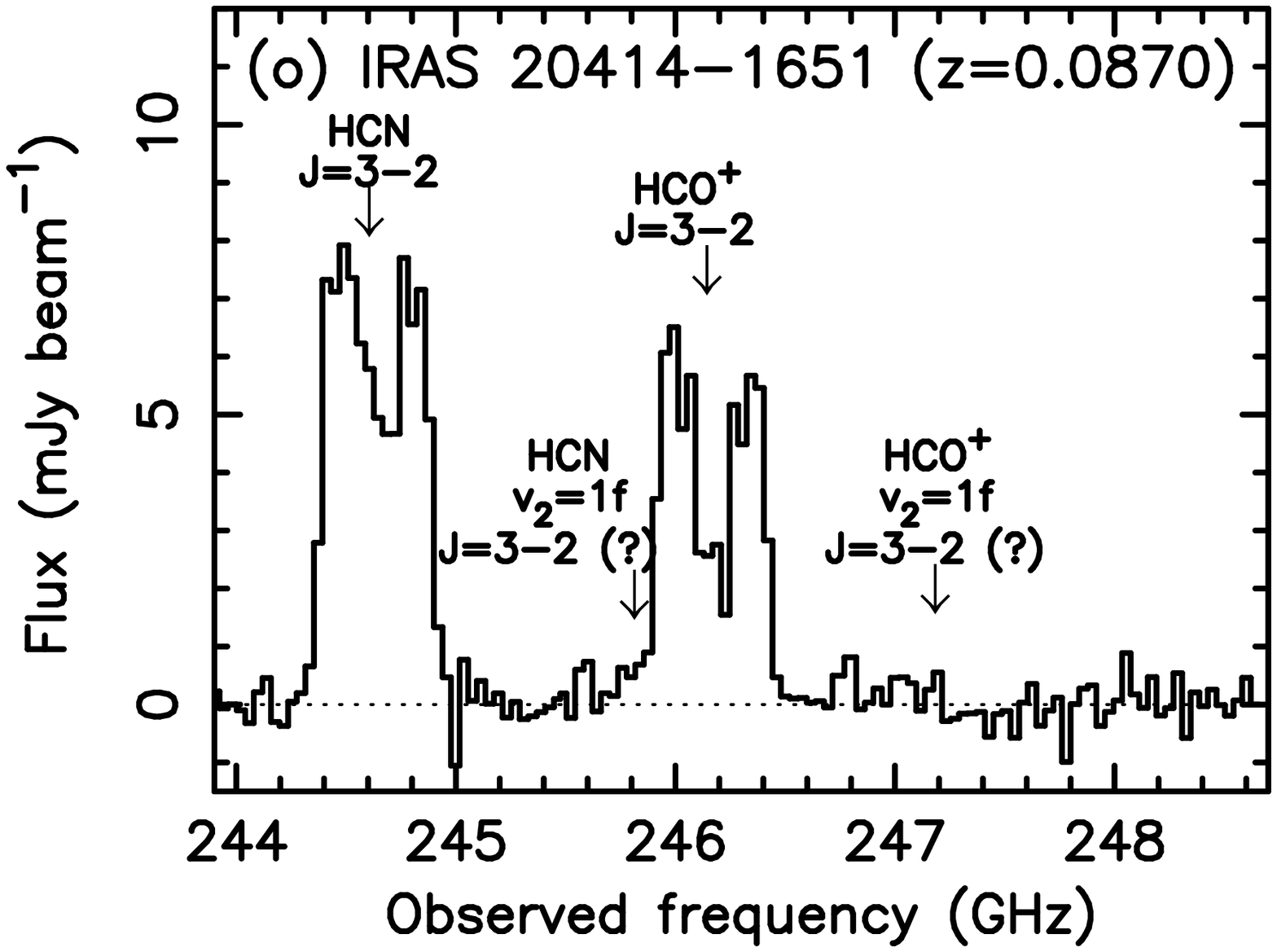} 
\vspace{-0.4cm}
\end{center}
\caption{Full frequency coverage spectra of LIRGs at interesting regions.
Spectra are taken within the beam size, except for (h).
The SB1 and SB2 of NGC 1614 are defined from the HCN J=3--2 emission
peak (see text in $\S$4).
Down arrows are shown at the expected observed frequency of some emission
lines for the redshifts shown in Table 1. 
The binned spectrum of the NE source detected in the continuum map of
IRAS 13509$+$0442 is also shown.  
}
\end{figure}

\begin{figure}
\begin{center}
\includegraphics[angle=0,scale=.6]{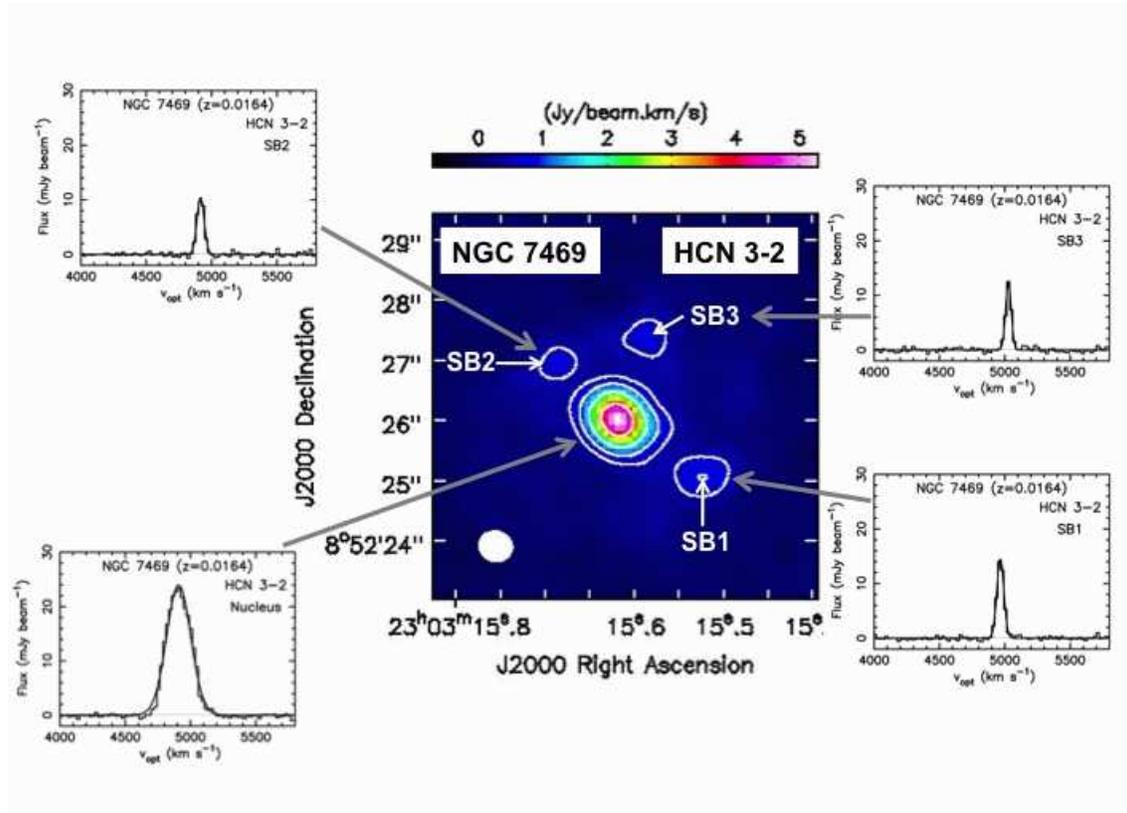} \\
\vspace{0.4cm}
\includegraphics[angle=0,scale=.6]{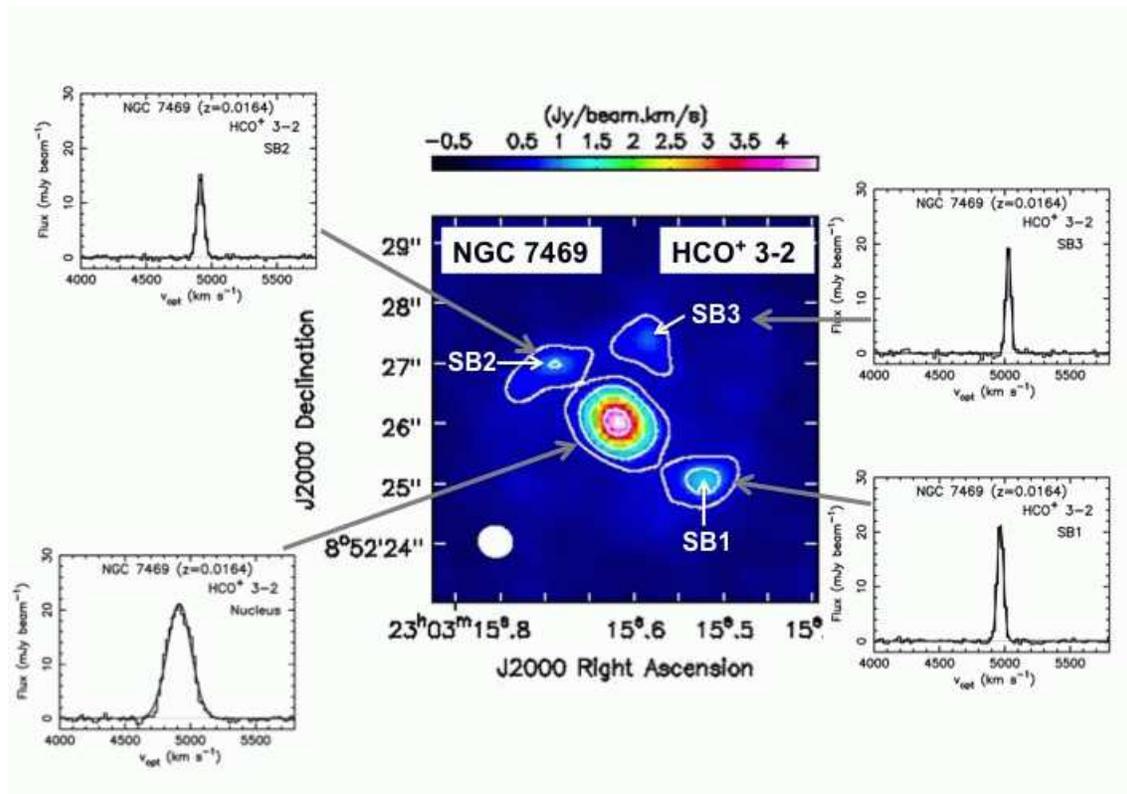} 
\end{center}
\vspace{-0.7cm}
\caption{
Integrated intensity (moment 0) maps of the HCN J=3--2 and HCO$^{+}$ J=3--2
emission lines, and their Gaussian fits in the spectra at individual
locations within the beam size, for NGC 7469. 
For moment 0 maps of HCN J=3--2 and HCO$^{+}$ J=3--2, the contours
represent the 5$\sigma$, 10$\sigma$, 20$\sigma$, and 40$\sigma$ levels.}
\end{figure}

\begin{figure}
\begin{center}
\includegraphics[angle=0,scale=.6]{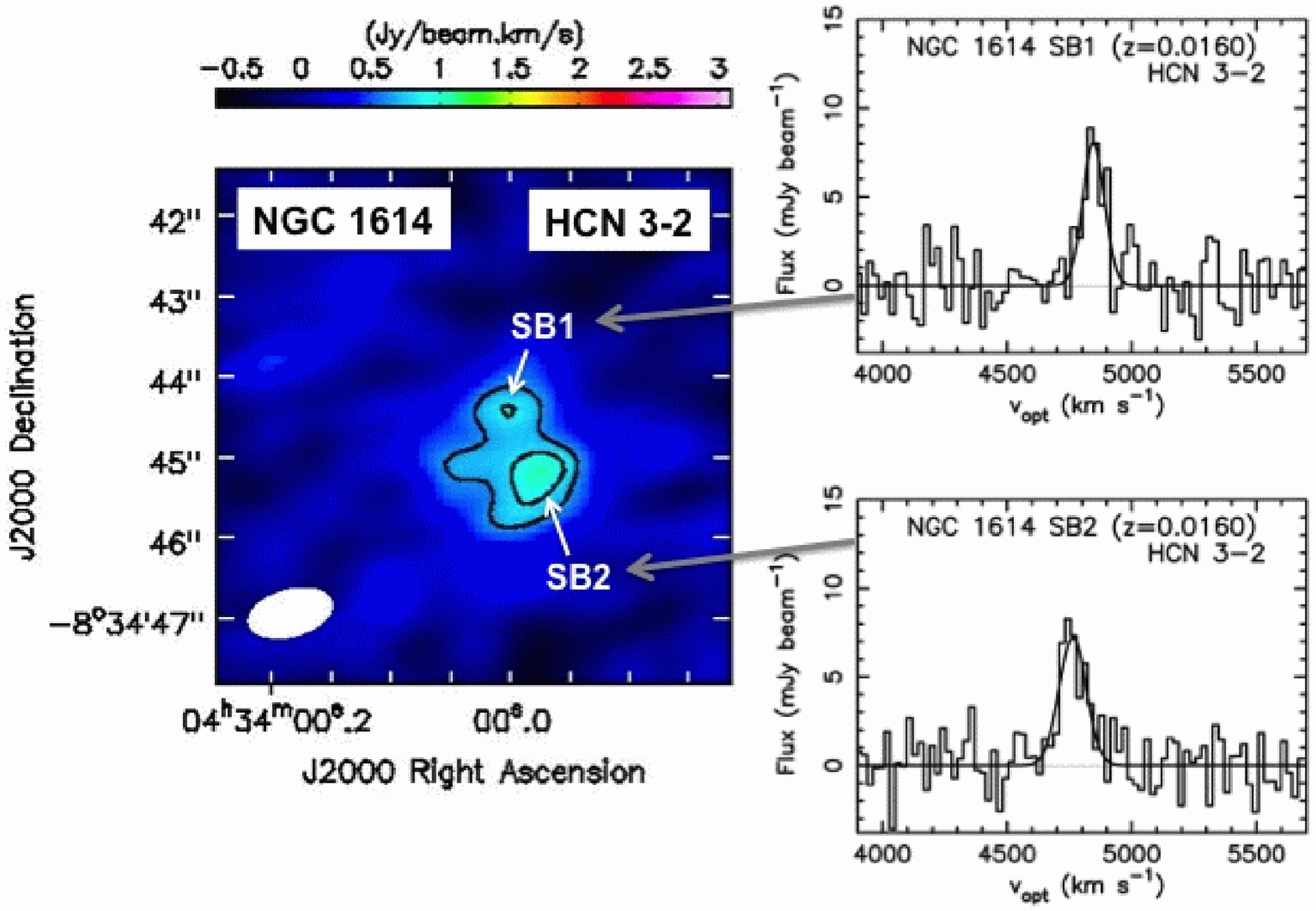} \\
\vspace{0.4cm}
\includegraphics[angle=0,scale=.6]{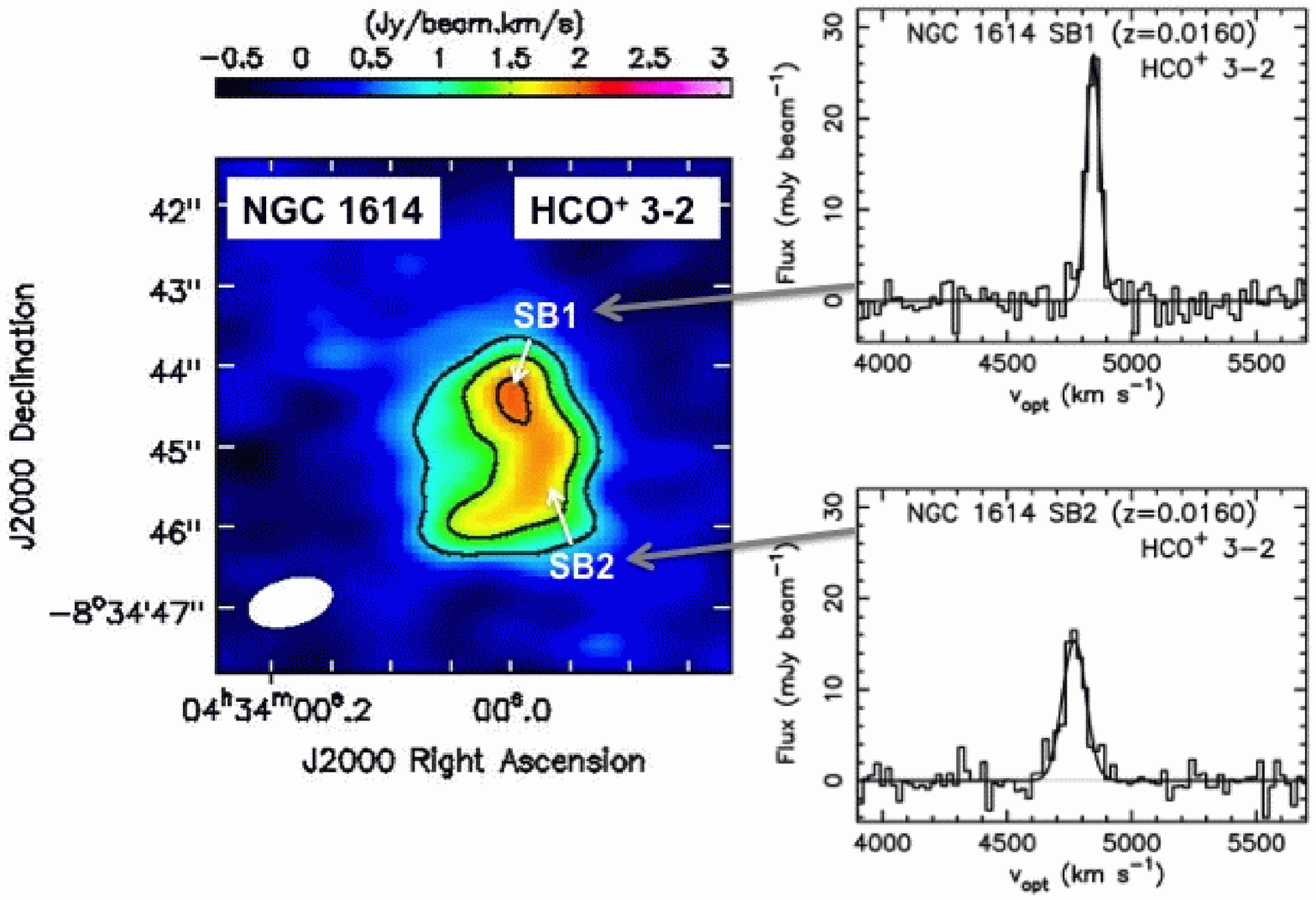} 
\end{center}
\vspace{-0.7cm}
\caption{
Integrated intensity (moment 0) maps of the HCN J=3--2 and HCO$^{+}$ J=3--2
emission lines, and their Gaussian fits in the spectra at individual
locations within the beam size, for NGC 1614. 
The contours of the moment 0 maps are 3$\sigma$, 4$\sigma$ for HCN J=3--2, 
and 3$\sigma$, 5$\sigma$, 7$\sigma$ for HCO$^{+}$ J=3--2.}
\end{figure}

\begin{figure}
\begin{center}
\includegraphics[angle=0,scale=.41]{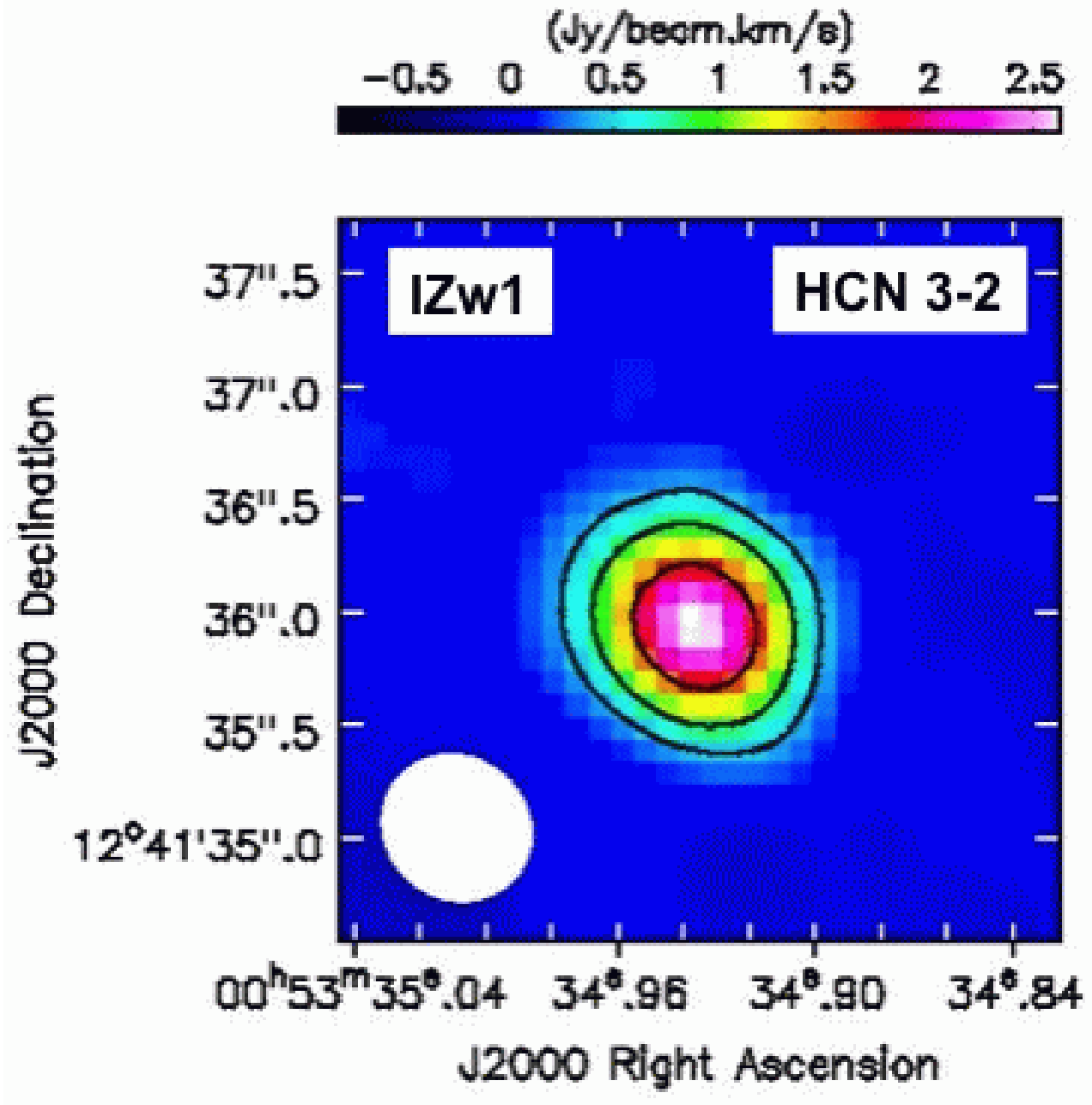} 
\includegraphics[angle=0,scale=.41]{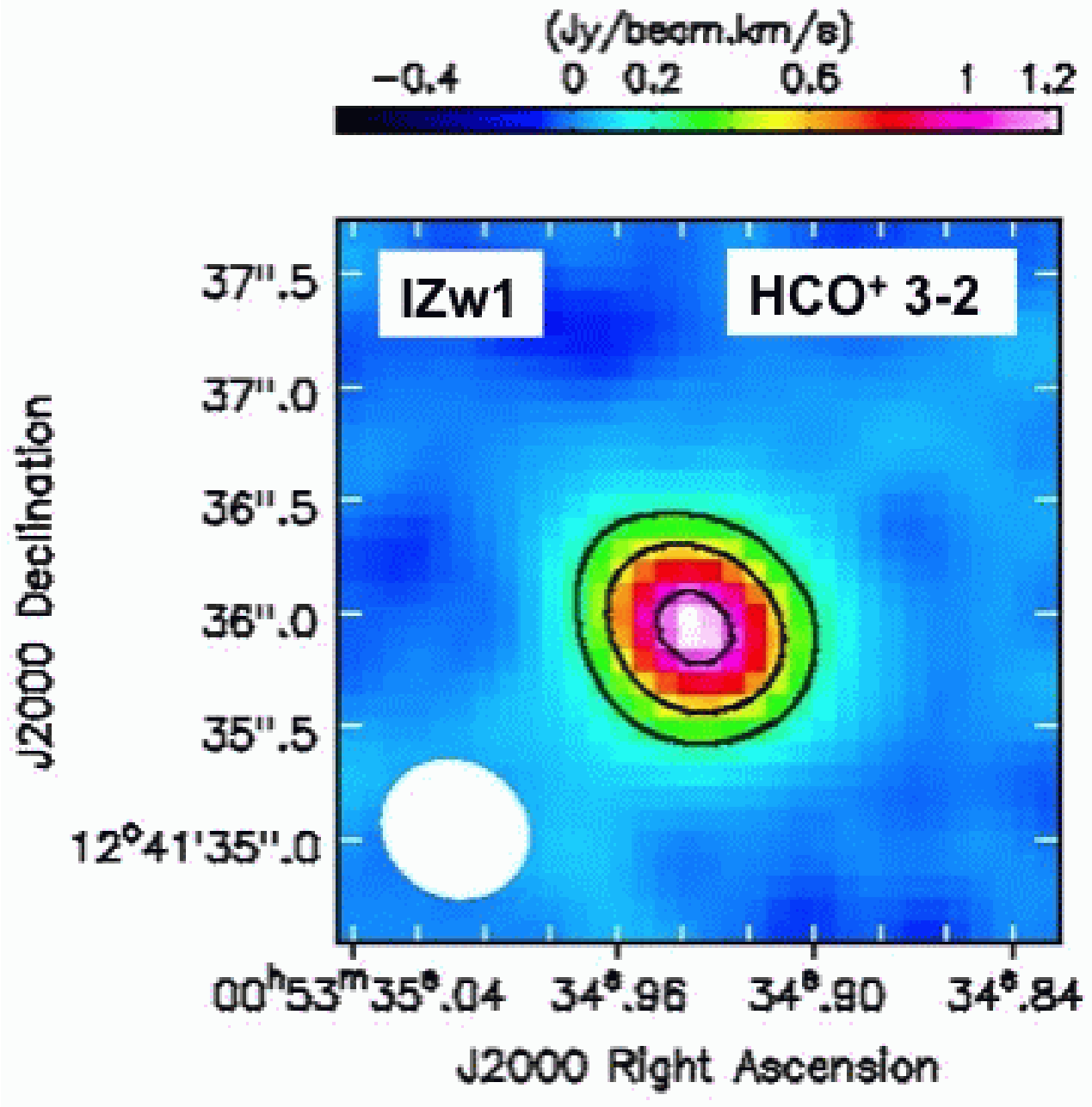} \\ 
\vspace*{-1.3cm}
\includegraphics[angle=0,scale=.41]{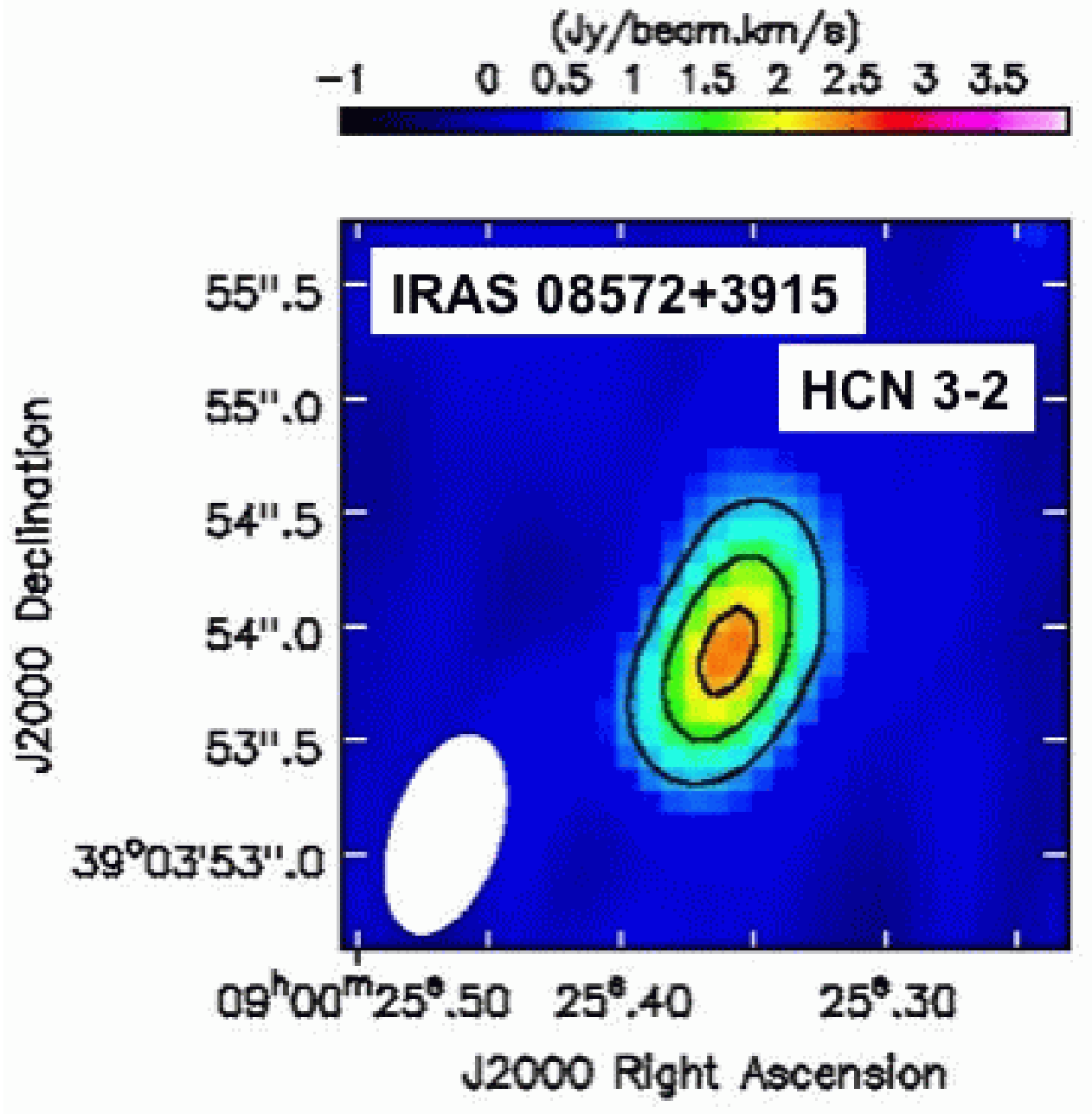} 
\includegraphics[angle=0,scale=.41]{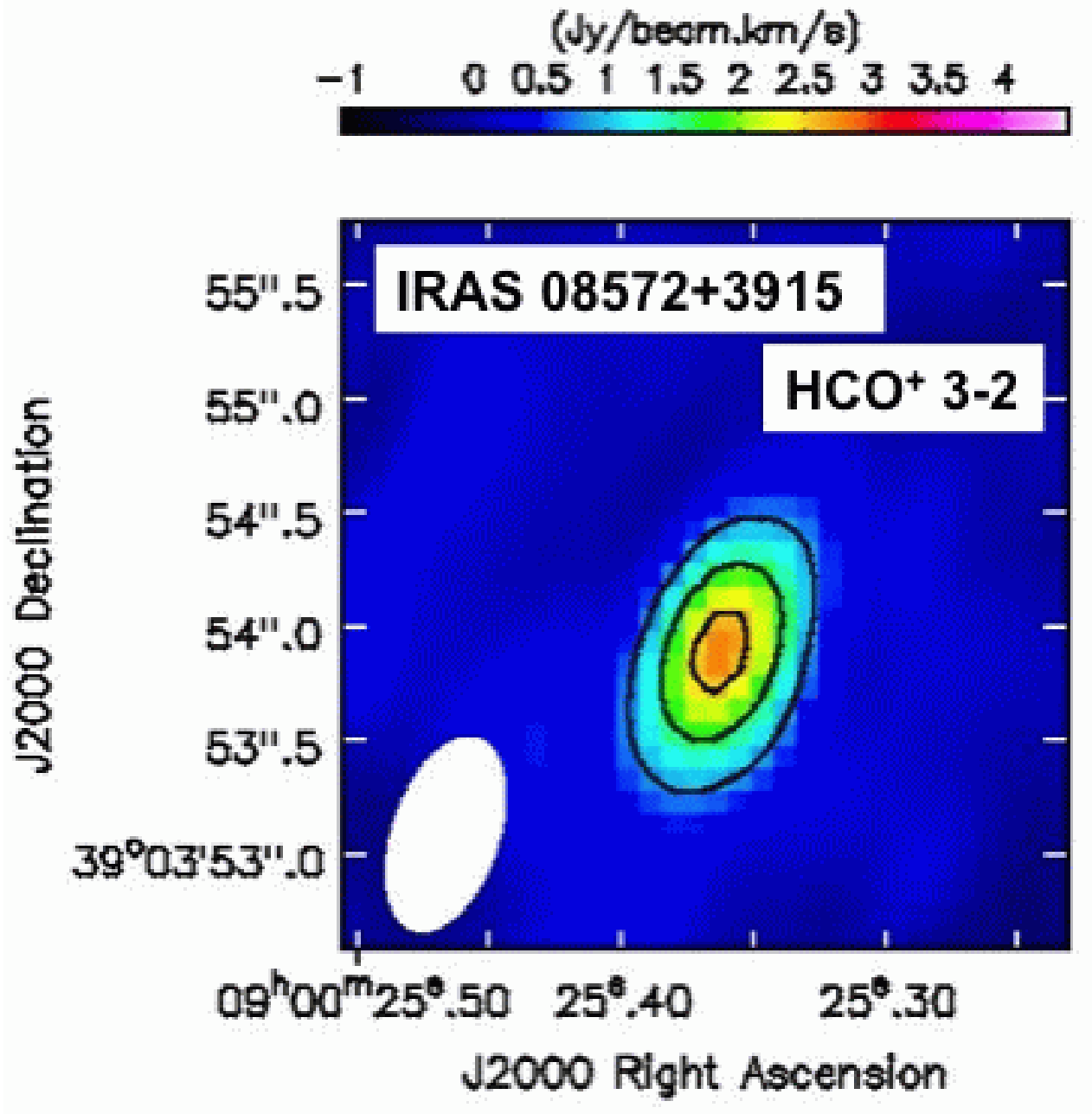}\\ 
\vspace*{-1.3cm}
\includegraphics[angle=0,scale=.41]{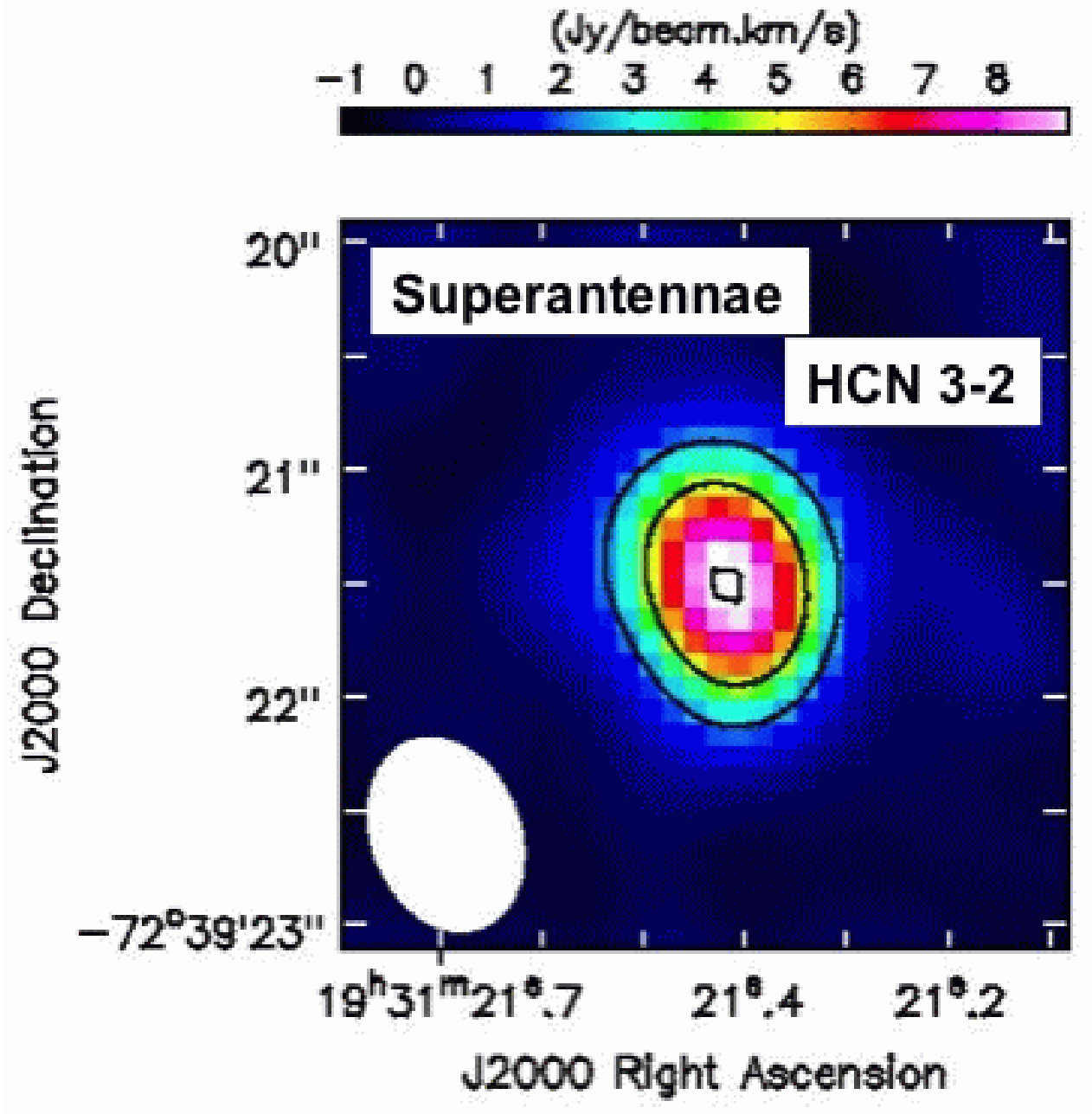}
\includegraphics[angle=0,scale=.41]{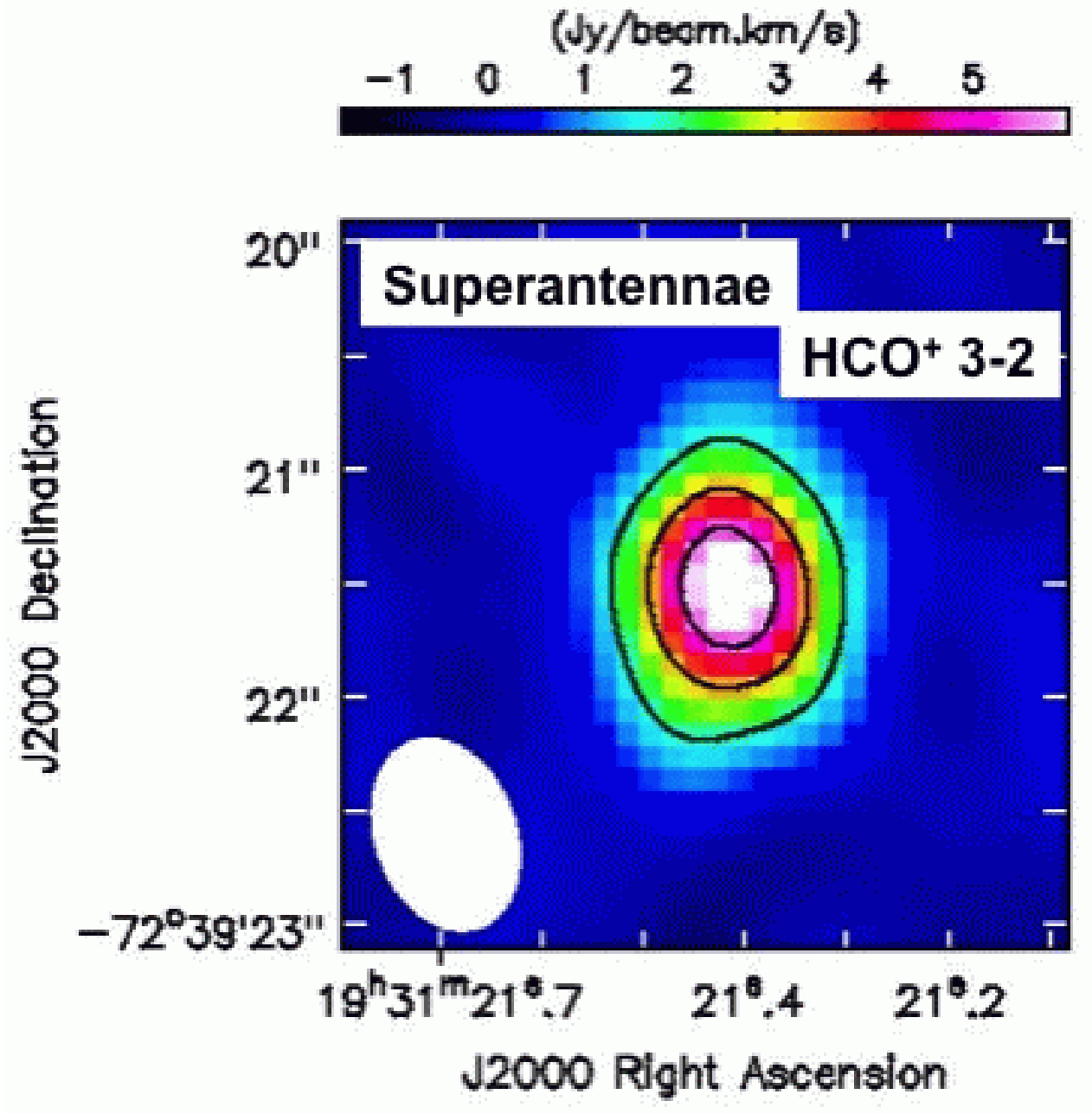} \\
\end{center}
\end{figure}

\clearpage

\begin{figure}
\begin{center}
\includegraphics[angle=0,scale=.41]{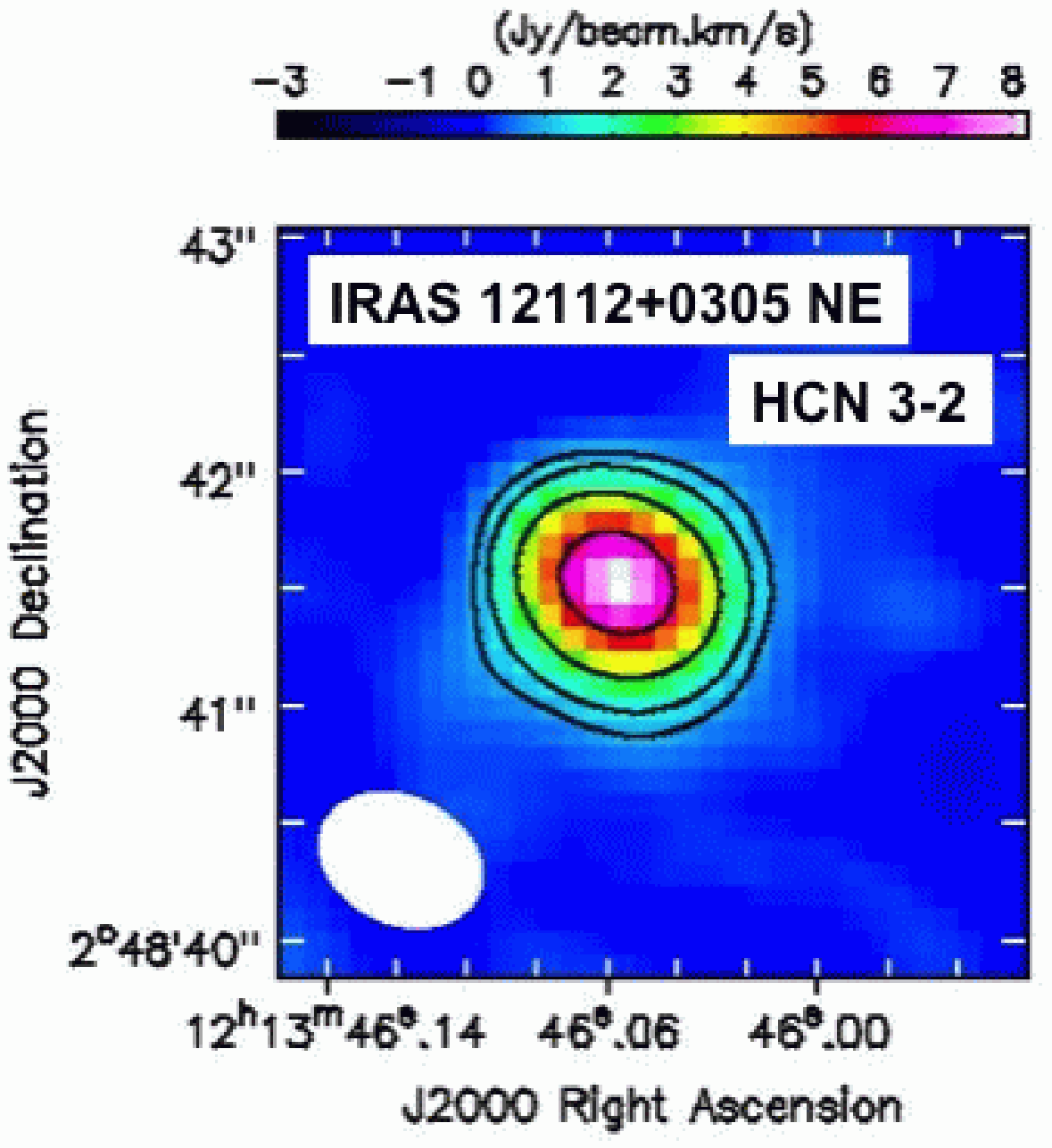} 
\includegraphics[angle=0,scale=.41]{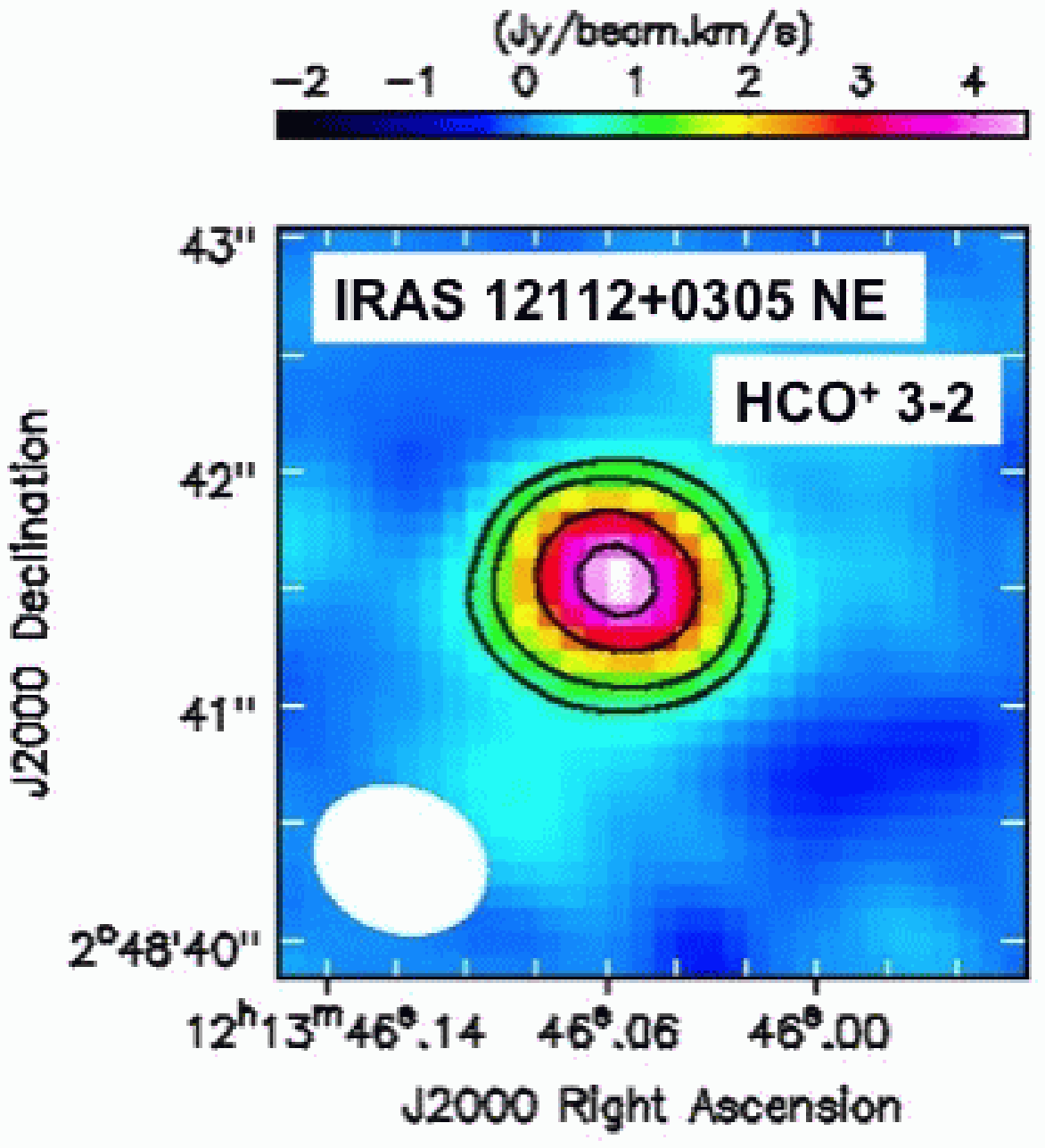} \\
\vspace*{-1.3cm}
\includegraphics[angle=0,scale=.41]{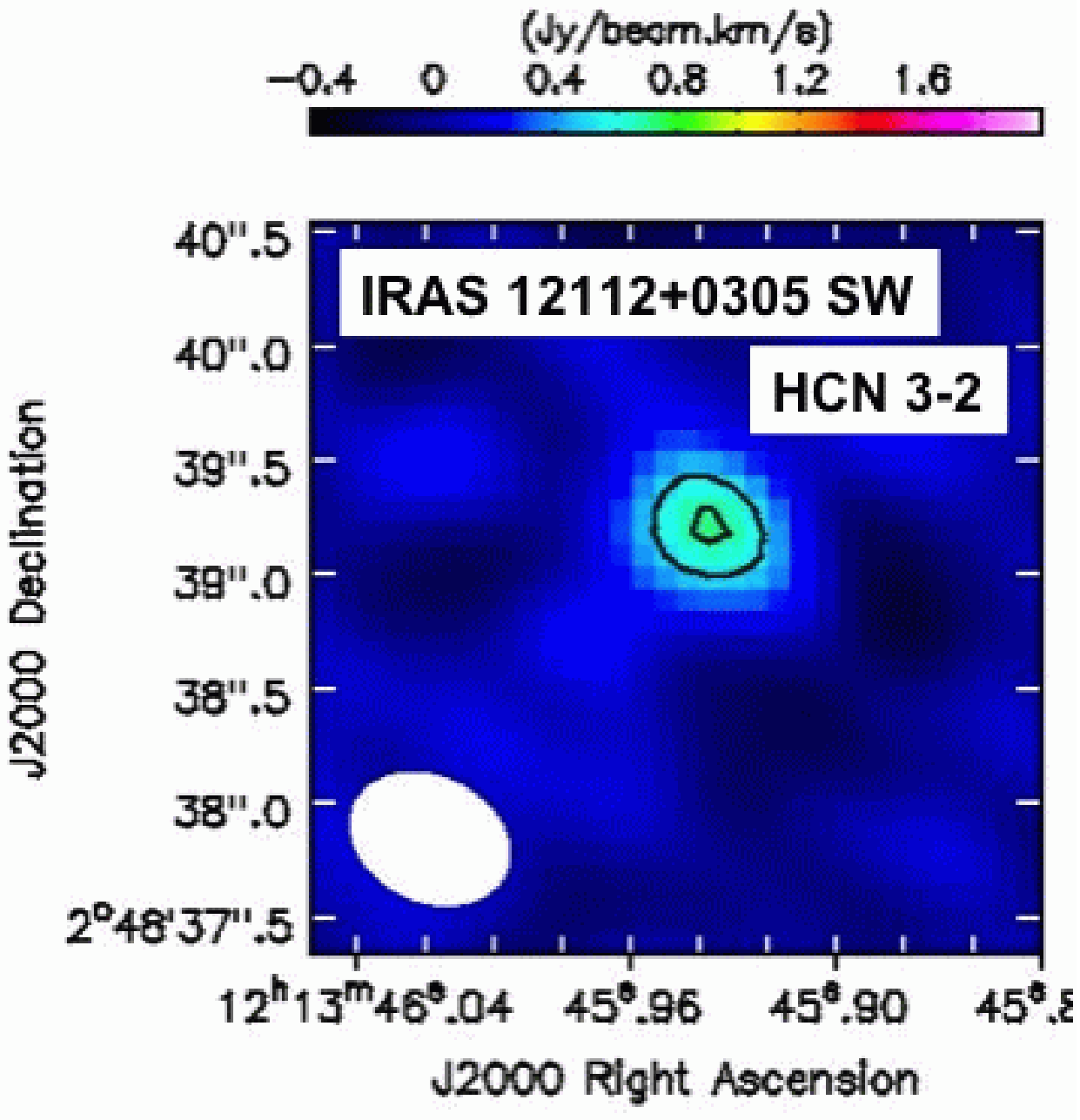} 
\includegraphics[angle=0,scale=.41]{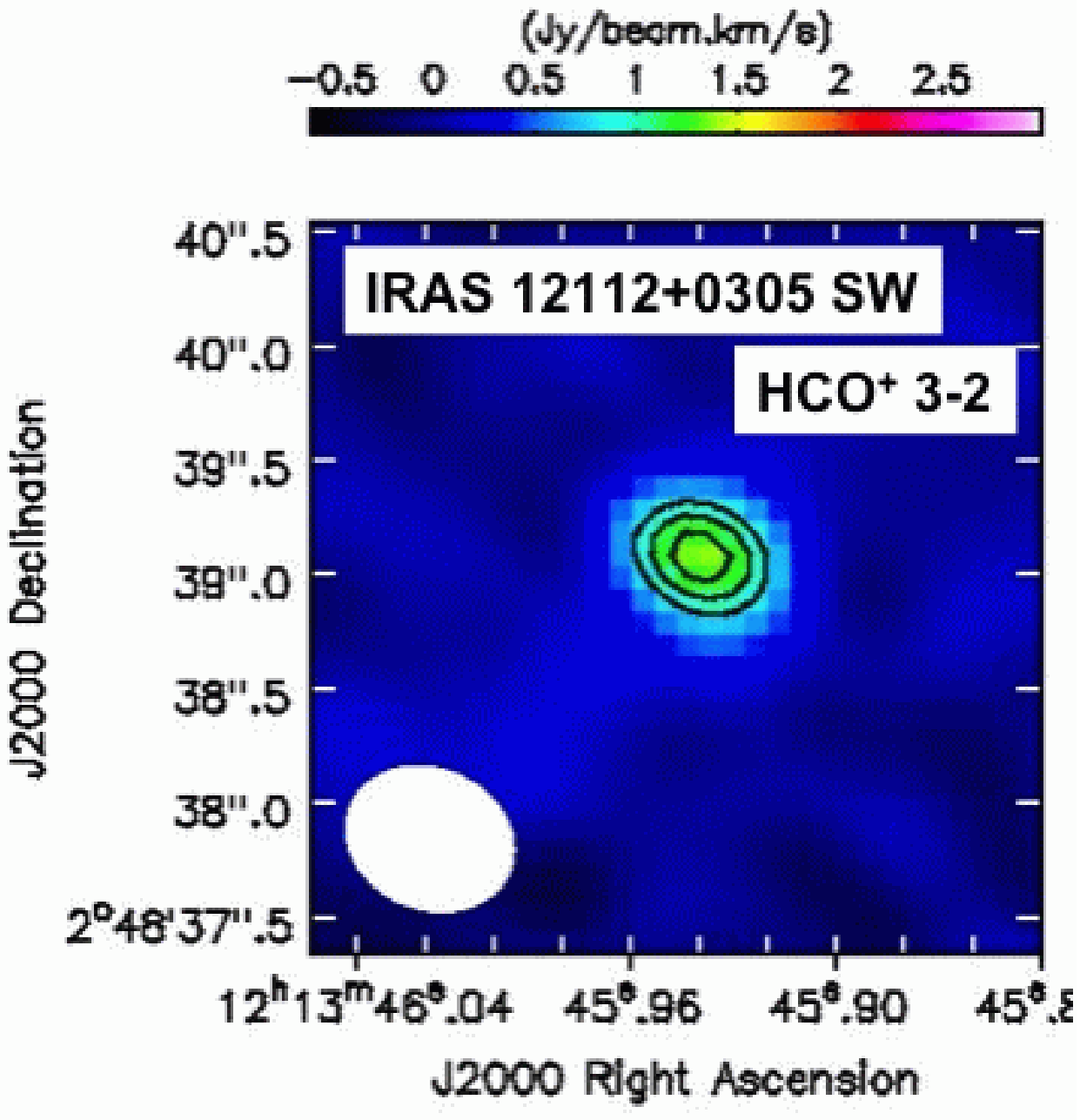} \\
\vspace*{-1.3cm}
\includegraphics[angle=0,scale=.41]{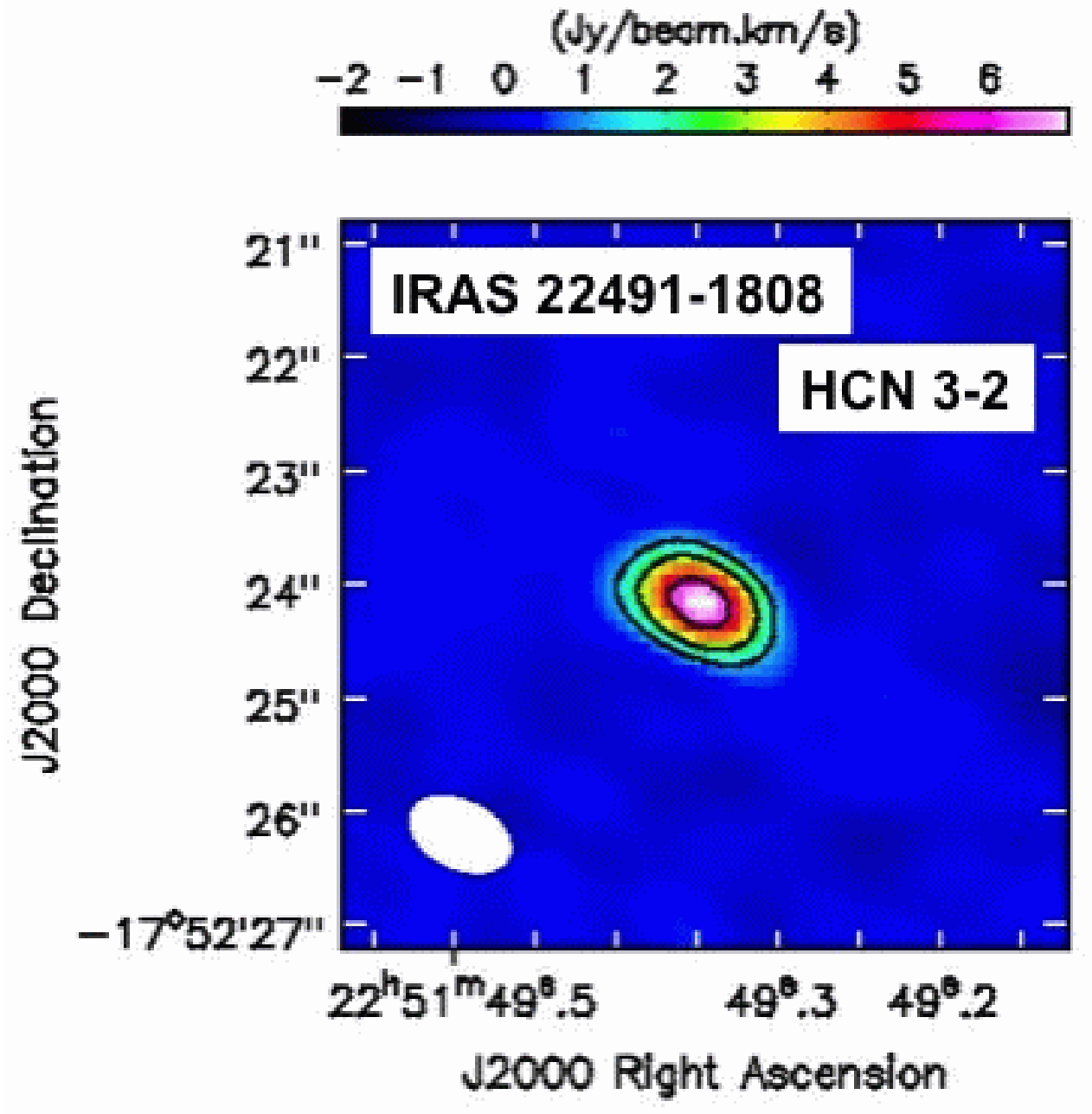} 
\includegraphics[angle=0,scale=.41]{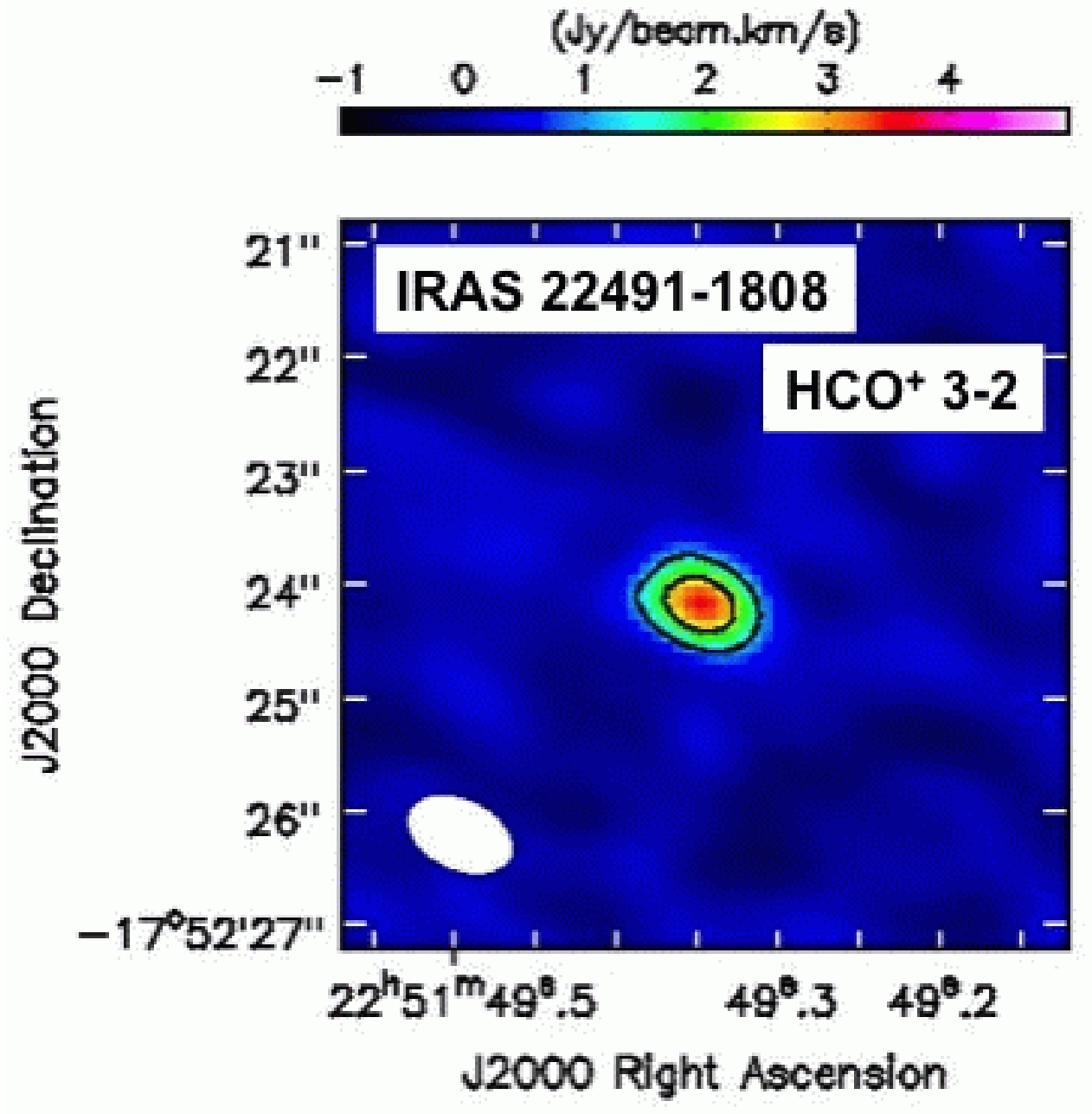} \\
\end{center}
\end{figure}

\clearpage

\begin{figure}
\begin{center}
\includegraphics[angle=0,scale=.43]{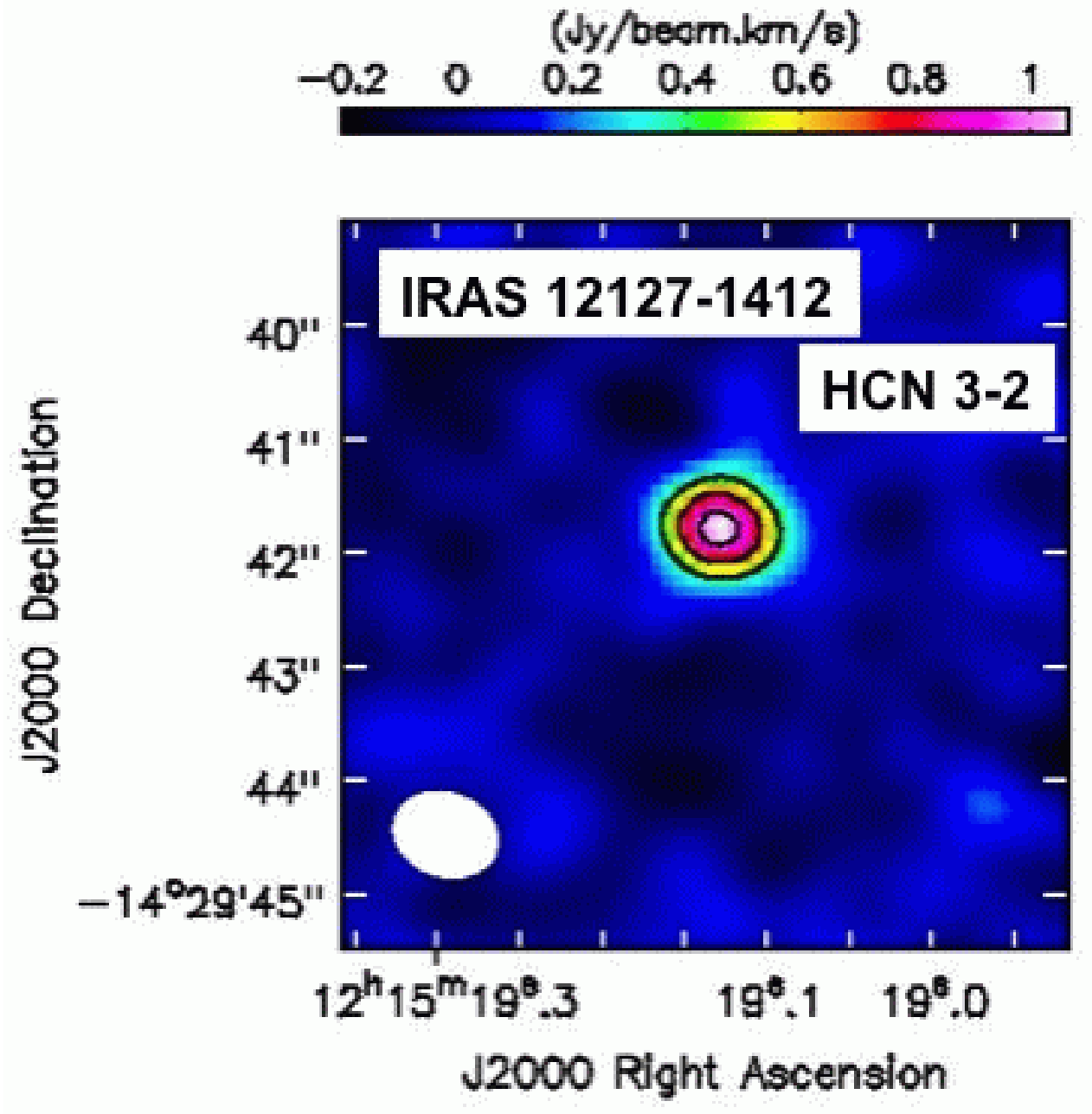} 
\includegraphics[angle=0,scale=.43]{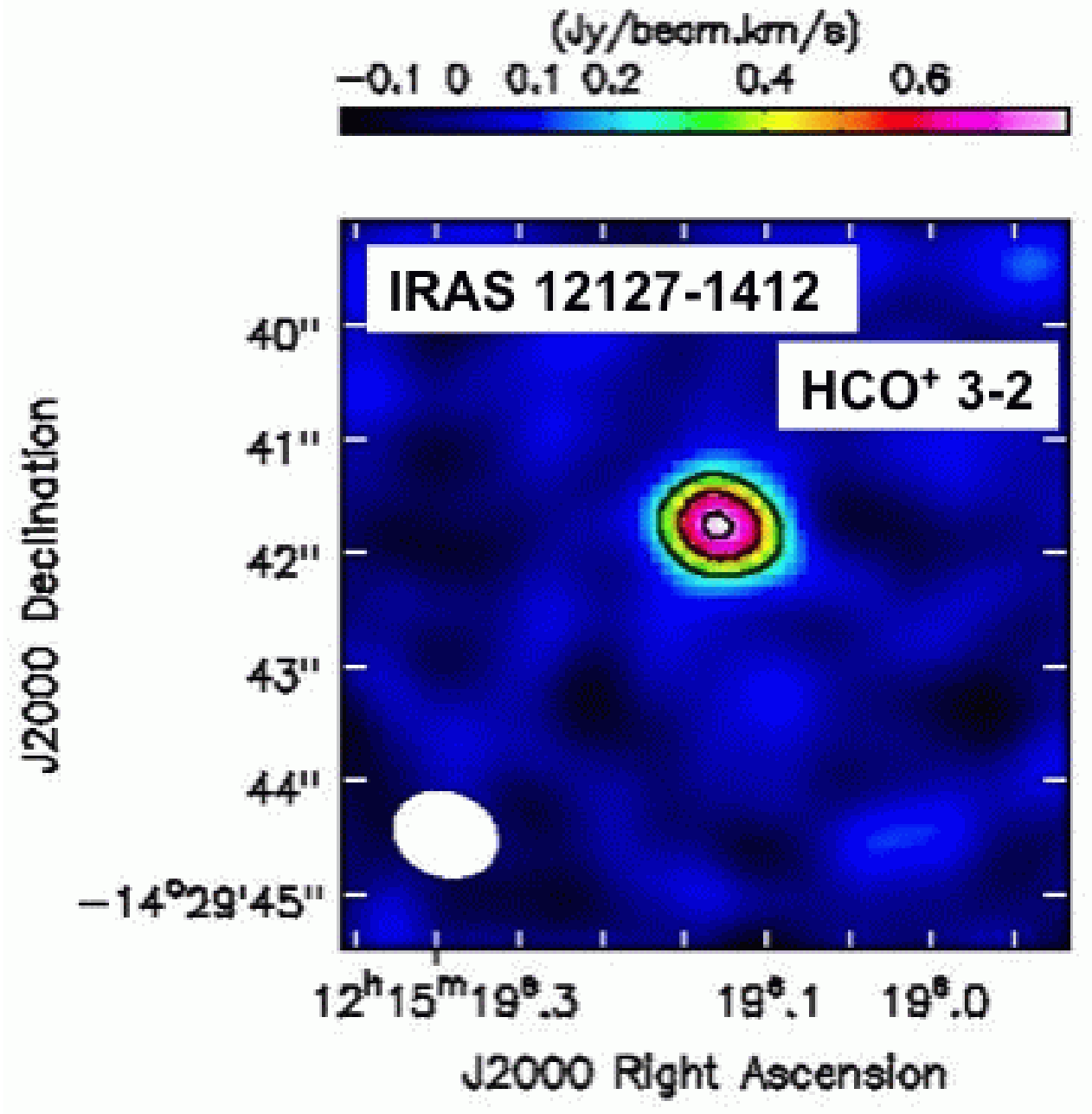} \\
\includegraphics[angle=0,scale=.41]{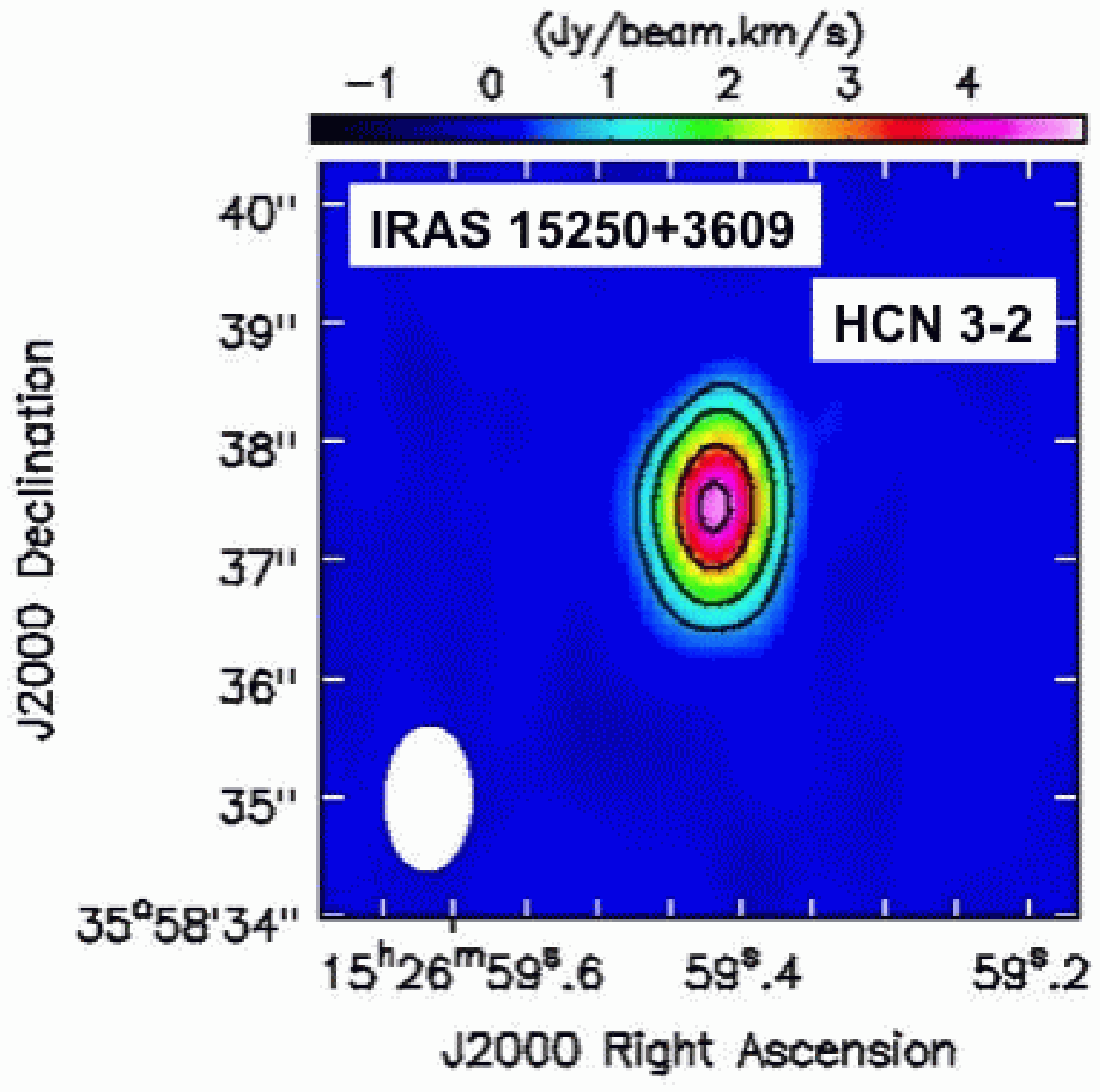} 
\includegraphics[angle=0,scale=.41]{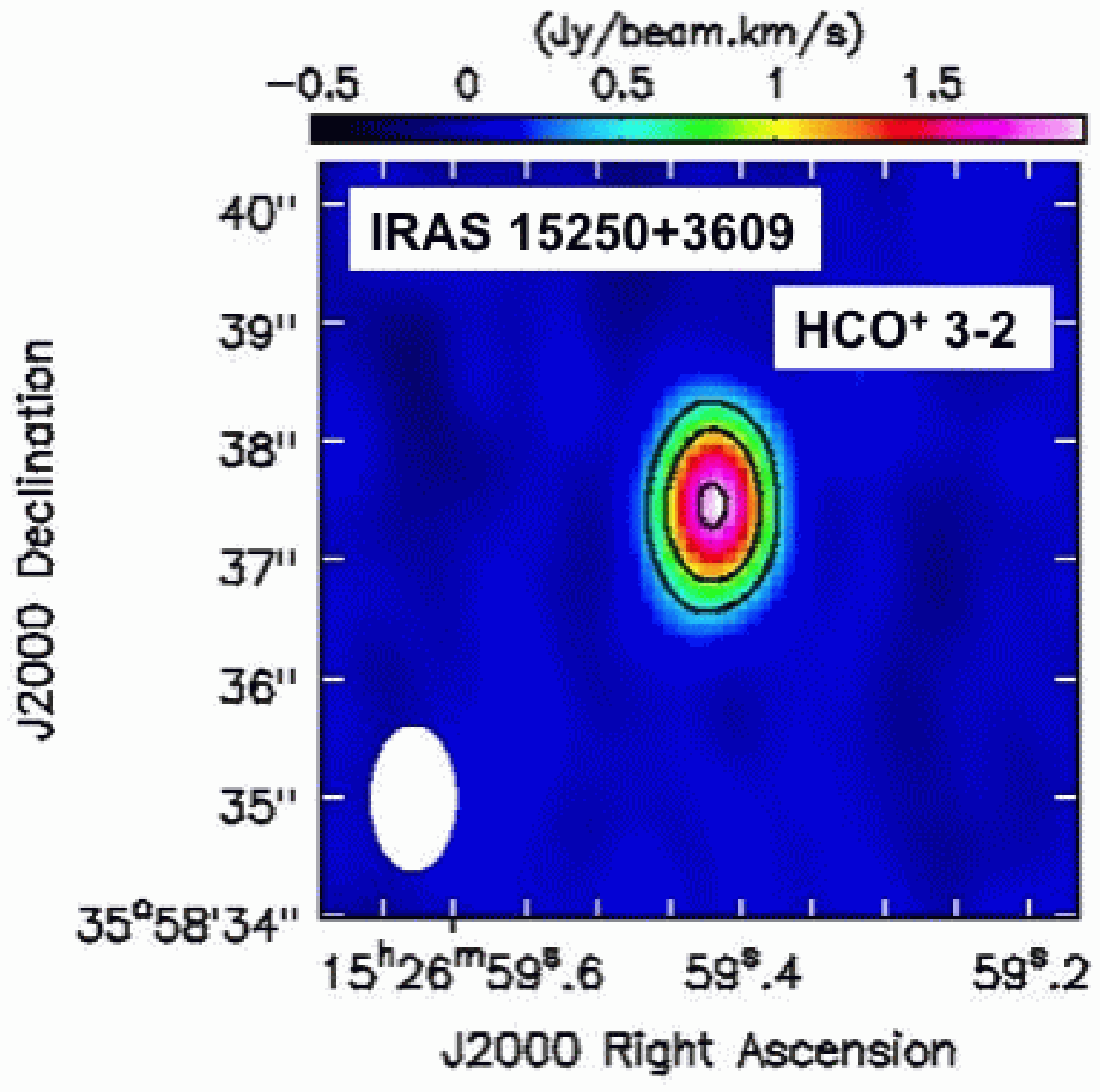} \\
\includegraphics[angle=0,scale=.41]{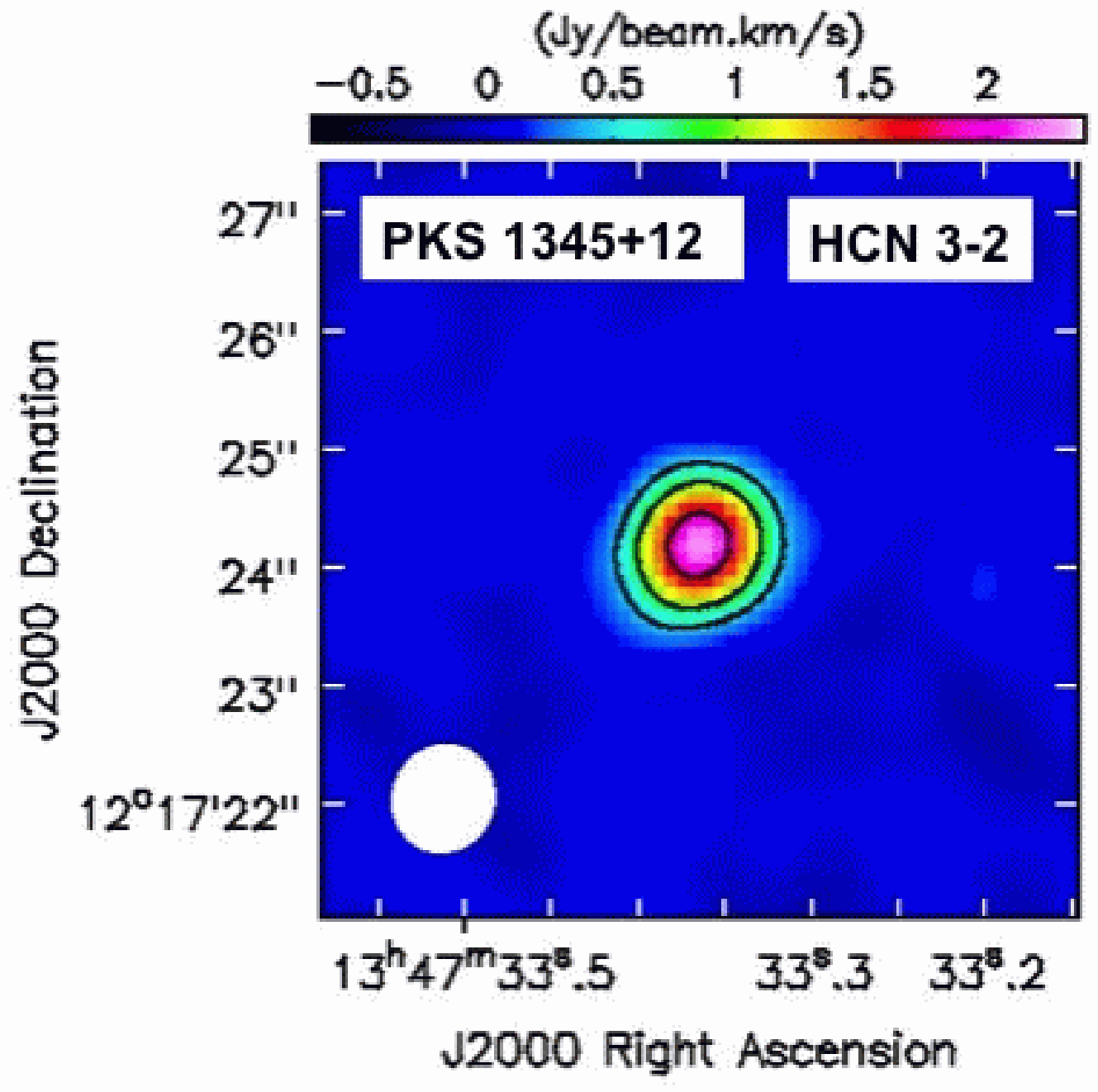} 
\includegraphics[angle=0,scale=.41]{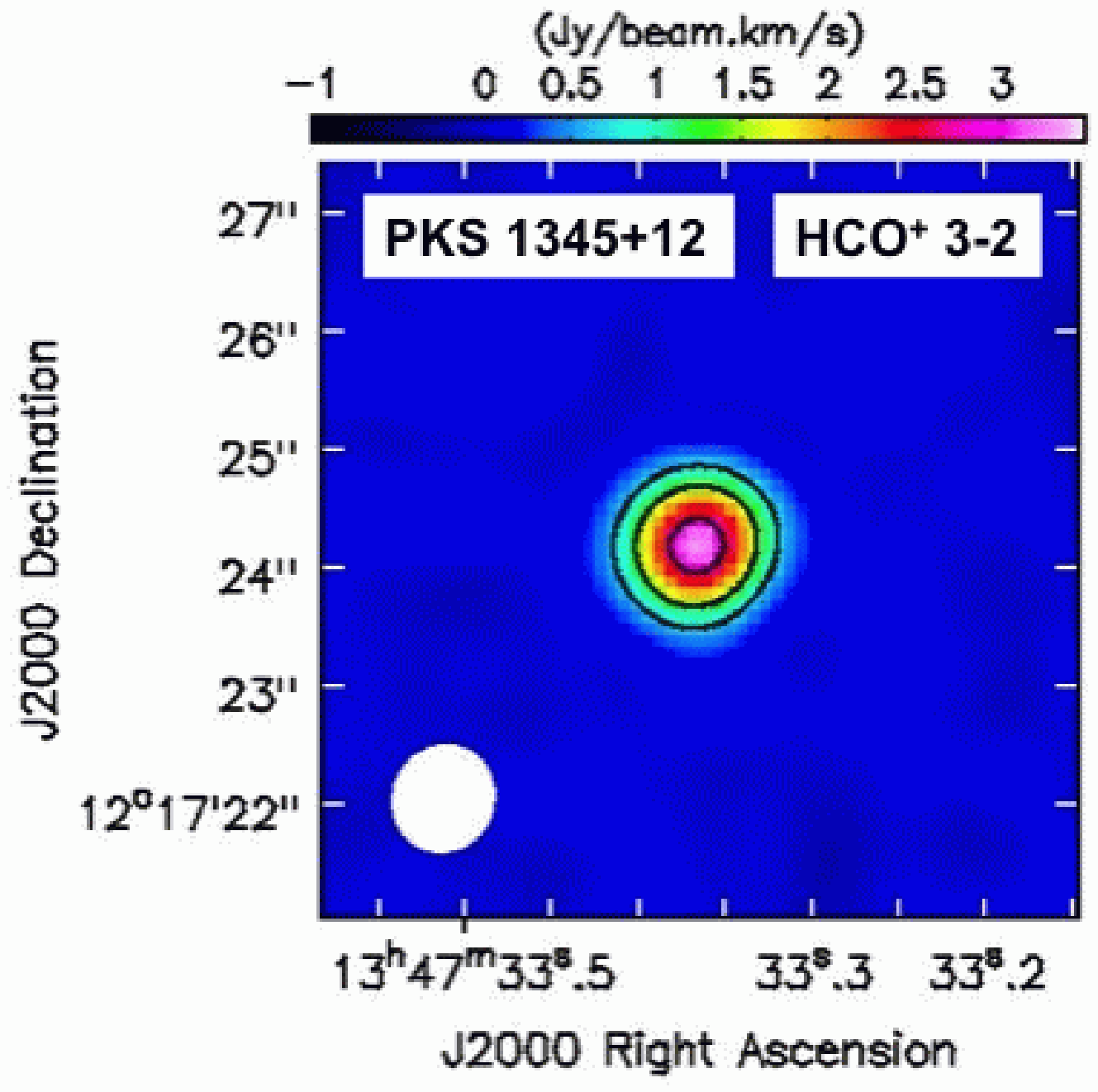} \\
\end{center}
\end{figure}

\clearpage

\begin{figure}
\begin{center}
\includegraphics[angle=0,scale=.41]{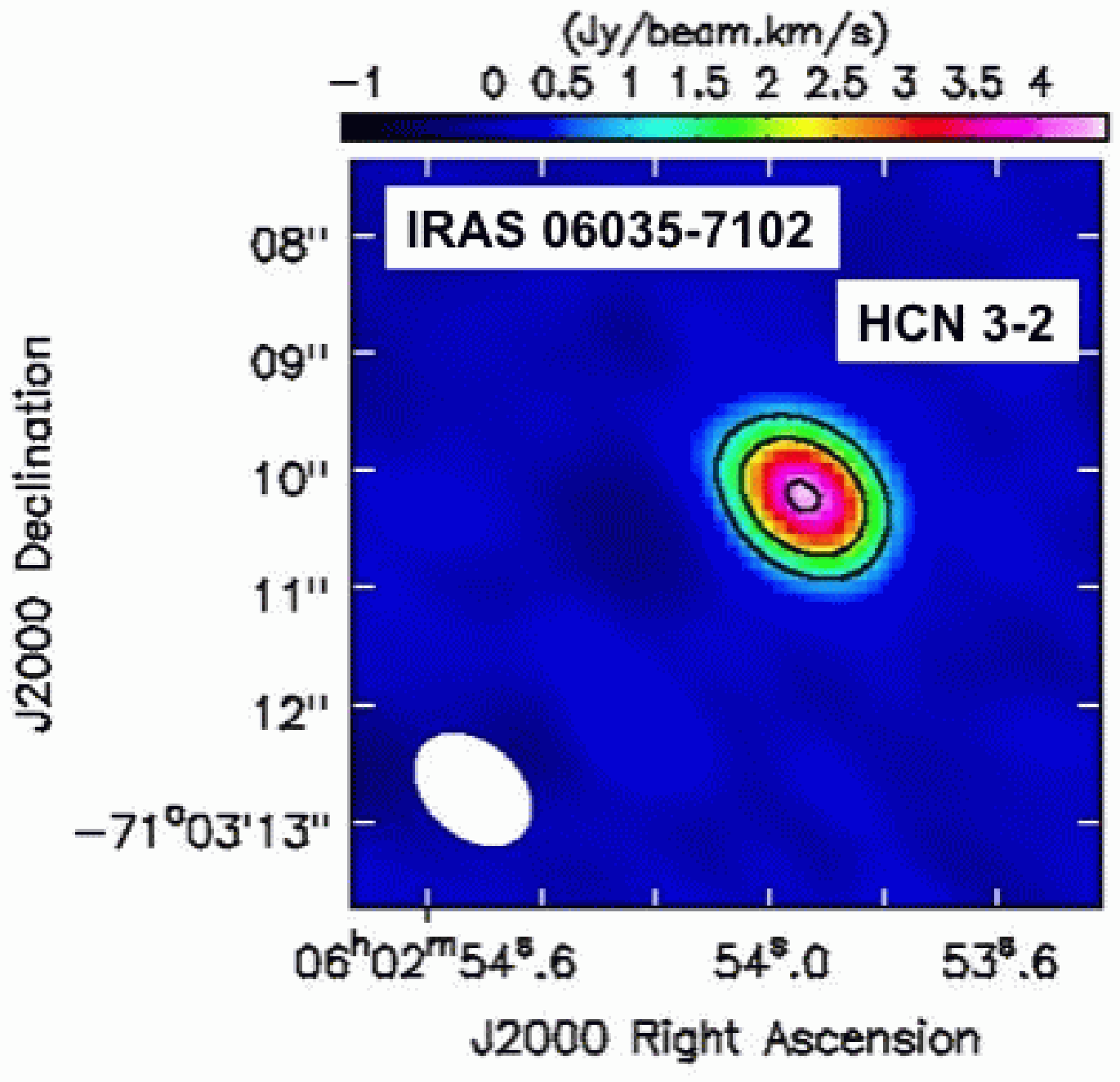} 
\includegraphics[angle=0,scale=.41]{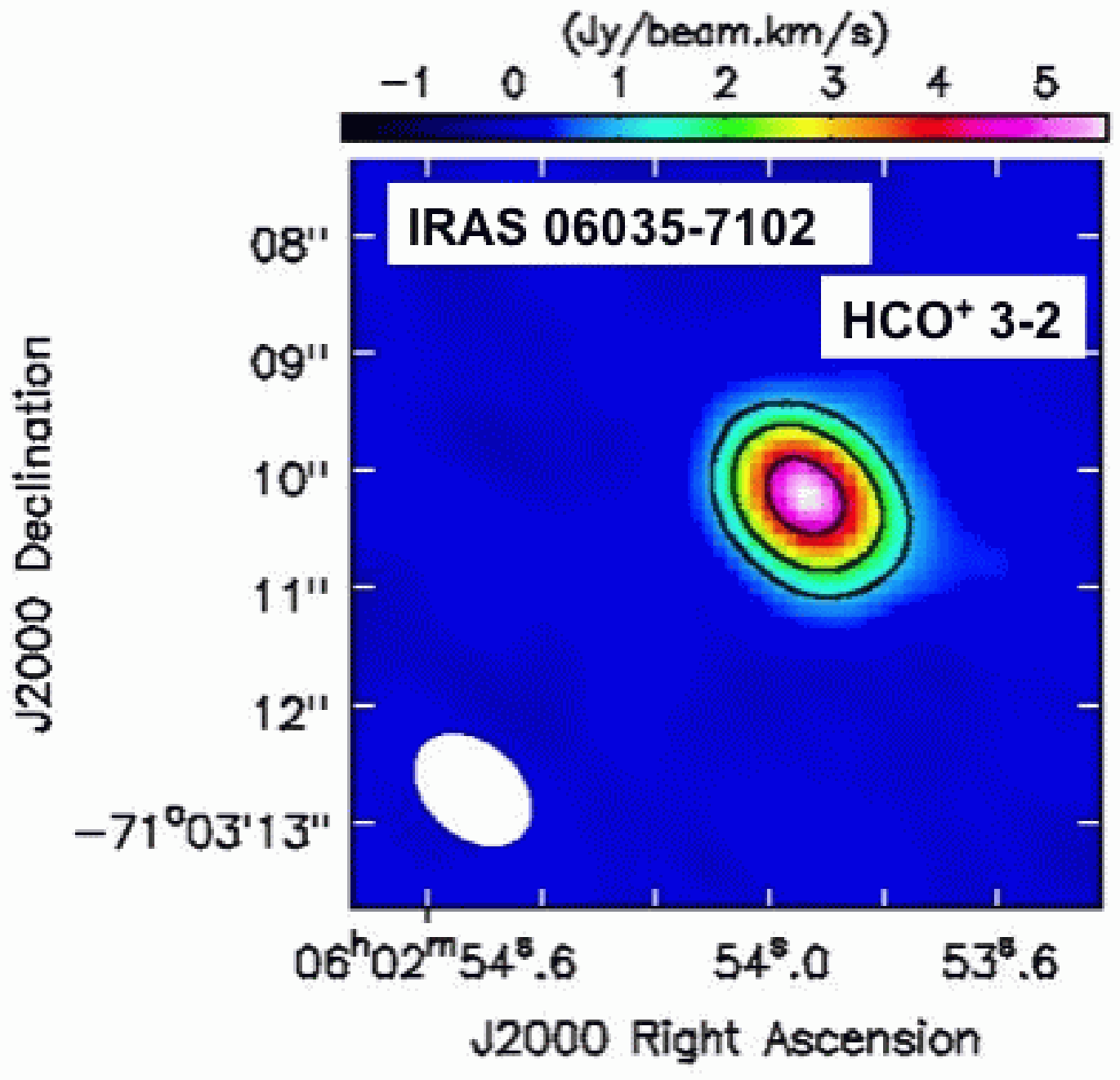} \\
\includegraphics[angle=0,scale=.41]{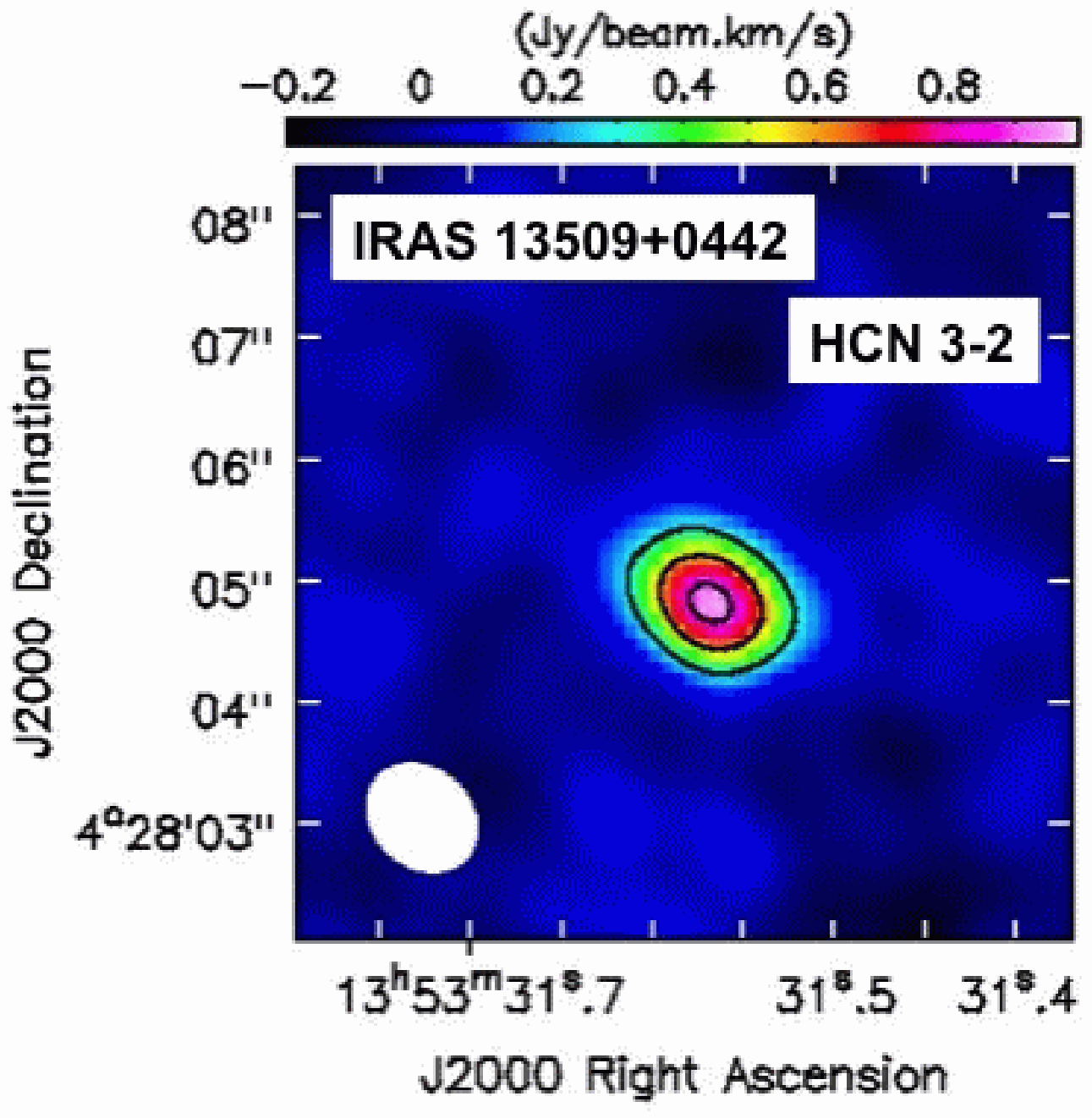} 
\includegraphics[angle=0,scale=.41]{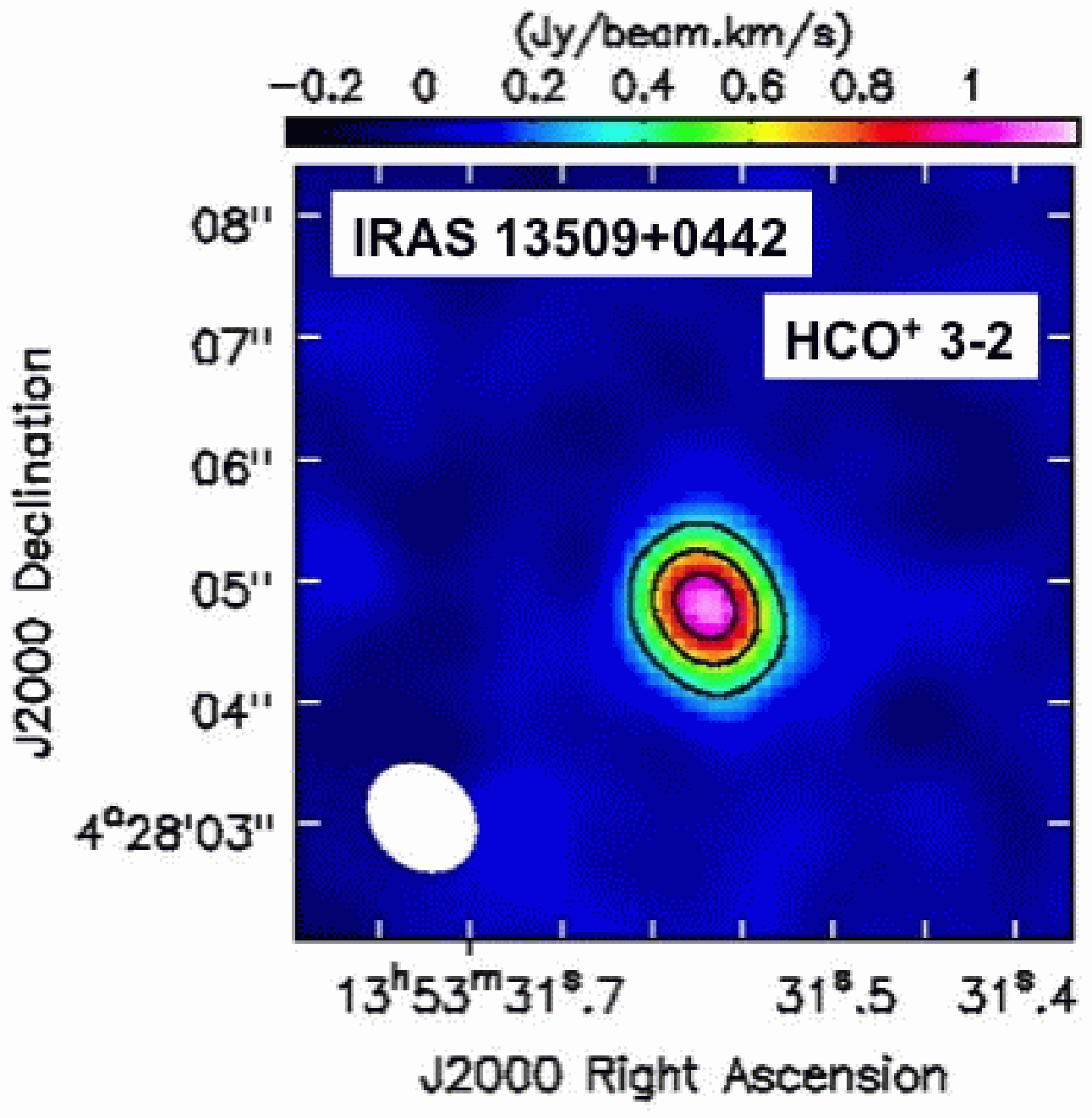} \\
\includegraphics[angle=0,scale=.43]{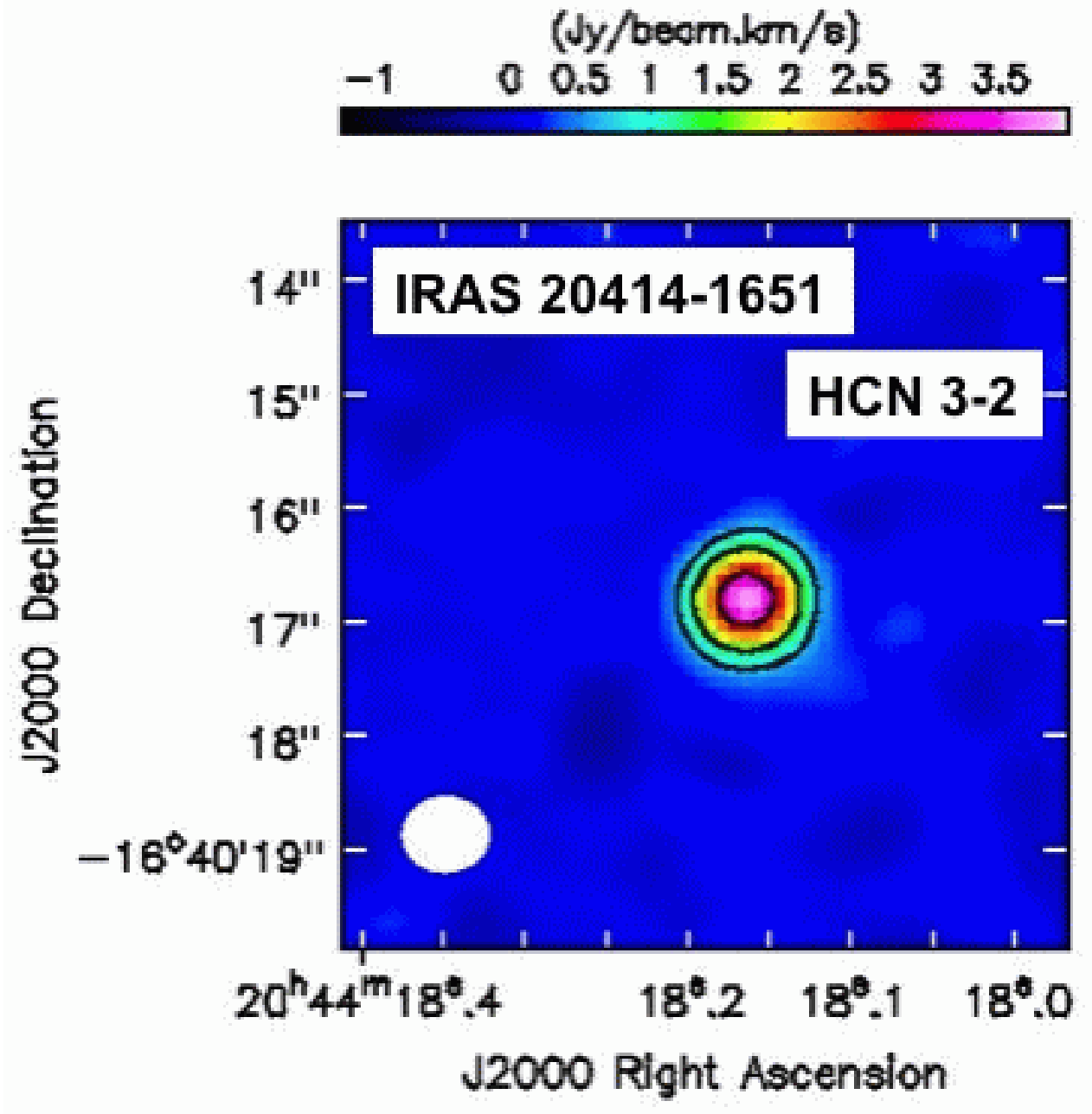} 
\includegraphics[angle=0,scale=.43]{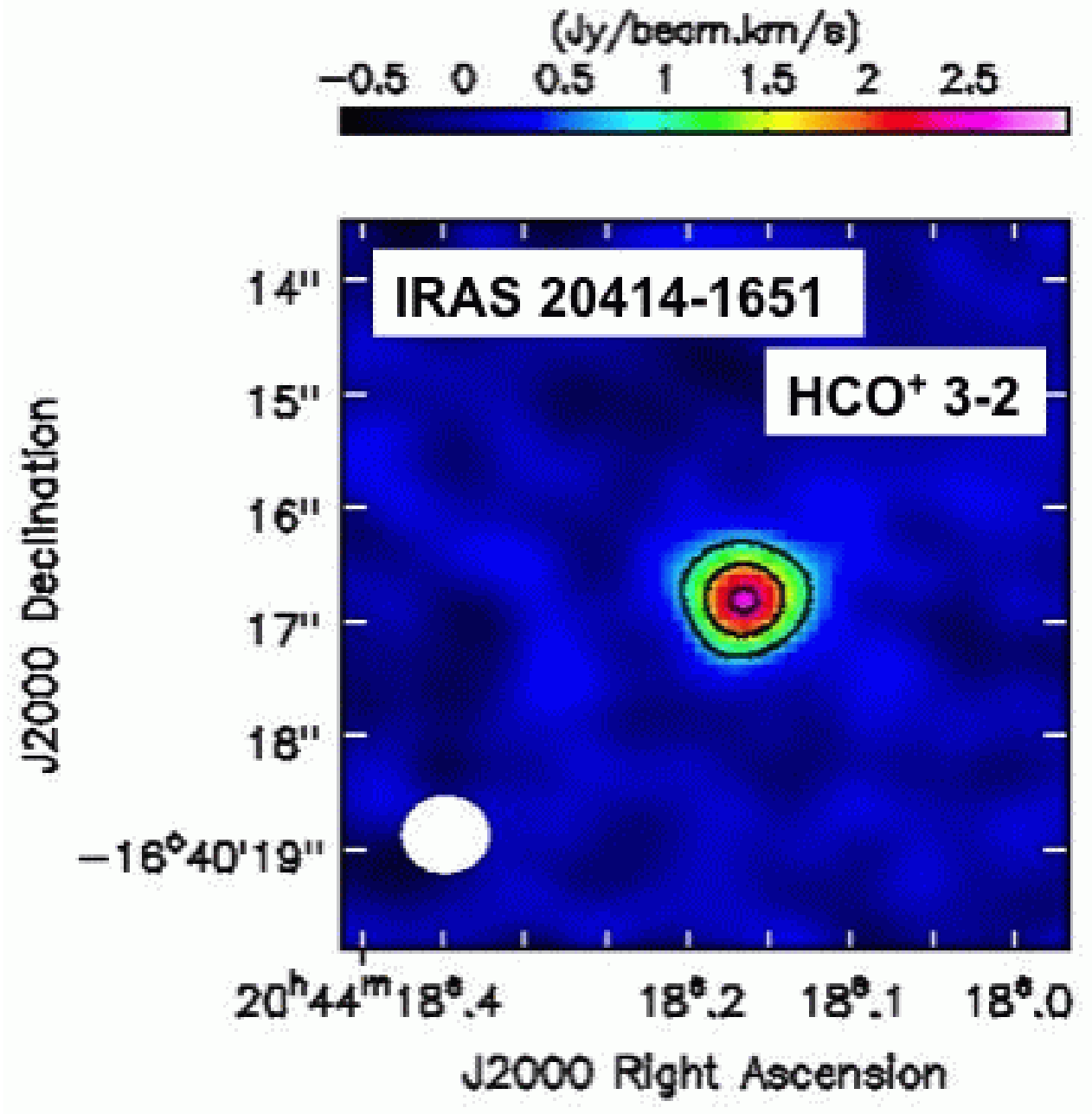} \\
\end{center}
\end{figure}

\clearpage

\begin{figure}
\caption{
Integrated intensity (moment 0) maps of HCN J=3--2 (left) and HCO$^{+}$
J=3--2 (right).
For I Zw 1, the contours are 5$\sigma$, 10$\sigma$, 20$\sigma$ for 
both HCN J=3--2 and HCO$^{+}$ J=3--2. 
For IRAS 08572$+$3915, the contours are 4$\sigma$, 8$\sigma$, 12$\sigma$
for both HCN J=3--2 and HCO$^{+}$ J=3--2.
For The Superantennae, the contours are 5$\sigma$, 10$\sigma$, 20$\sigma$
for HCN J=3--2, and 5$\sigma$, 10$\sigma$, 15$\sigma$ for HCO$^{+}$ J=3--2.
For IRAS 12112$+$0305 NE, the contours are 3$\sigma$, 5$\sigma$, 10$\sigma$,
20$\sigma$ for HCN J=3--2, and 3$\sigma$, 5$\sigma$, 10$\sigma$,
15$\sigma$ for HCO$^{+}$ J=3--2.
For IRAS 12112$+$0305 SW, the contours are 3$\sigma$, 4$\sigma$ for HCN 
J=3--2, and 4$\sigma$, 5$\sigma$, 6$\sigma$ for HCO$^{+}$ J=3--2.
For IRAS 22491$-$1808, the contours are 5$\sigma$, 10$\sigma$,
20$\sigma$ for HCN J=3--2, and 5$\sigma$, 10$\sigma$ for HCO$^{+}$ J=3--2.
For IRAS 12127$-$1412, the contours are 5$\sigma$, 8$\sigma$,
11$\sigma$ for HCN J=3--2, and 4$\sigma$, 7$\sigma$,
10$\sigma$ for HCO$^{+}$ J=3--2.
For IRAS 15250$+$3609, the contours are 5$\sigma$, 10$\sigma$,
20$\sigma$, 30$\sigma$ for HCN J=3--2, and 5$\sigma$, 10$\sigma$,
20$\sigma$ for HCO$^{+}$ J=3--2.
For PKS 1345$+$12, the contours are 4$\sigma$, 8$\sigma$,
16$\sigma$ for HCN J=3--2, and 6$\sigma$, 12$\sigma$, 24$\sigma$ for
HCO$^{+}$ J=3--2.
For IRAS 06035$-$7102, the contours are 6$\sigma$, 12$\sigma$,
24$\sigma$ for both HCN J=3--2 and HCO$^{+}$ J=3--2.
For IRAS 13509$+$0442, the contours are 4$\sigma$, 8$\sigma$,
12$\sigma$ for HCN J=3--2, and 5$\sigma$, 10$\sigma$, 15$\sigma$ for
HCO$^{+}$ J=3--2.
For IRAS 20414$-$1651, the contours are 5$\sigma$, 10$\sigma$,
20$\sigma$ for HCN J=3--2, and 4$\sigma$, 8$\sigma$, 12$\sigma$ for
HCO$^{+}$ J=3--2.
} 
\end{figure}

\begin{figure}
\begin{center}
\includegraphics[angle=0,scale=.35]{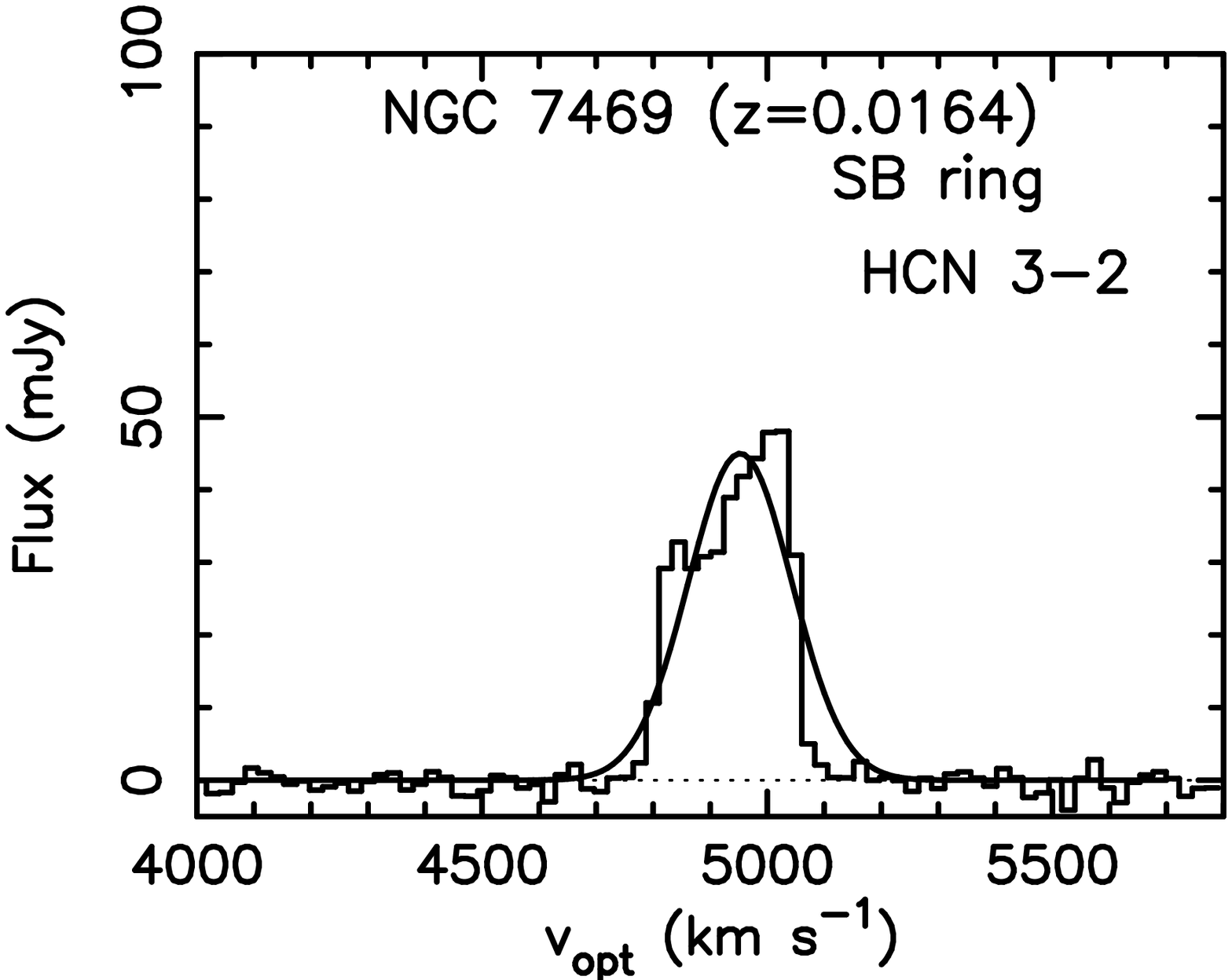} 
\includegraphics[angle=0,scale=.35]{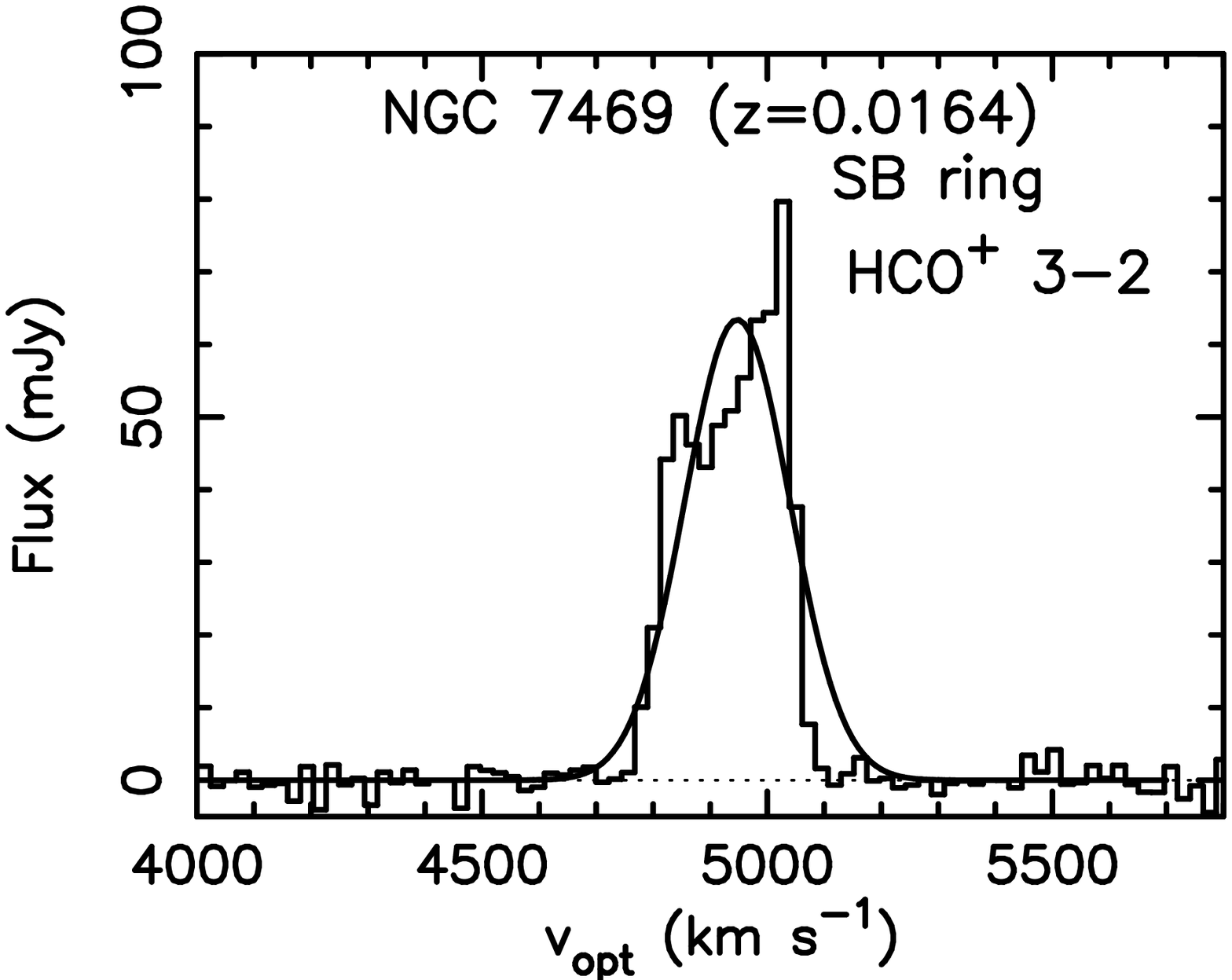} \\
\includegraphics[angle=0,scale=.35]{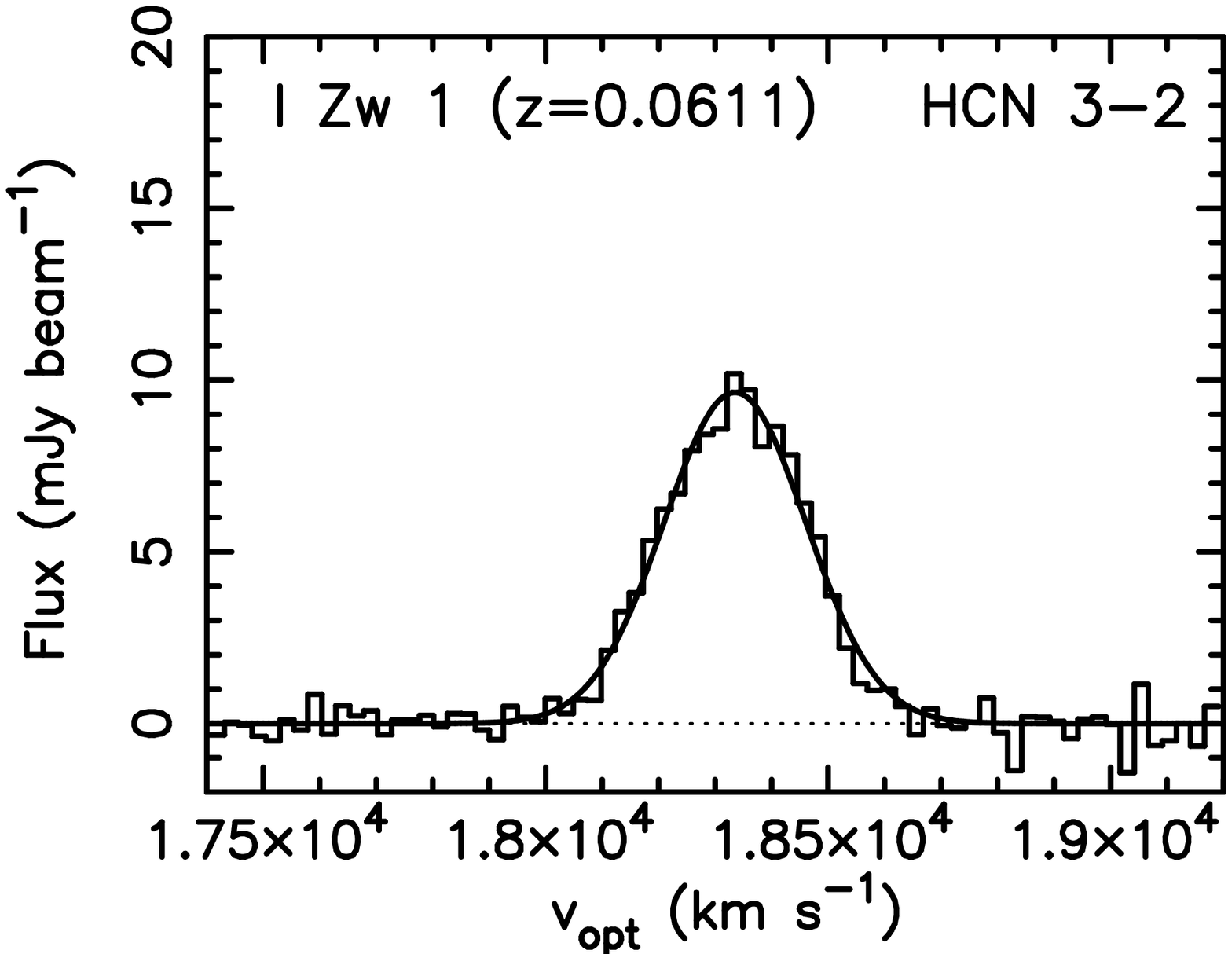} 
\includegraphics[angle=0,scale=.35]{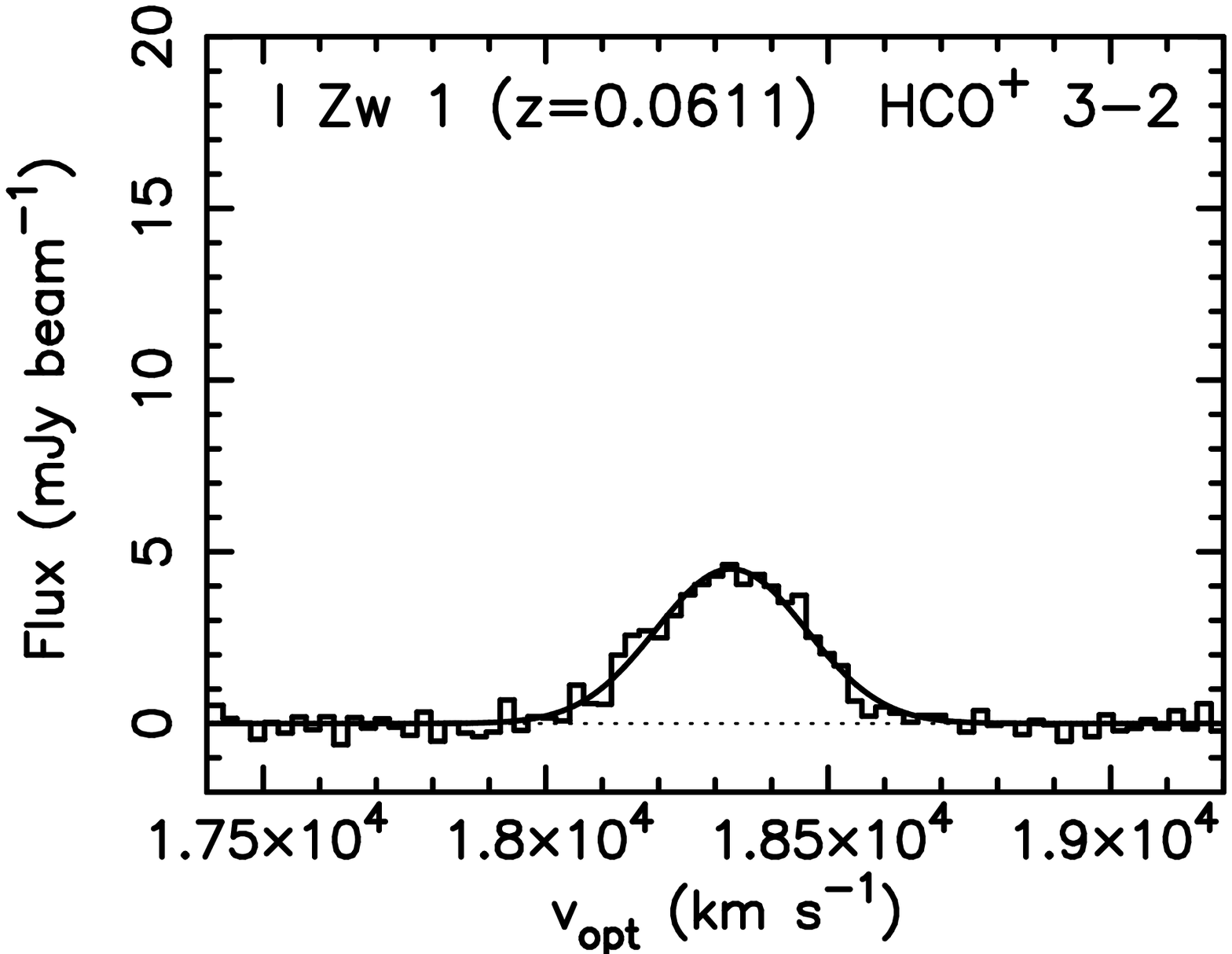} \\
\includegraphics[angle=0,scale=.35]{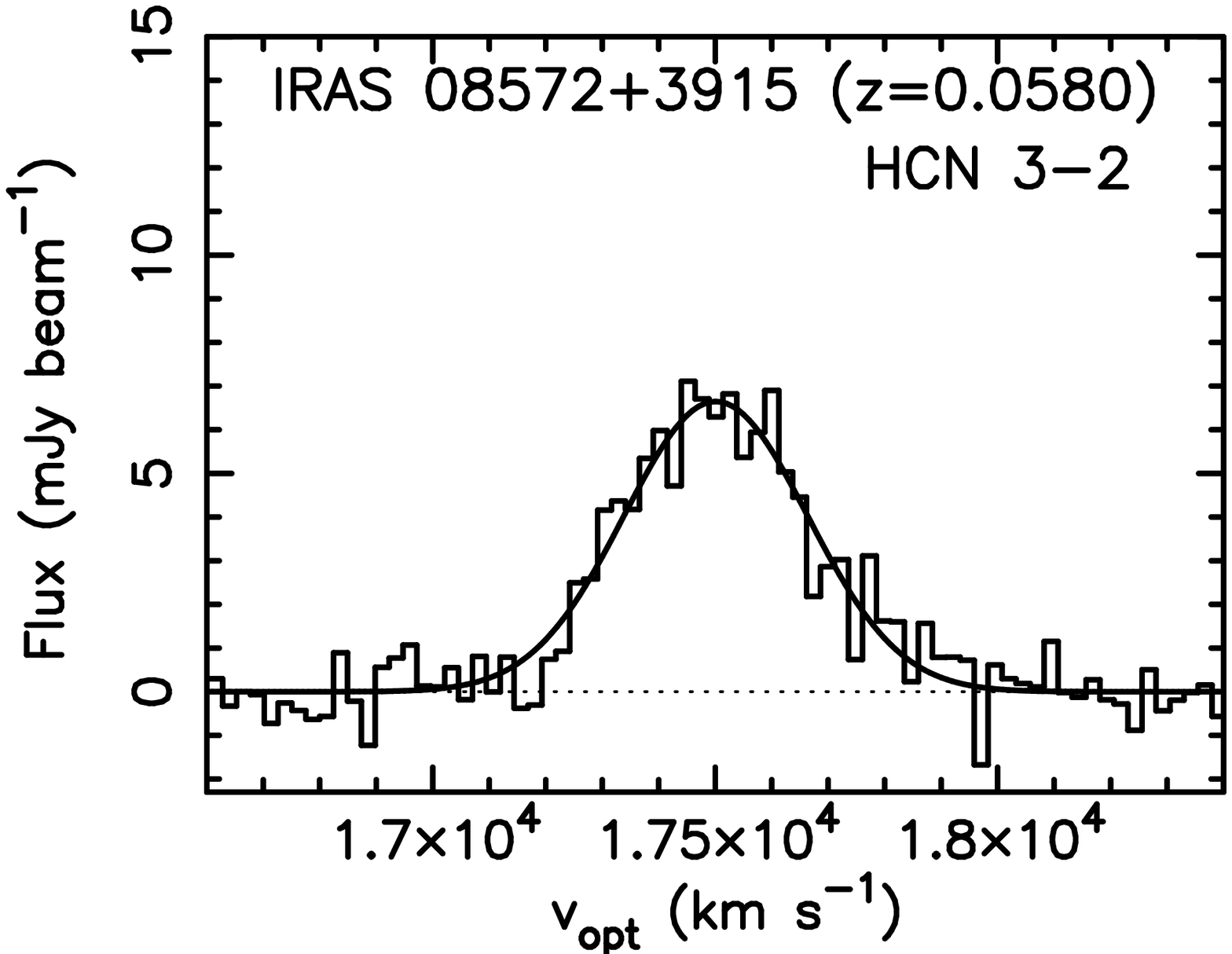} 
\includegraphics[angle=0,scale=.35]{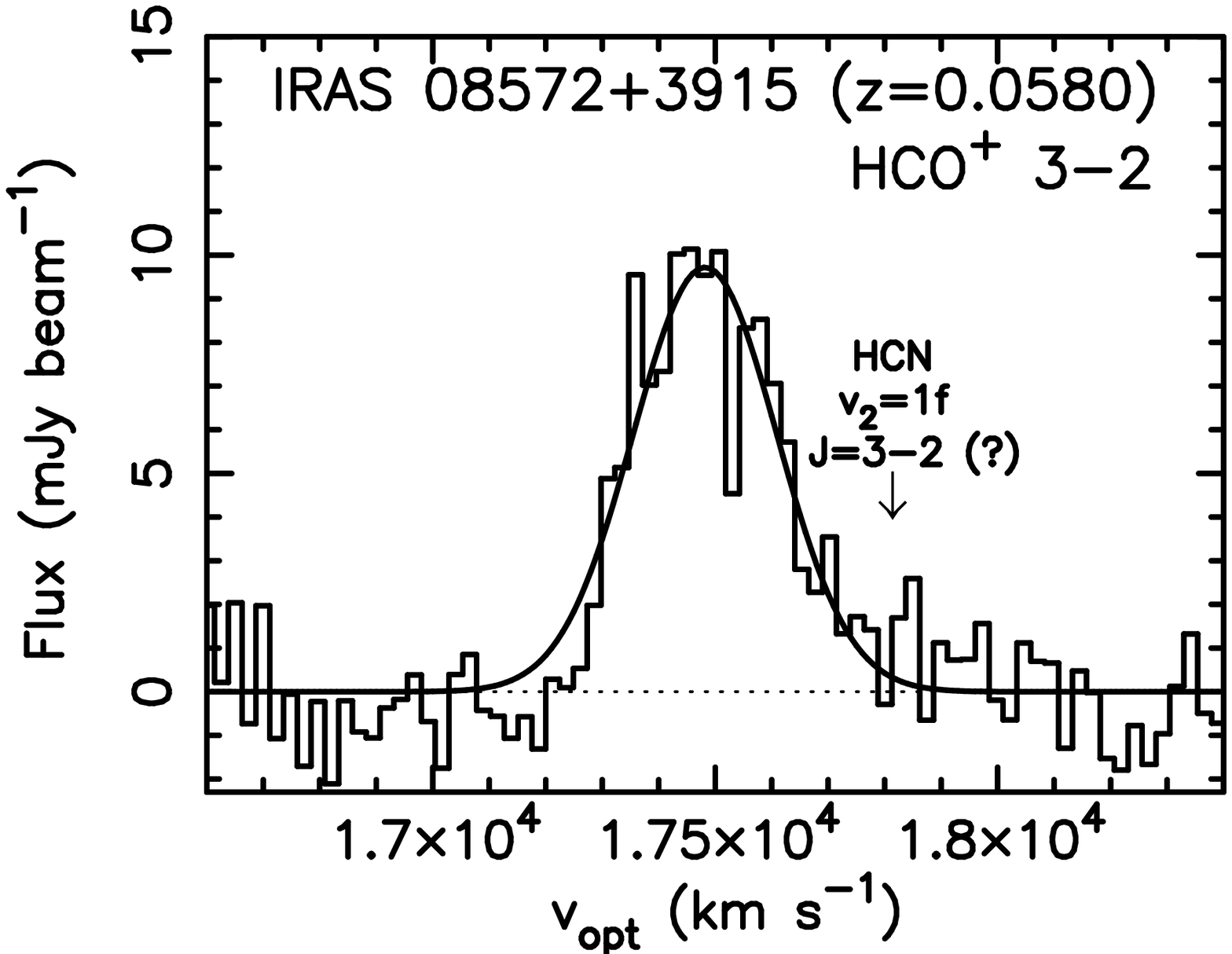} \\
\includegraphics[angle=0,scale=.35]{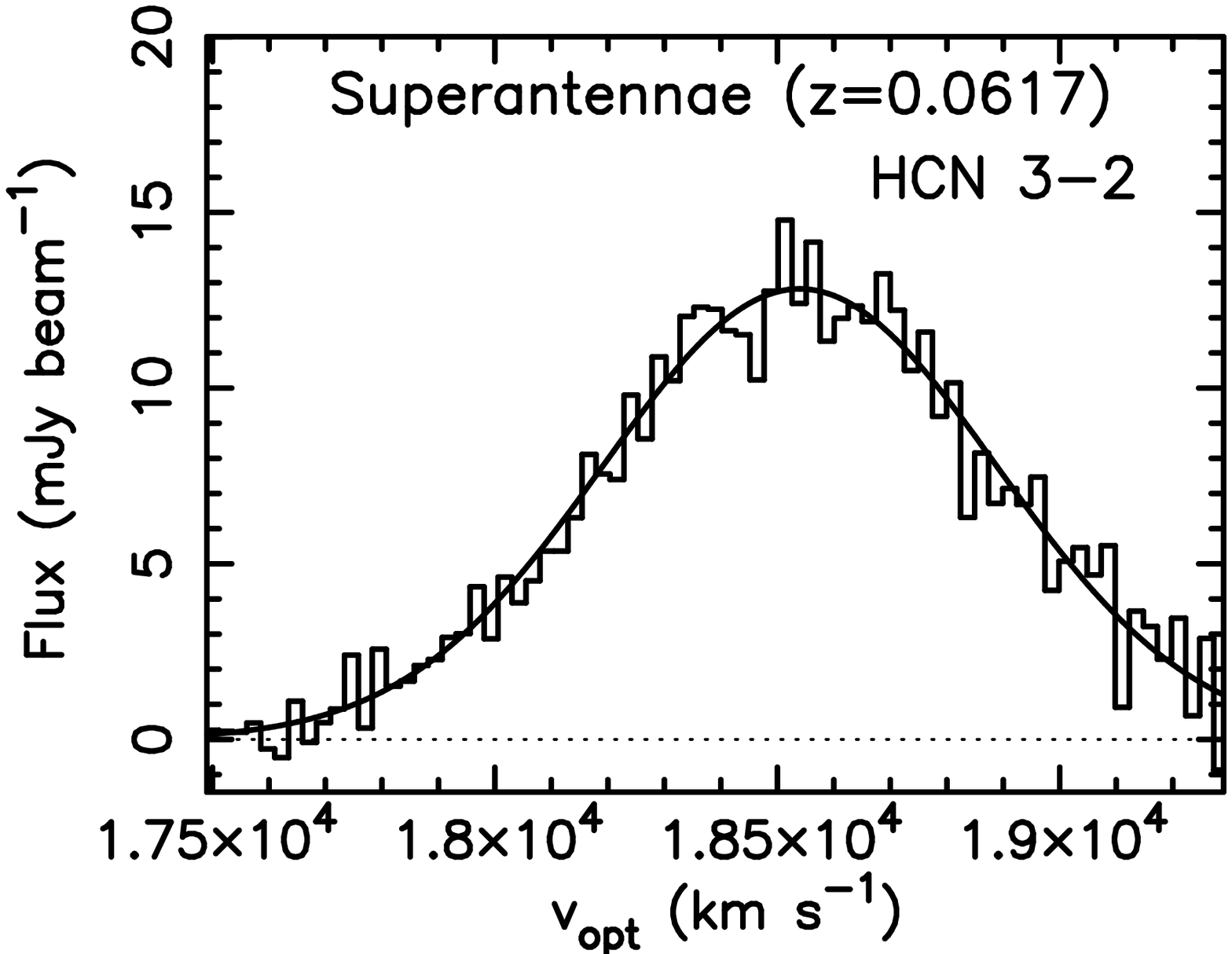}  
\includegraphics[angle=0,scale=.35]{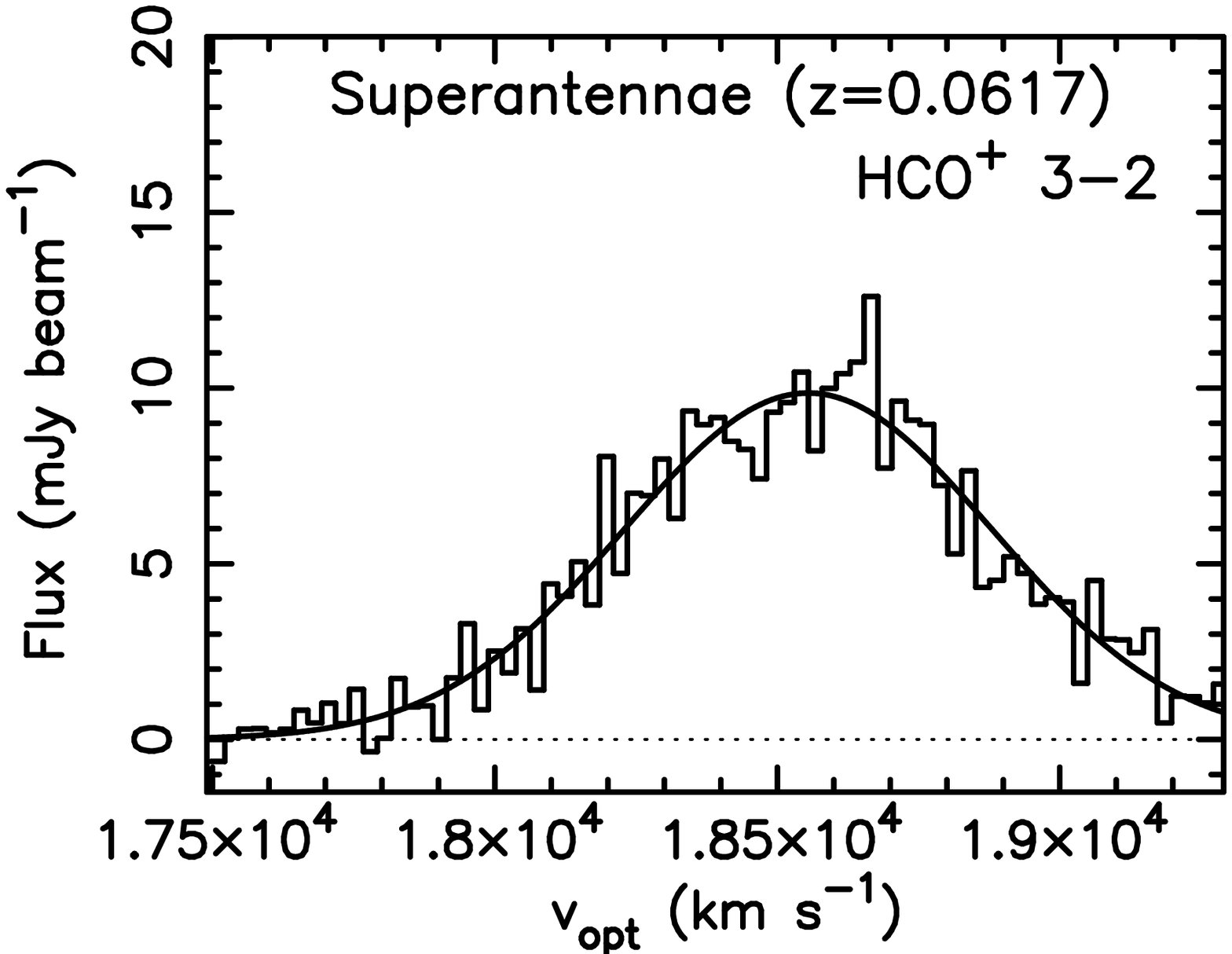} \\
\end{center}
\end{figure}

\clearpage

\begin{figure}
\begin{center}
\includegraphics[angle=0,scale=.35]{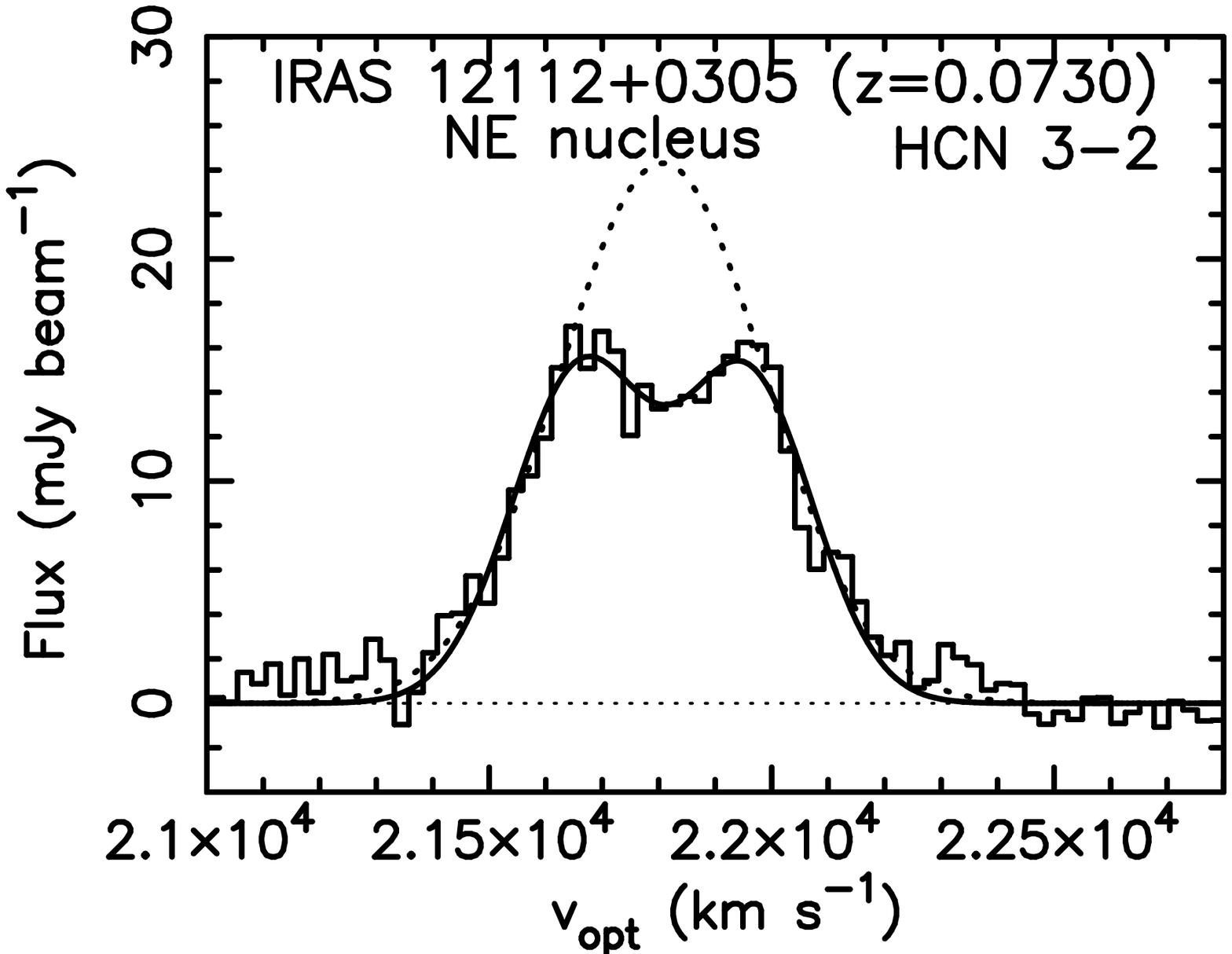} 
\includegraphics[angle=0,scale=.35]{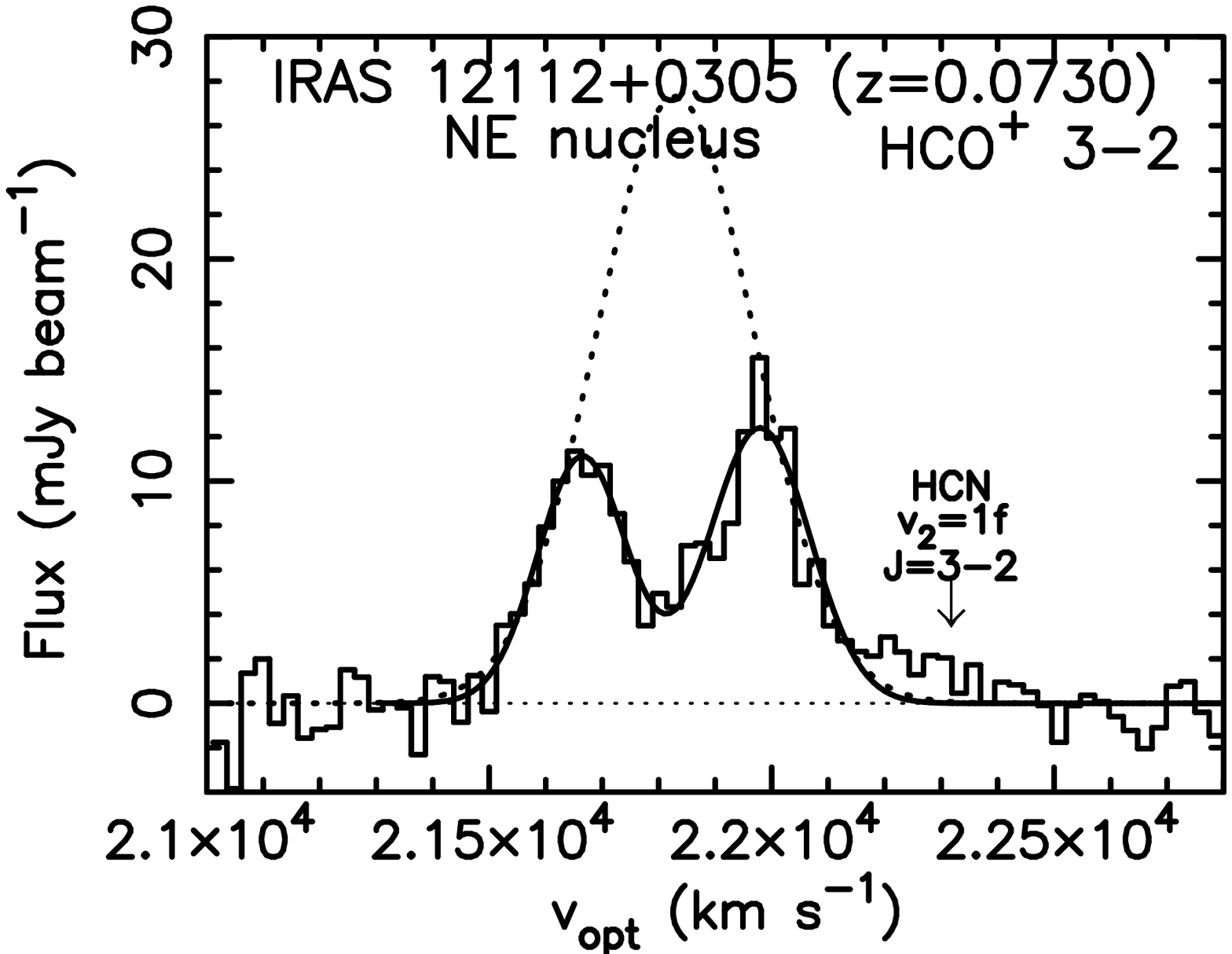} \\
\vspace*{-0.3cm}
\includegraphics[angle=0,scale=.35]{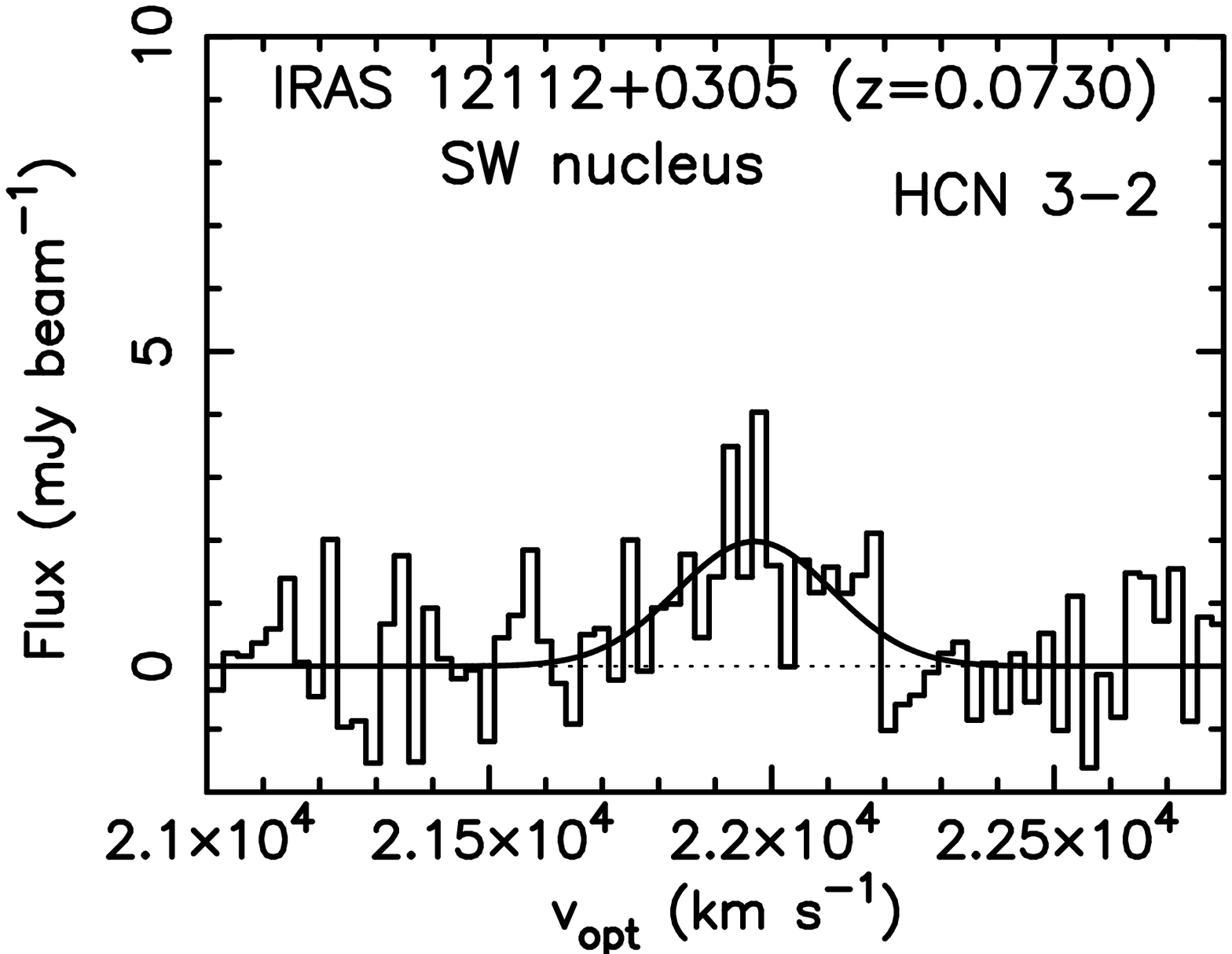} 
\includegraphics[angle=0,scale=.35]{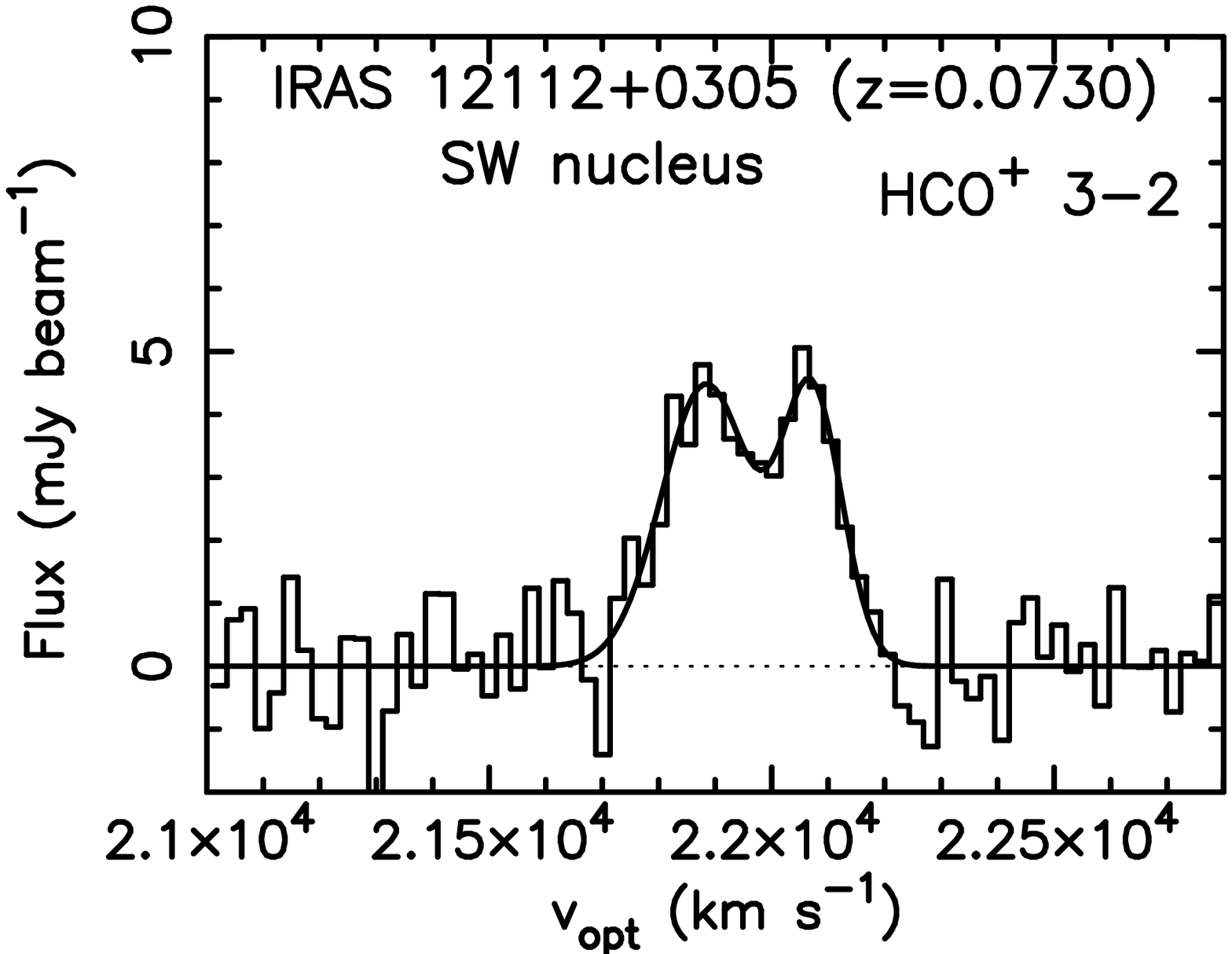} \\
\vspace*{-0.3cm}
\includegraphics[angle=0,scale=.35]{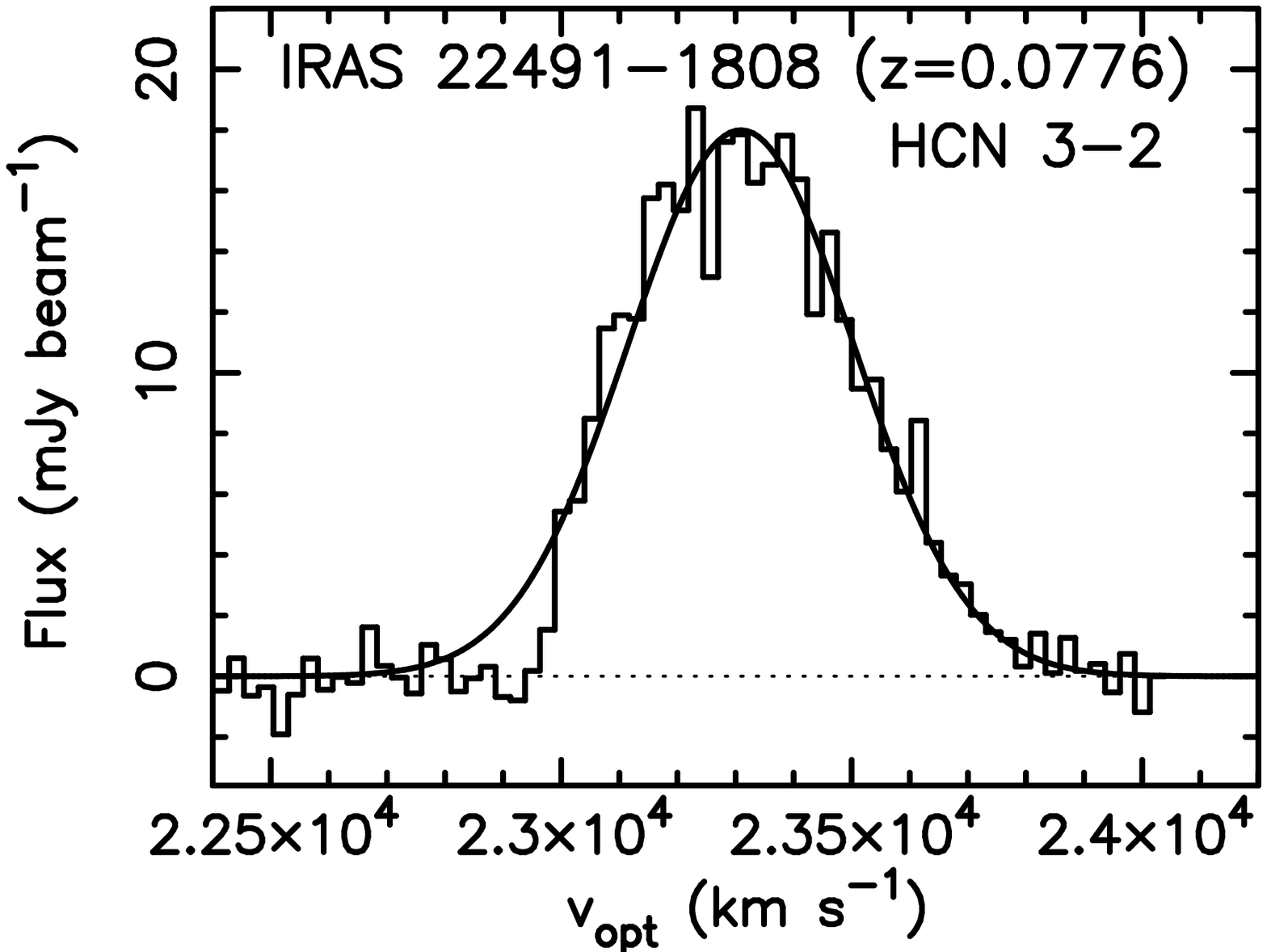} 
\includegraphics[angle=0,scale=.35]{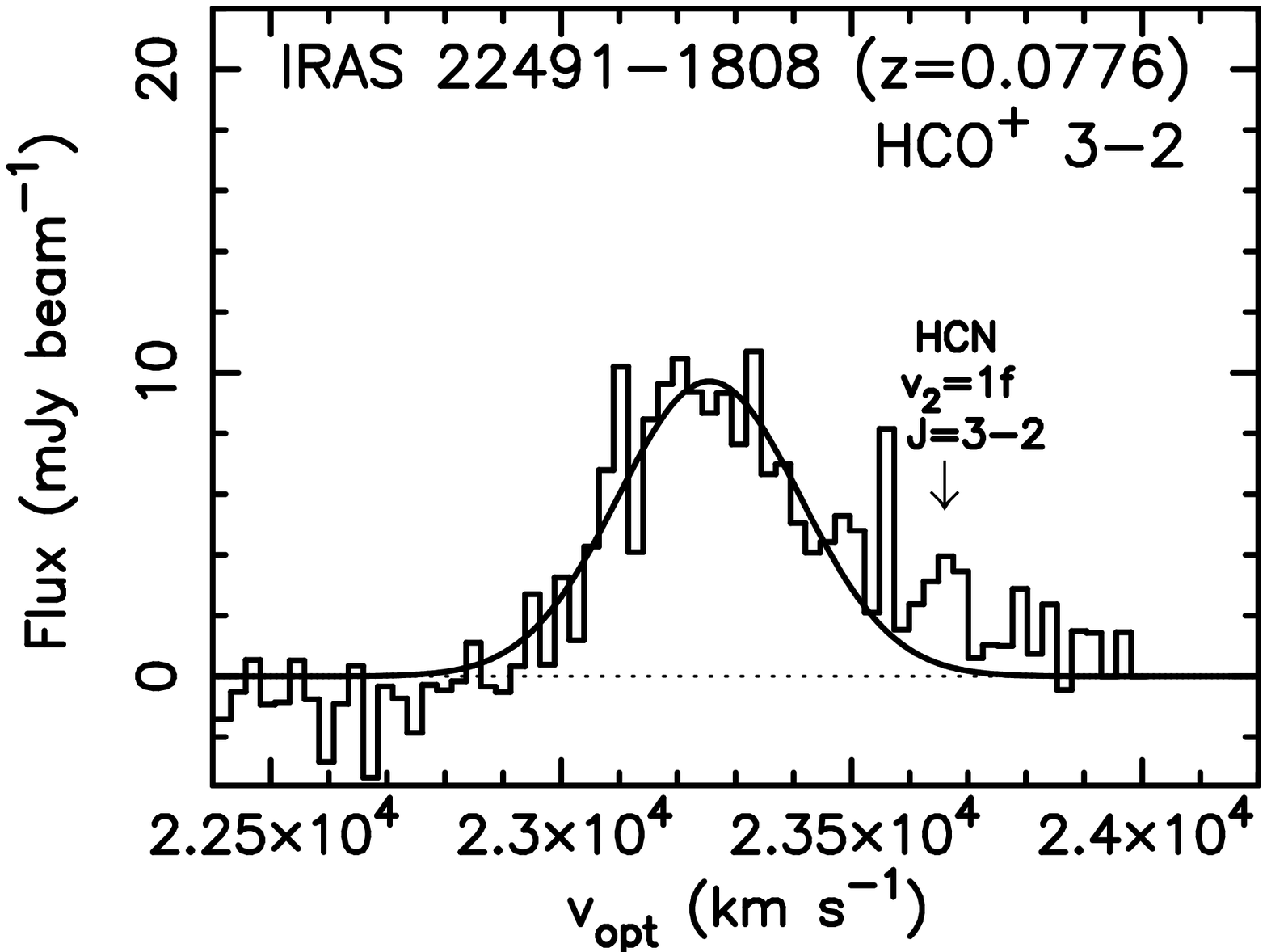} \\
\vspace*{-0.3cm}
\includegraphics[angle=0,scale=.35]{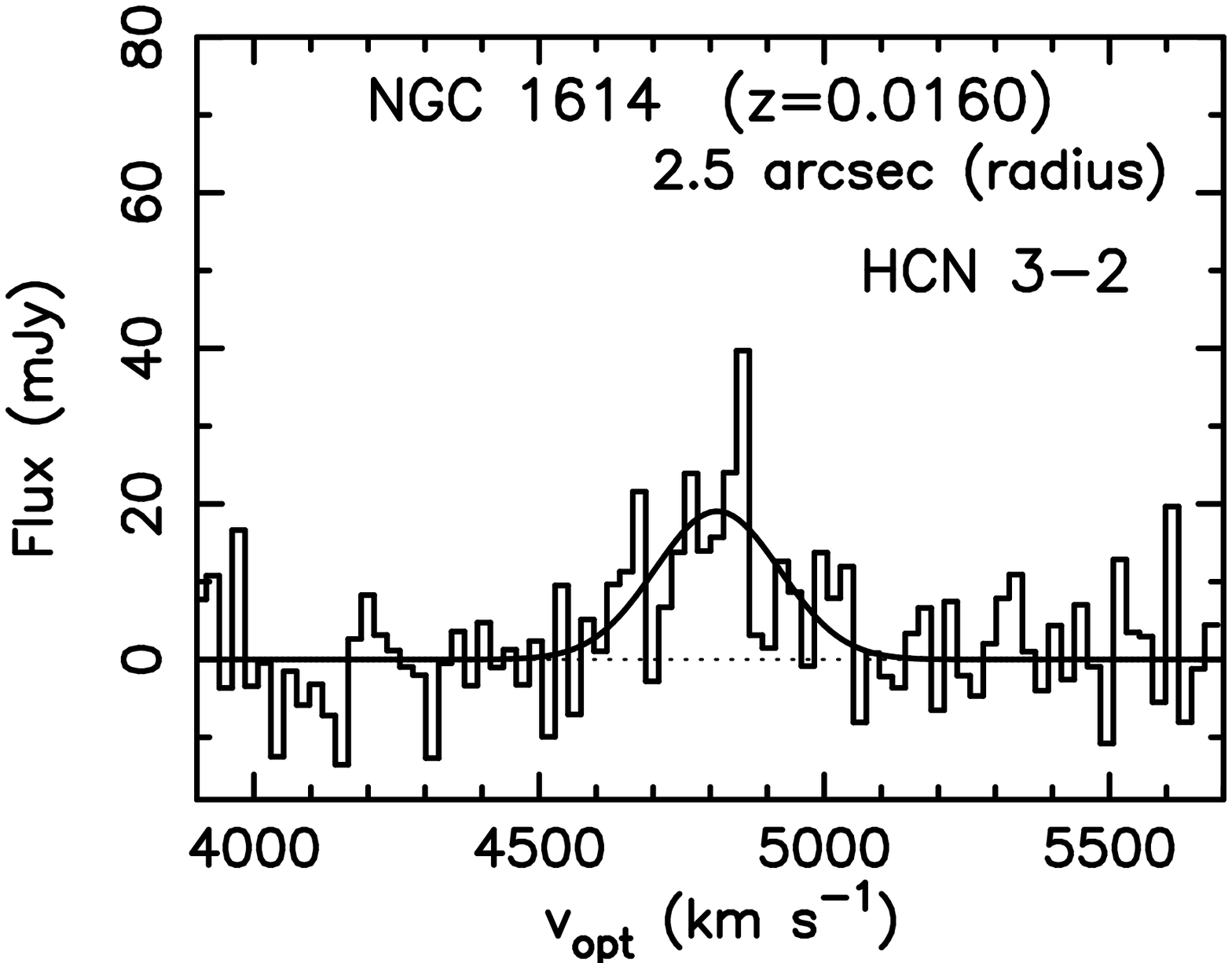} 
\includegraphics[angle=0,scale=.35]{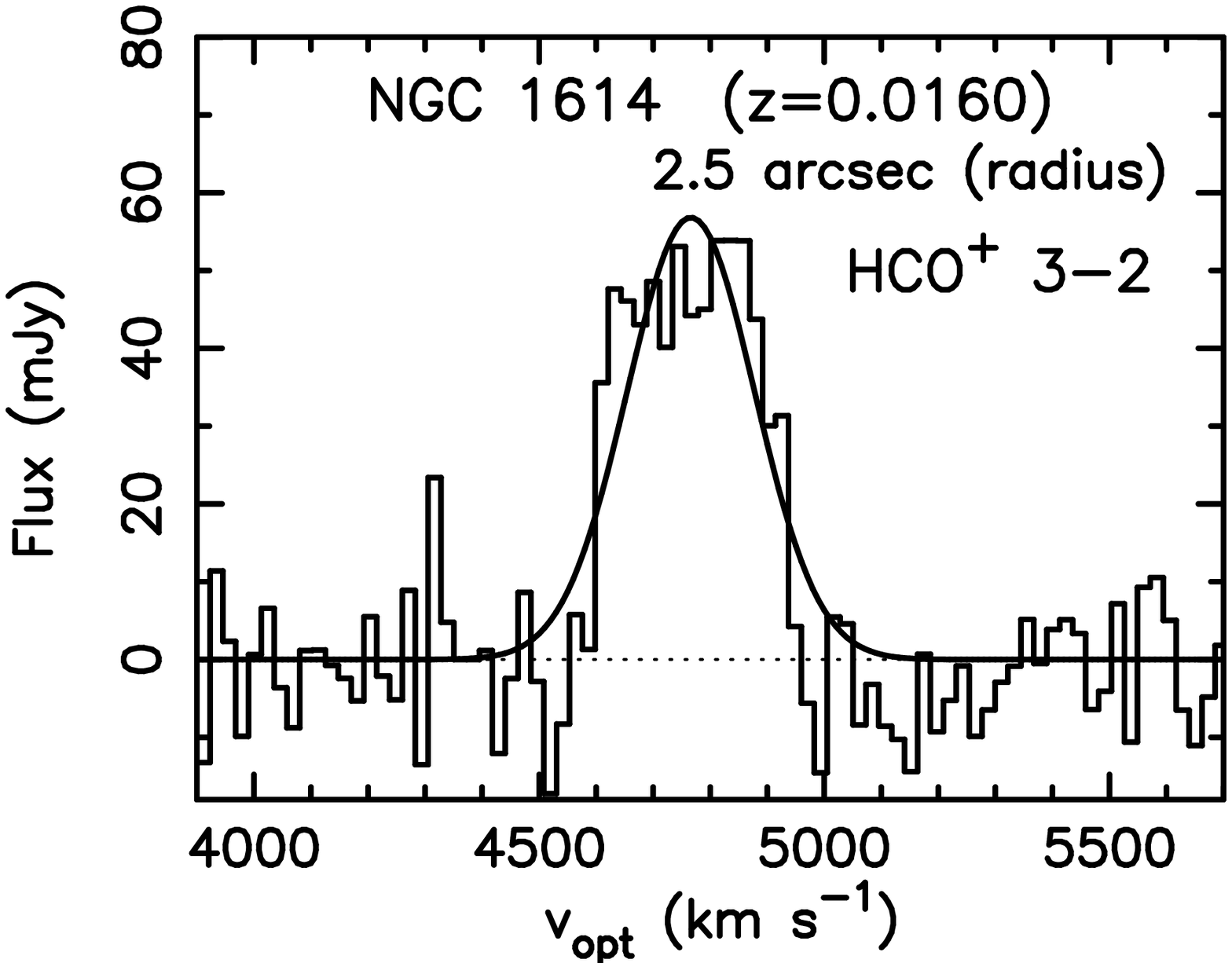} \\
\end{center}
\end{figure}

\clearpage

\begin{figure}
\begin{center}
\includegraphics[angle=0,scale=.35]{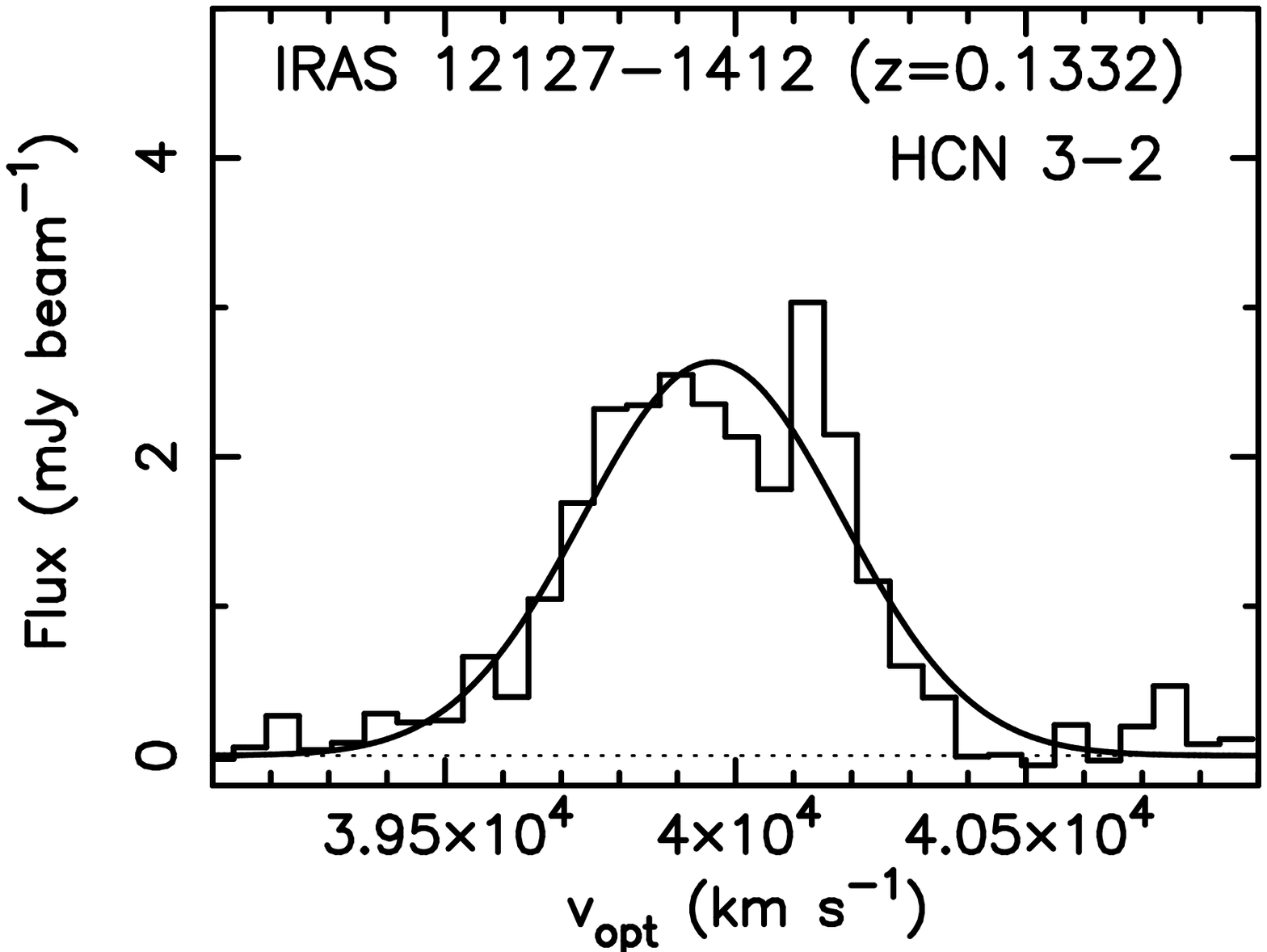} 
\includegraphics[angle=0,scale=.35]{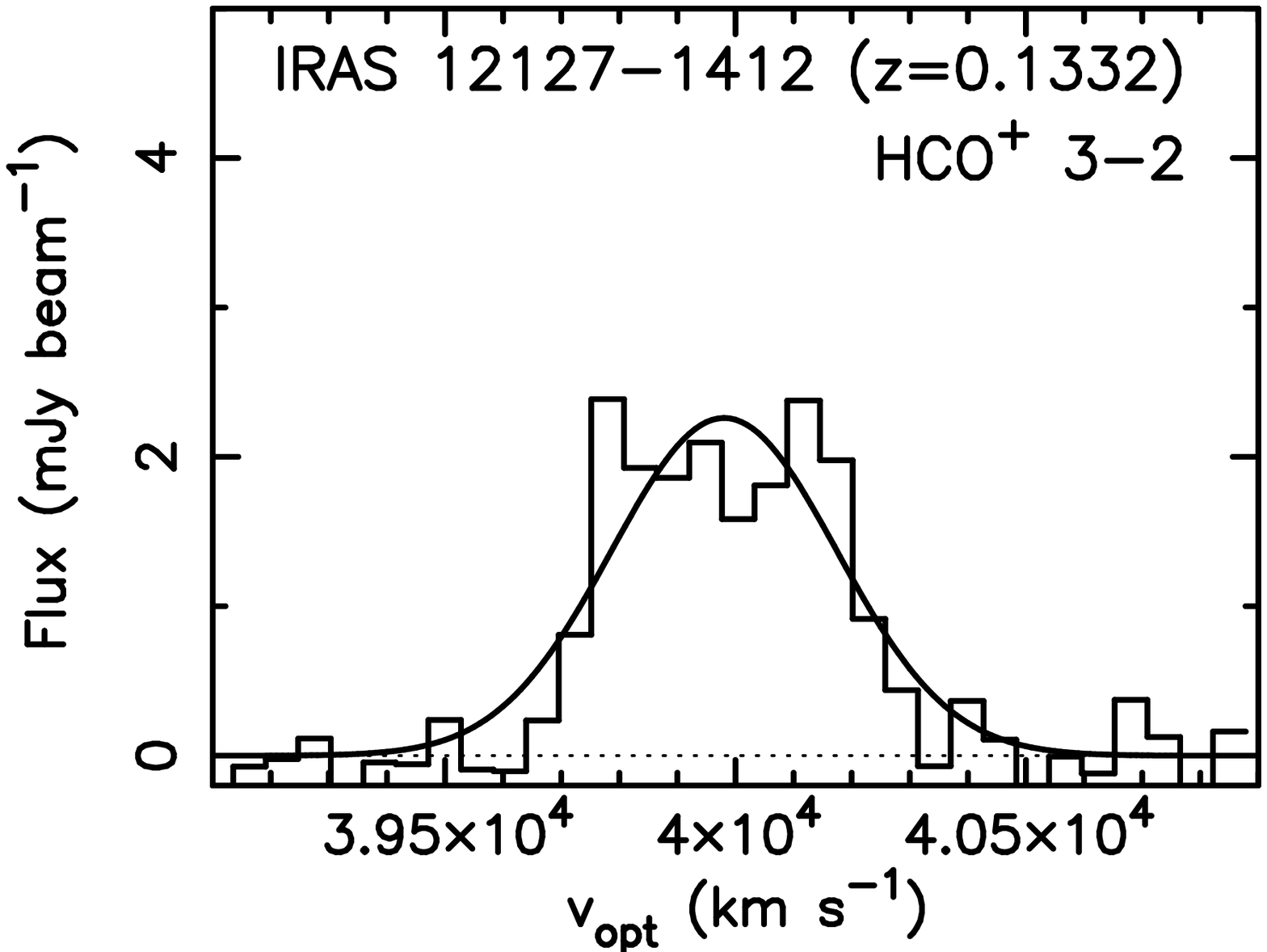} \\
\vspace*{-0.3cm}
\includegraphics[angle=0,scale=.35]{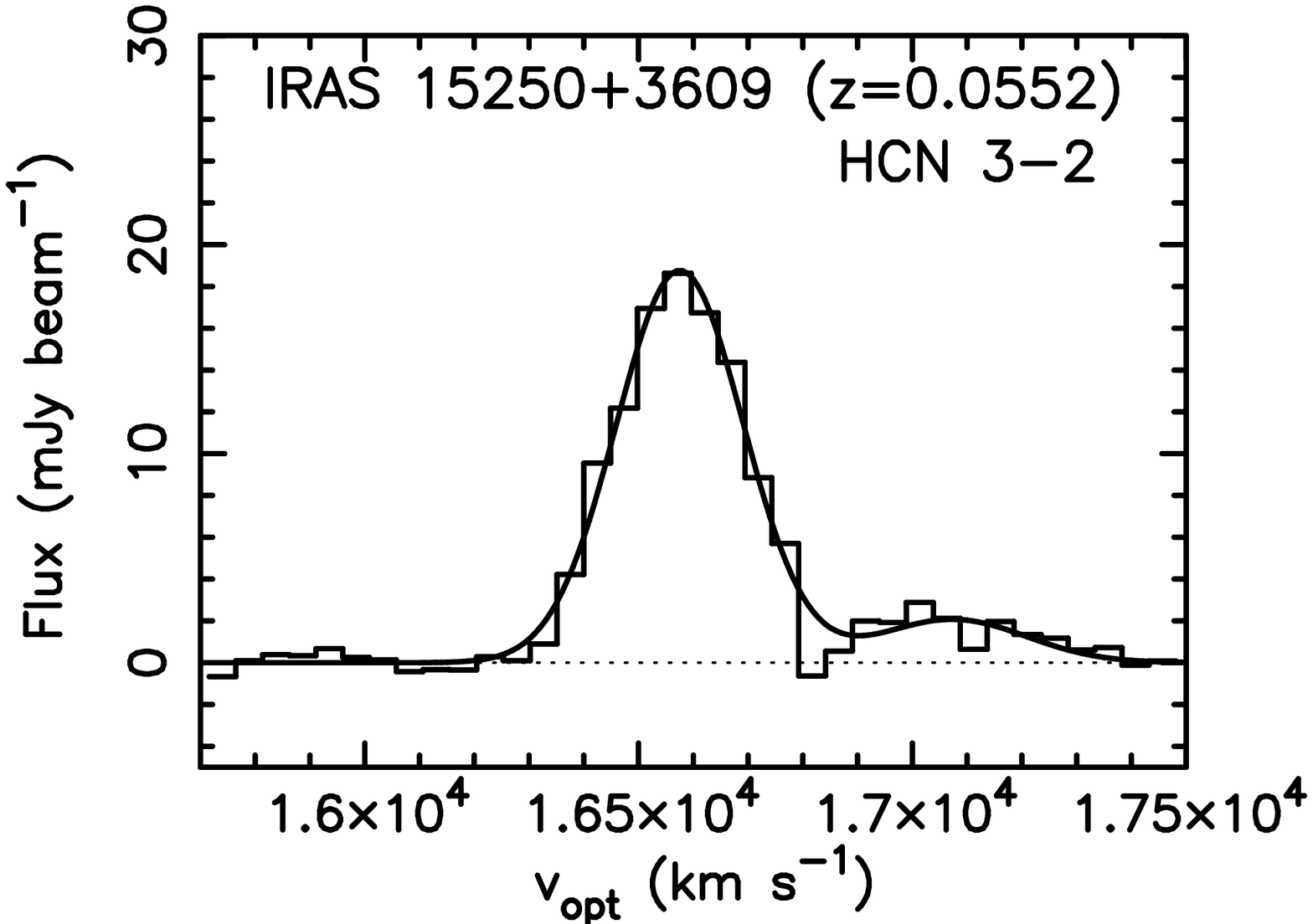} 
\includegraphics[angle=0,scale=.35]{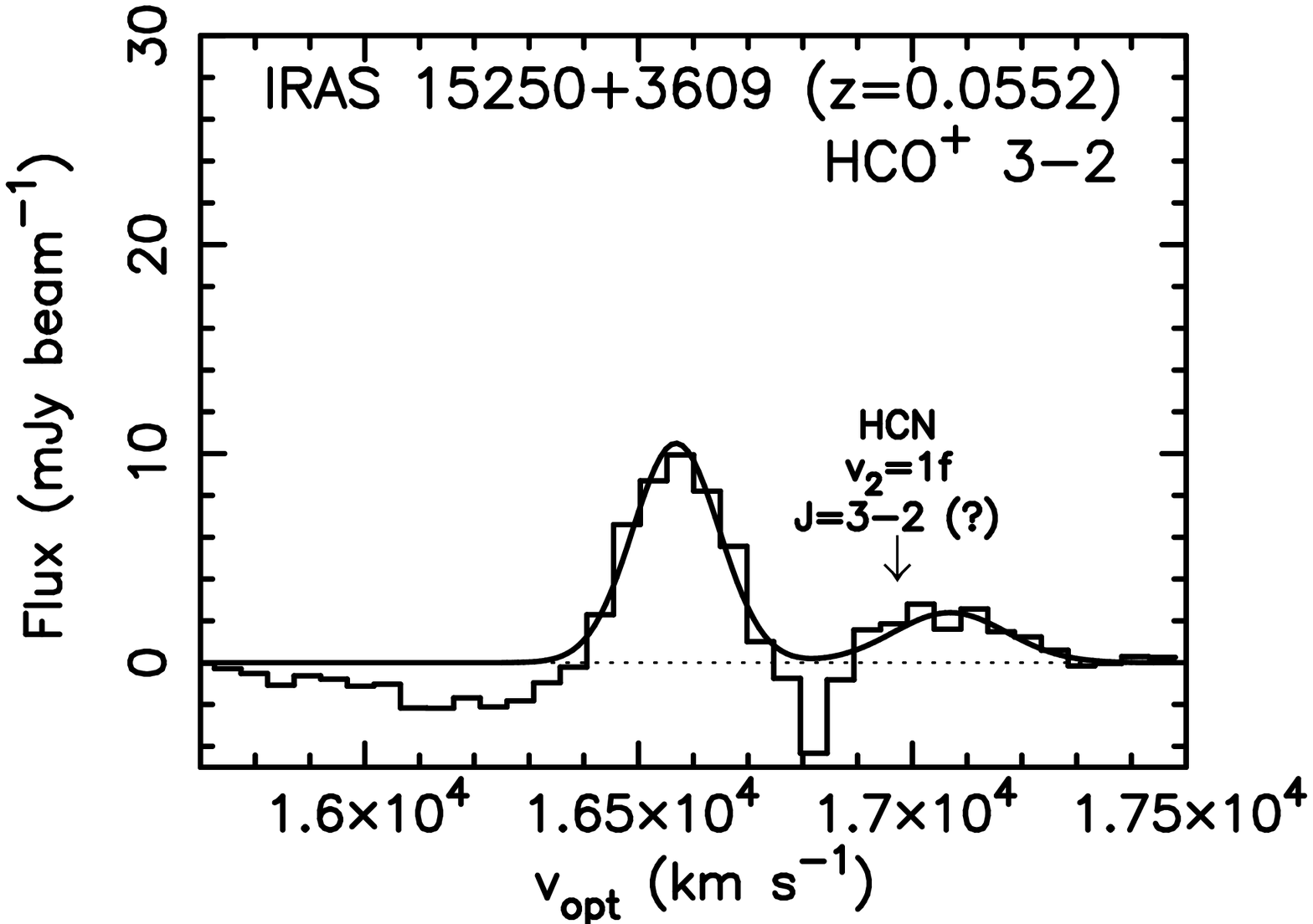} \\
\vspace*{-0.3cm}
\includegraphics[angle=0,scale=.35]{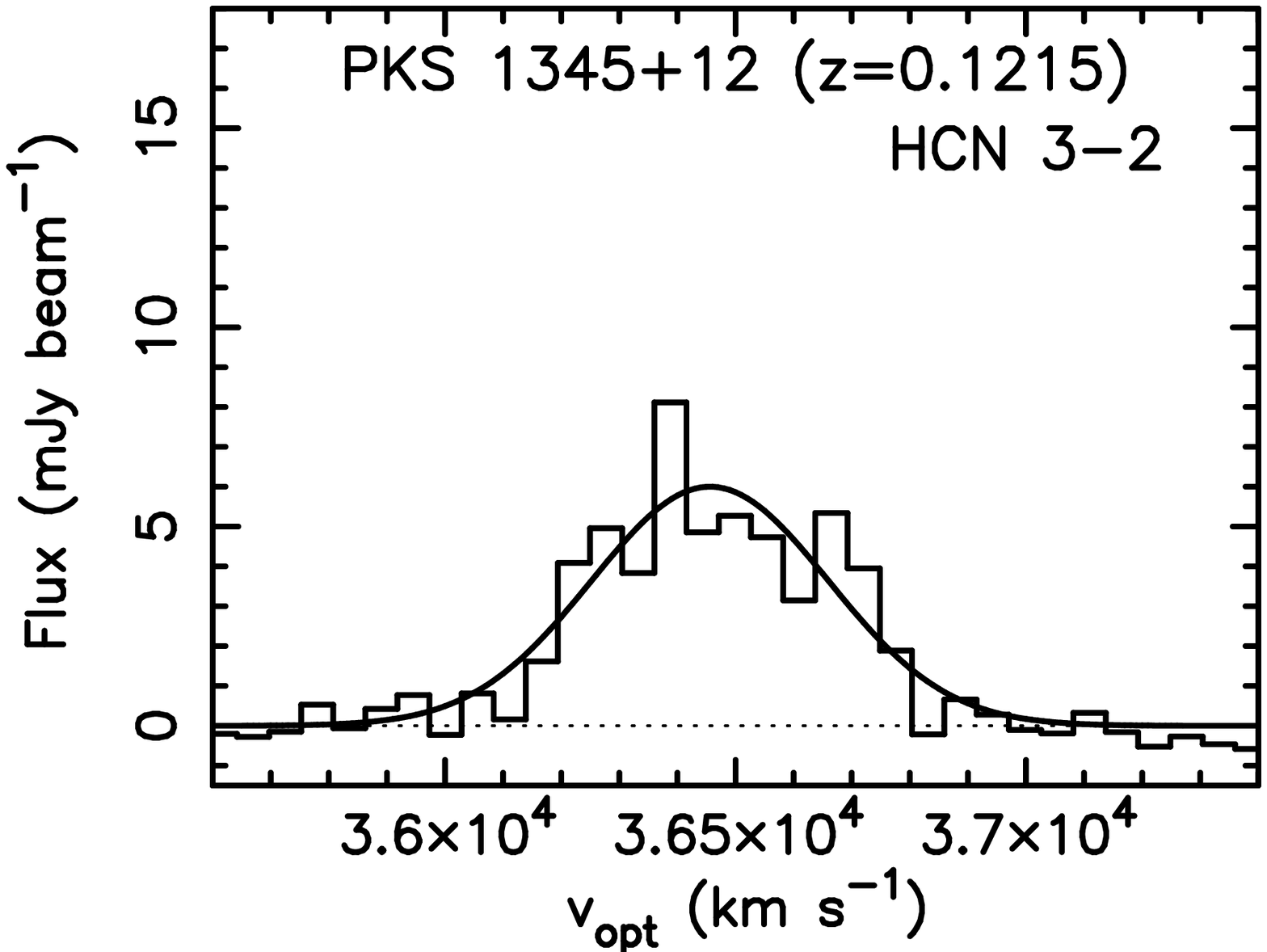} 
\includegraphics[angle=0,scale=.35]{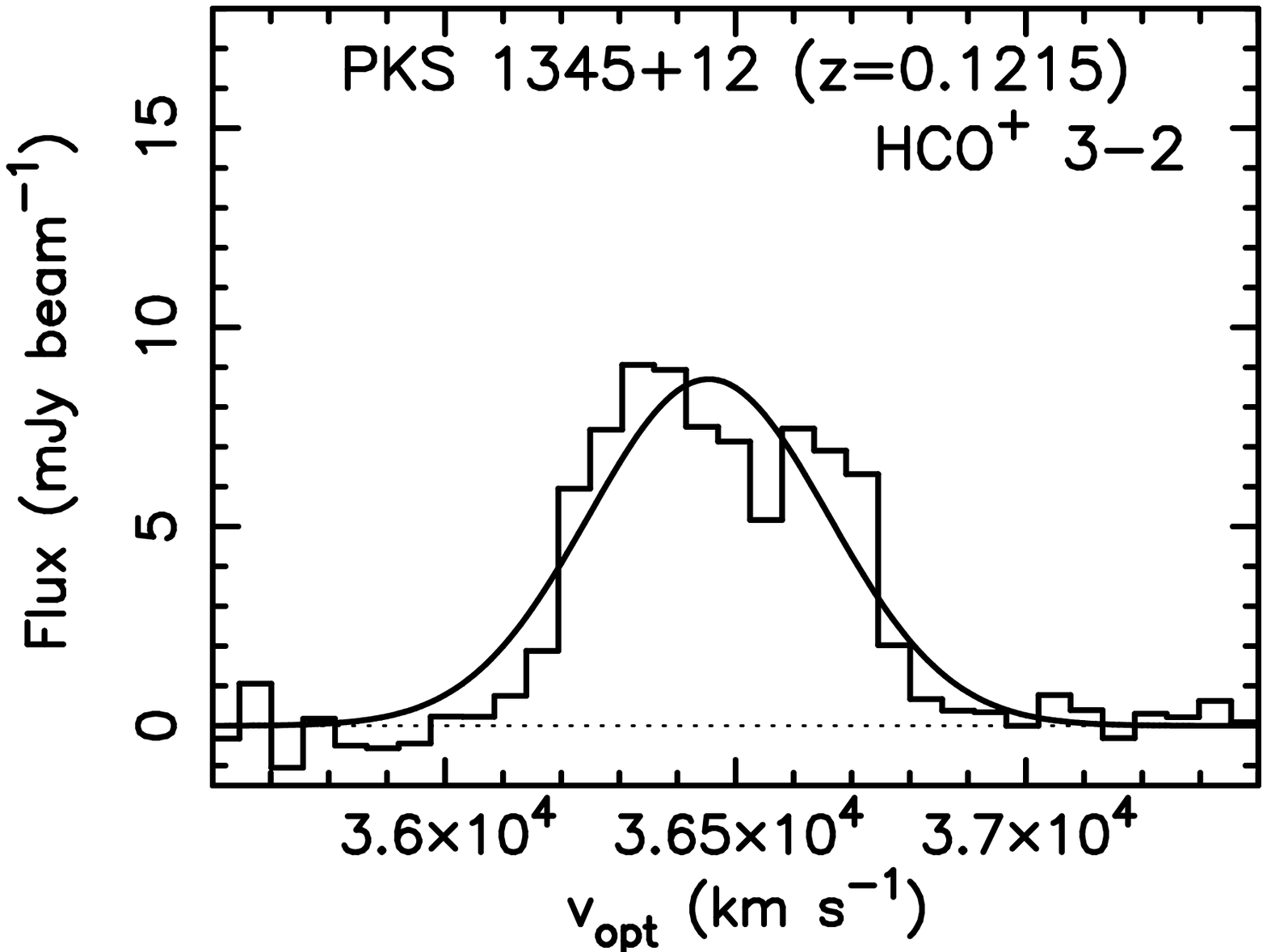} \\
\vspace*{-0.3cm}
\includegraphics[angle=0,scale=.35]{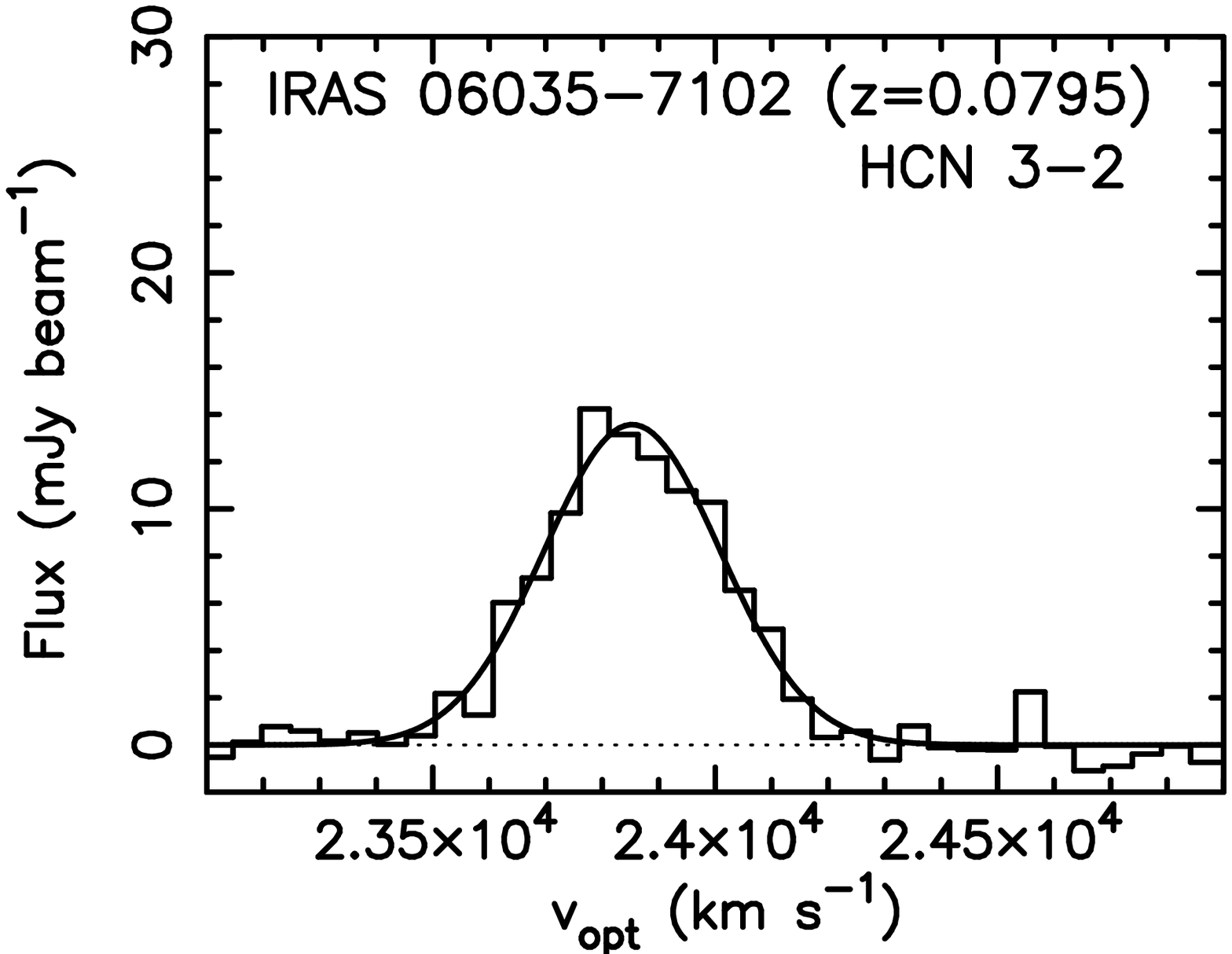} 
\includegraphics[angle=0,scale=.35]{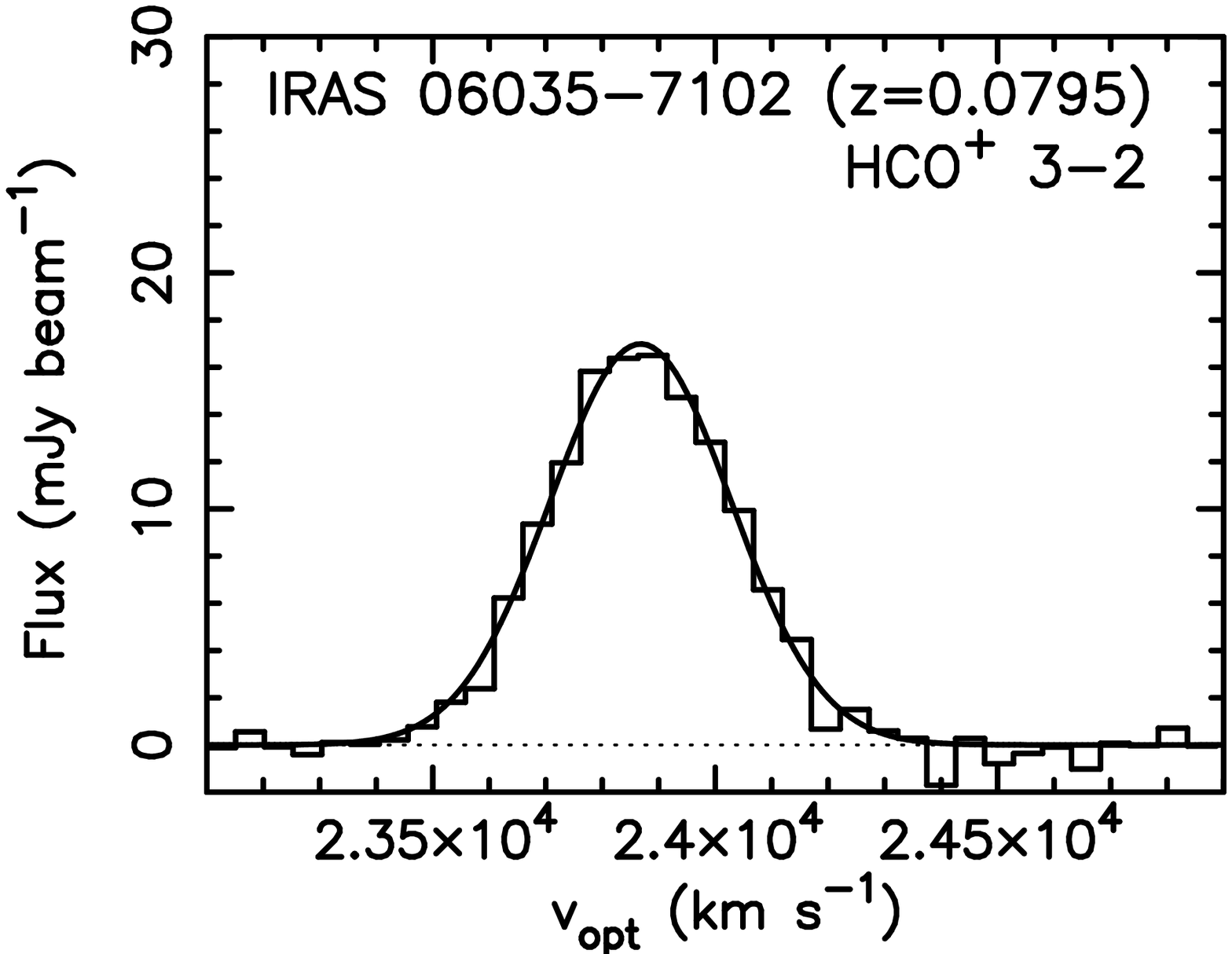} \\
\end{center}
\end{figure}

\clearpage

\begin{figure}
\begin{center}
\includegraphics[angle=0,scale=.35]{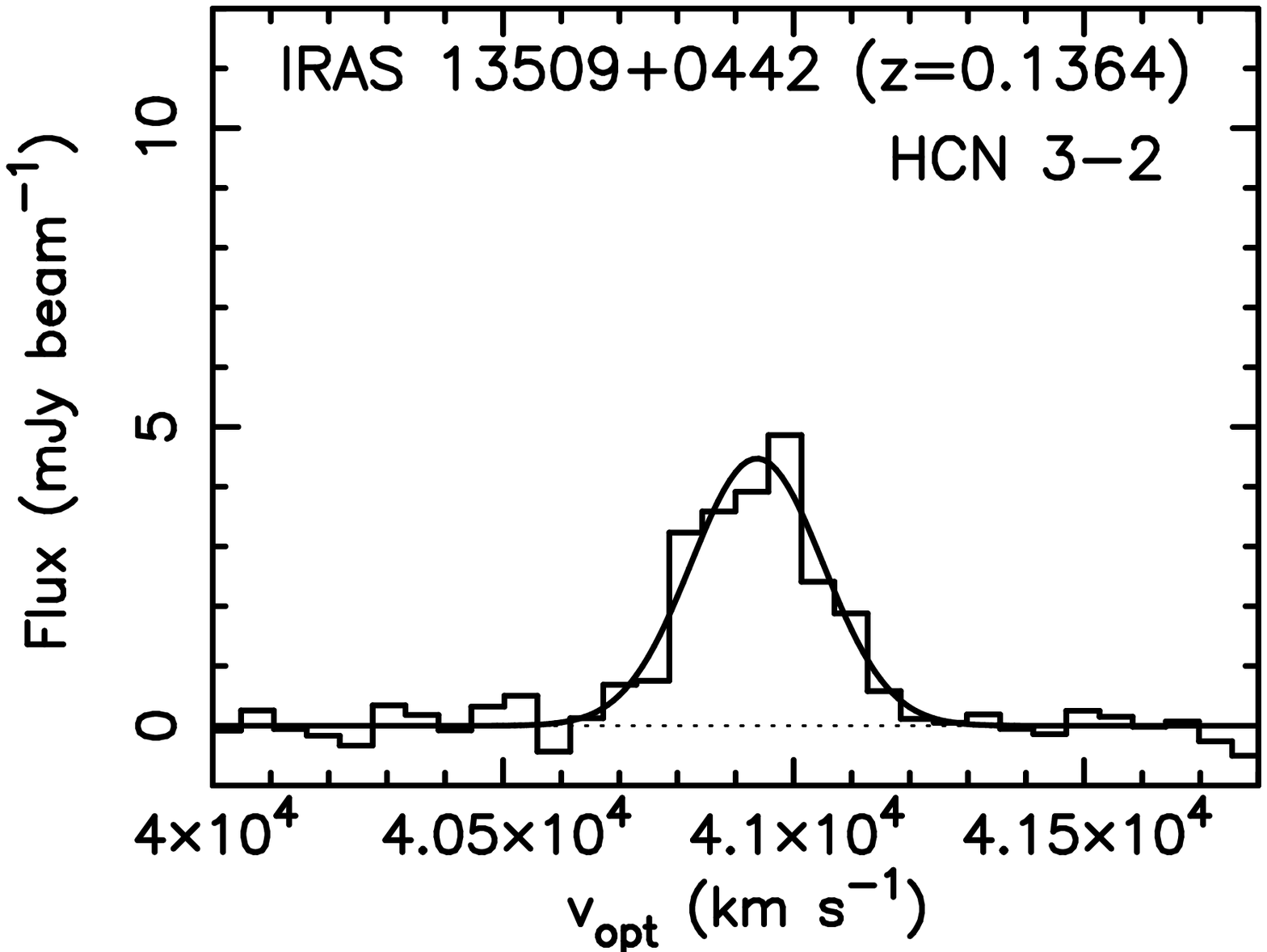} 
\includegraphics[angle=0,scale=.35]{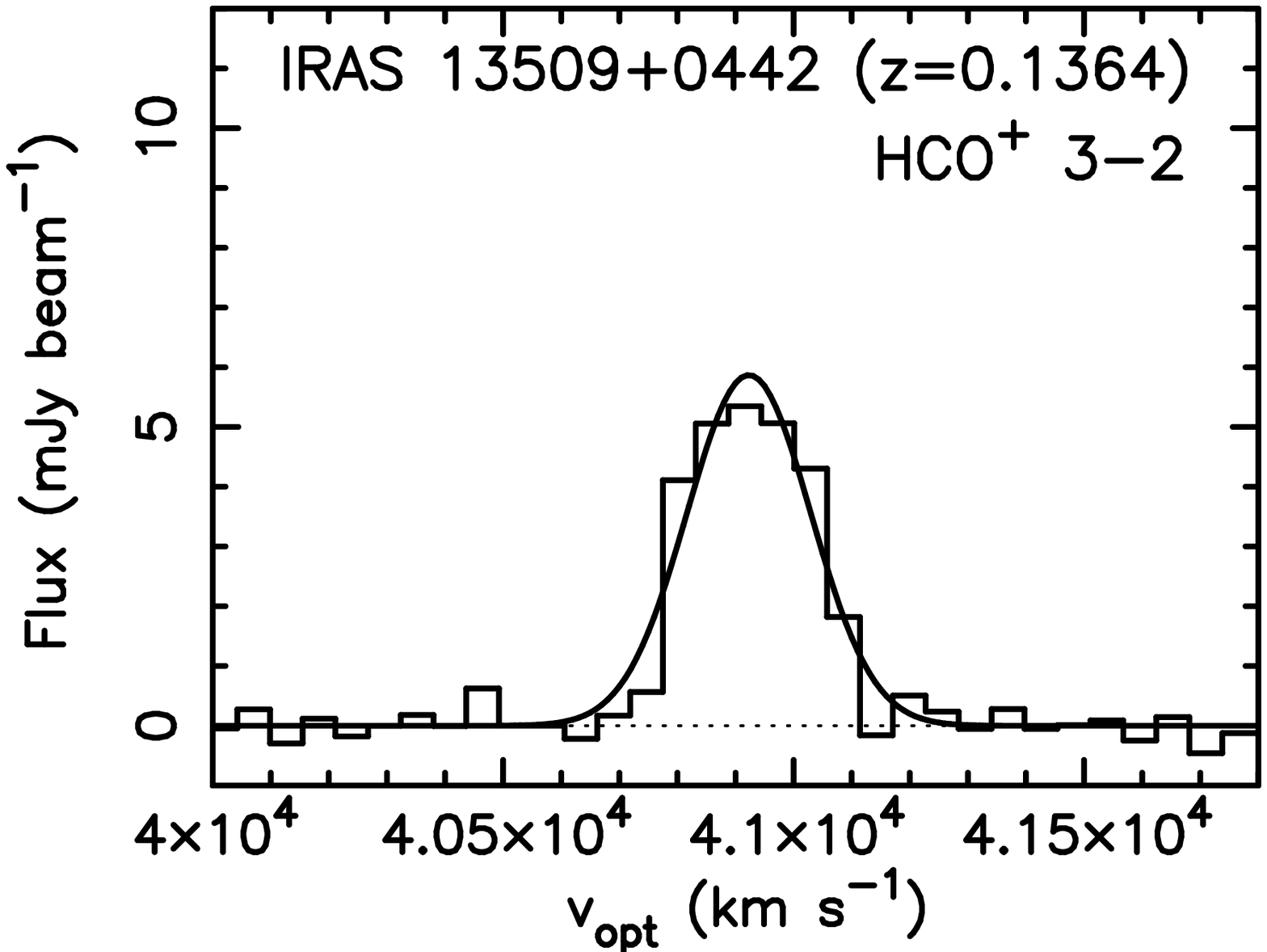} 
\vspace*{-0.3cm}
\includegraphics[angle=0,scale=.35]{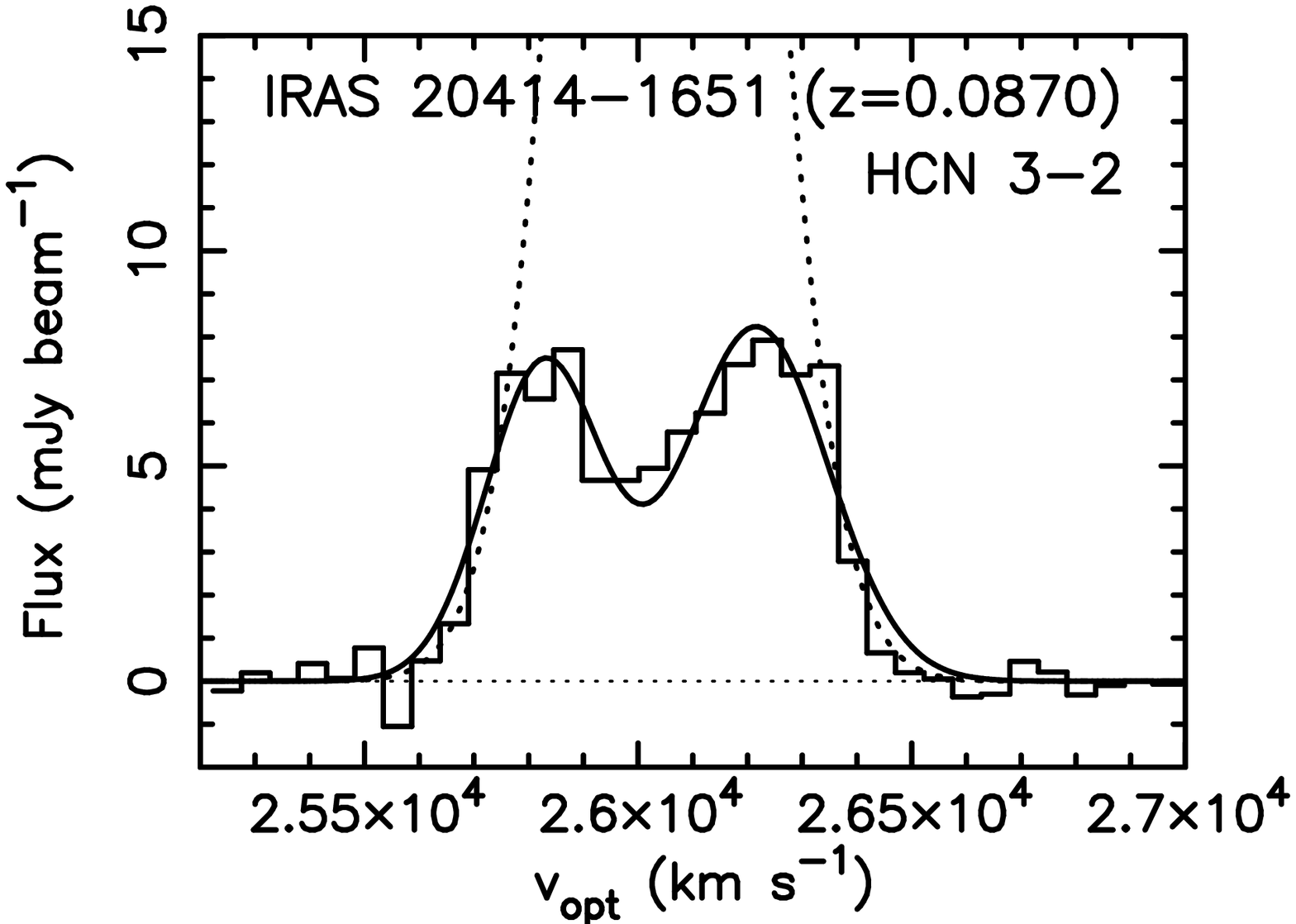} 
\includegraphics[angle=0,scale=.35]{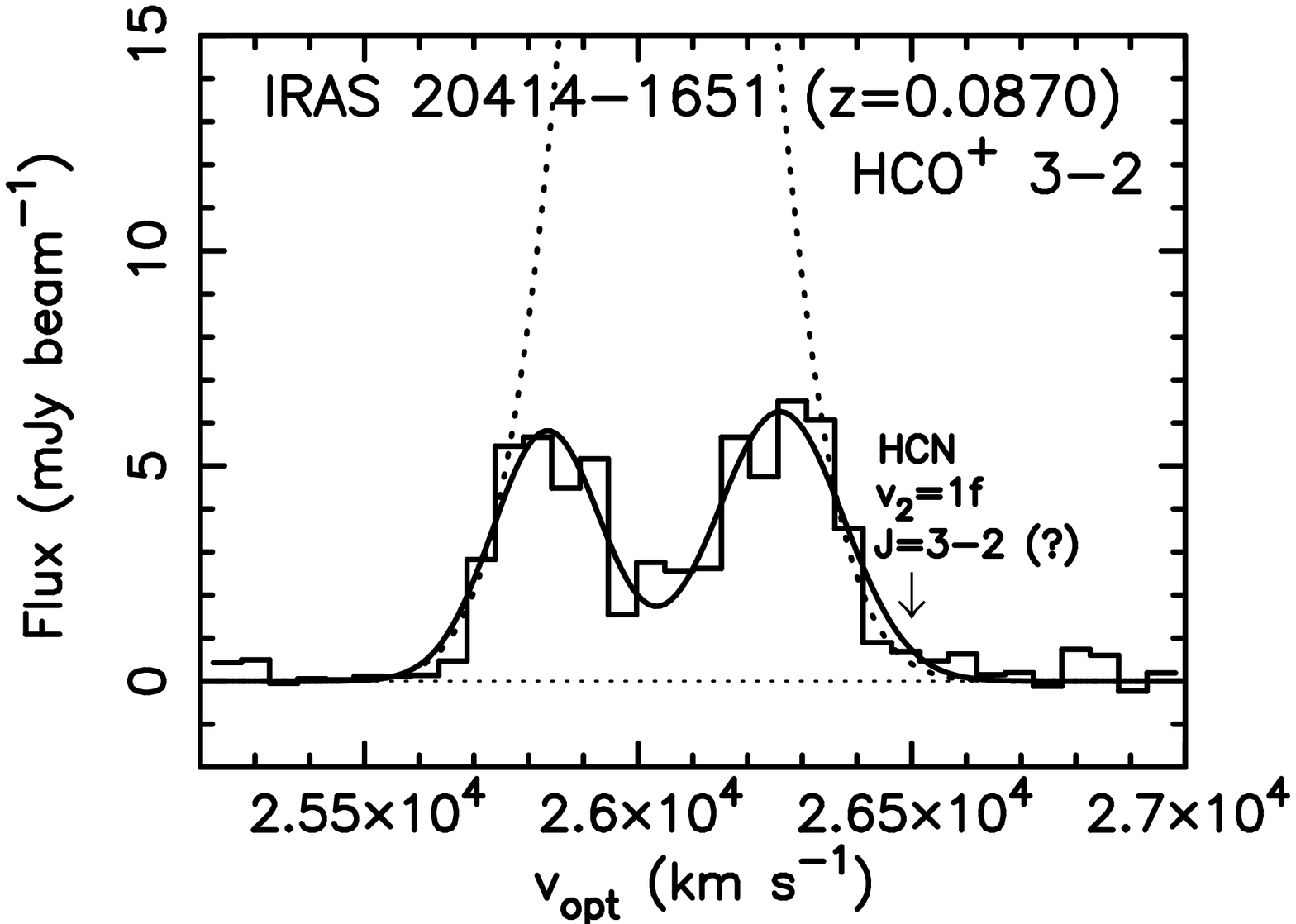} 
\end{center}
\caption{
Gaussian fits to the detected HCN J=3--2 (left) and HCO$^{+}$J=3--2
(v=0) (right) emission lines.
For NGC 7469 and NGC 1614, Gaussian fits to the detected HCN J=3--2 and
HCO$^{+}$J=3--2 (v=0) emission lines from the integrated starburst
regions, 0$\farcs$8--2$\farcs$5 radius annulus around the nucleus for
NGC 7469, and circular region with 2$\farcs$5 radius around (04 34
00.03, $-$08 34 44.6)J2000 for NGC 1614, are shown here.  
For the IRAS 22491$-$1808 HCO$^{+}$ J=3--2 emission line, the frequency which
can largely be contaminated by the HCN v$_{2}$=1f J=3--2 emission line
is excluded from the Gaussian fit.
For IRAS 15250$+$3609, a Gaussian fit to the fainter sub-peak component 
is added, after fixing the best Gaussian fit to the main component. 
For objects with clear double-peaked HCN J=3--2 and HCO$^{+}$ J=3--2
emission line profiles (IRAS 12112$+$0305 NE and IRAS 20414$-$1651), single
Gaussian fits using data points at the edge of the emission tail, after
removing data possibly affected by the central dip (based on the
assumption of self-absorption), are also shown with curved dashed lines.   
} 
\end{figure}

\clearpage

\begin{figure}
\begin{center}
\includegraphics[angle=0,scale=.35]{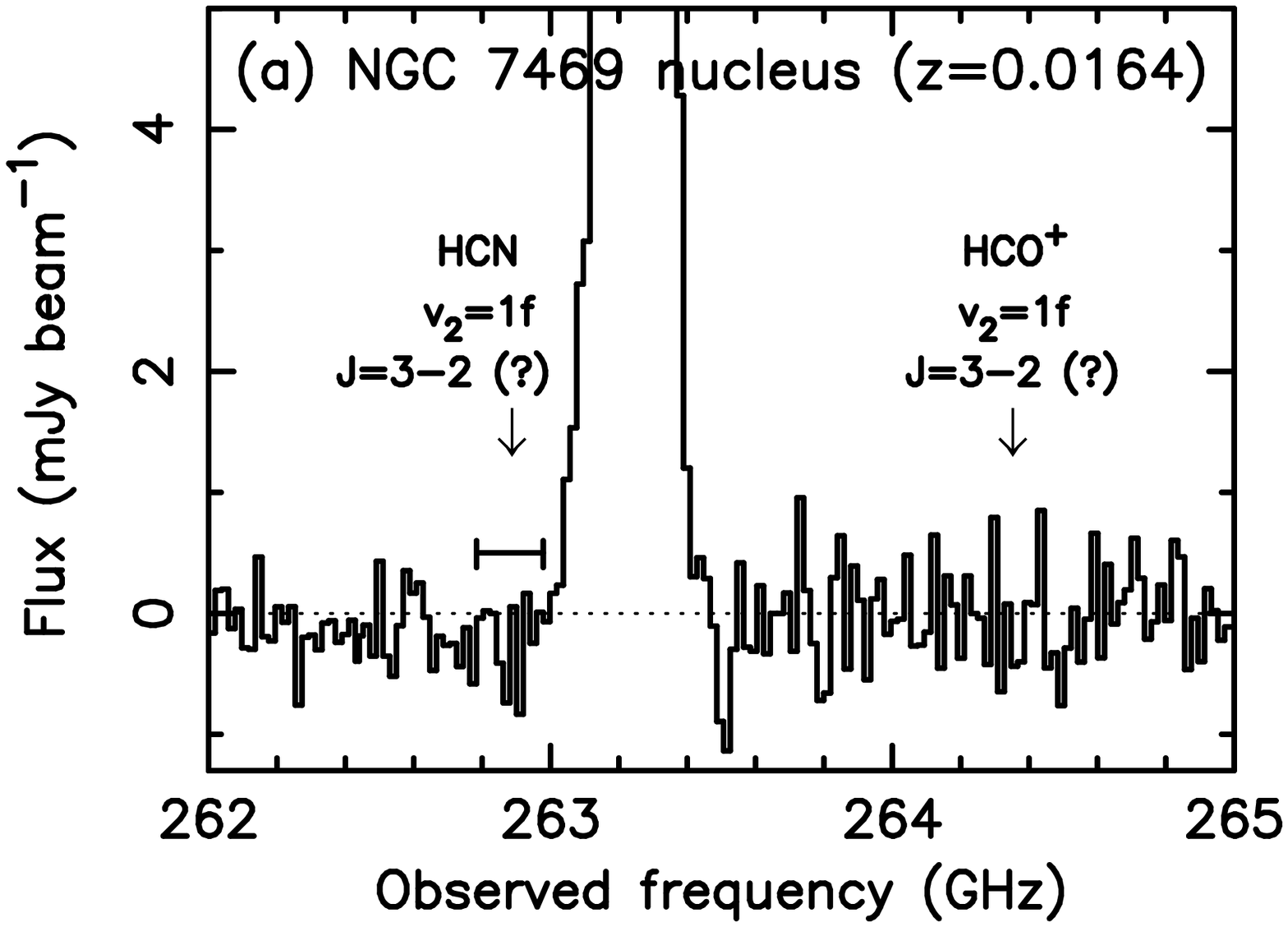} 
\includegraphics[angle=0,scale=.35]{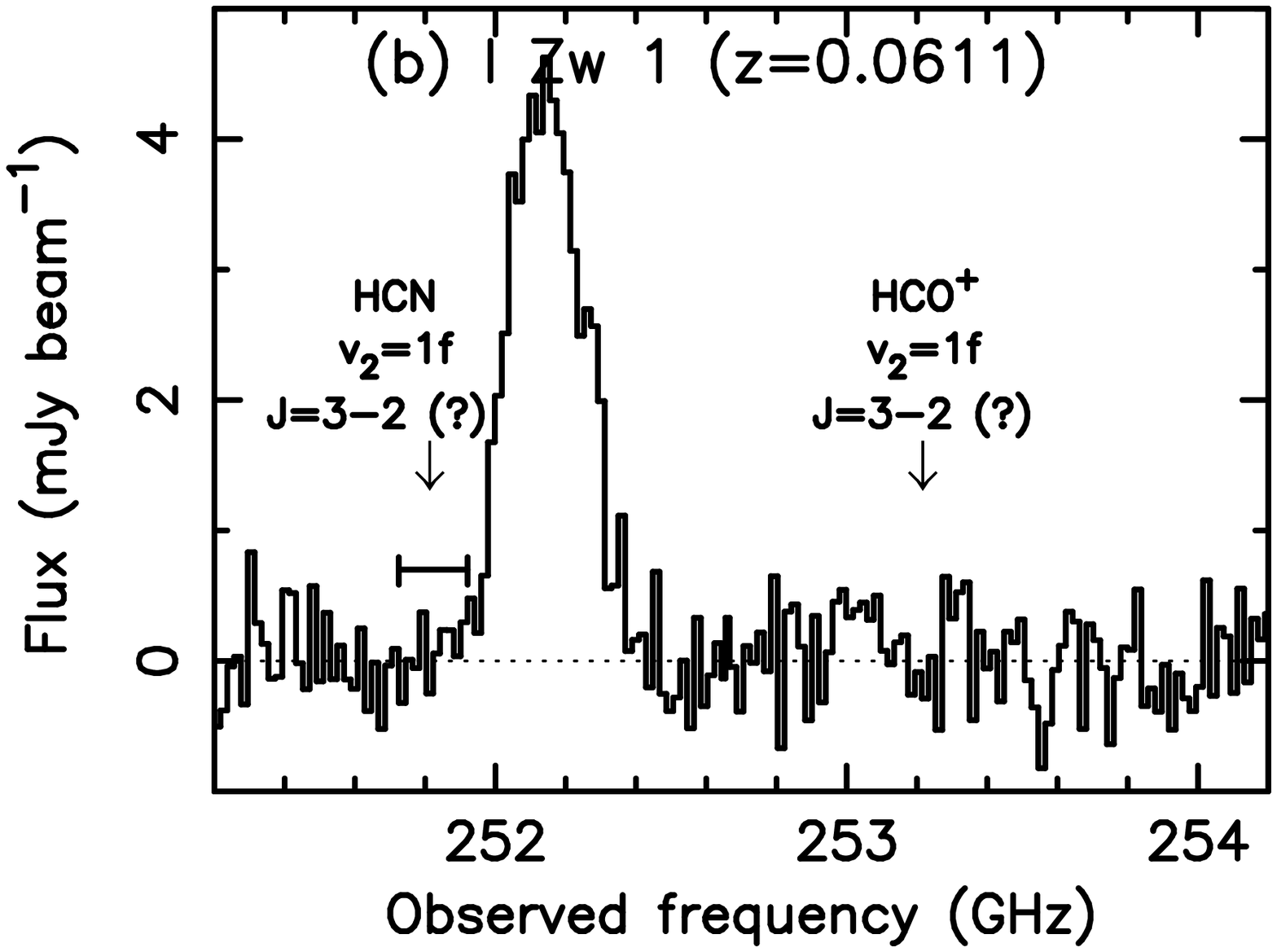} \\
\vspace{-0.4cm}
\includegraphics[angle=0,scale=.35]{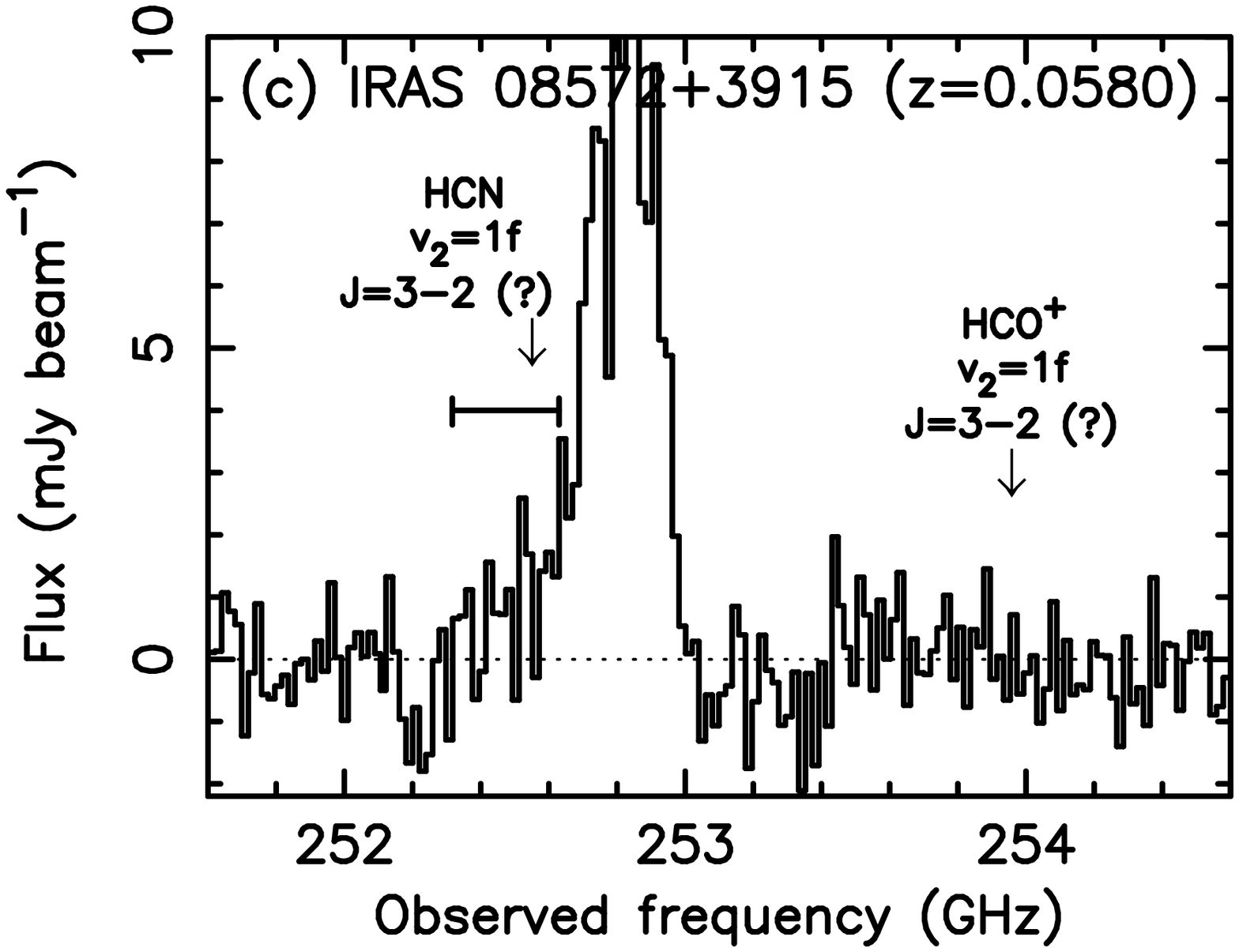} 
\includegraphics[angle=0,scale=.35]{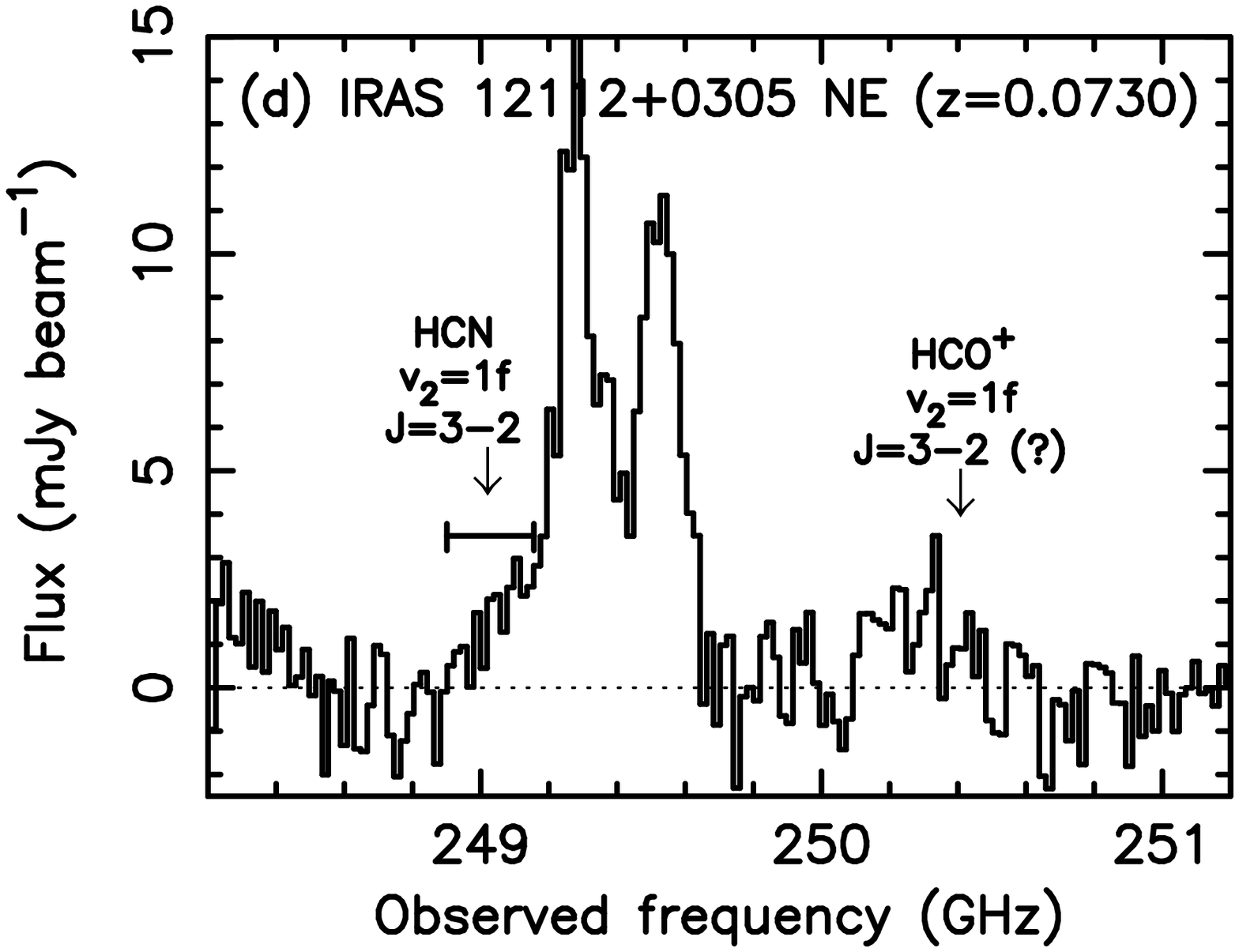} \\ 
\vspace{-0.4cm}
\includegraphics[angle=0,scale=.35]{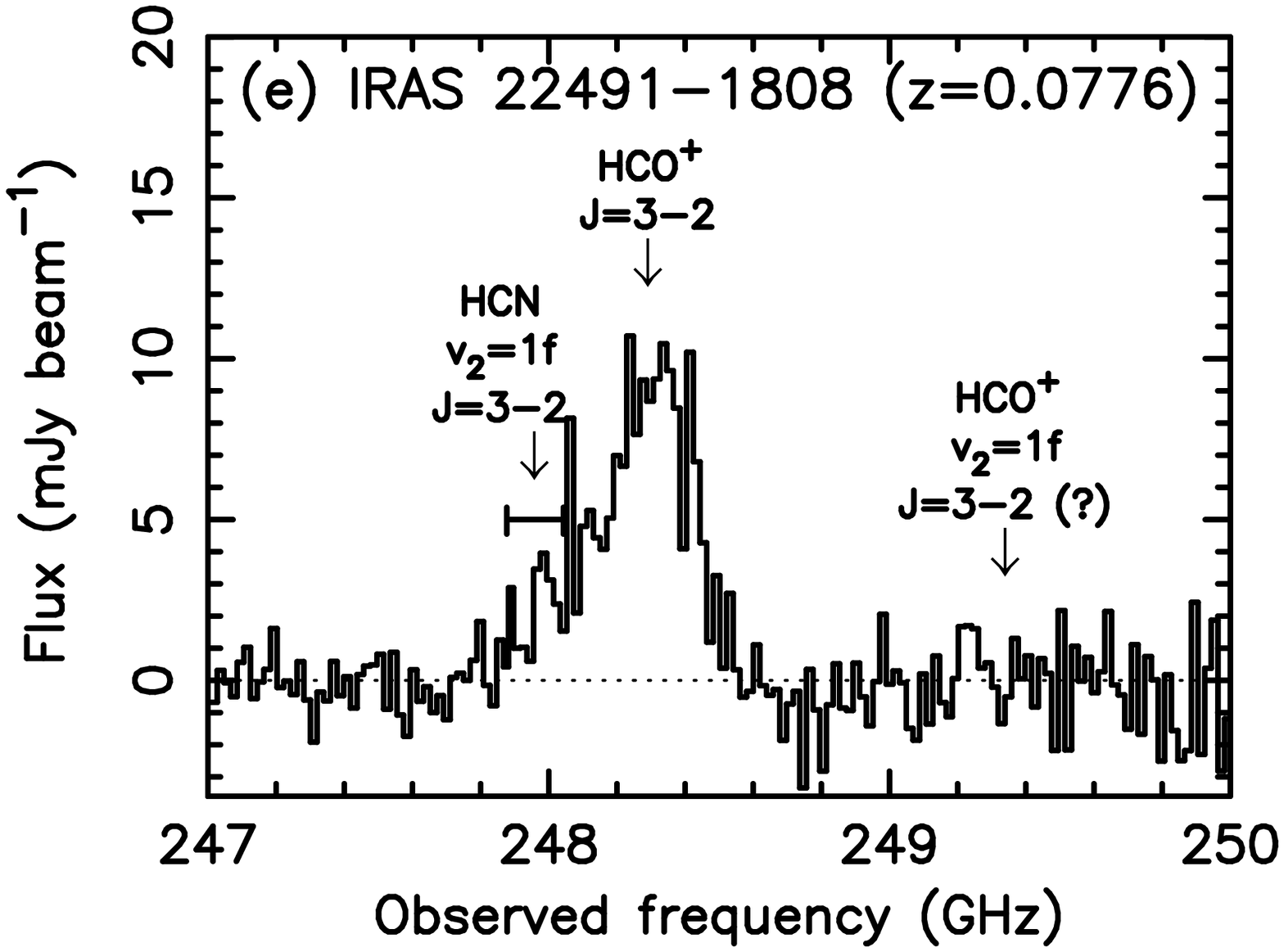} 
\includegraphics[angle=0,scale=.35]{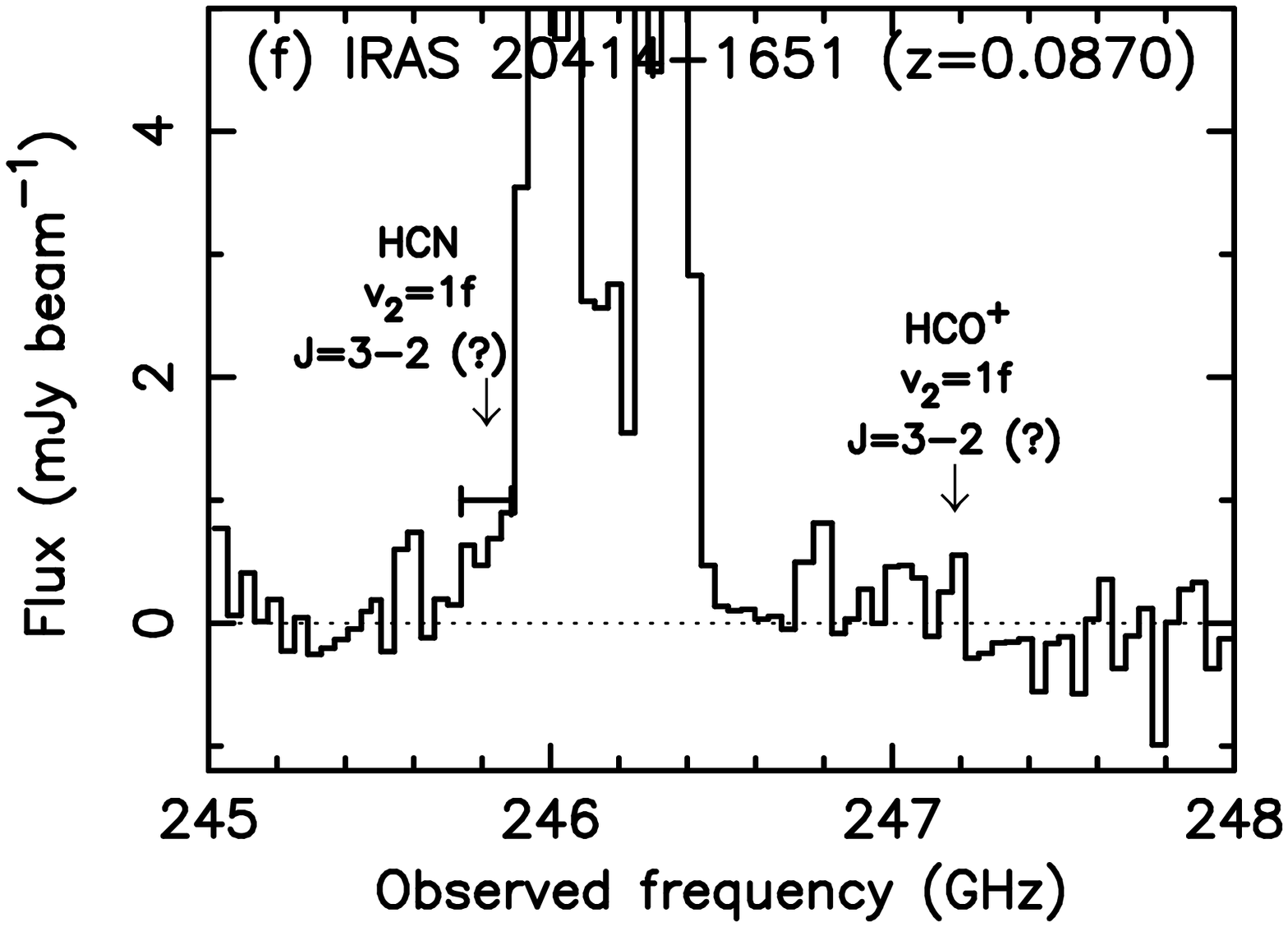} \\ 
\end{center}
\vspace{-0.6cm}
\caption{Zoomed-in spectra for selected sources to better display possible
vibrationally excited (v$_{2}$=1f) HCN J=3--2 and HCO$^{+}$ J=3--2
emission lines. 
The horizontal solid straight lines, inserted by short vertical solid
lines, are frequency ranges used to estimate the flux of the HCN
v$_{2}$=1f J=3--2 emission line.}
\end{figure}

\clearpage

\begin{figure}
\begin{center}
\includegraphics[angle=0,scale=.8]{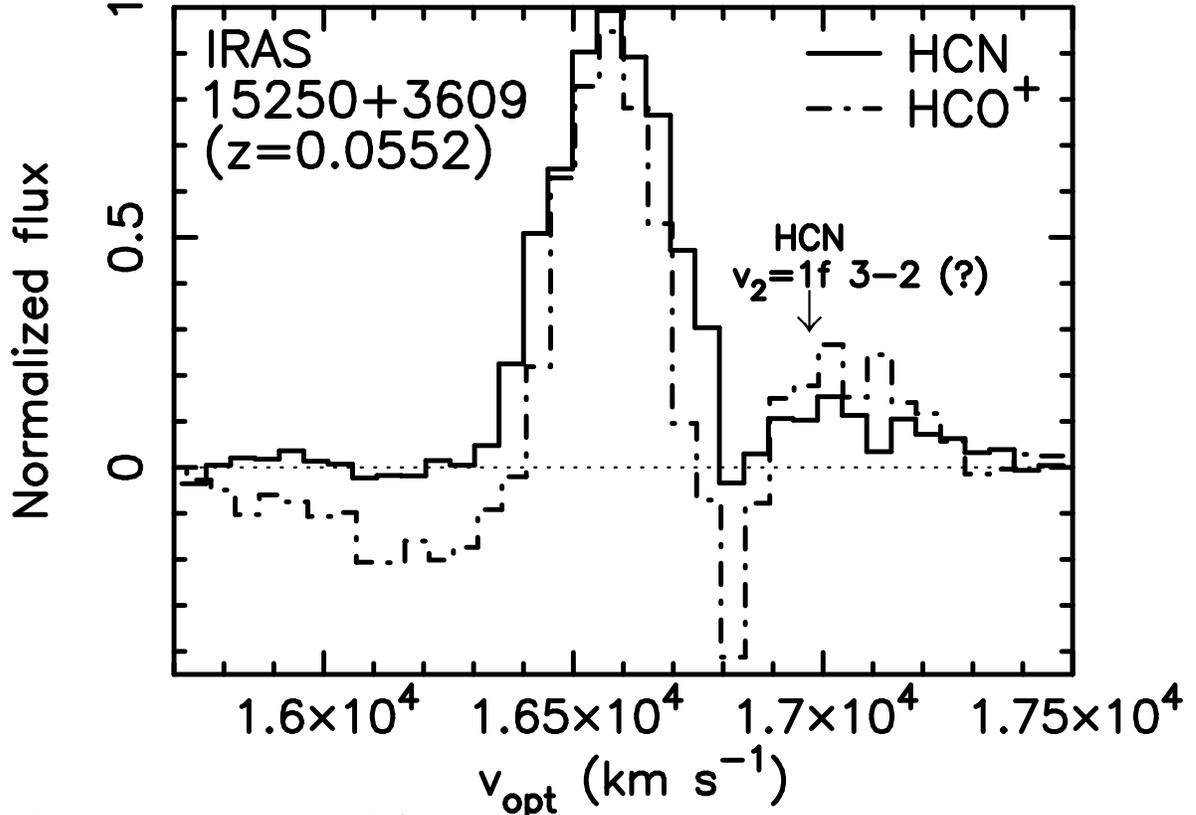} \\
\end{center}
\vspace{-0.6cm}
\caption{
Comparison of HCN J=3--2 and HCO$^{+}$ J=3--2 emission line profiles for
IRAS 15250$+$3609, after normalizing the Gaussian-fit peak flux for the 
brighter main component. 
Solid and dash-dotted lines represent HCN J=3--2 and HCO$^{+}$ J=3--2
emission line profiles, respectively. 
The expected position in the abscissa, of the HCN v$_{2}$=1f J=3--2
emission at $z =$ 0.0552 is shown as a down arrow for the HCO$^{+}$
J=3--2 emission line profile (dash-dotted line). 
}
\end{figure}

\begin{figure}
\begin{center}
\includegraphics[angle=0,scale=.46]{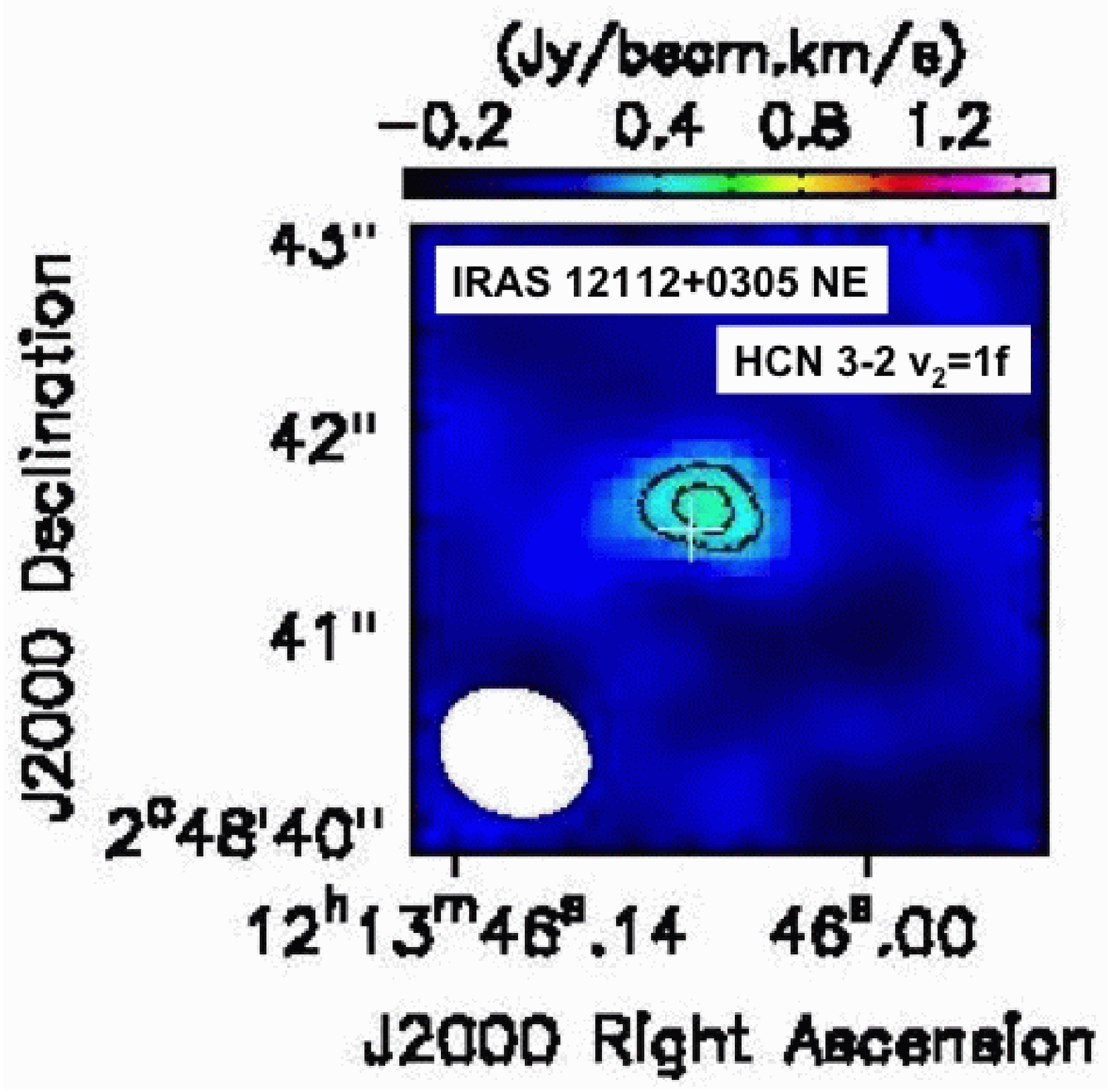} 
\includegraphics[angle=0,scale=.46]{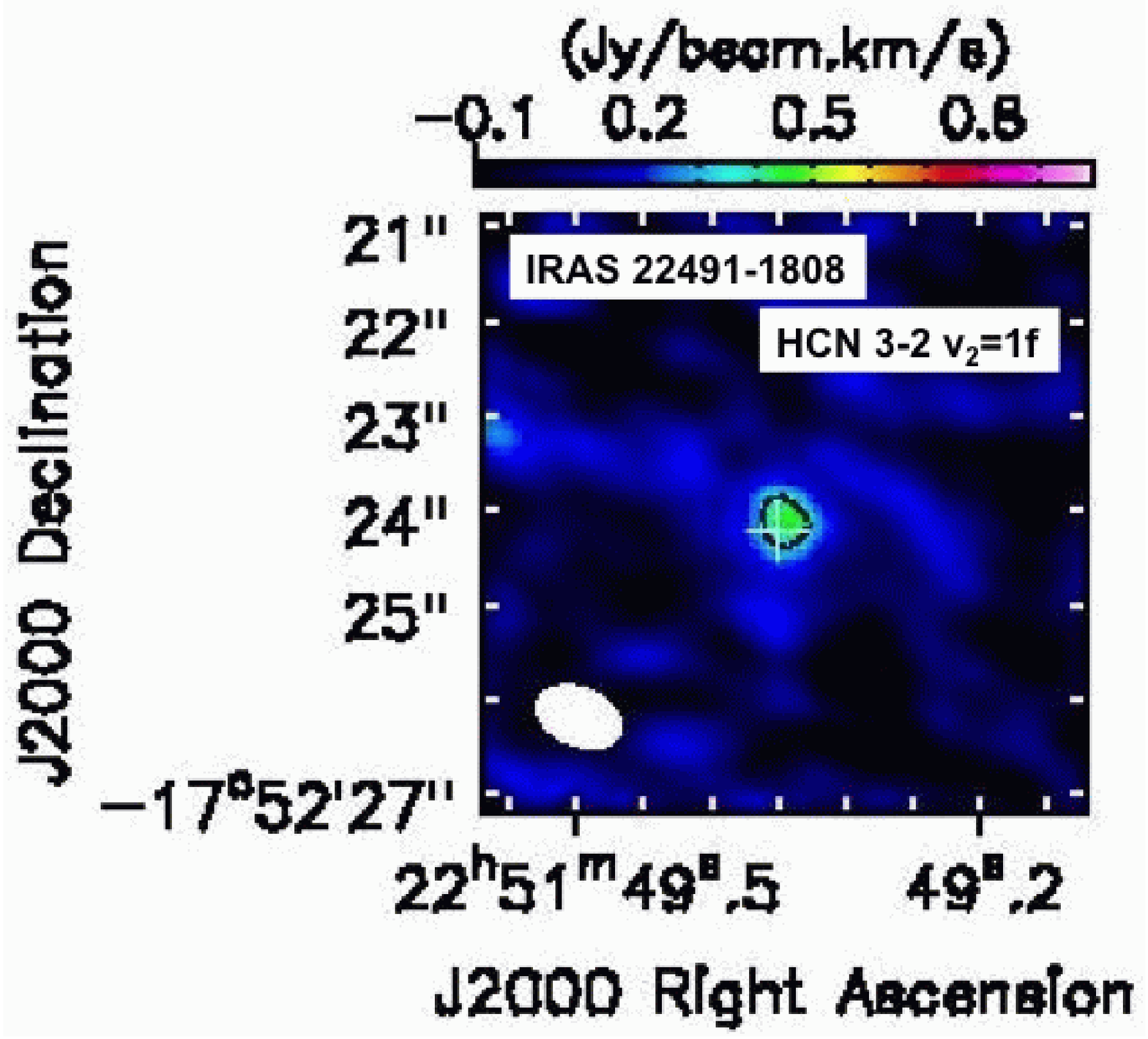} \\
\end{center}
\caption{
Integrated intensity (moment 0) maps of the HCN v$_{2}$=1f J=3--2 emission
line of IRAS 12112$+$0305 NE and IRAS 22491$-$1808. 
Data at the frequency marked with a solid straight line inserted by
two vertical lines in Figure 10, are used. 
The contours are 3$\sigma$ and 4$\sigma$ for IRAS 12112$+$0305 NE, 
and 3$\sigma$ for IRAS 22491$-$1808.
The continuum peak positions are indicated with white crosses.
} 
\end{figure}

\begin{figure}
\begin{center}
\includegraphics[angle=0,scale=.28]{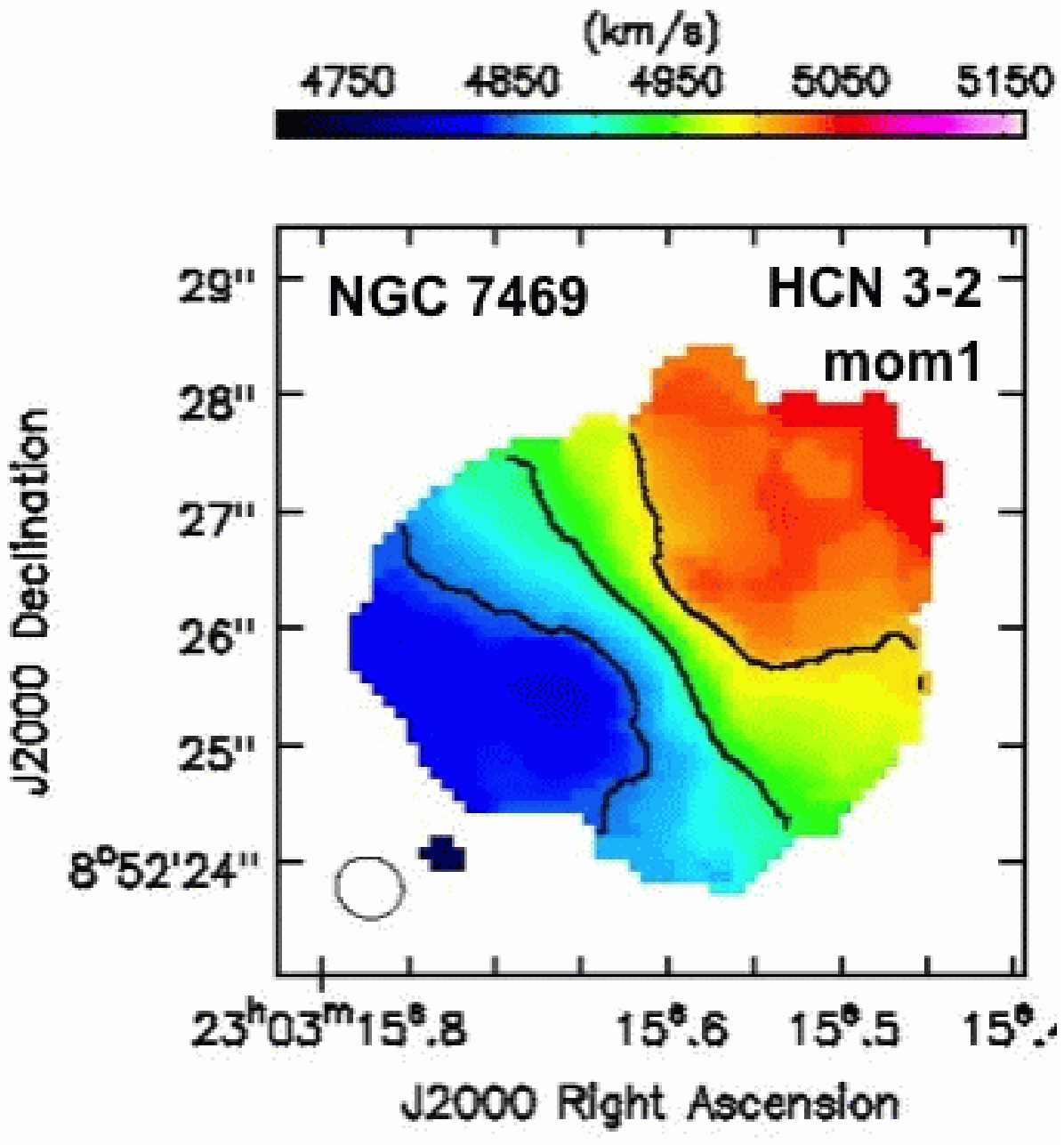}
\includegraphics[angle=0,scale=.28]{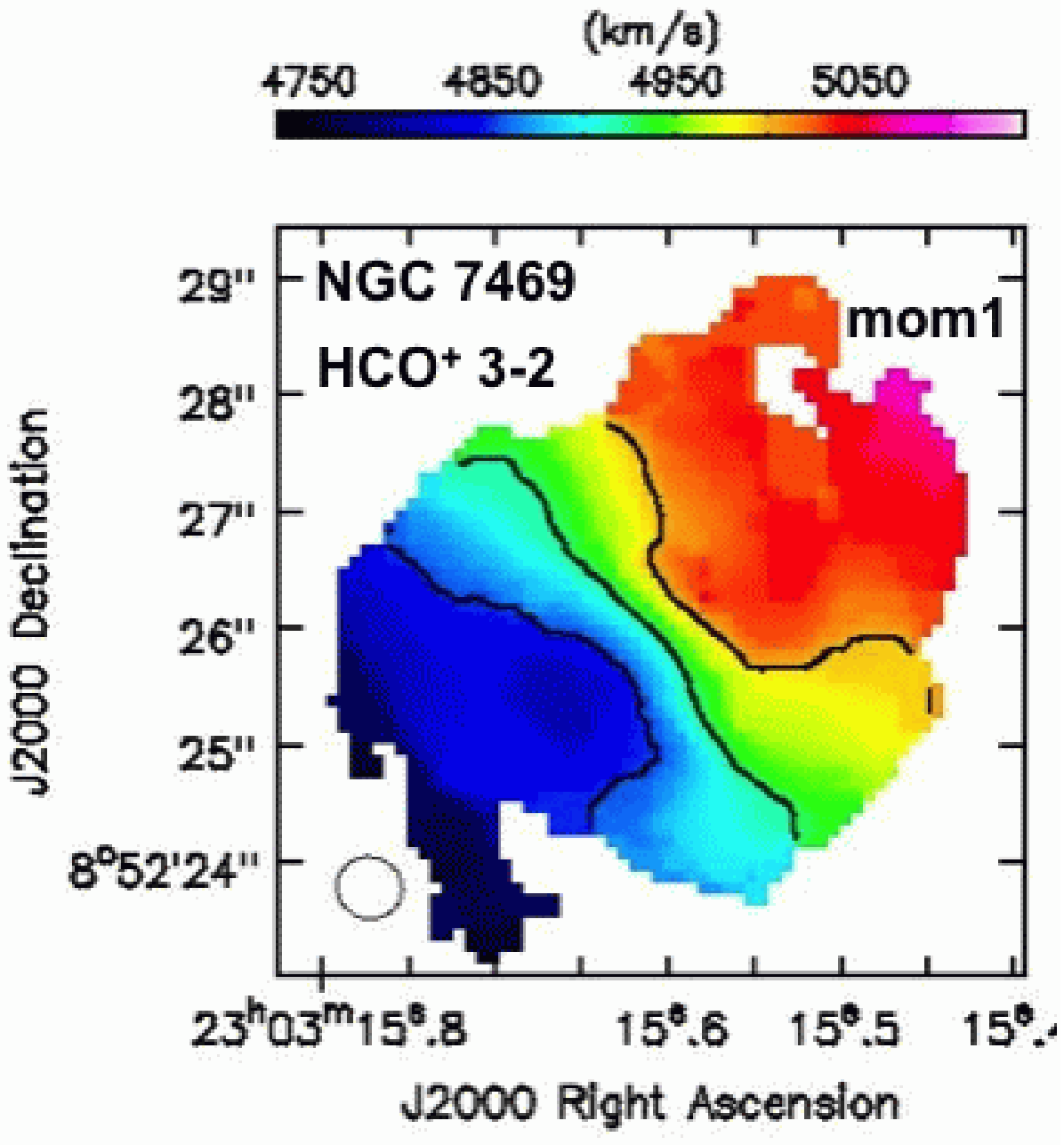} 
\includegraphics[angle=0,scale=.28]{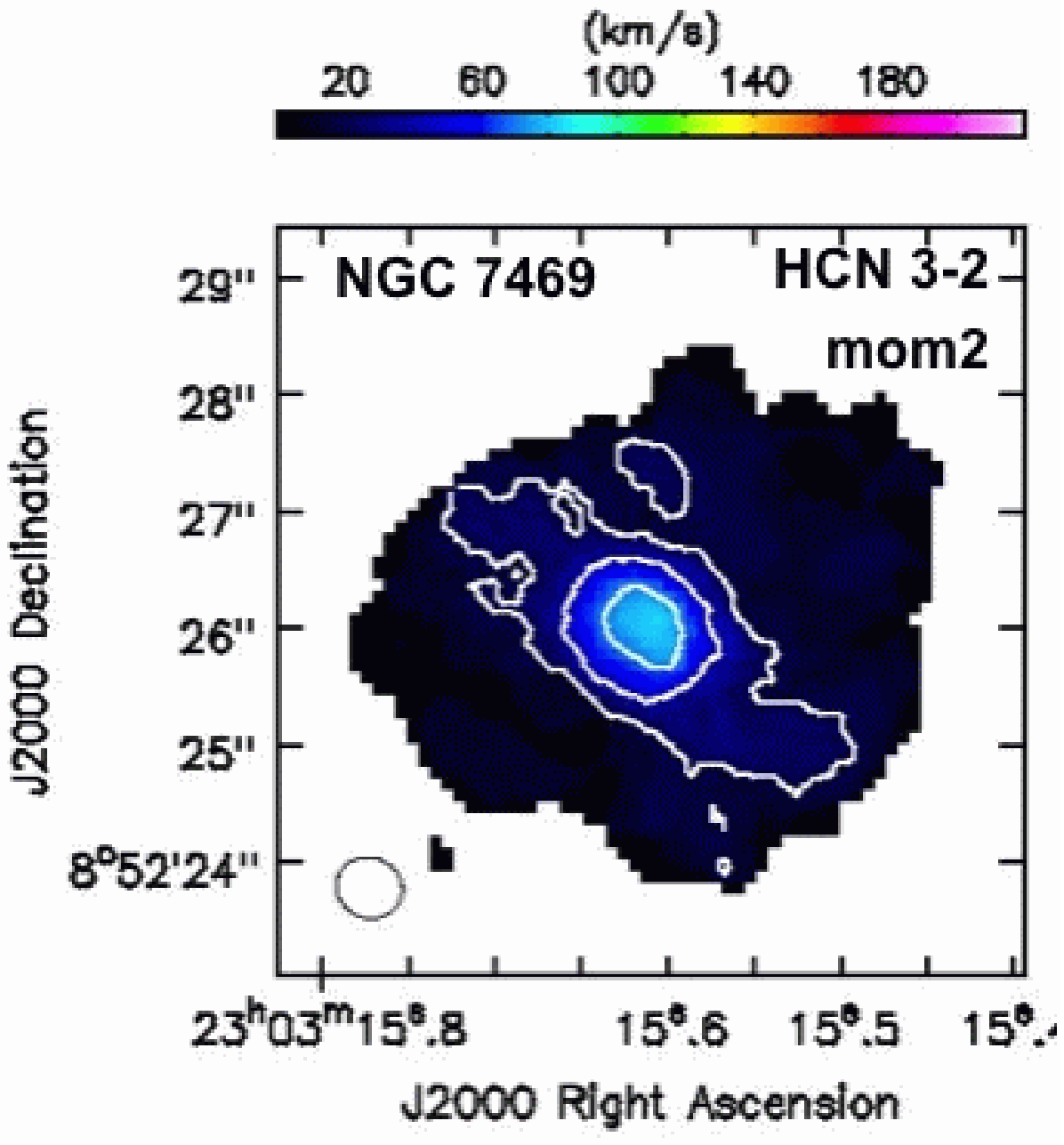} 
\includegraphics[angle=0,scale=.28]{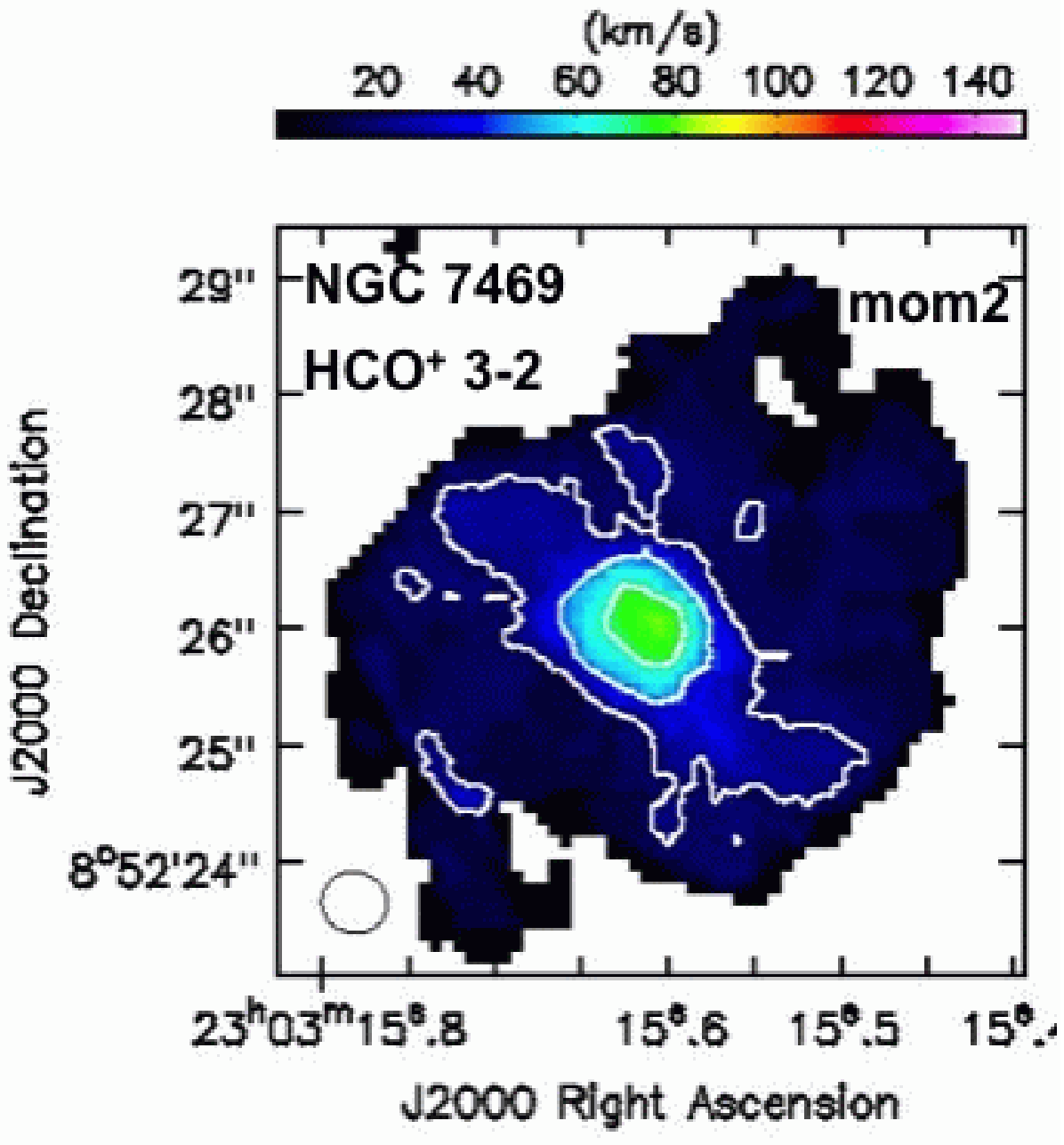} \\
\includegraphics[angle=0,scale=.28]{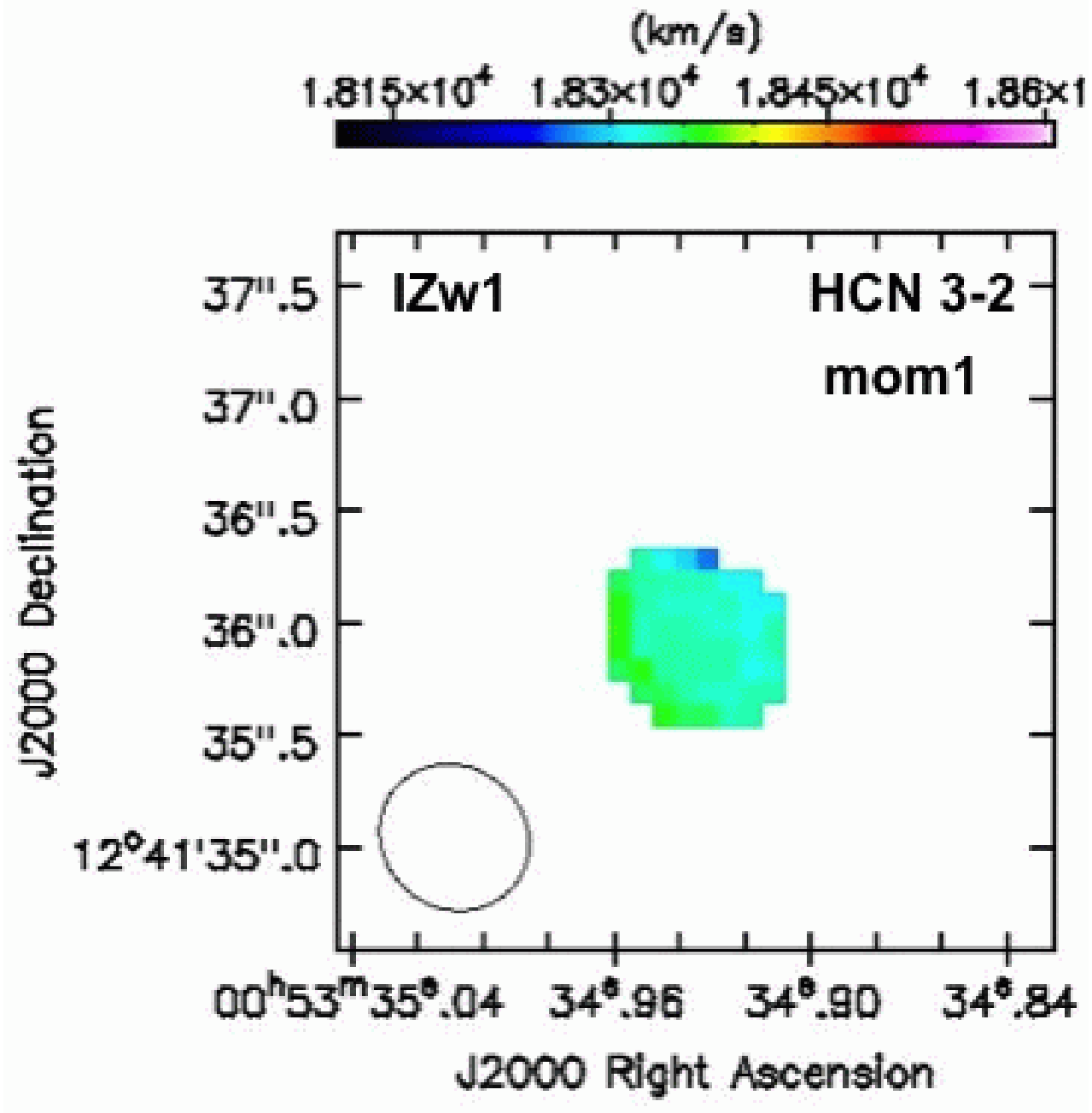} 
\includegraphics[angle=0,scale=.28]{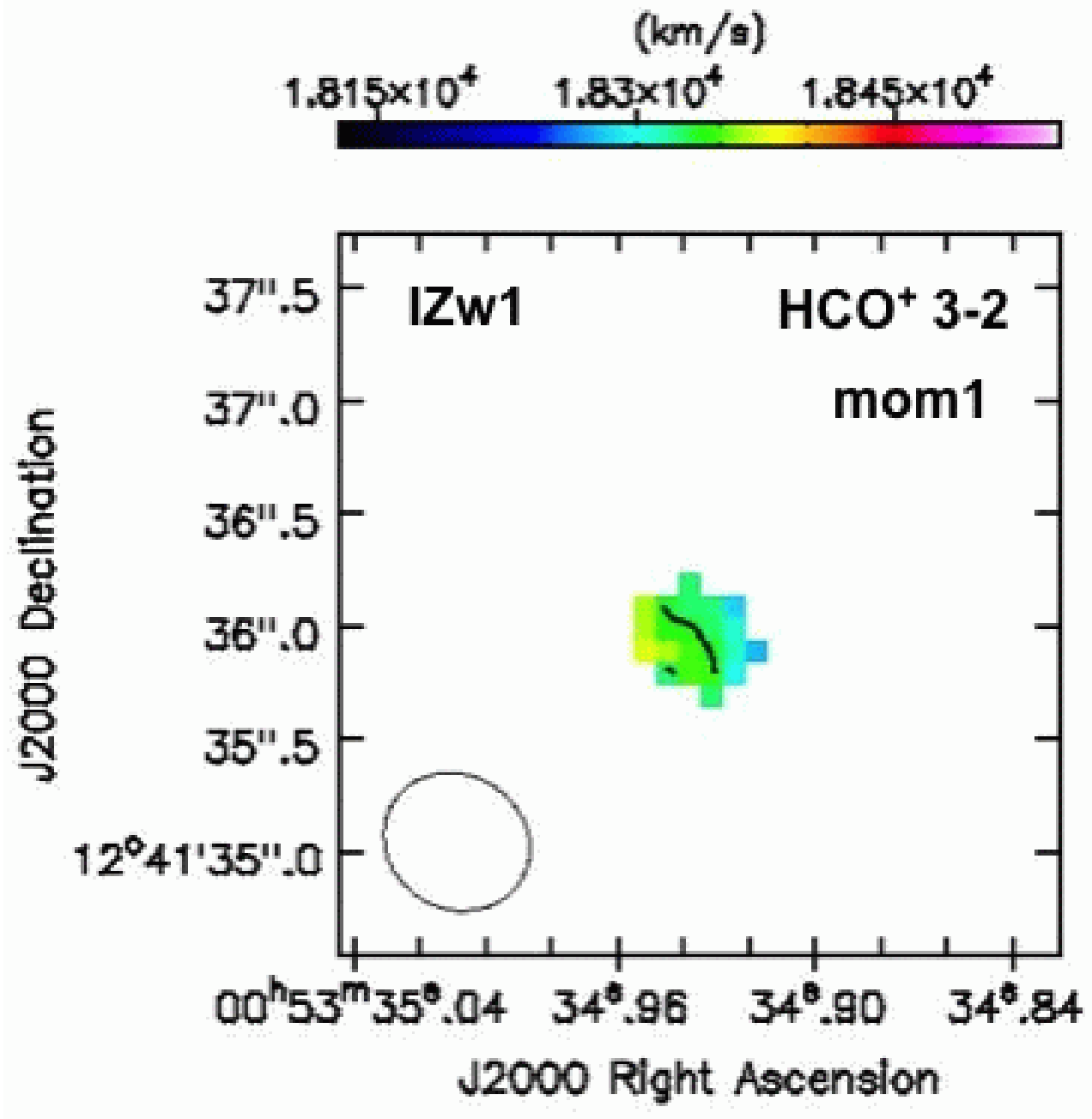} 
\includegraphics[angle=0,scale=.28]{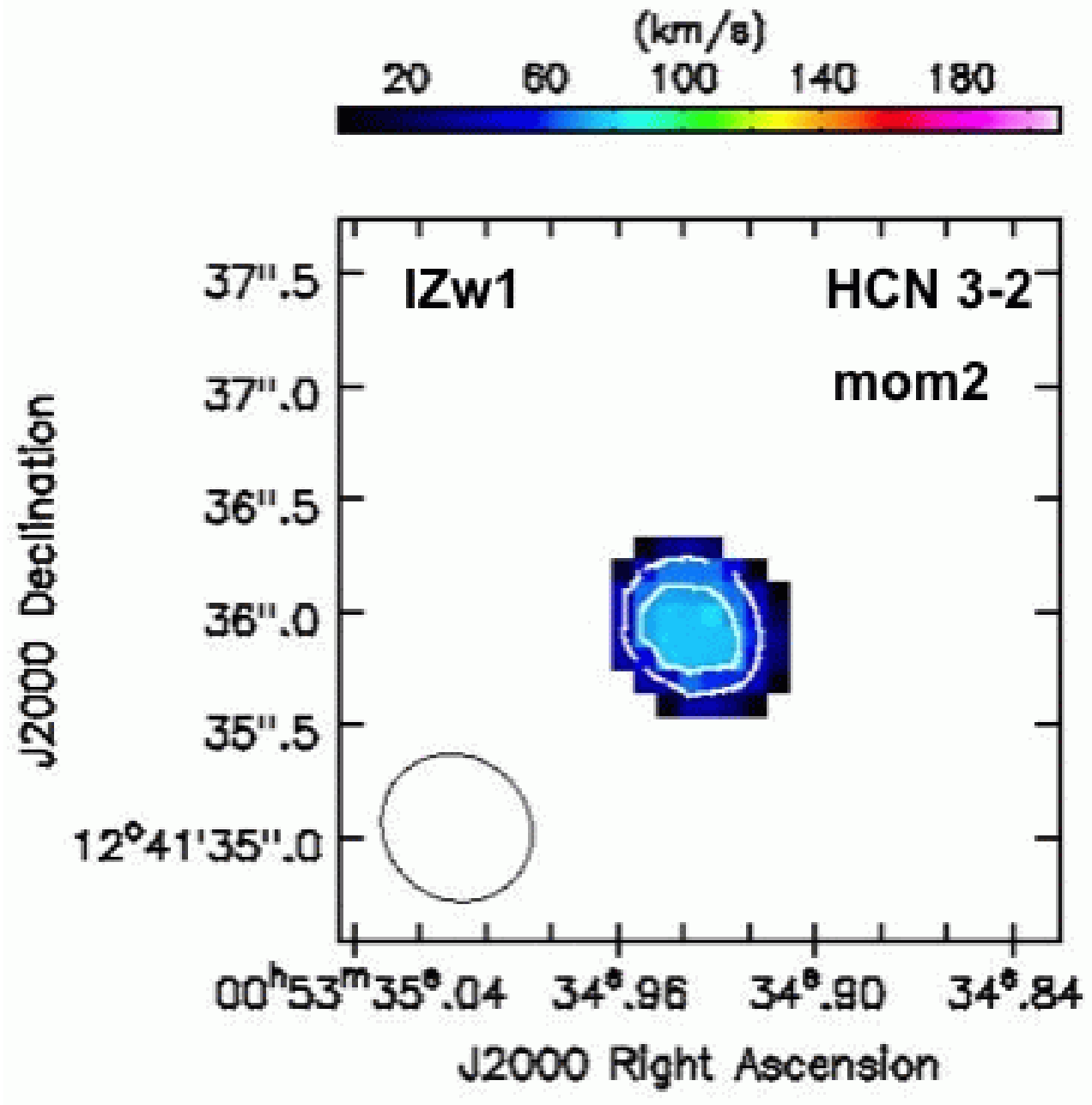} 
\includegraphics[angle=0,scale=.28]{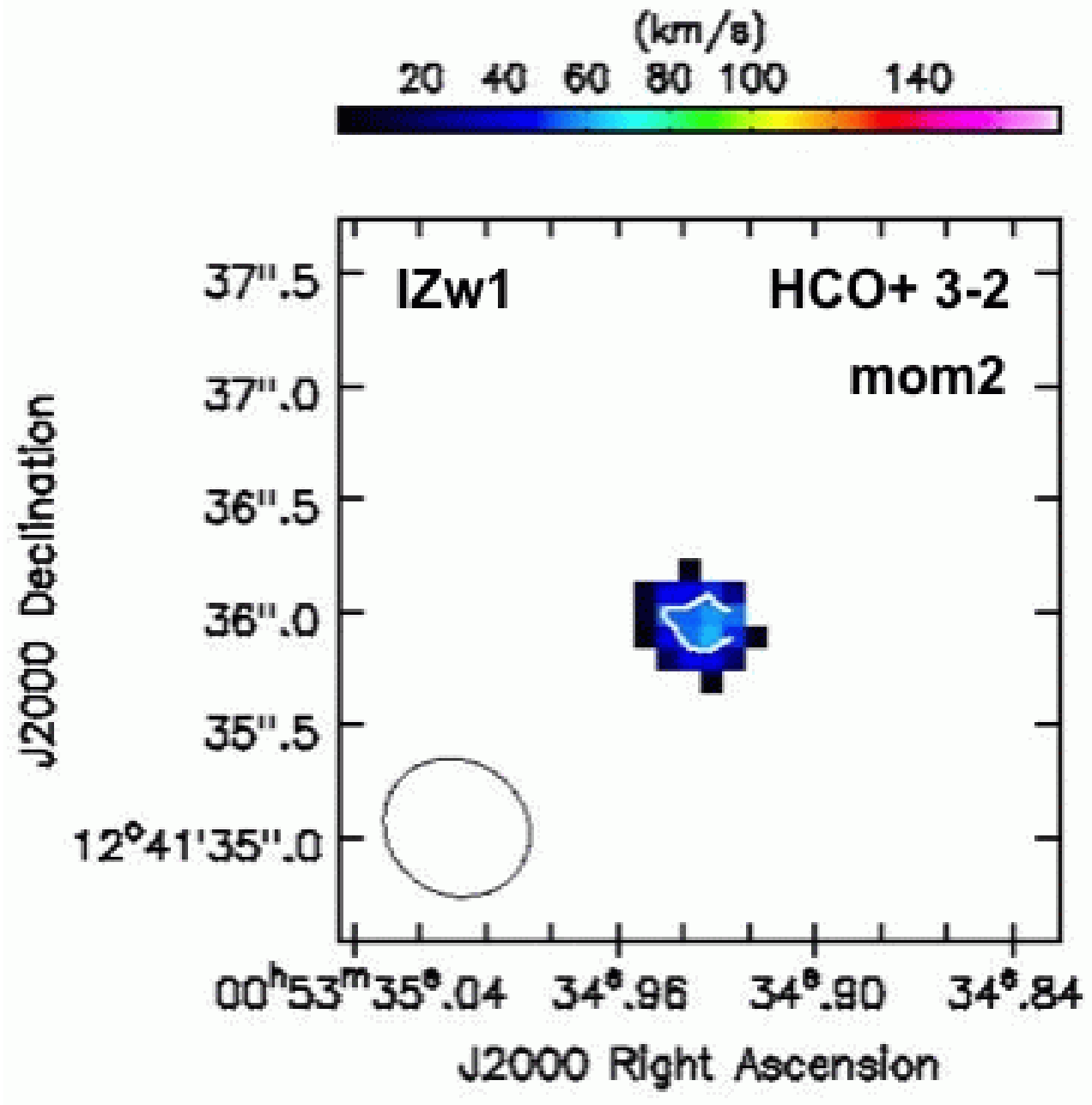} \\
\includegraphics[angle=0,scale=.28]{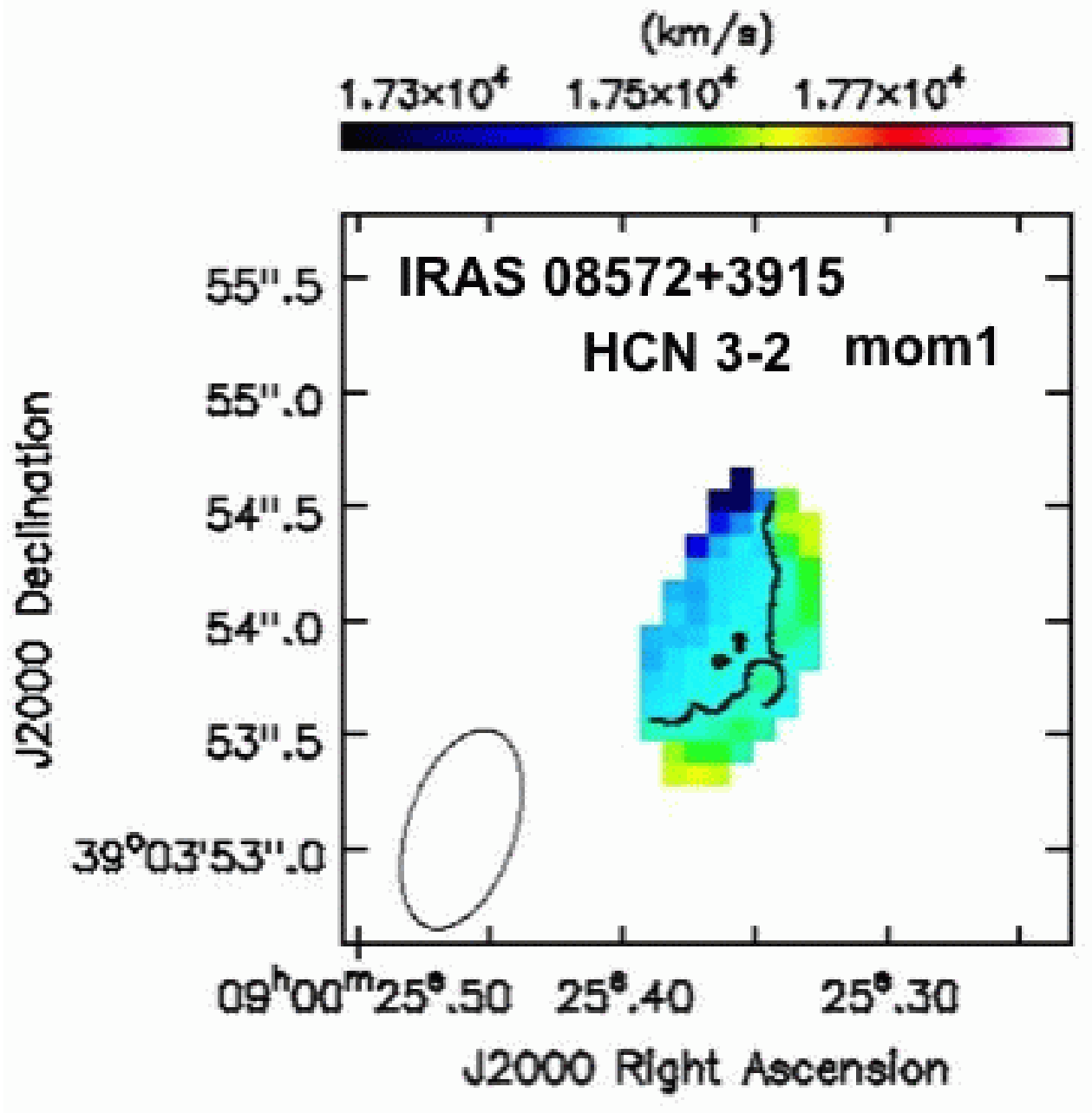} 
\includegraphics[angle=0,scale=.28]{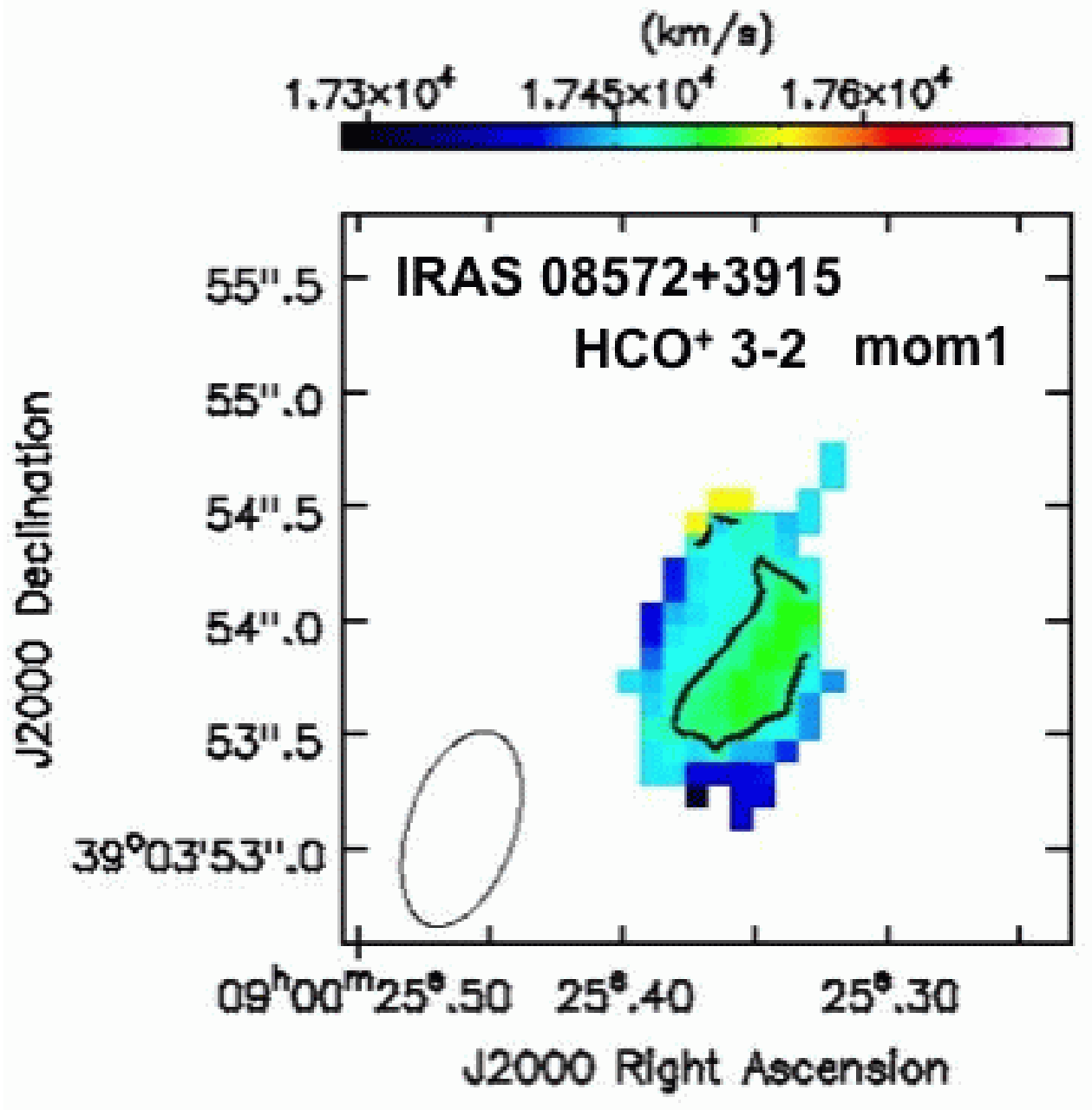}  
\includegraphics[angle=0,scale=.28]{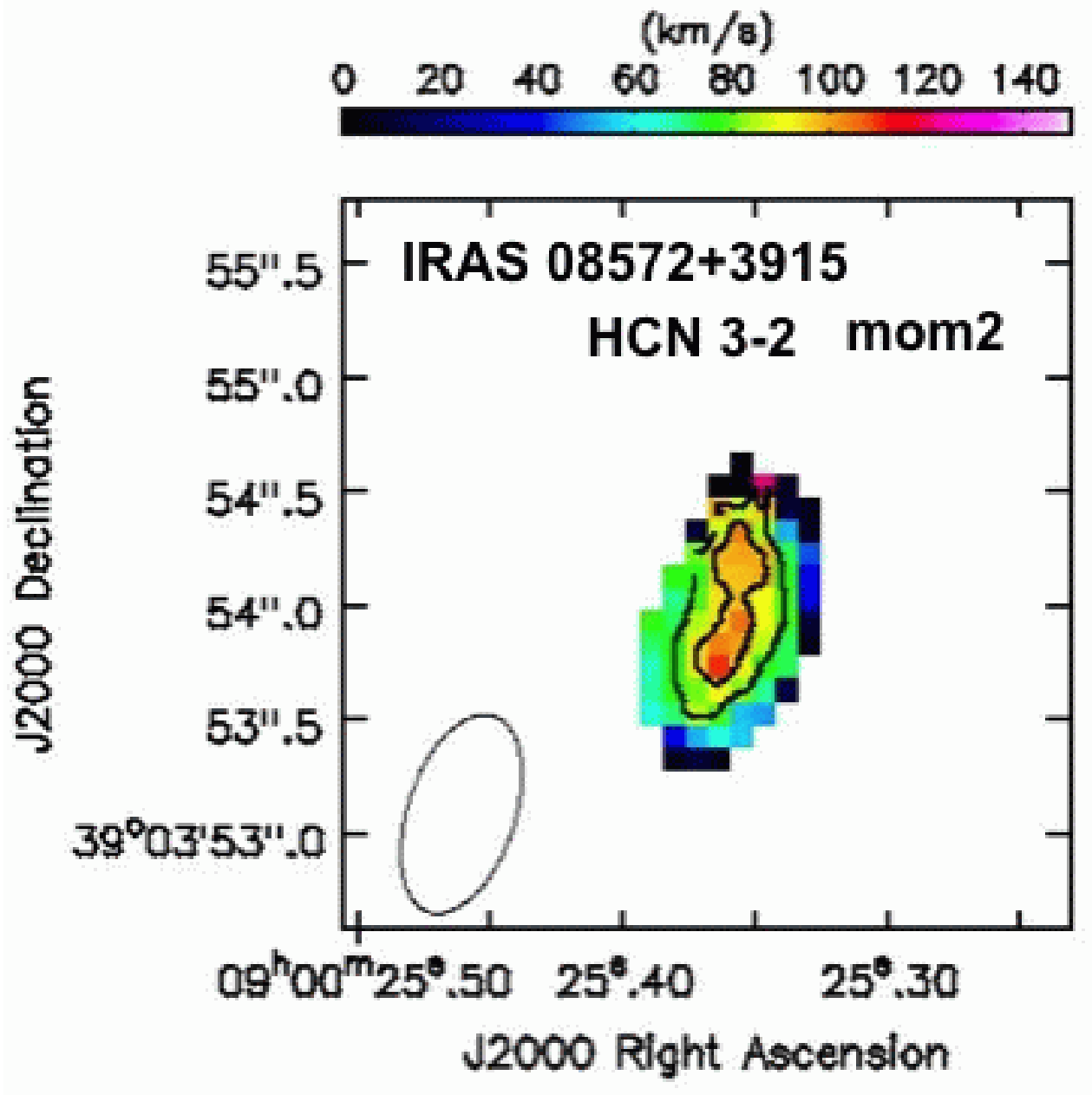} 
\includegraphics[angle=0,scale=.28]{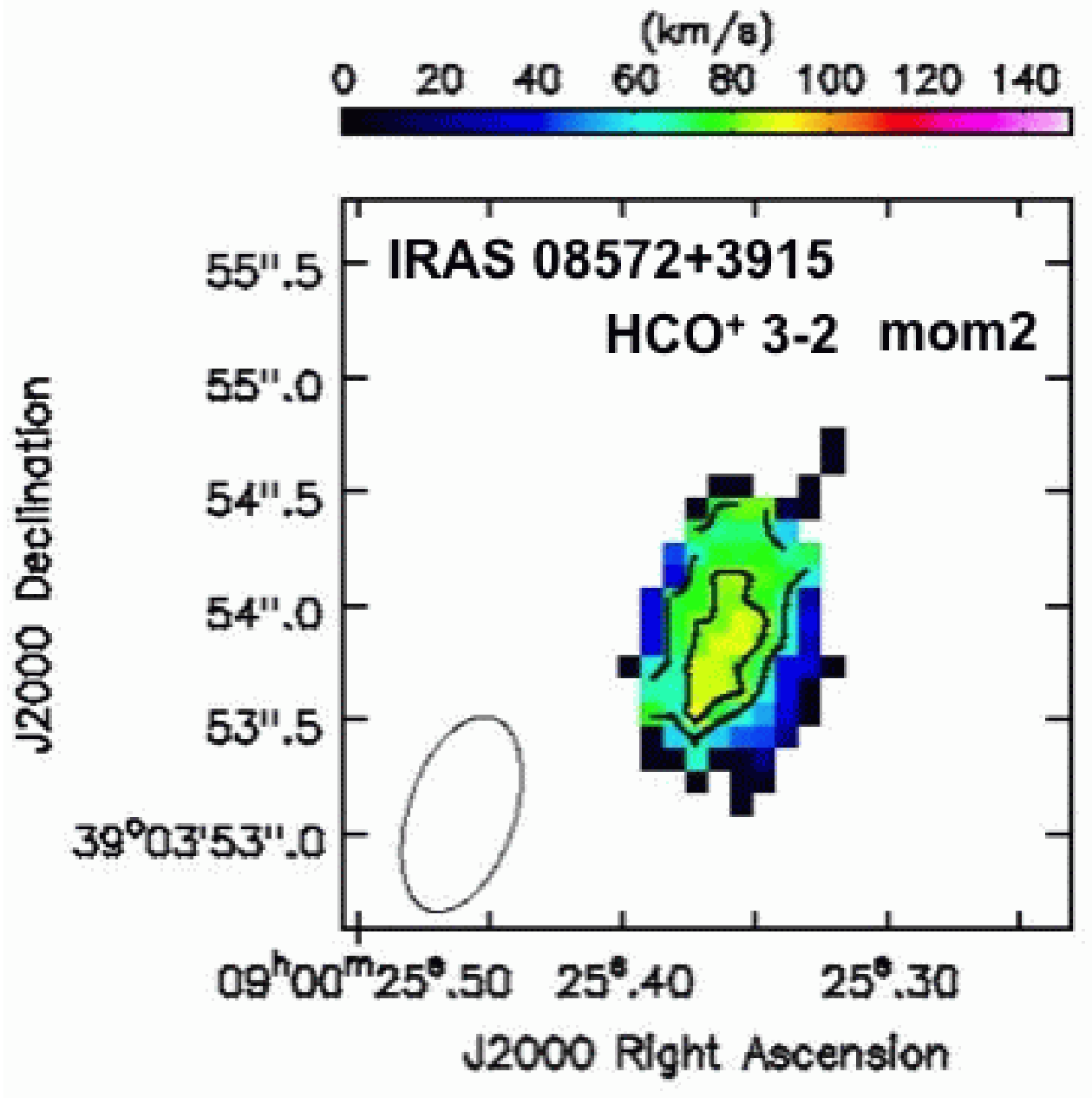}  \\
\includegraphics[angle=0,scale=.28]{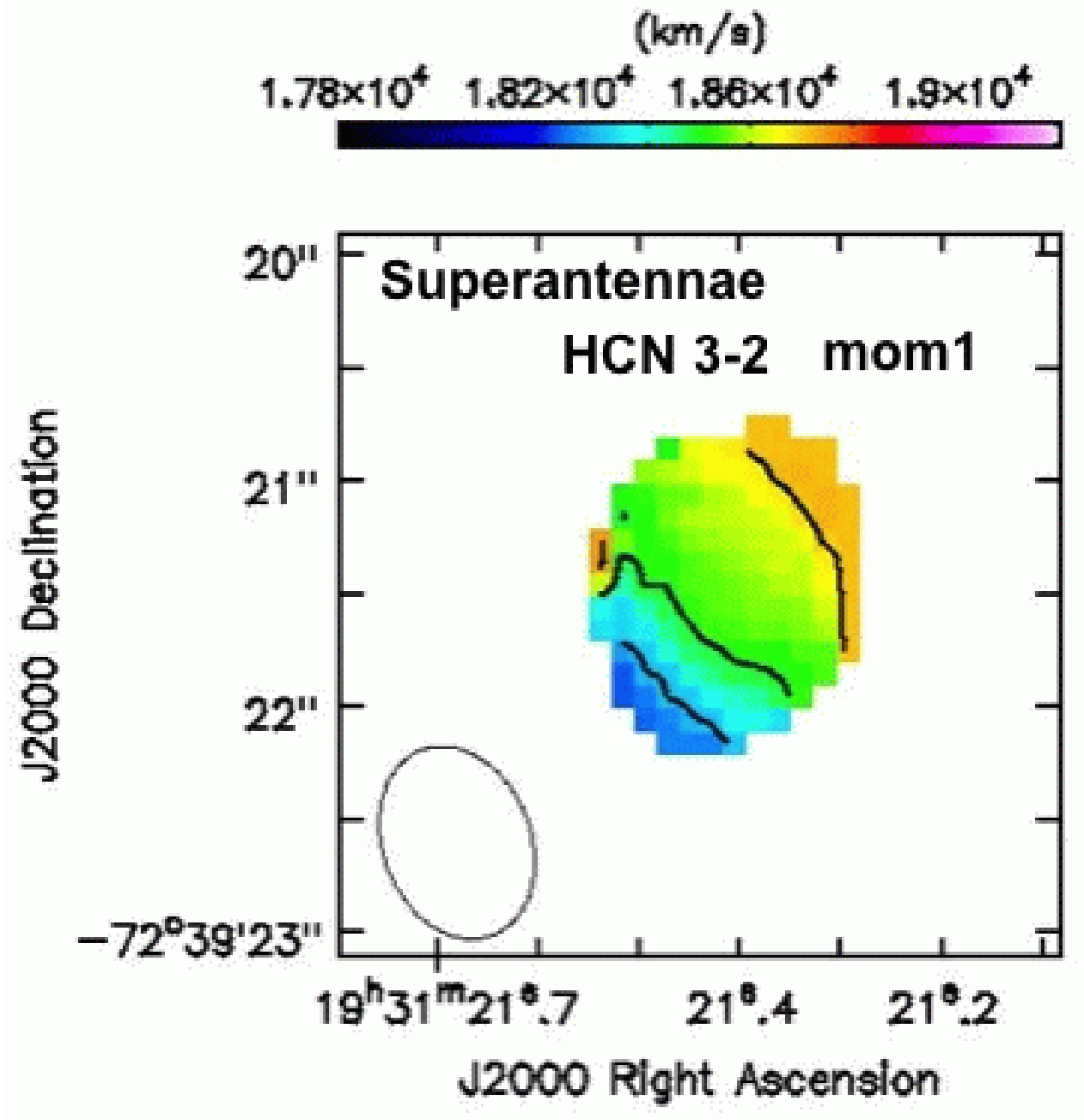} 
\includegraphics[angle=0,scale=.28]{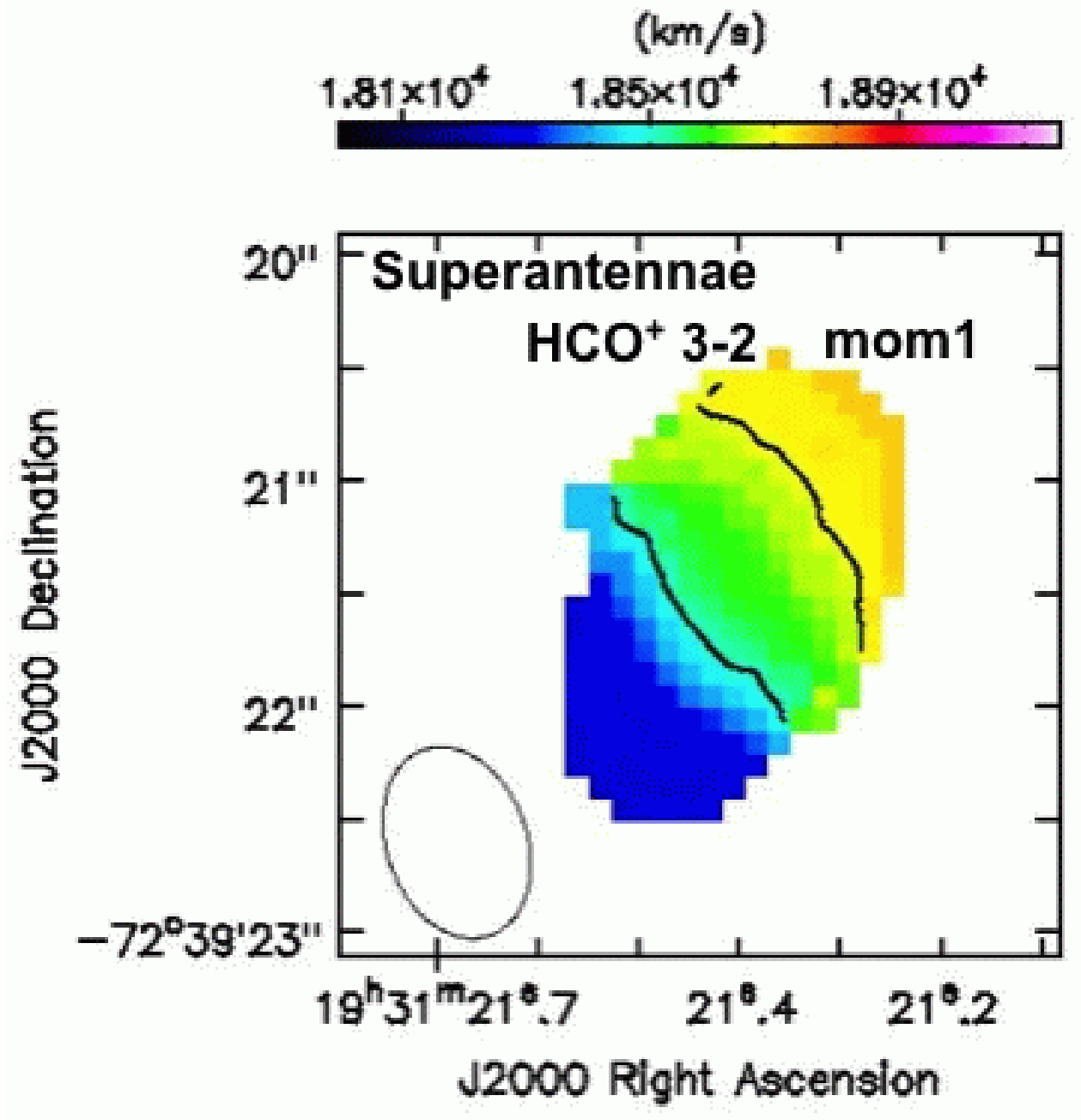}  
\includegraphics[angle=0,scale=.28]{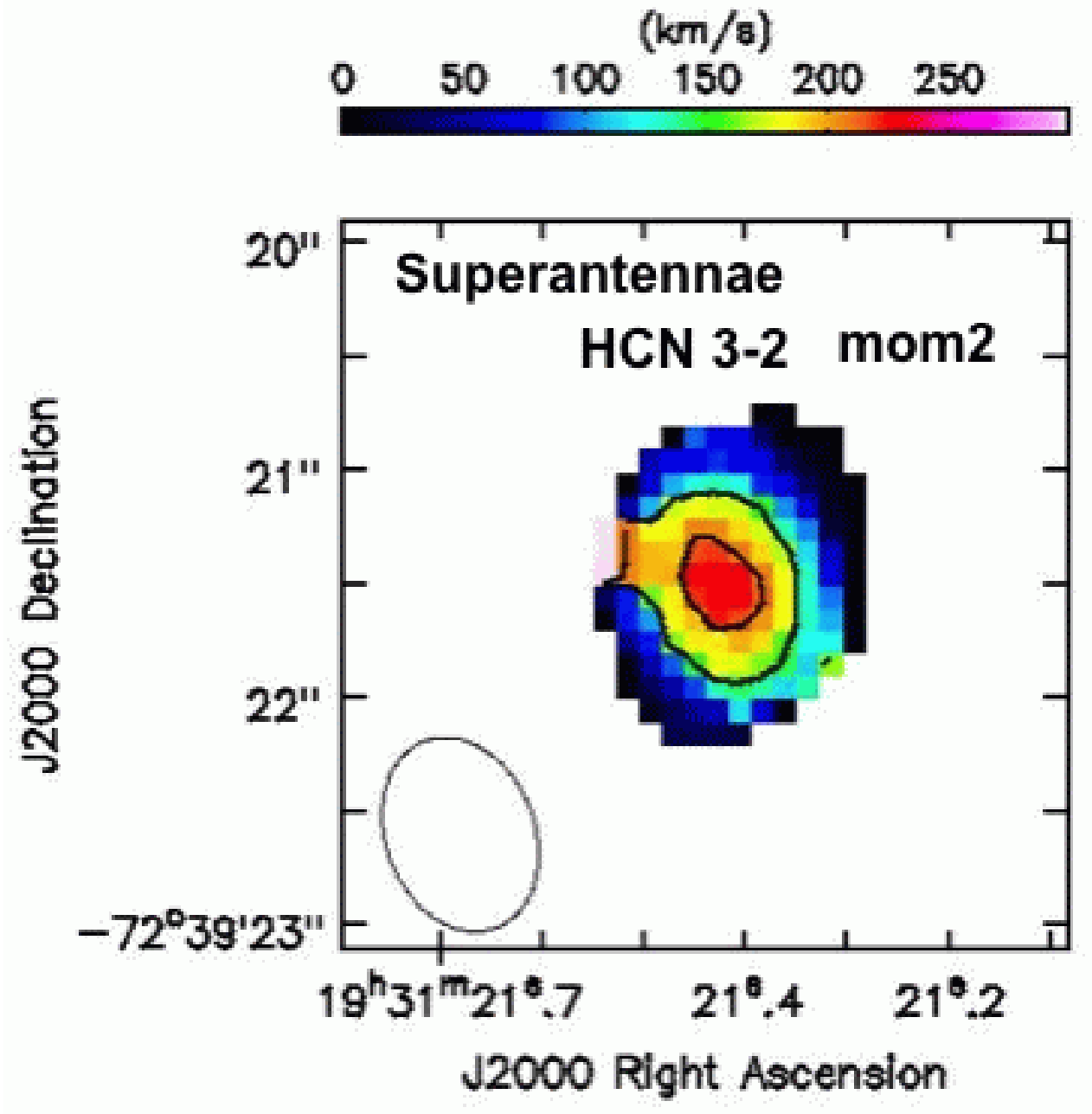} 
\includegraphics[angle=0,scale=.28]{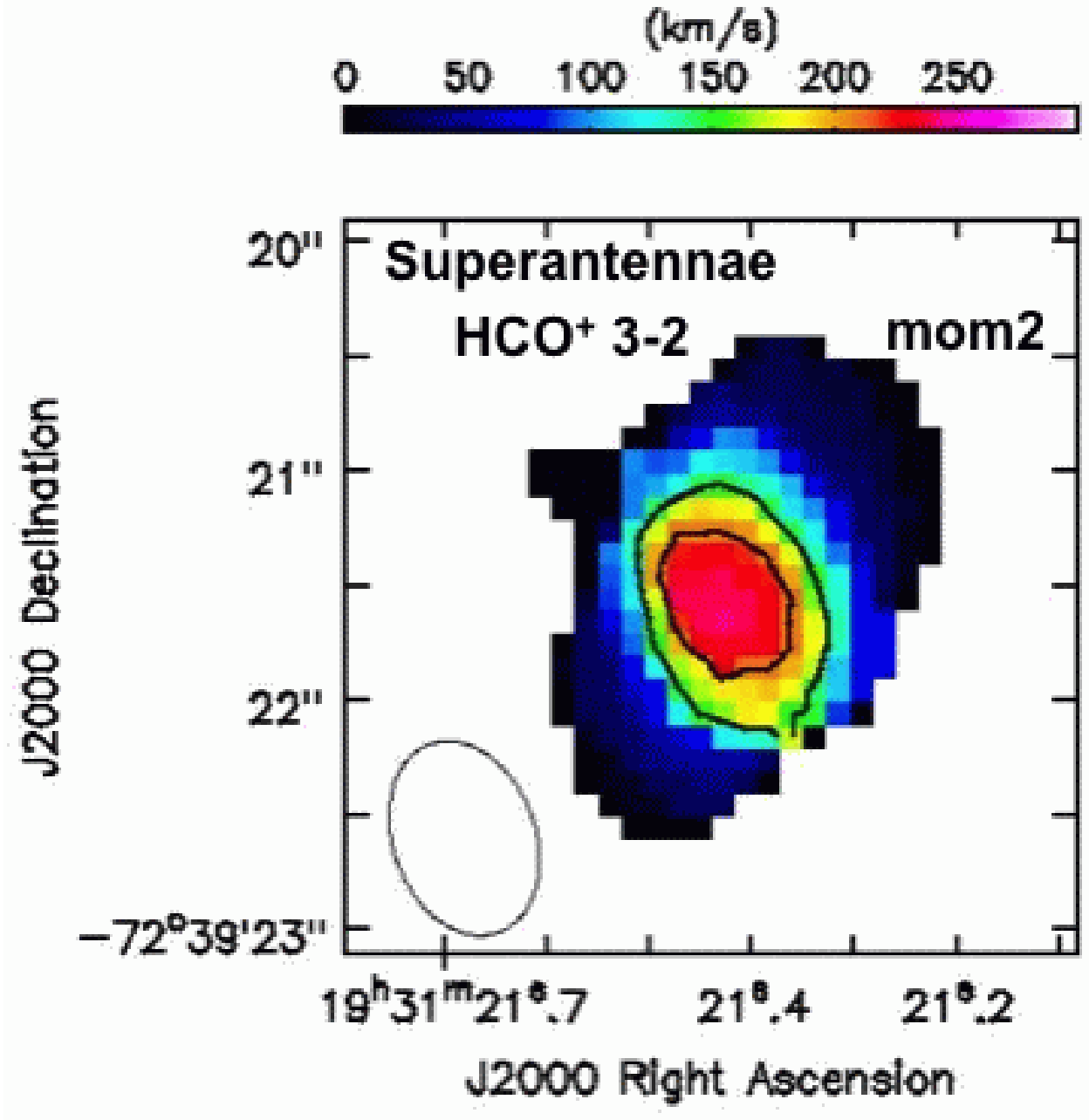}  \\
\end{center}
\end{figure}

\clearpage

\begin{figure}
\begin{center}
\includegraphics[angle=0,scale=.28]{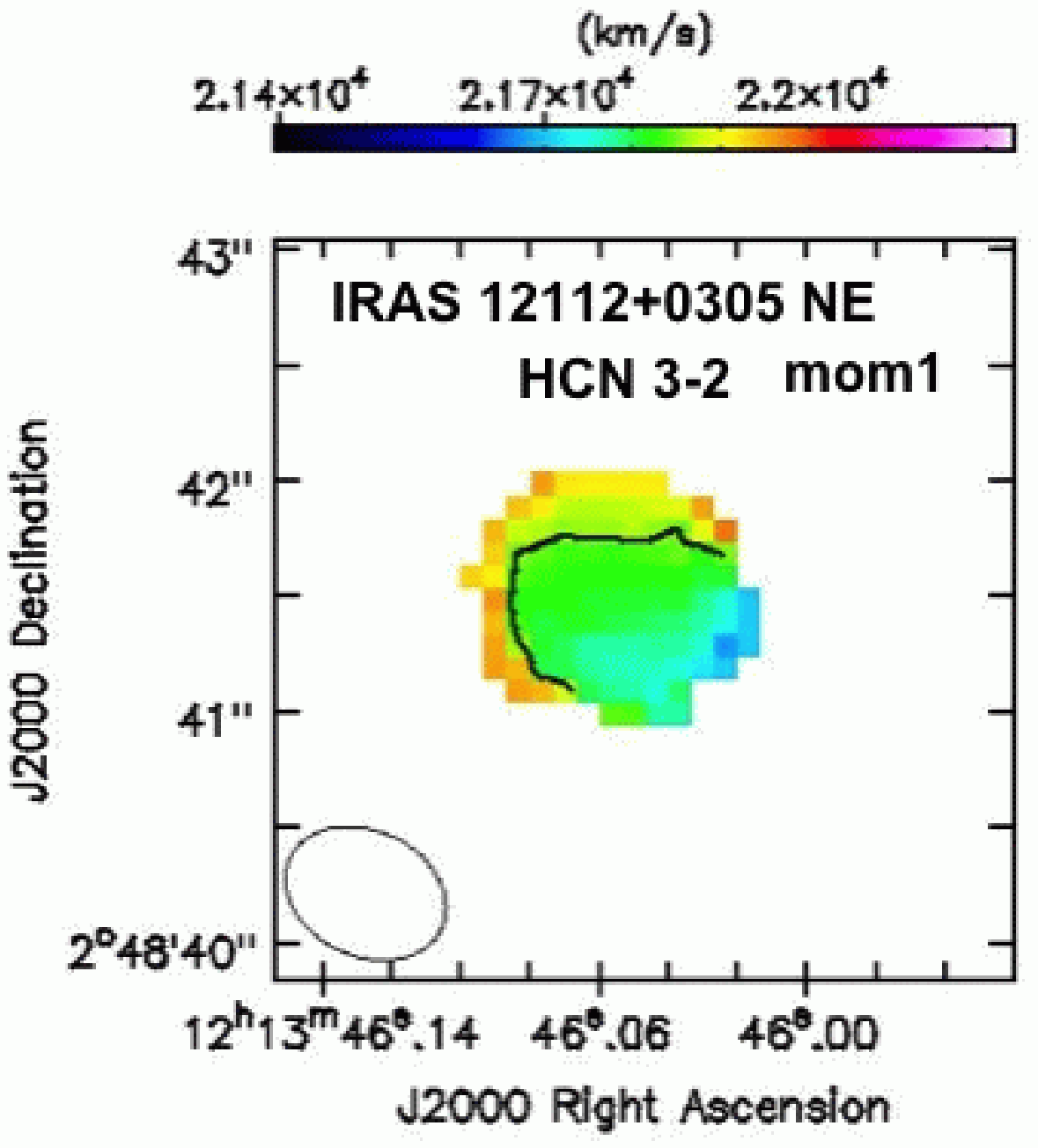} 
\includegraphics[angle=0,scale=.28]{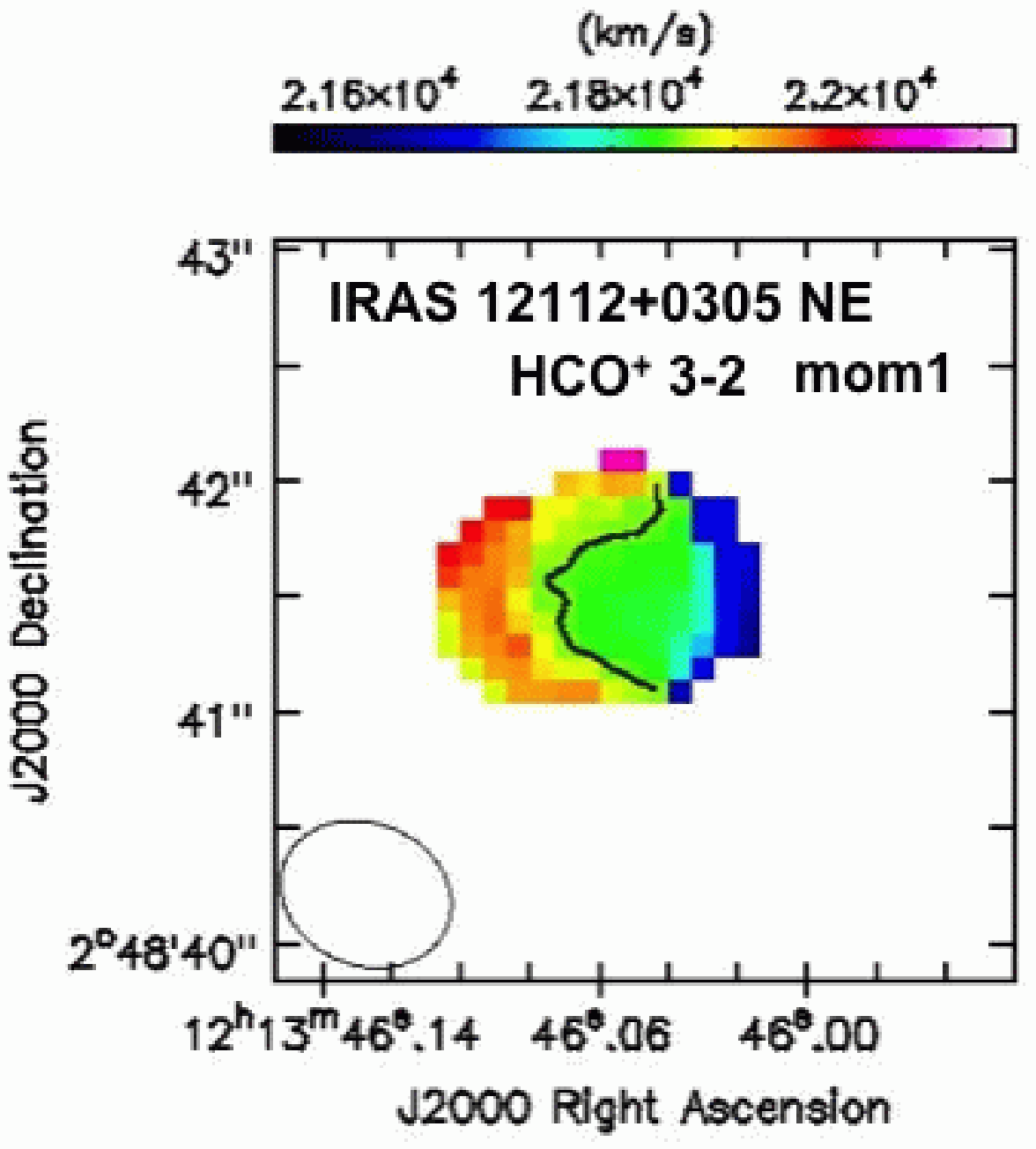}  
\includegraphics[angle=0,scale=.28]{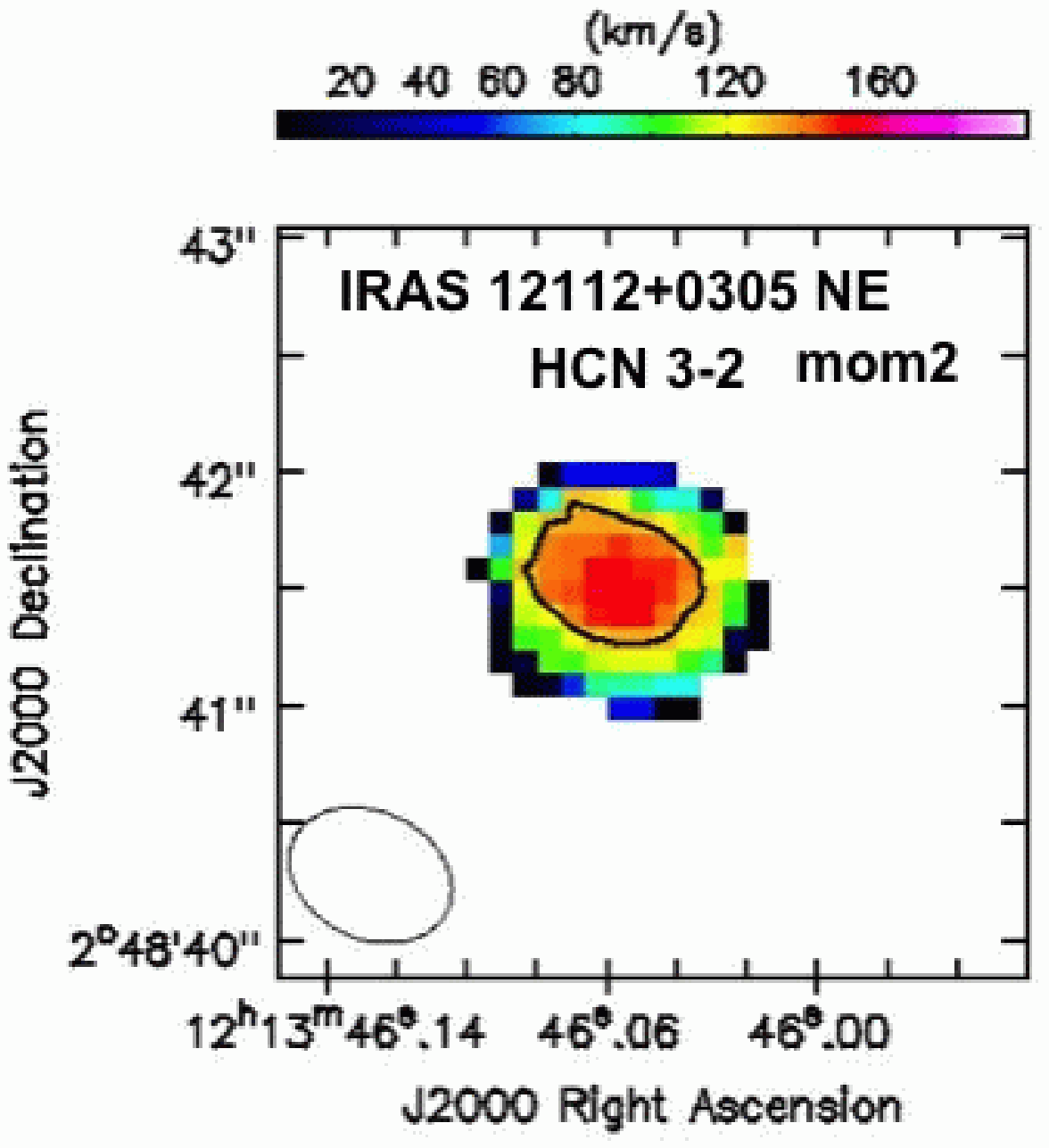} 
\includegraphics[angle=0,scale=.28]{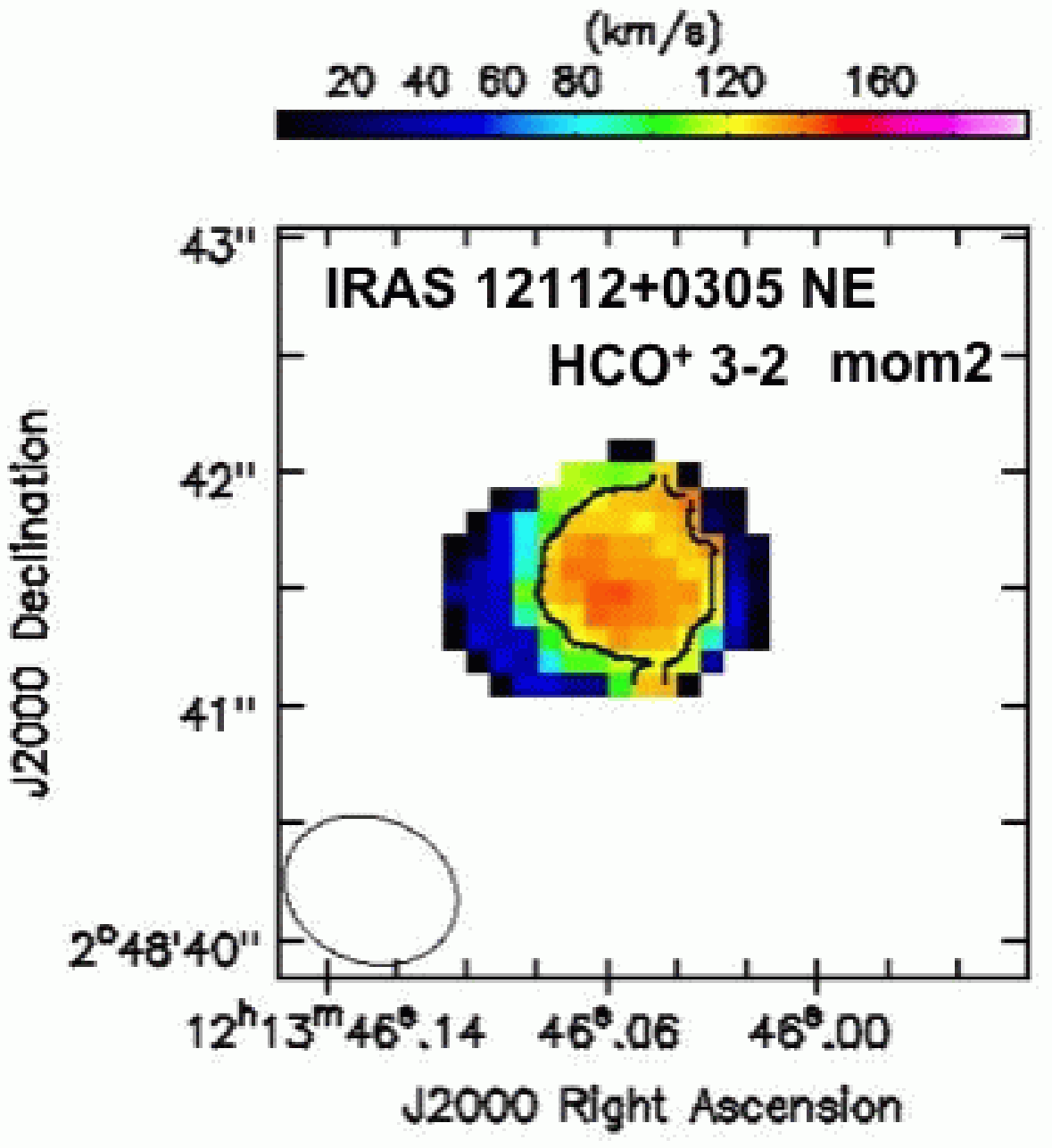}  \\
\includegraphics[angle=0,scale=.28]{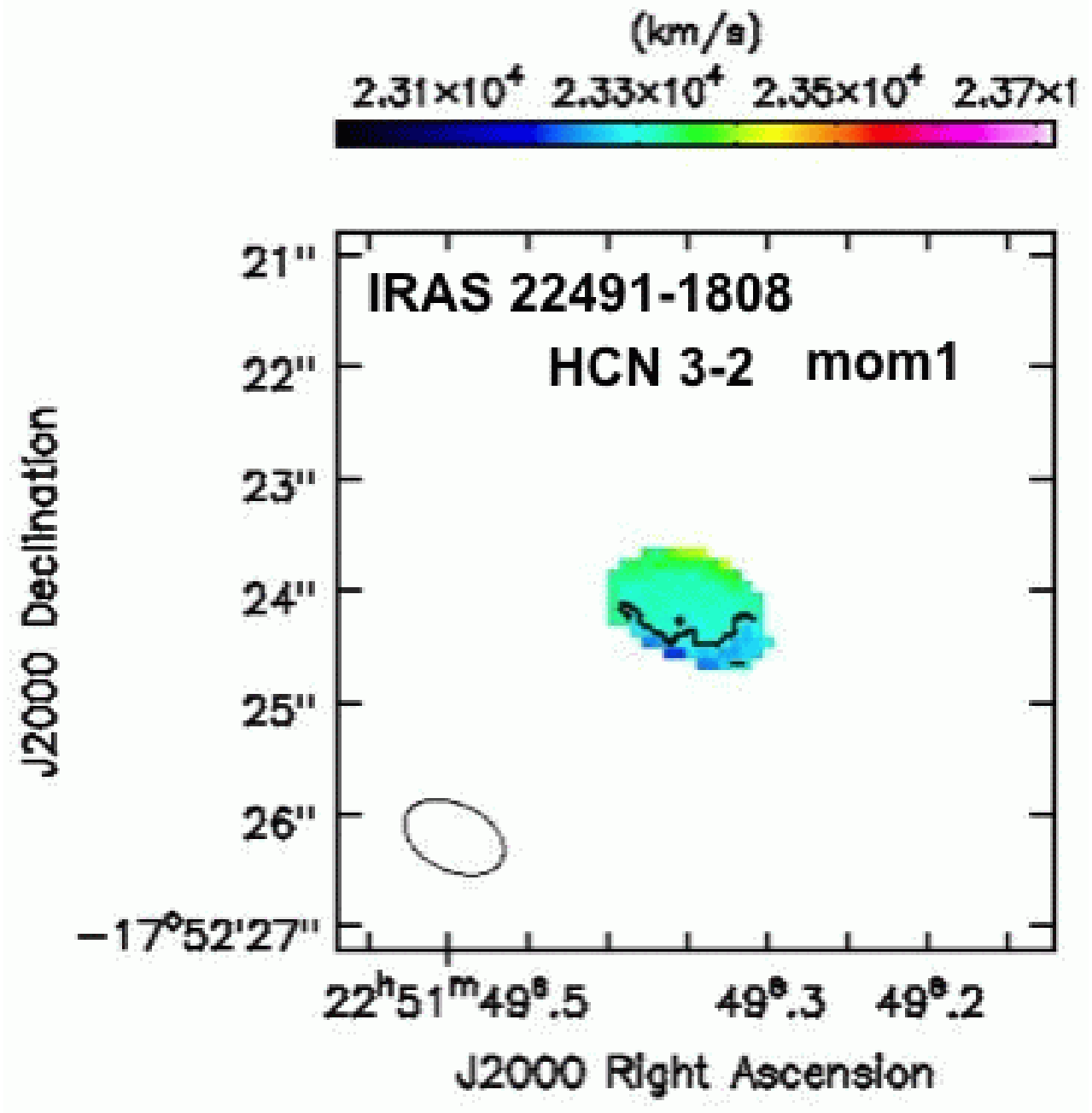} 
\includegraphics[angle=0,scale=.28]{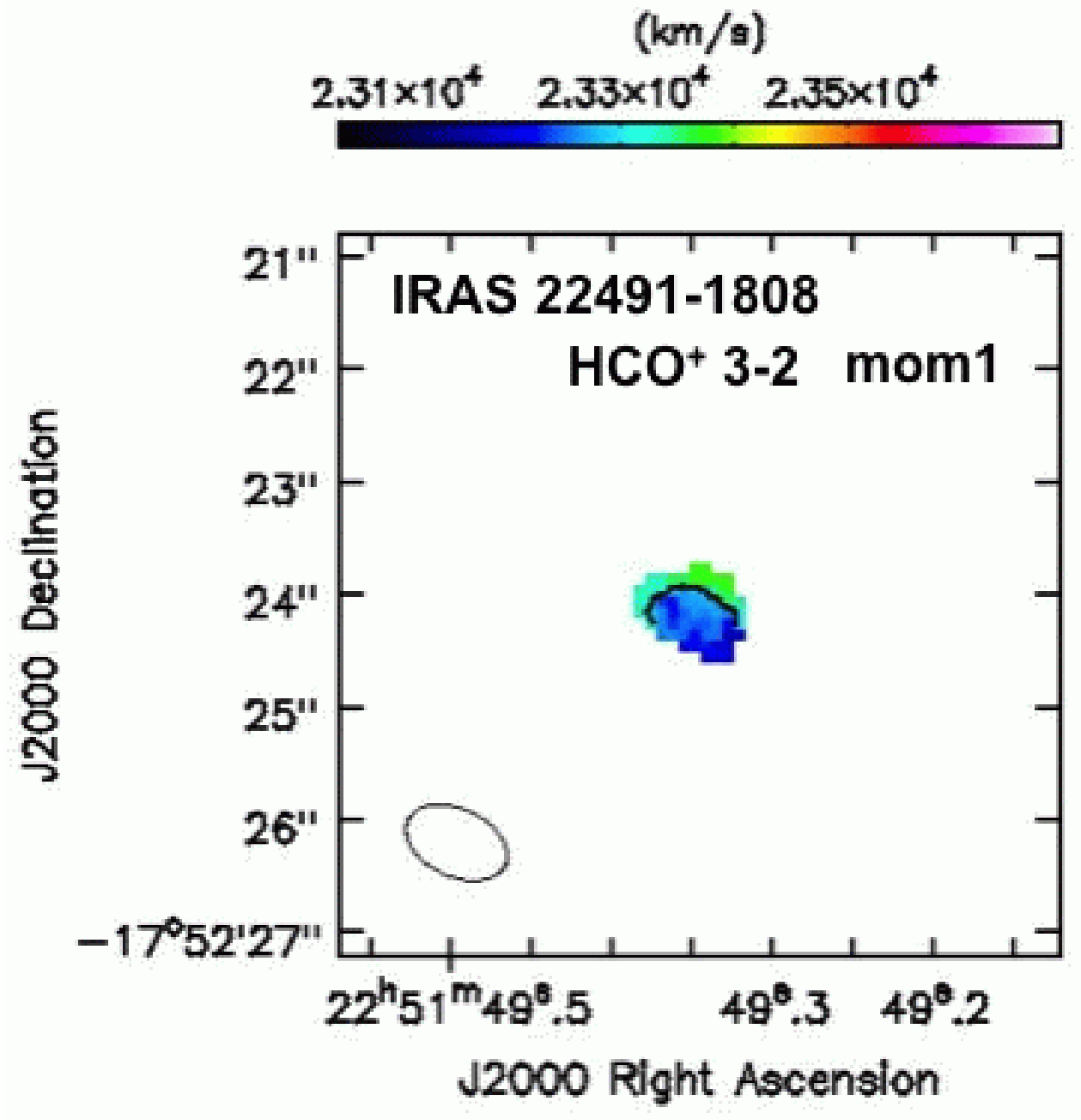}  
\includegraphics[angle=0,scale=.28]{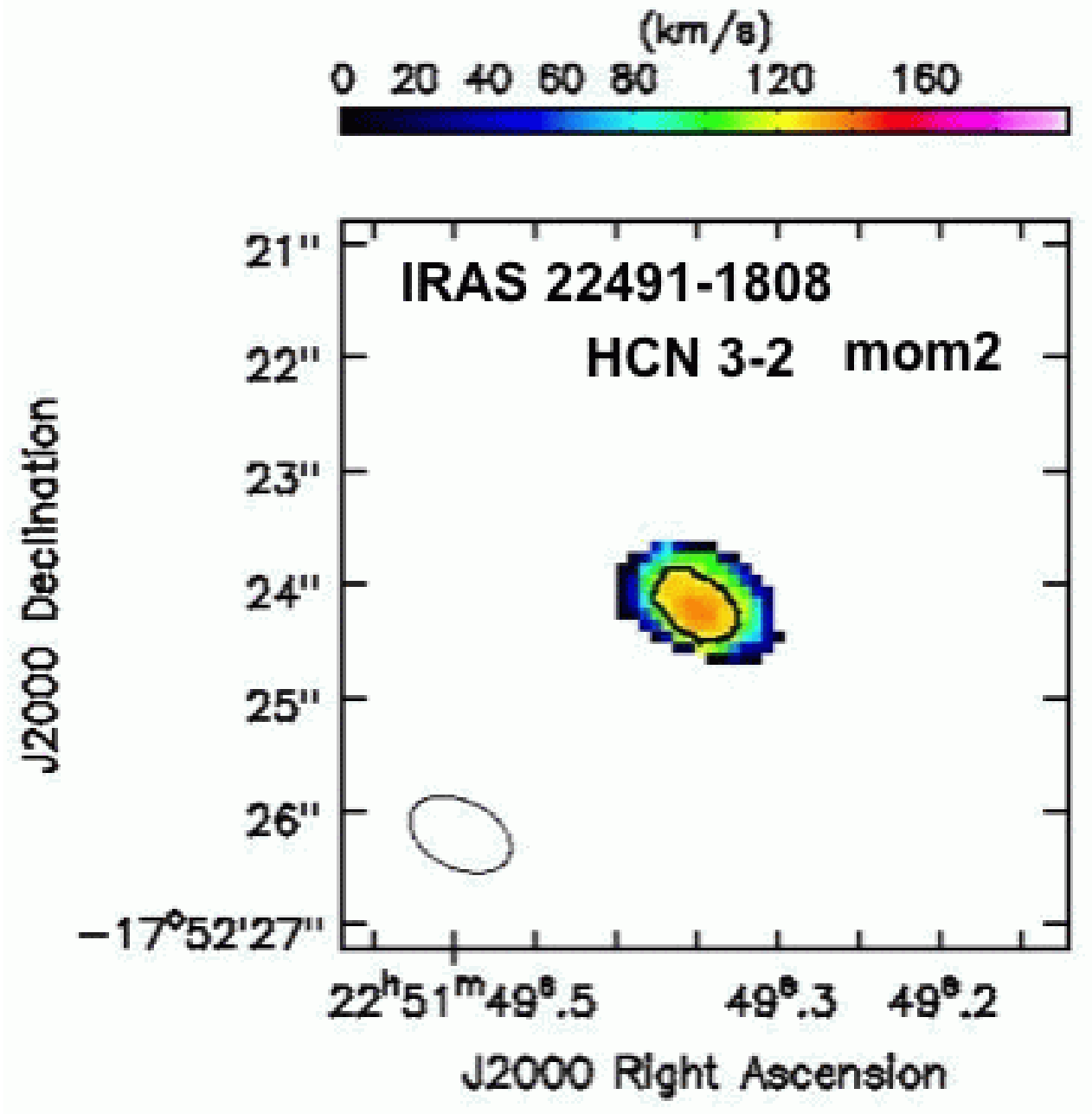} 
\includegraphics[angle=0,scale=.28]{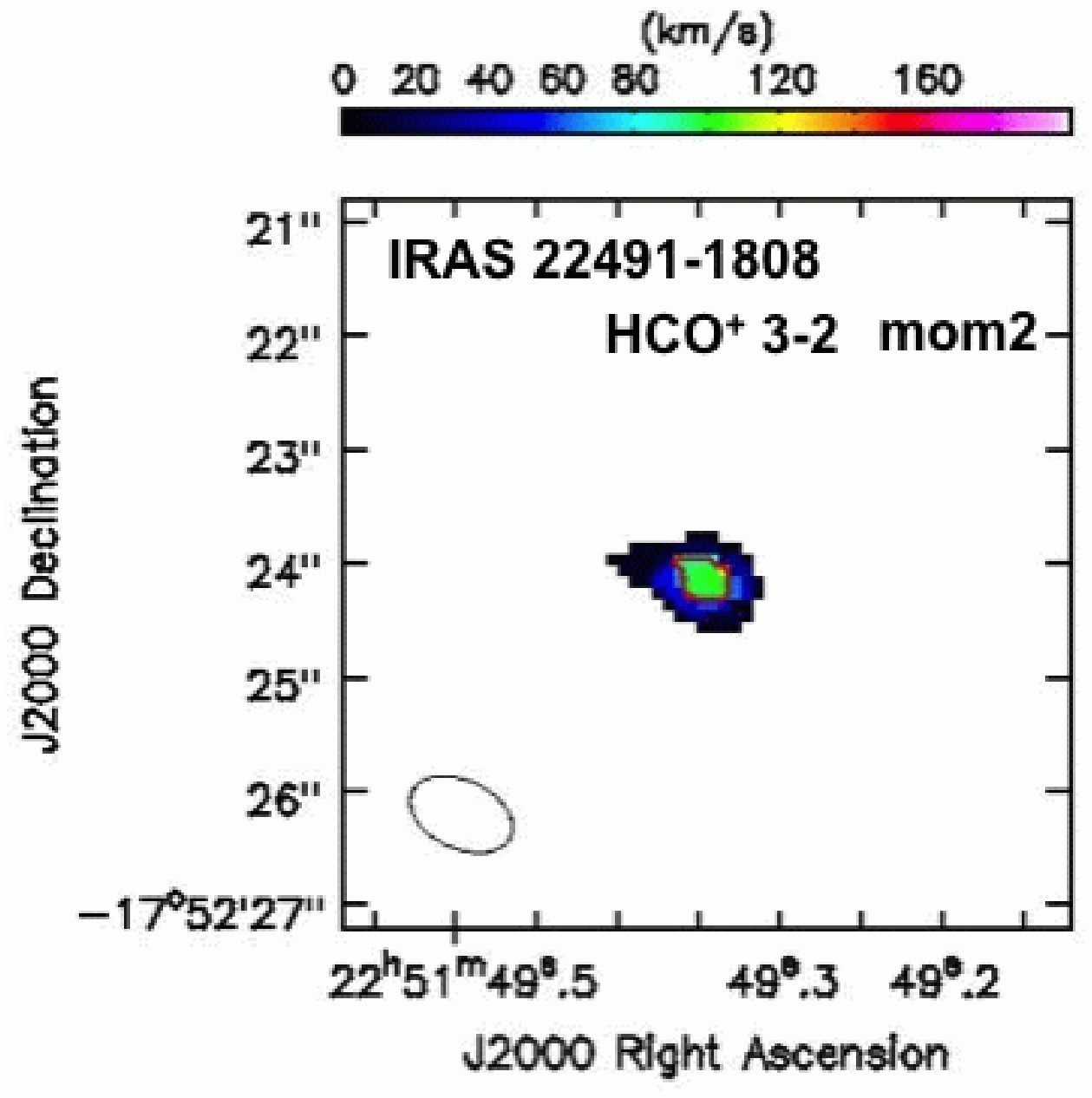}  \\
\includegraphics[angle=0,scale=.28]{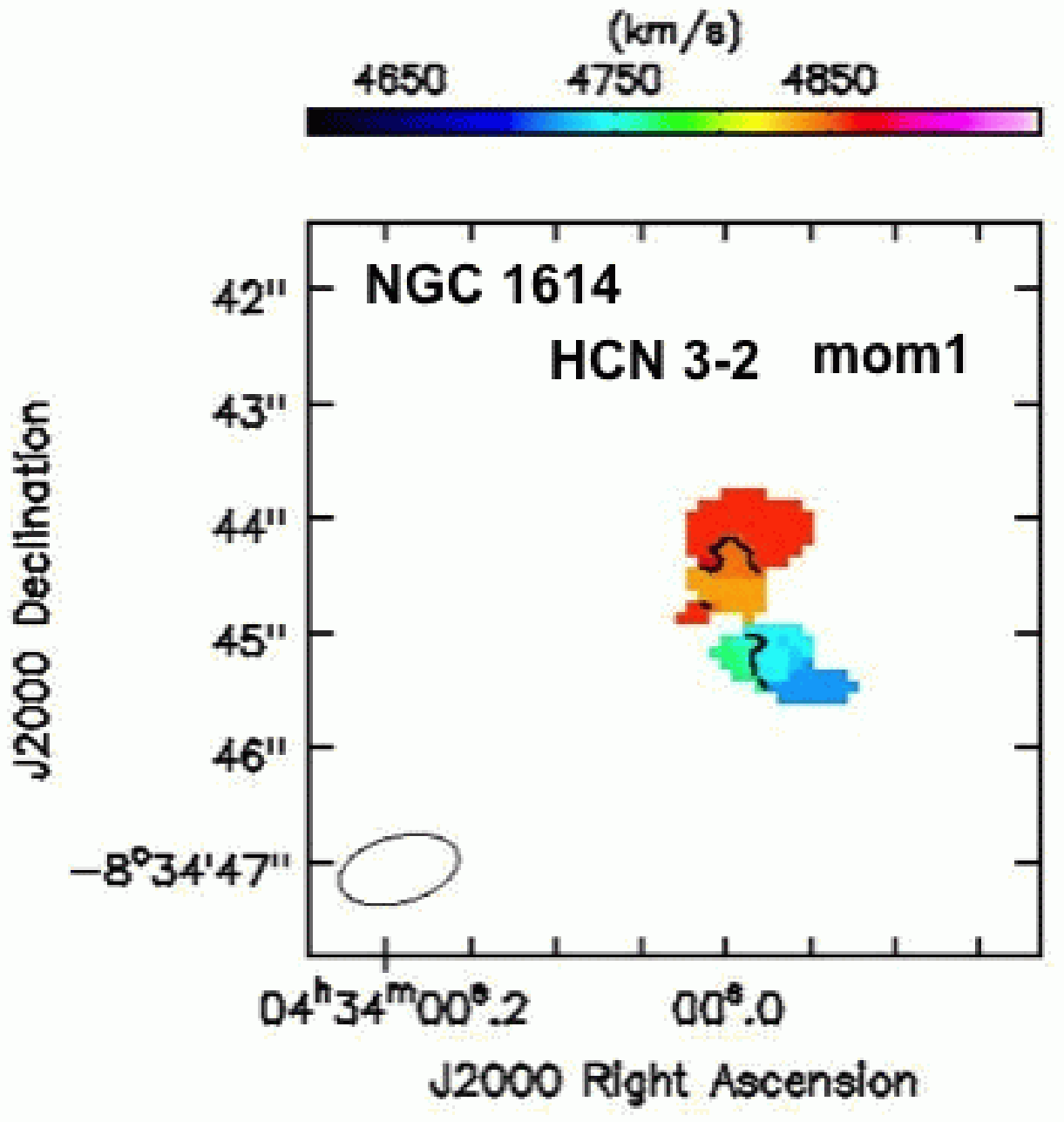} 
\includegraphics[angle=0,scale=.28]{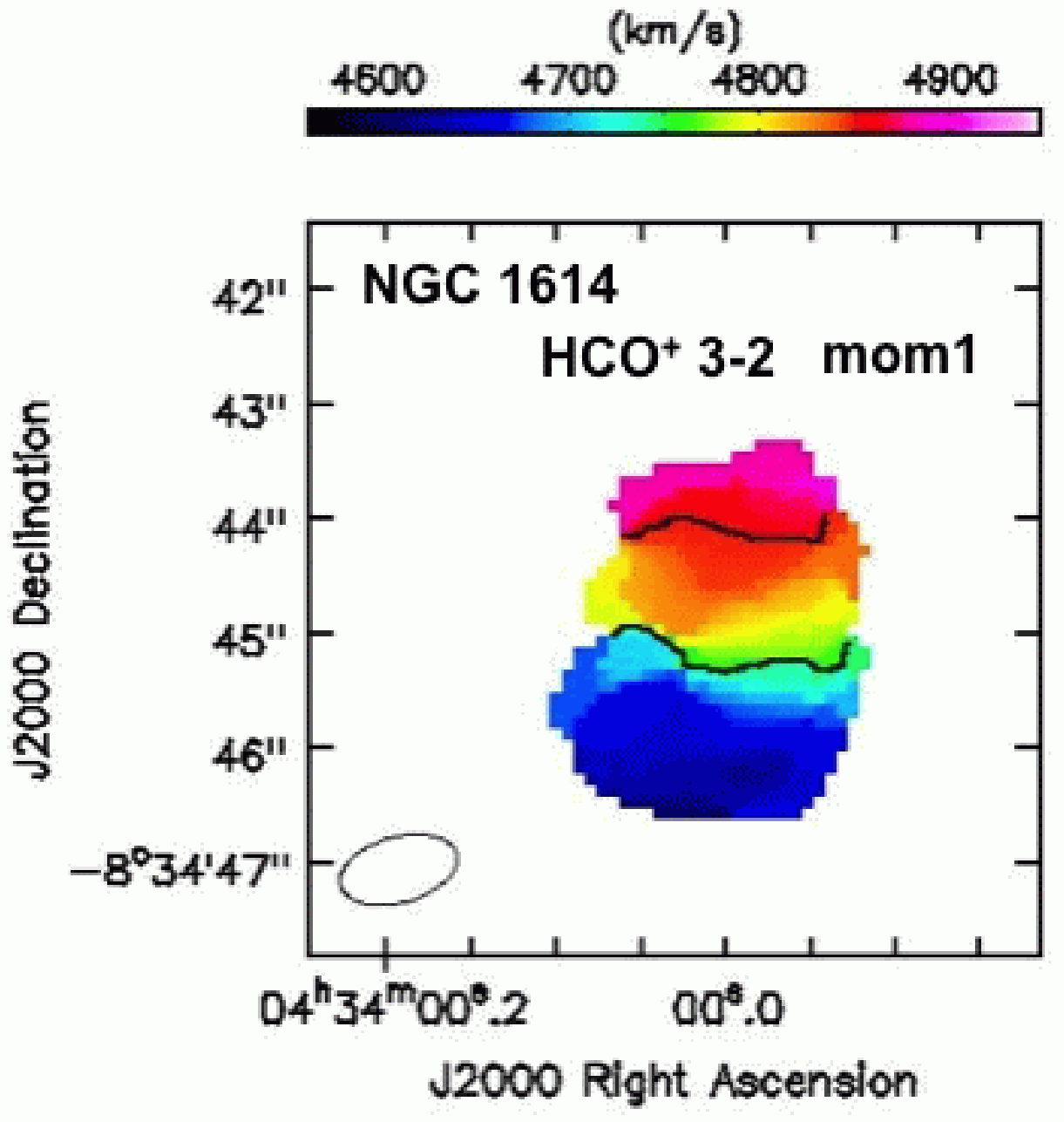}  
\includegraphics[angle=0,scale=.28]{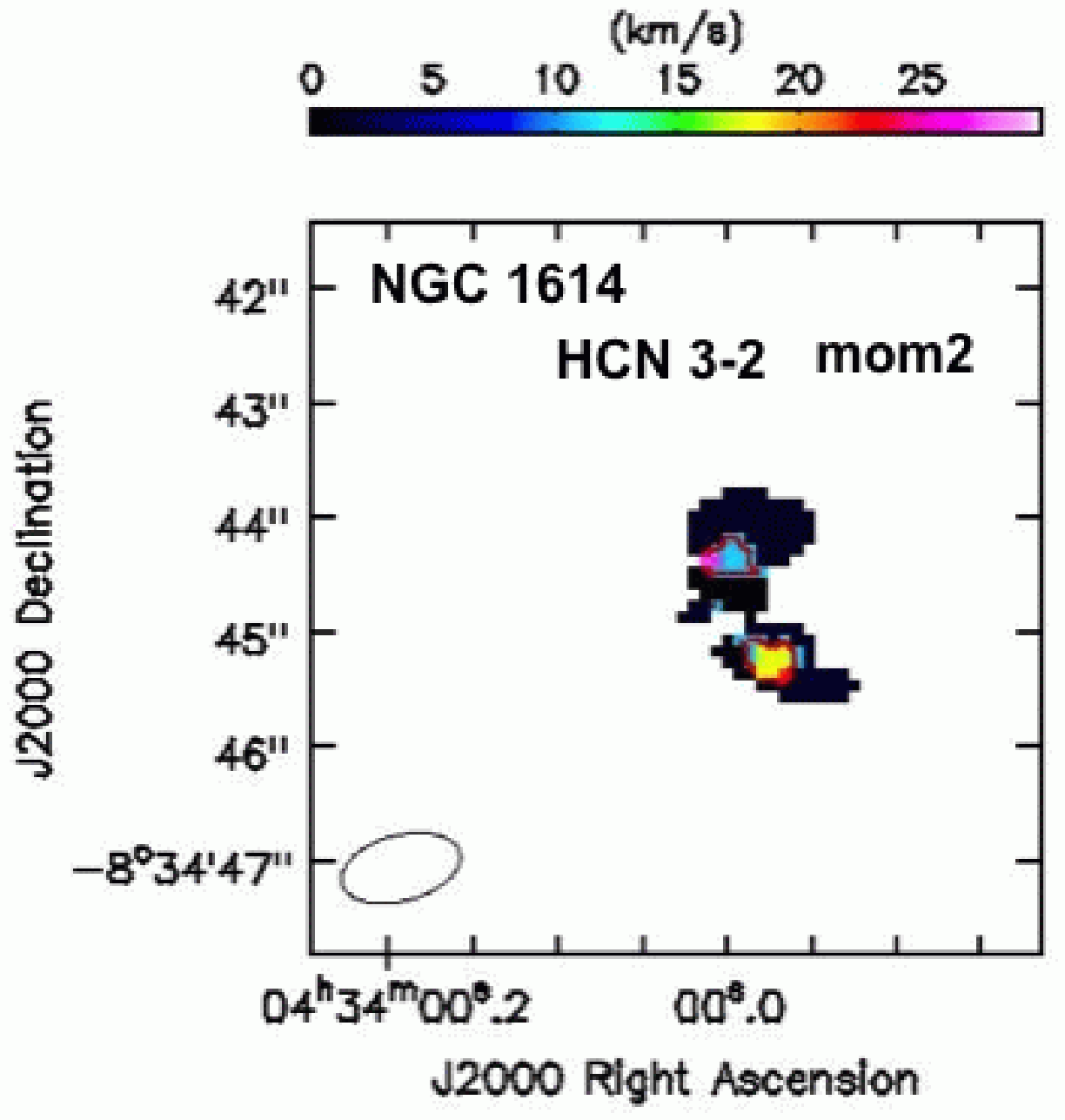} 
\includegraphics[angle=0,scale=.28]{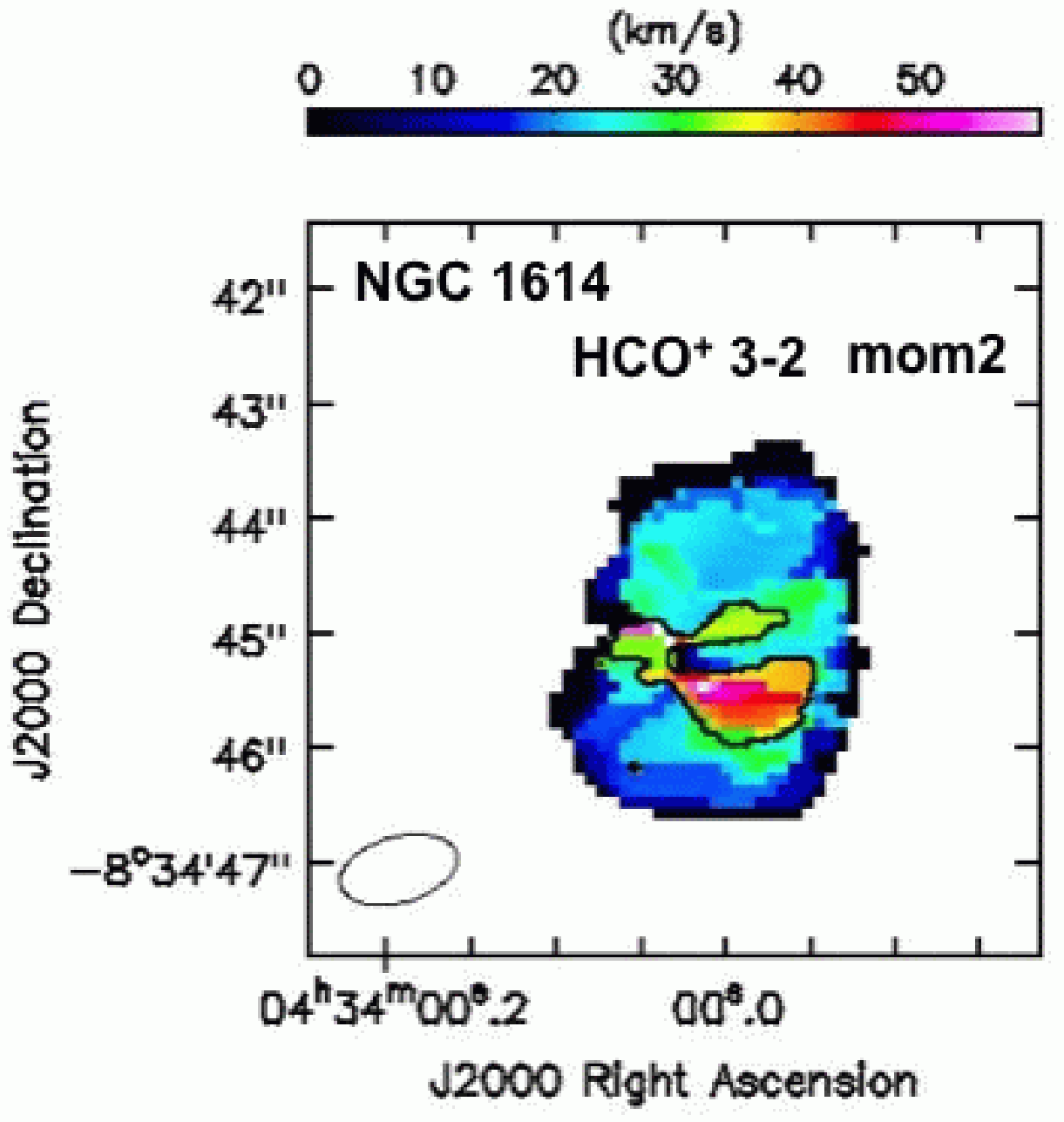}  \\
\includegraphics[angle=0,scale=.28]{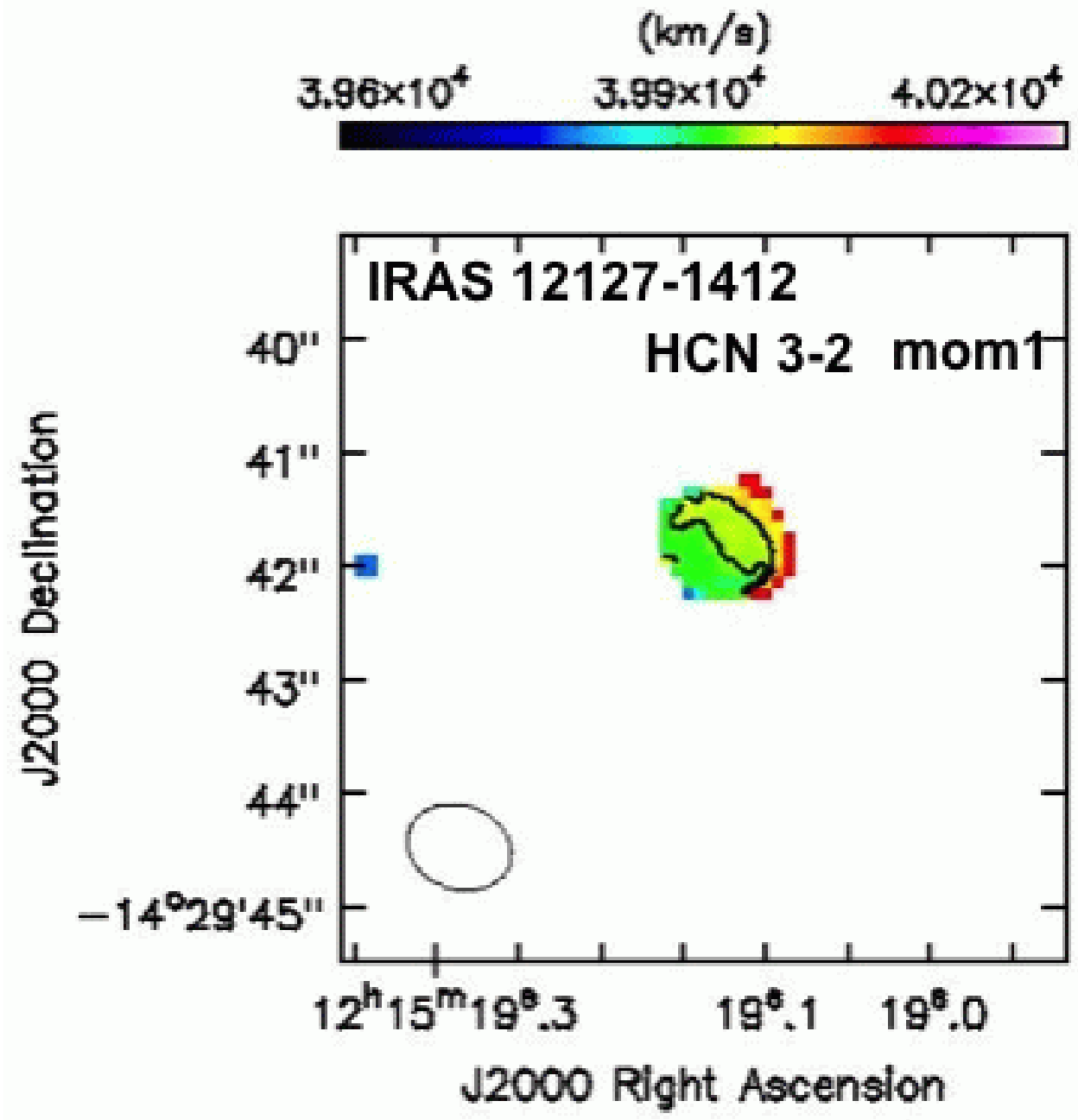} 
\includegraphics[angle=0,scale=.28]{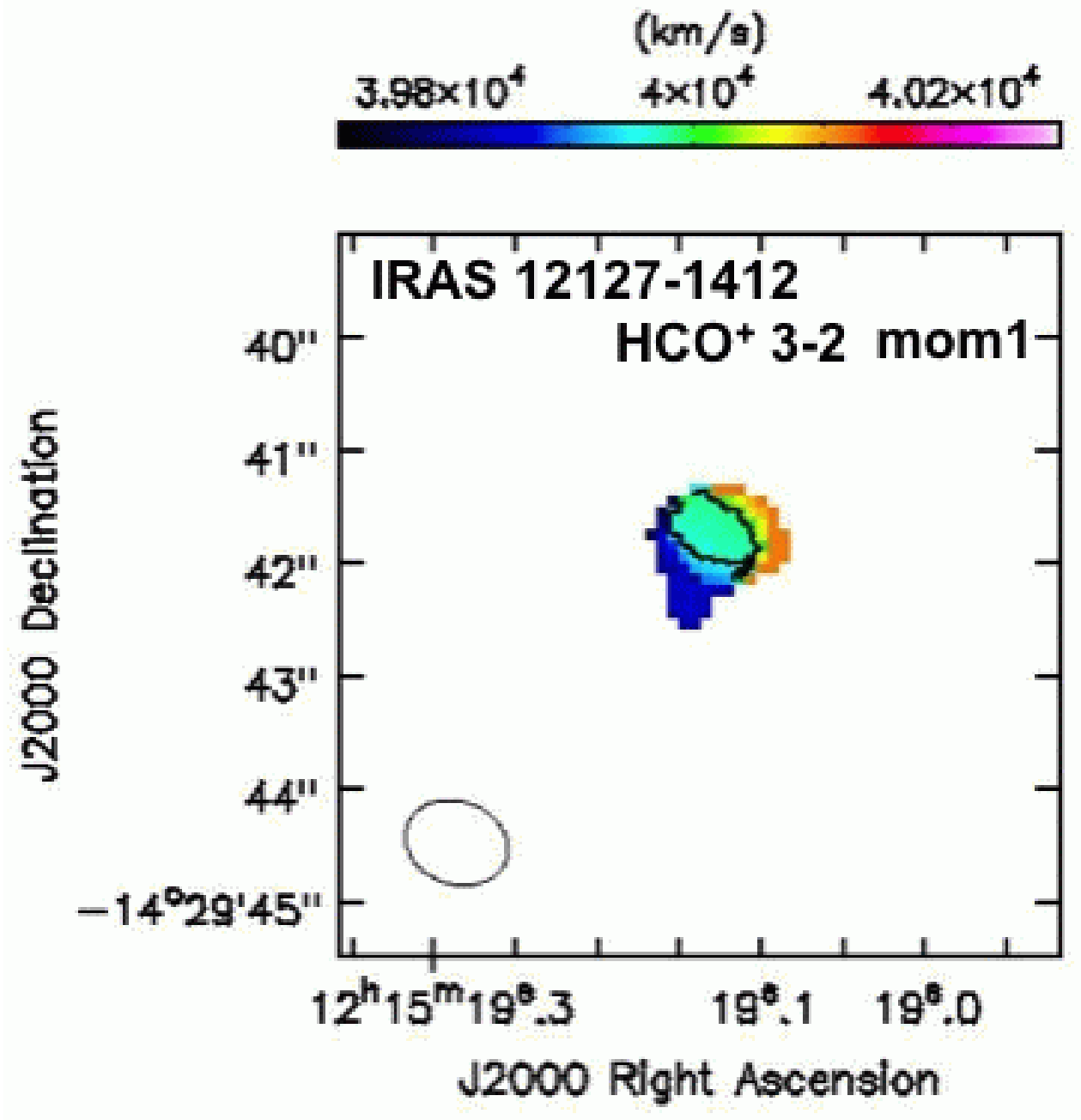} 
\includegraphics[angle=0,scale=.28]{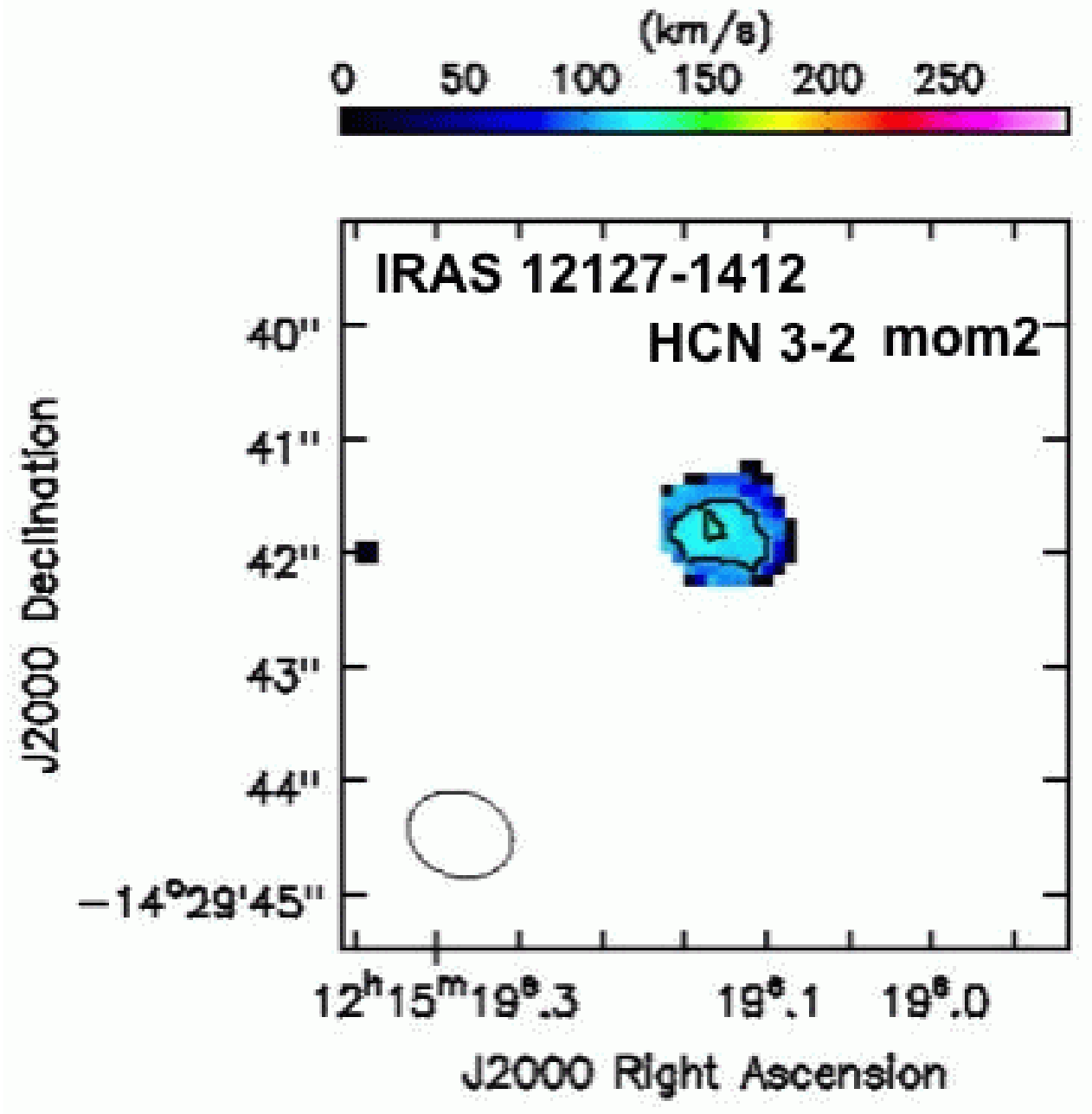} 
\includegraphics[angle=0,scale=.28]{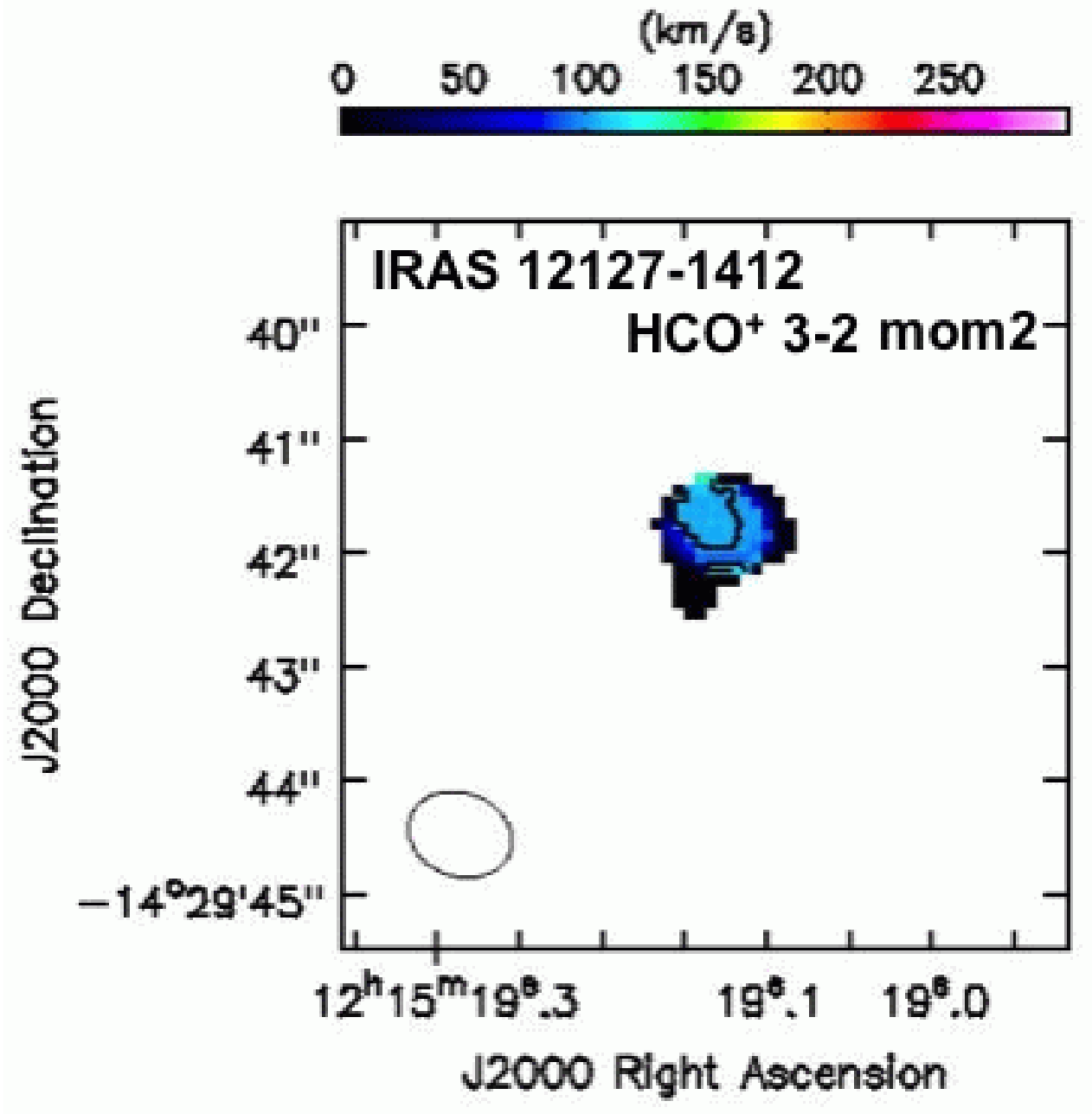} \\
\end{center}
\end{figure}

\clearpage

\begin{figure}
\begin{center}
\includegraphics[angle=0,scale=.25]{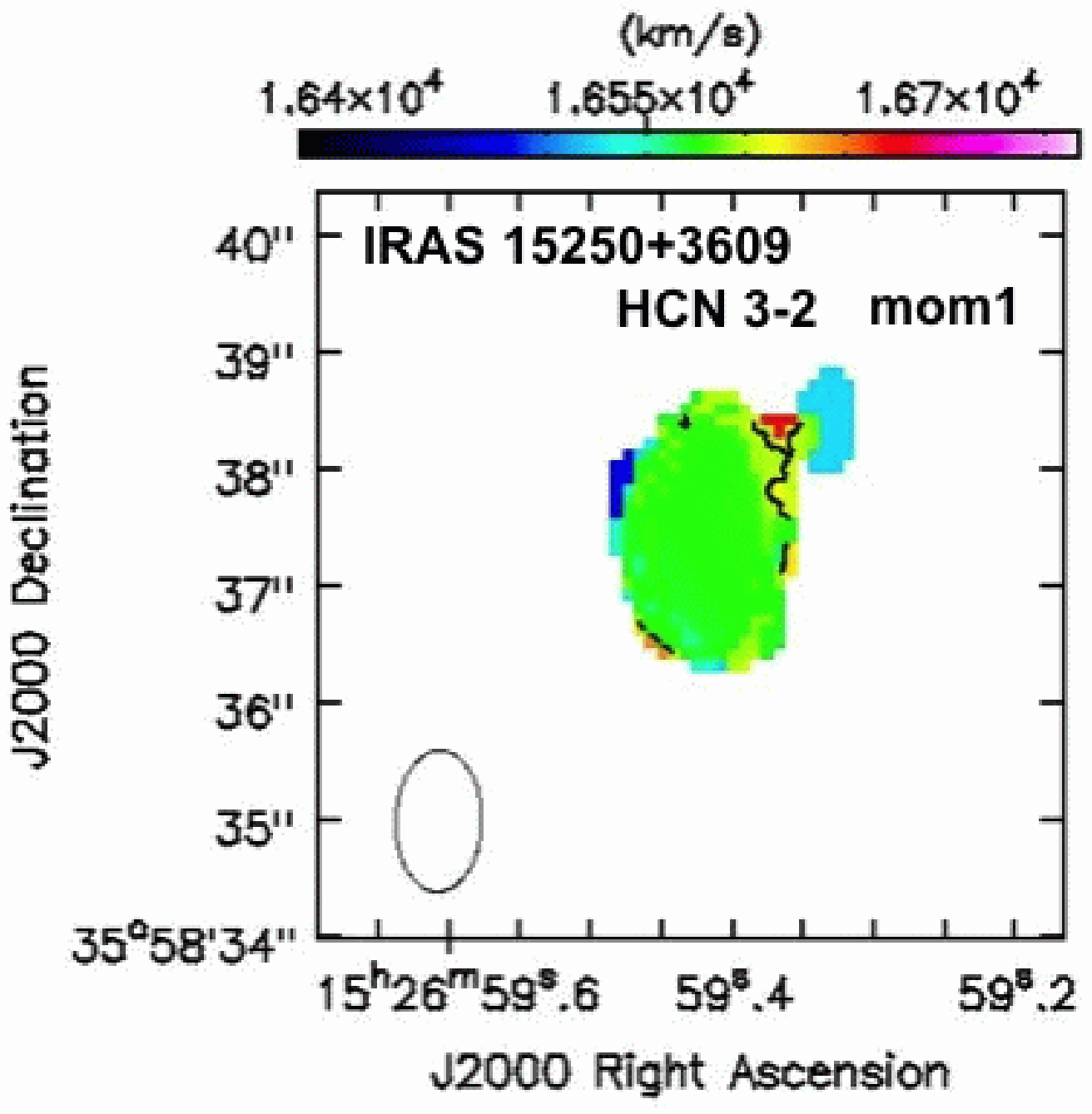} 
\includegraphics[angle=0,scale=.25]{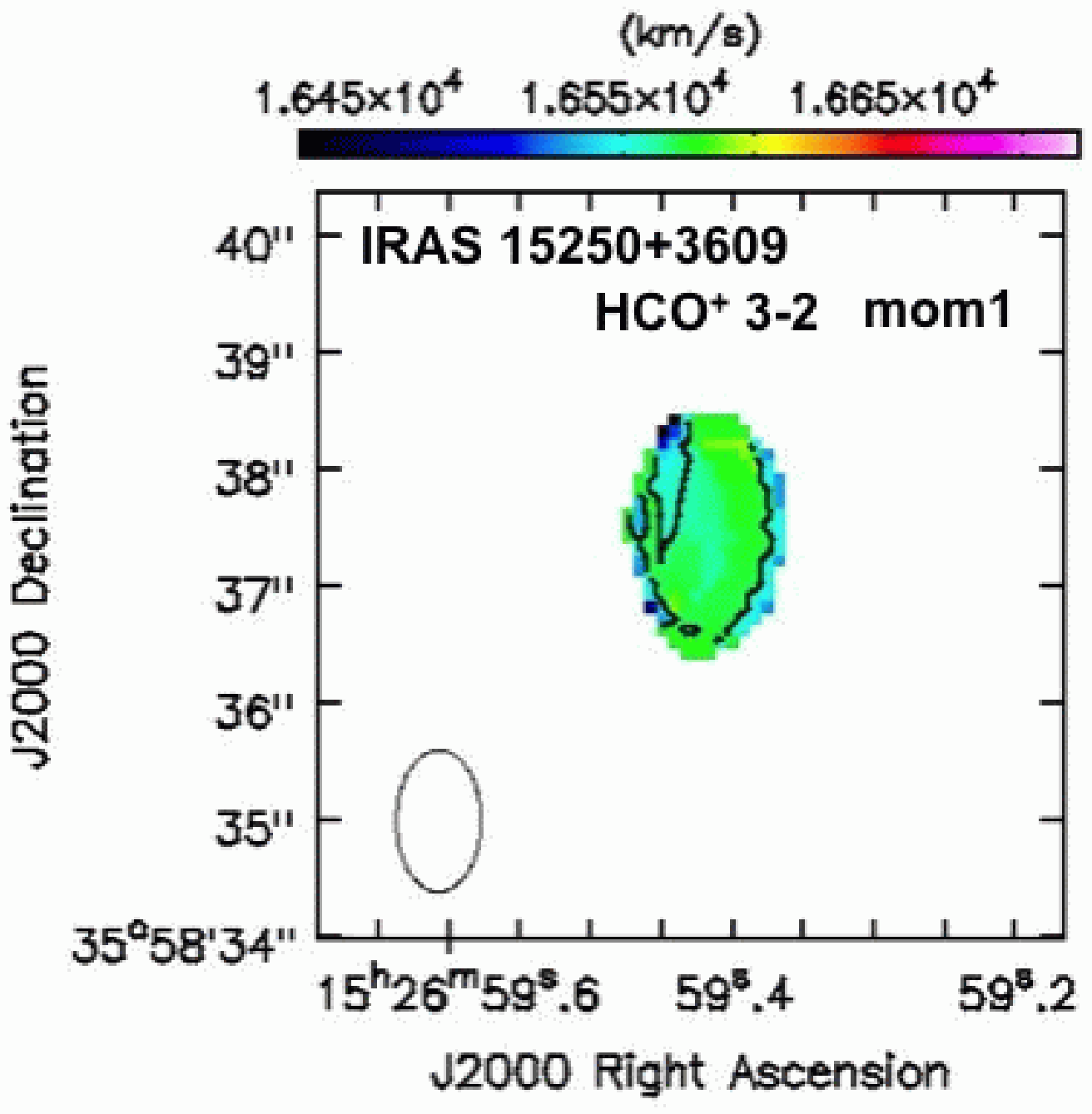} 
\includegraphics[angle=0,scale=.25]{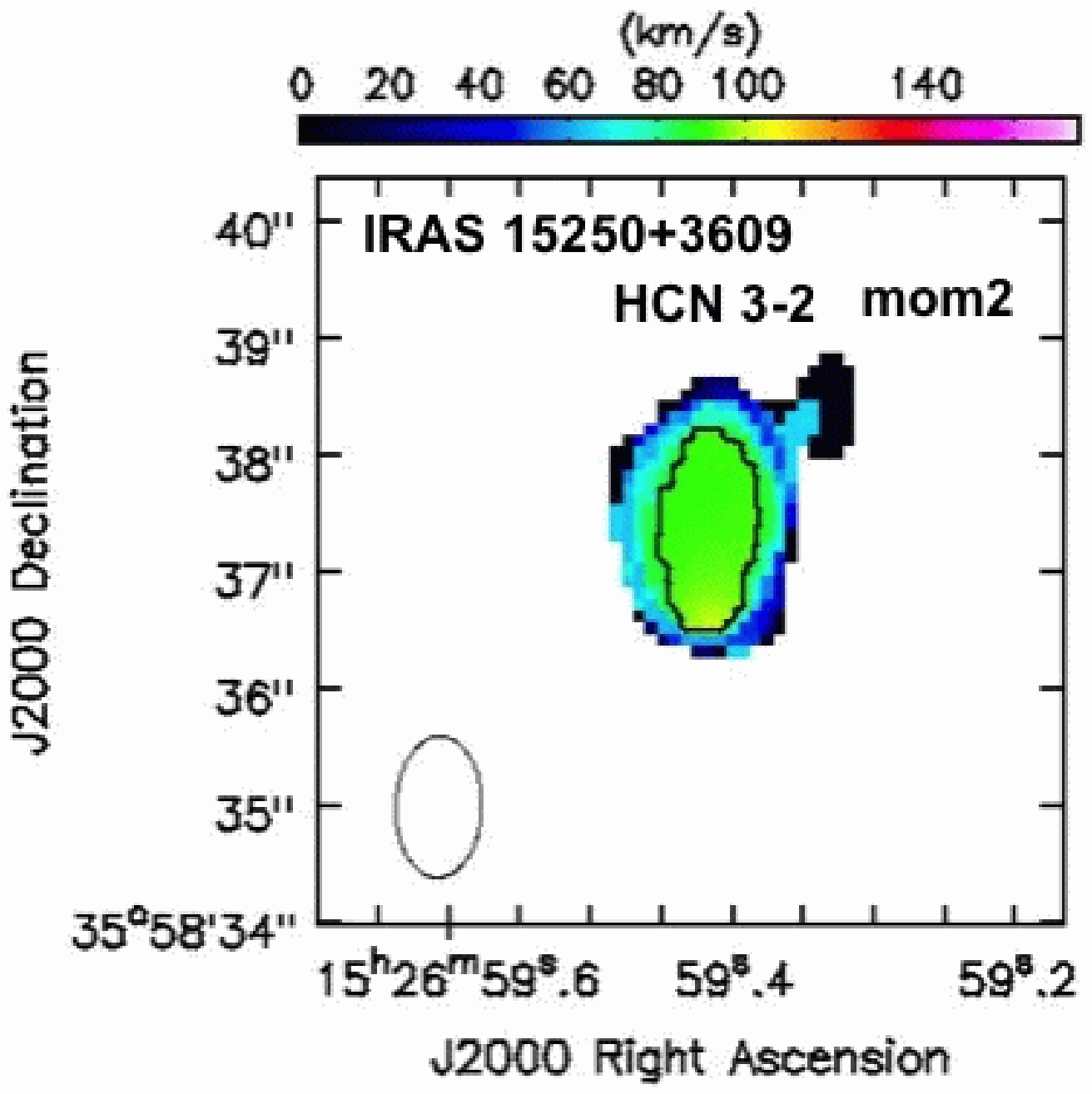} 
\includegraphics[angle=0,scale=.25]{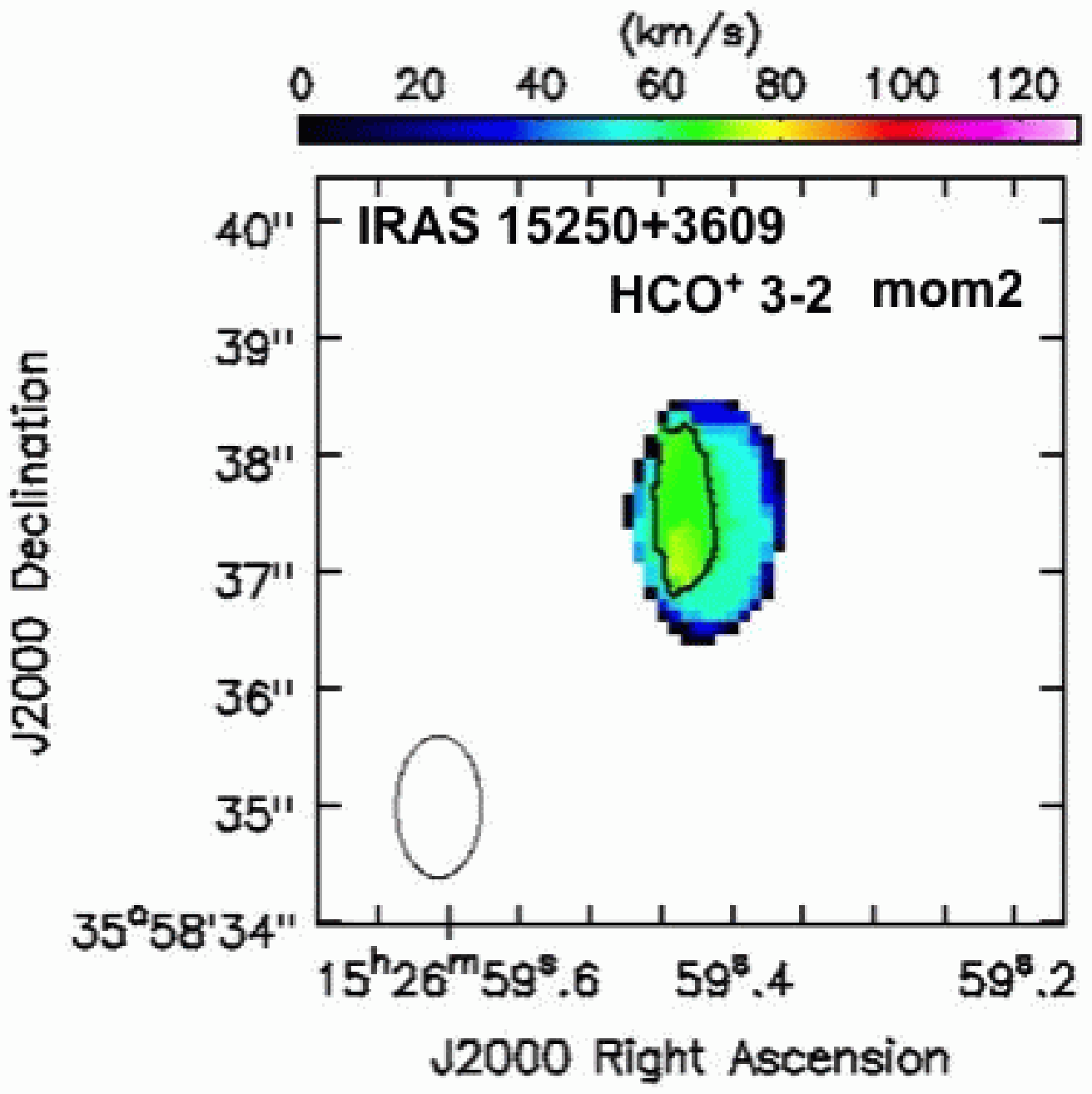} \\
\includegraphics[angle=0,scale=.25]{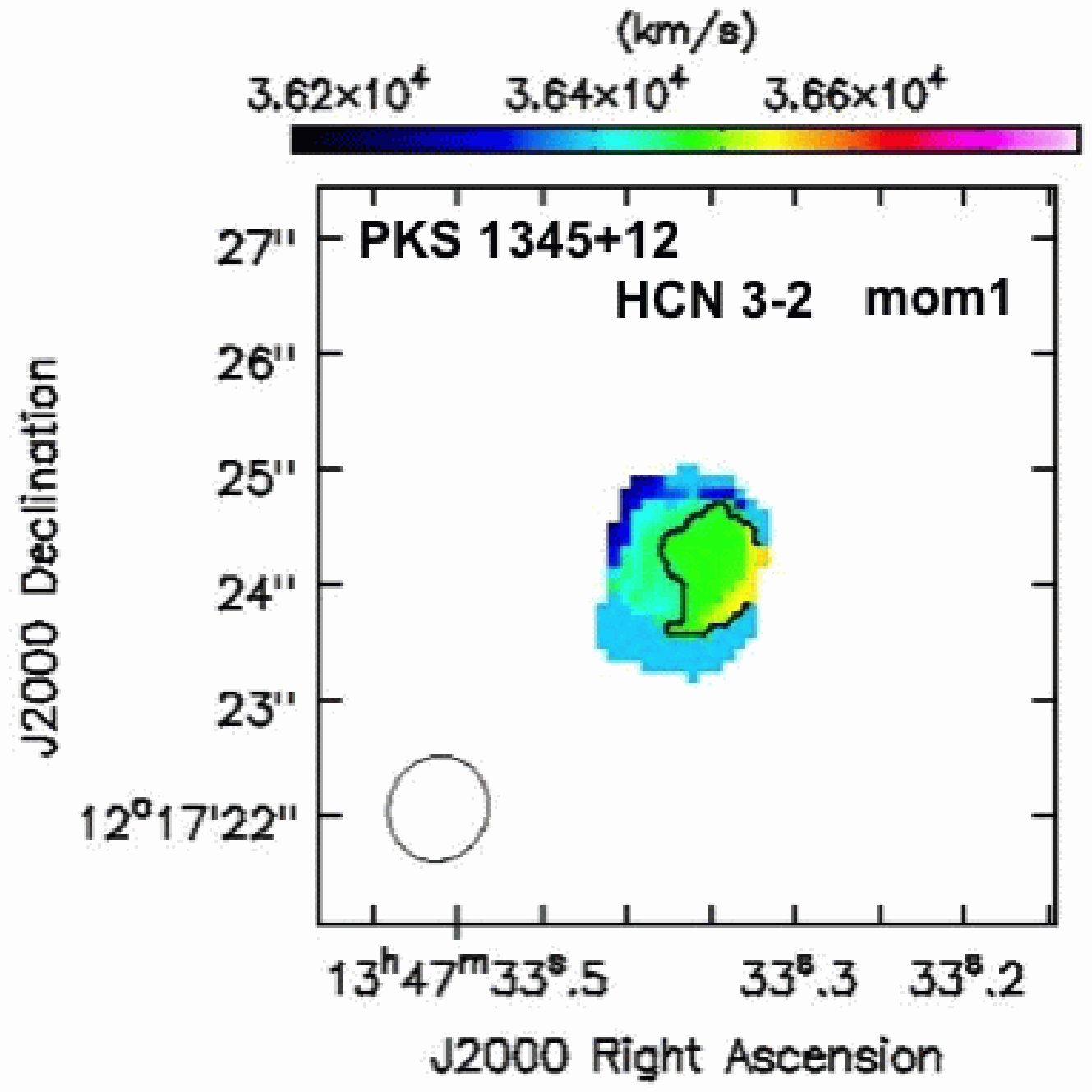} 
\includegraphics[angle=0,scale=.25]{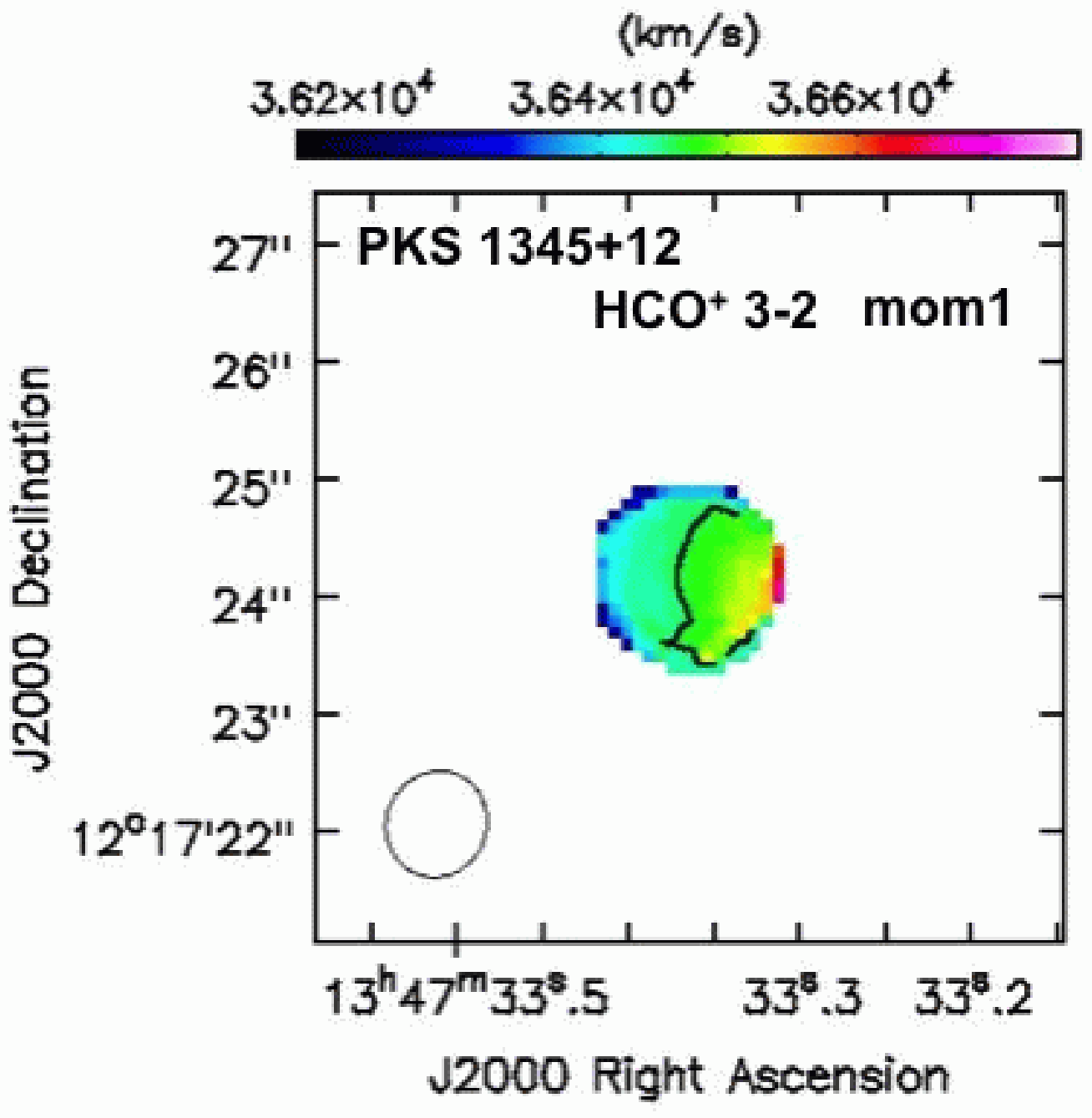} 
\includegraphics[angle=0,scale=.25]{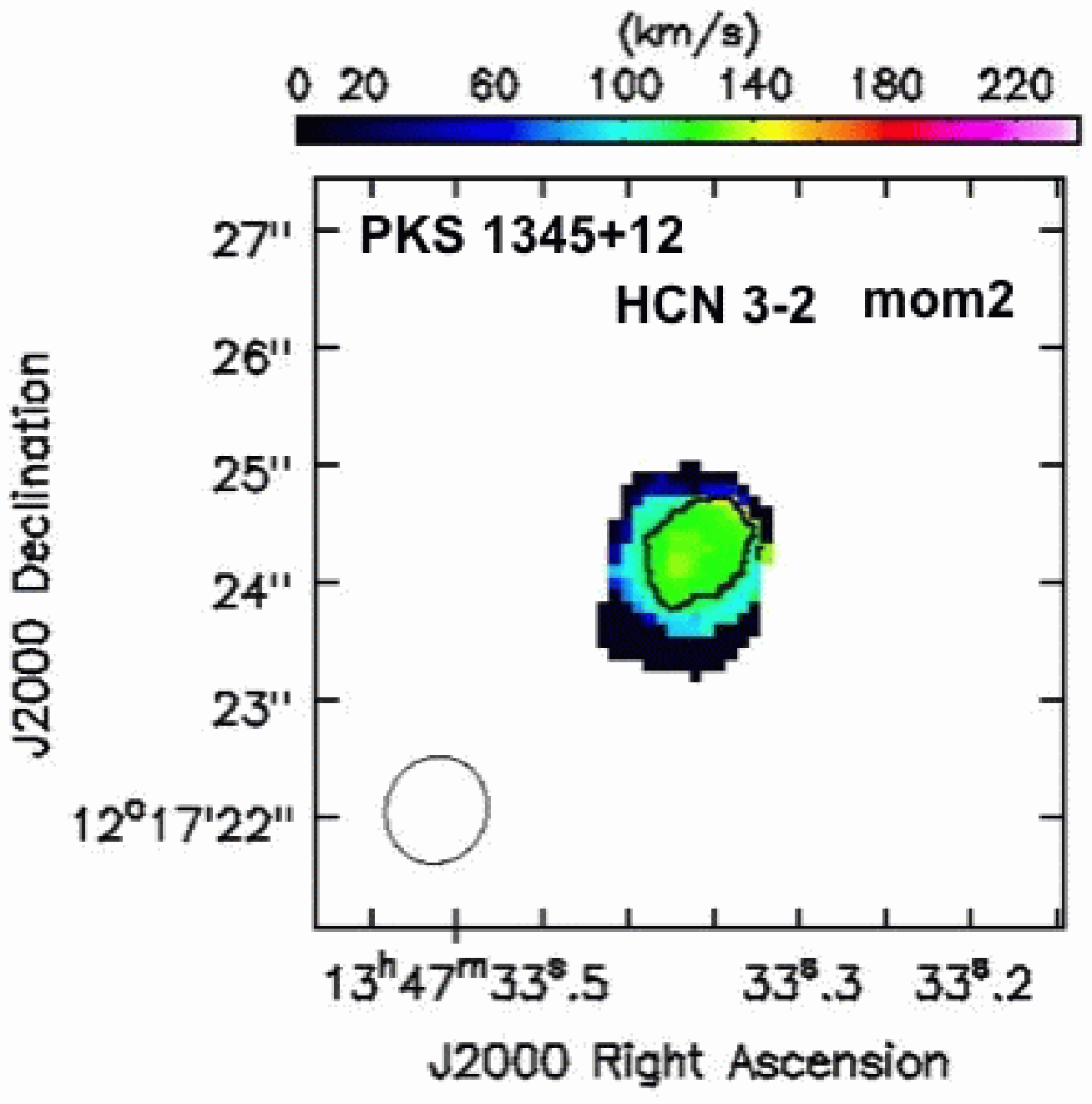} 
\includegraphics[angle=0,scale=.25]{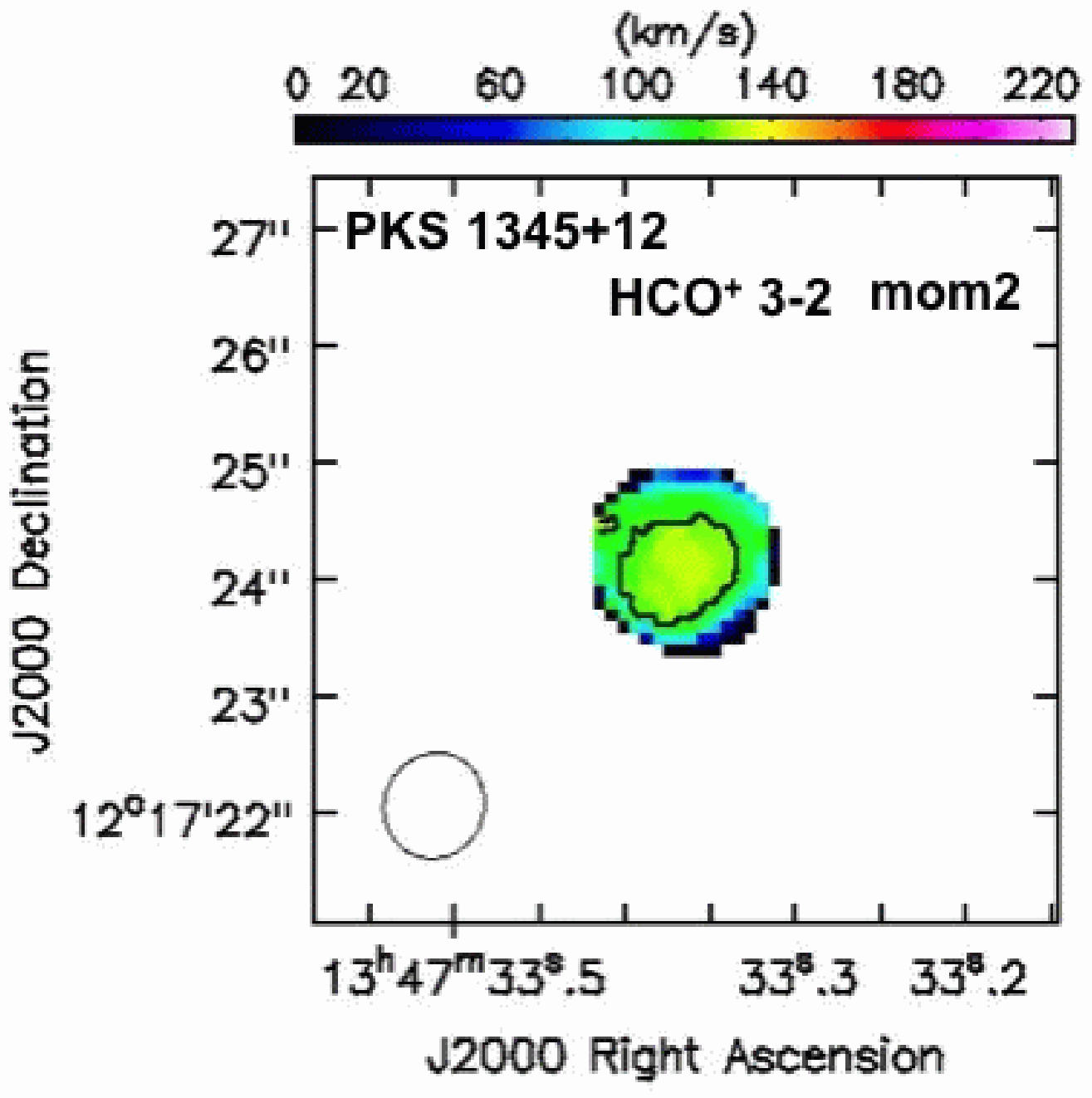} \\
\includegraphics[angle=0,scale=.25]{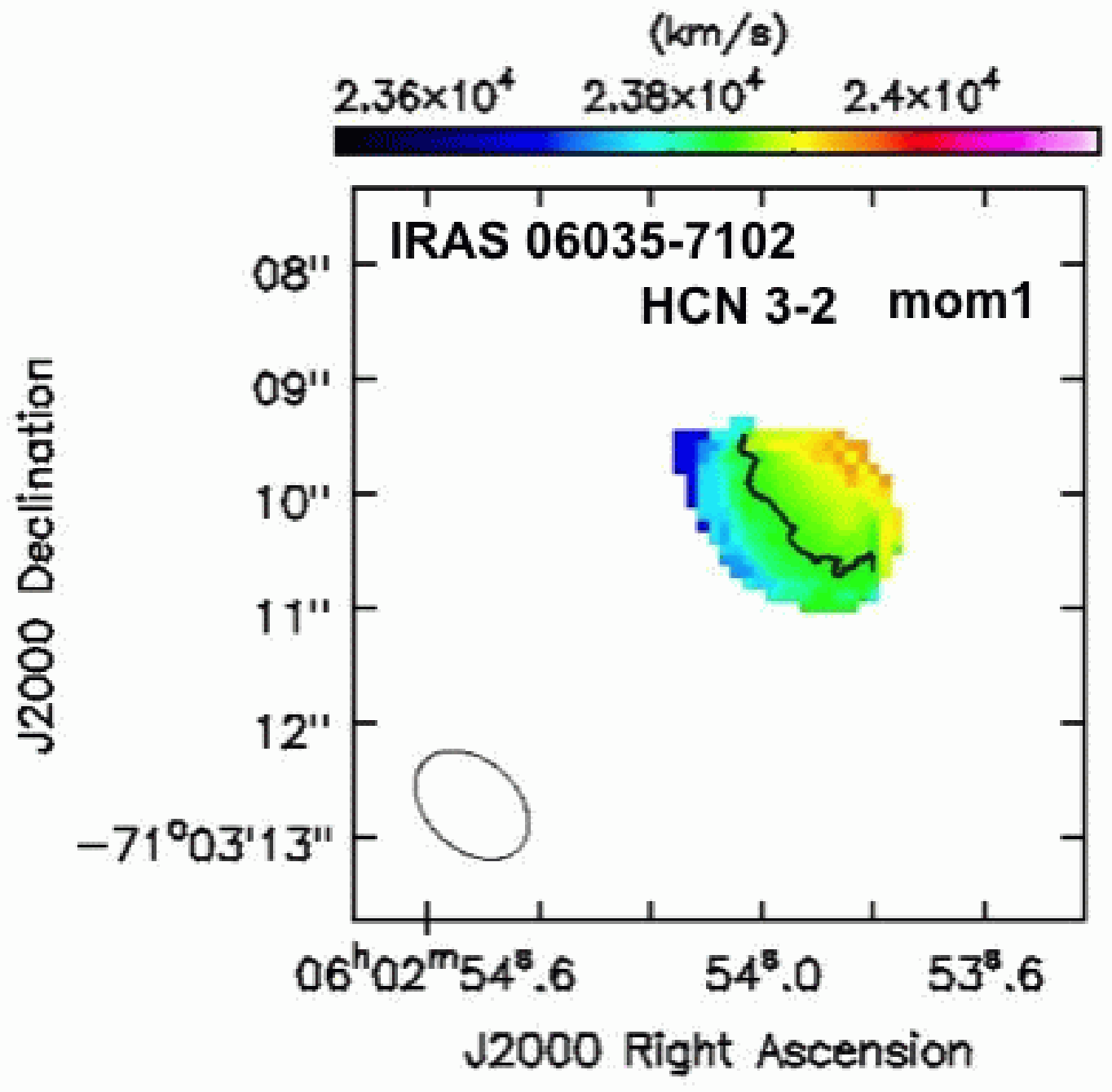} 
\includegraphics[angle=0,scale=.25]{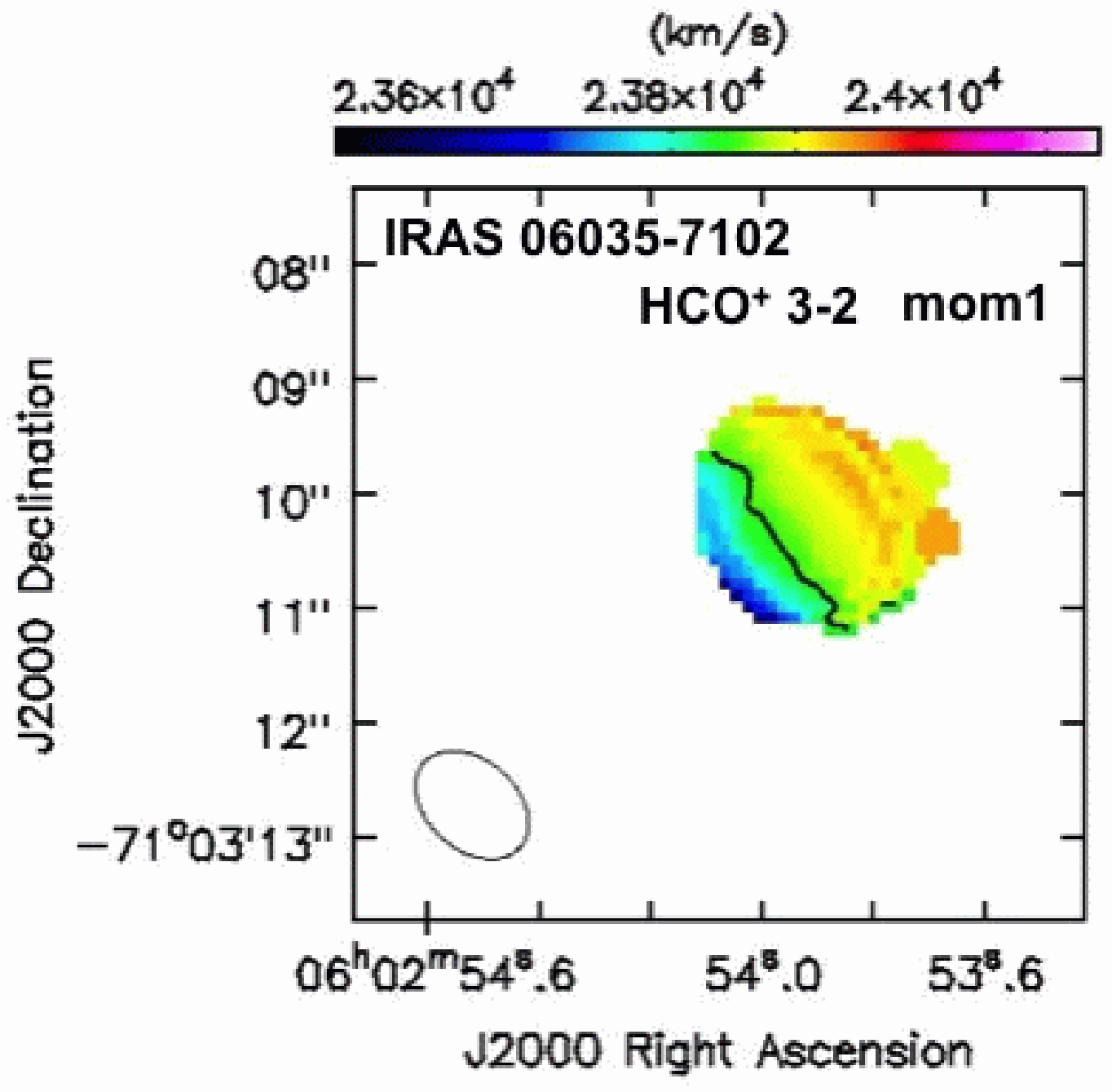} 
\includegraphics[angle=0,scale=.25]{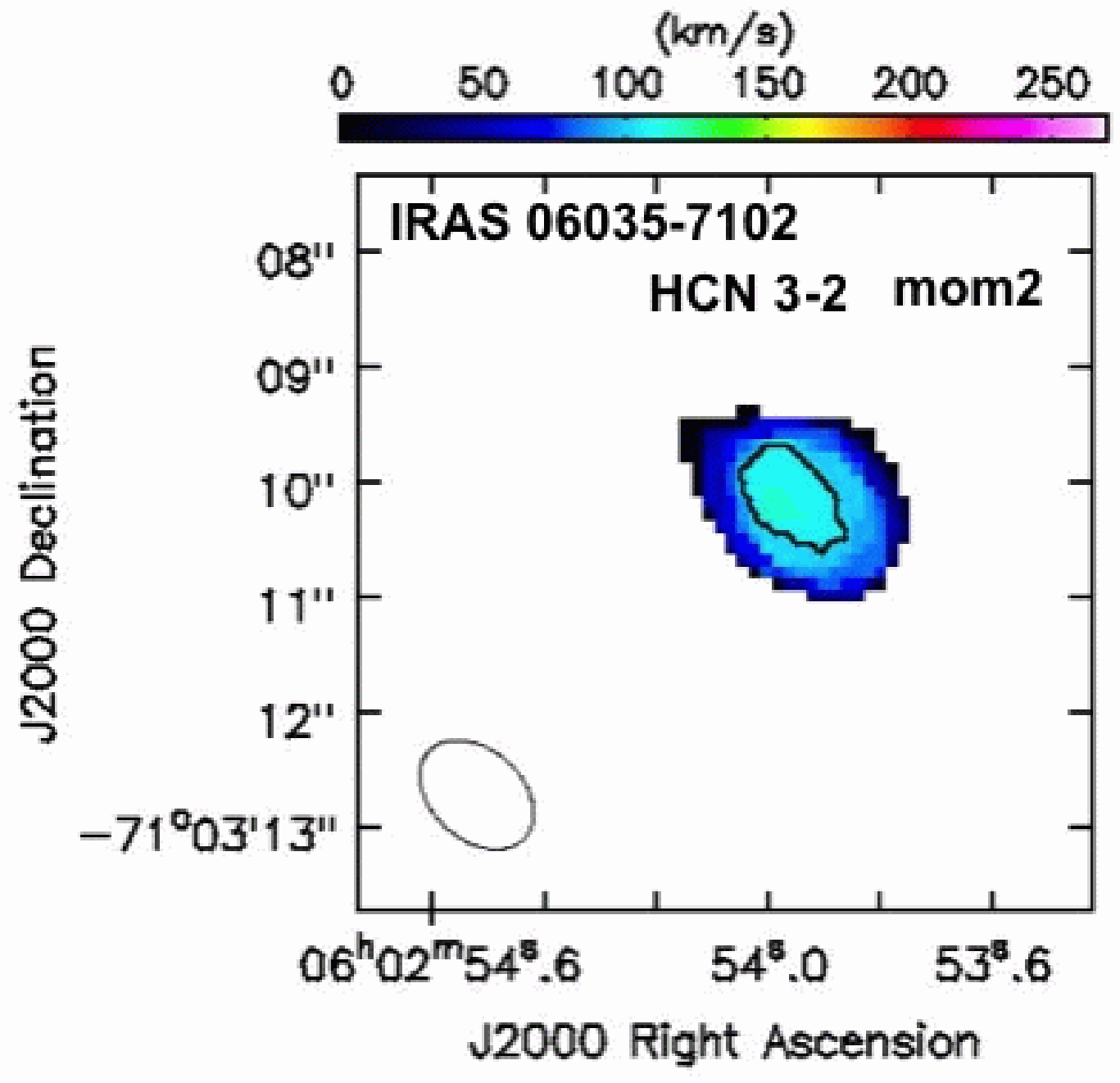} 
\includegraphics[angle=0,scale=.25]{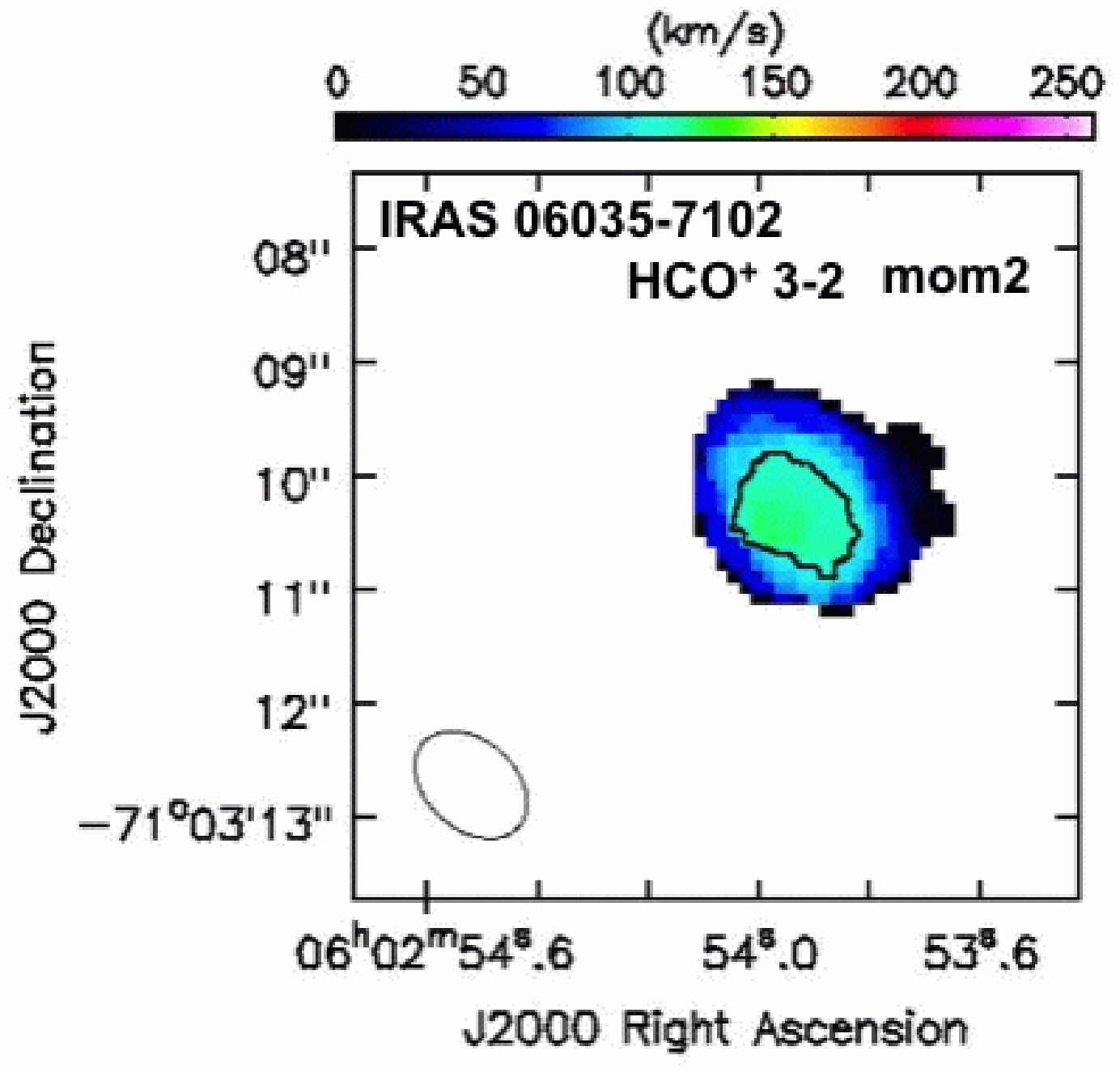} \\
\includegraphics[angle=0,scale=.25]{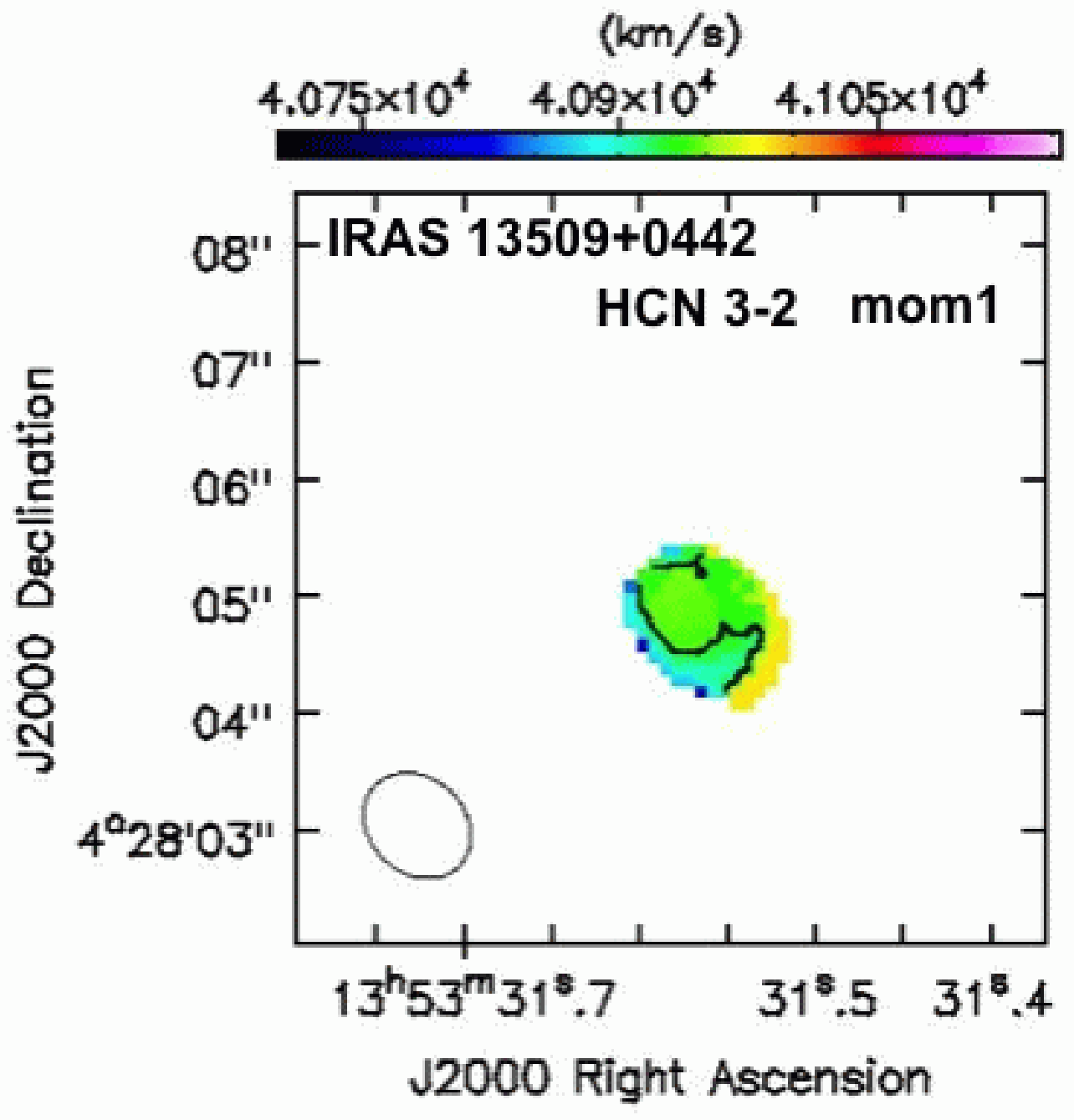} 
\includegraphics[angle=0,scale=.25]{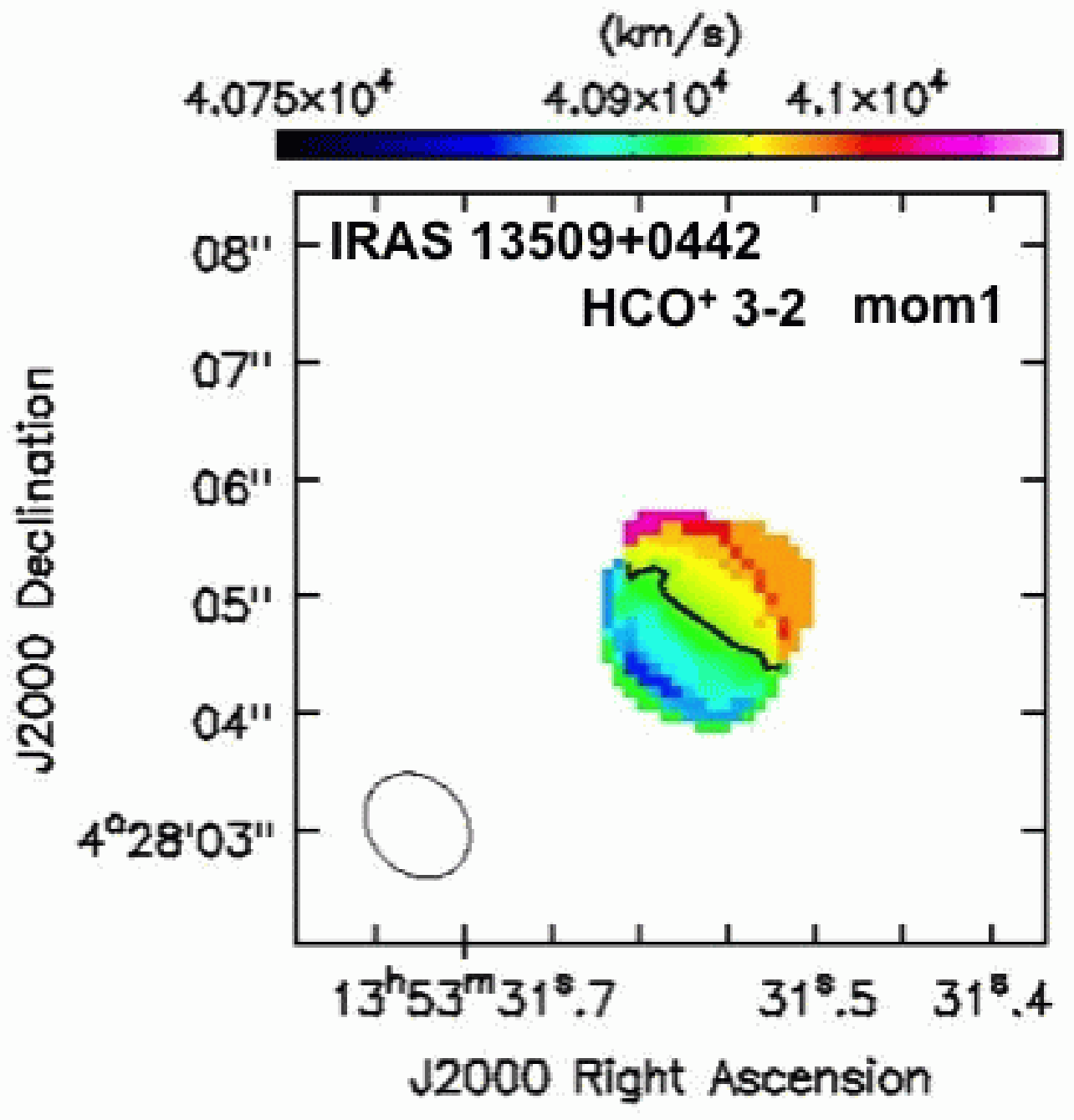} 
\includegraphics[angle=0,scale=.25]{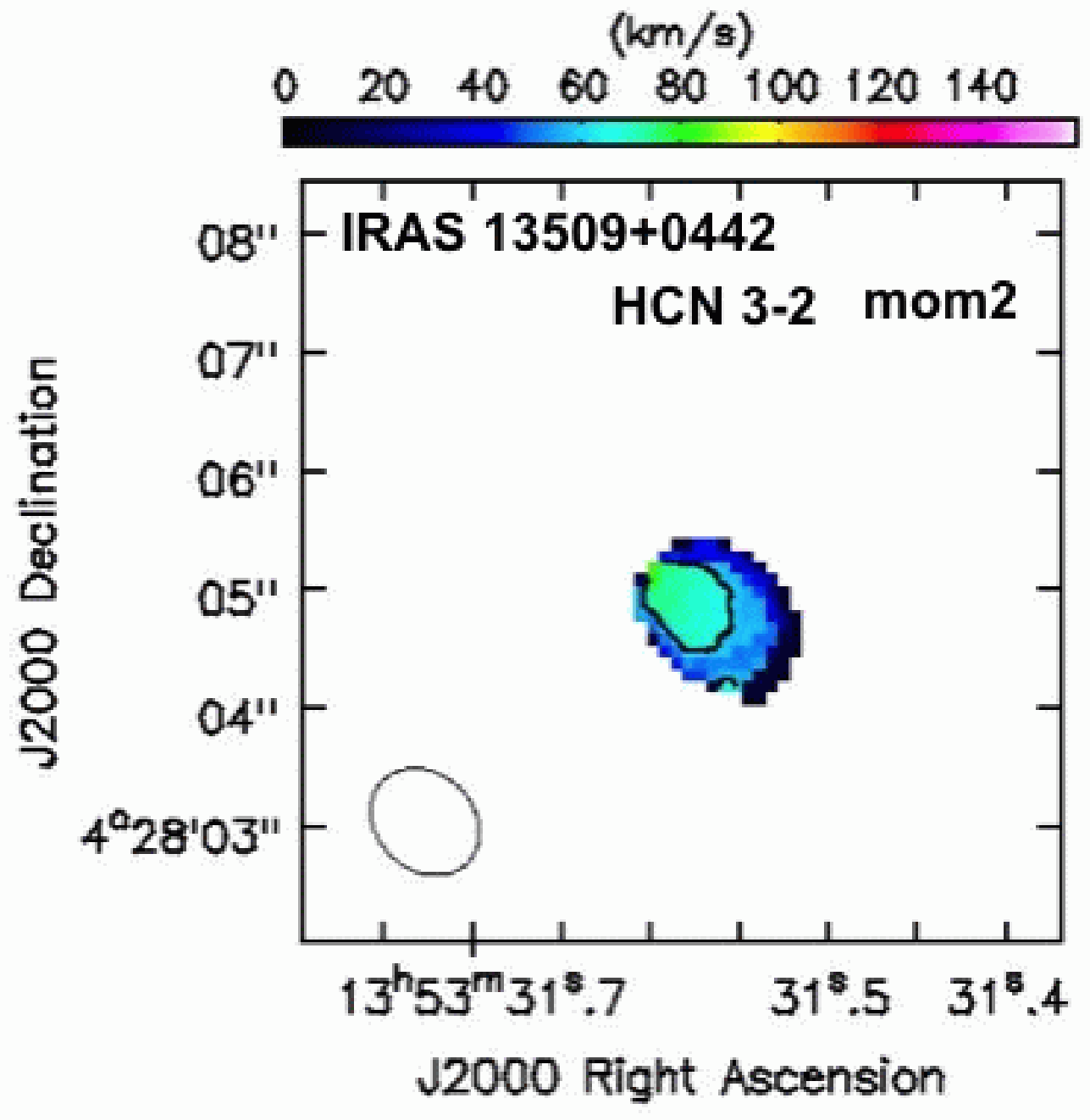} 
\includegraphics[angle=0,scale=.25]{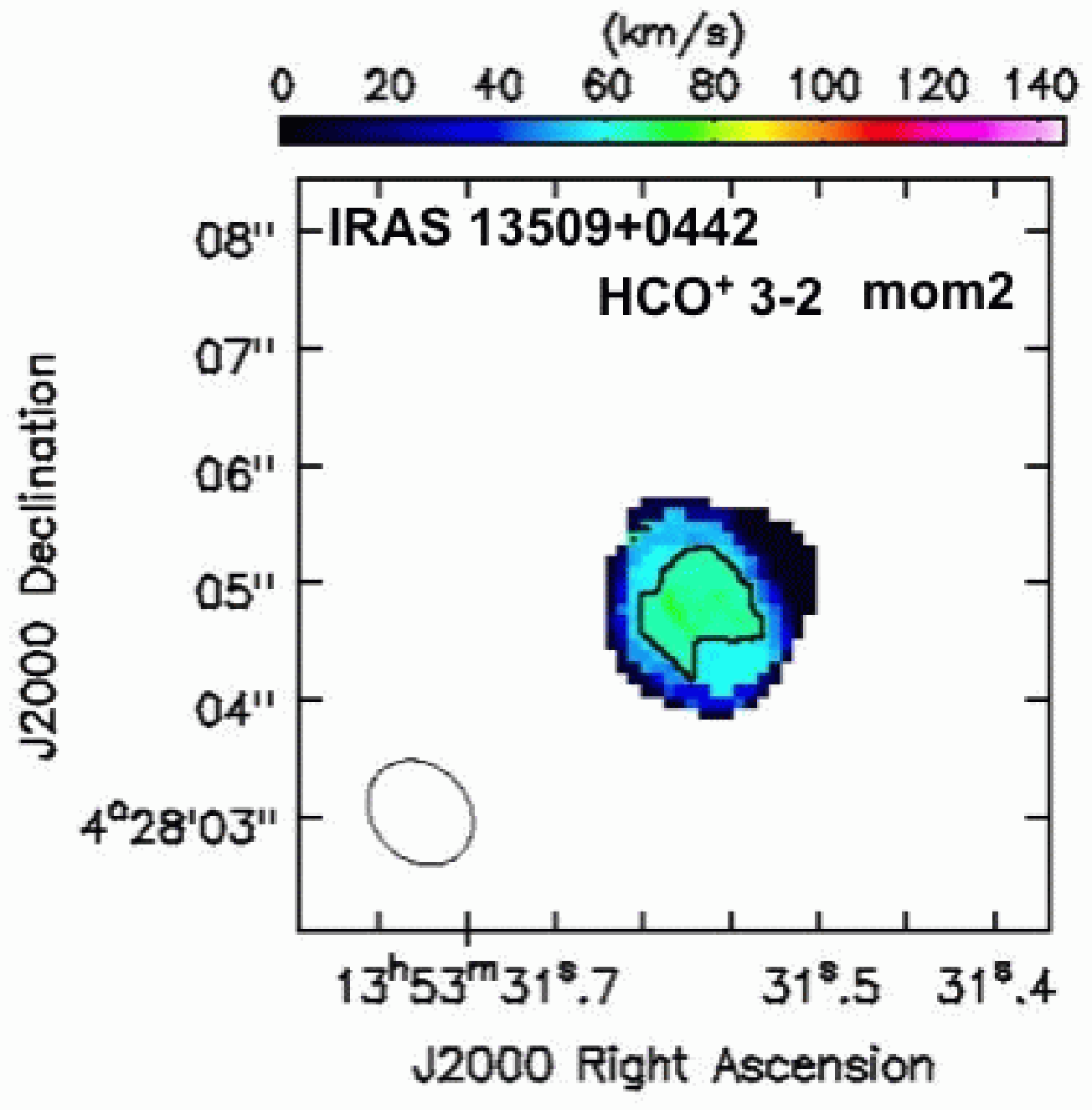} \\
\end{center}
\end{figure}

\begin{figure}
\begin{center}
\includegraphics[angle=0,scale=.25]{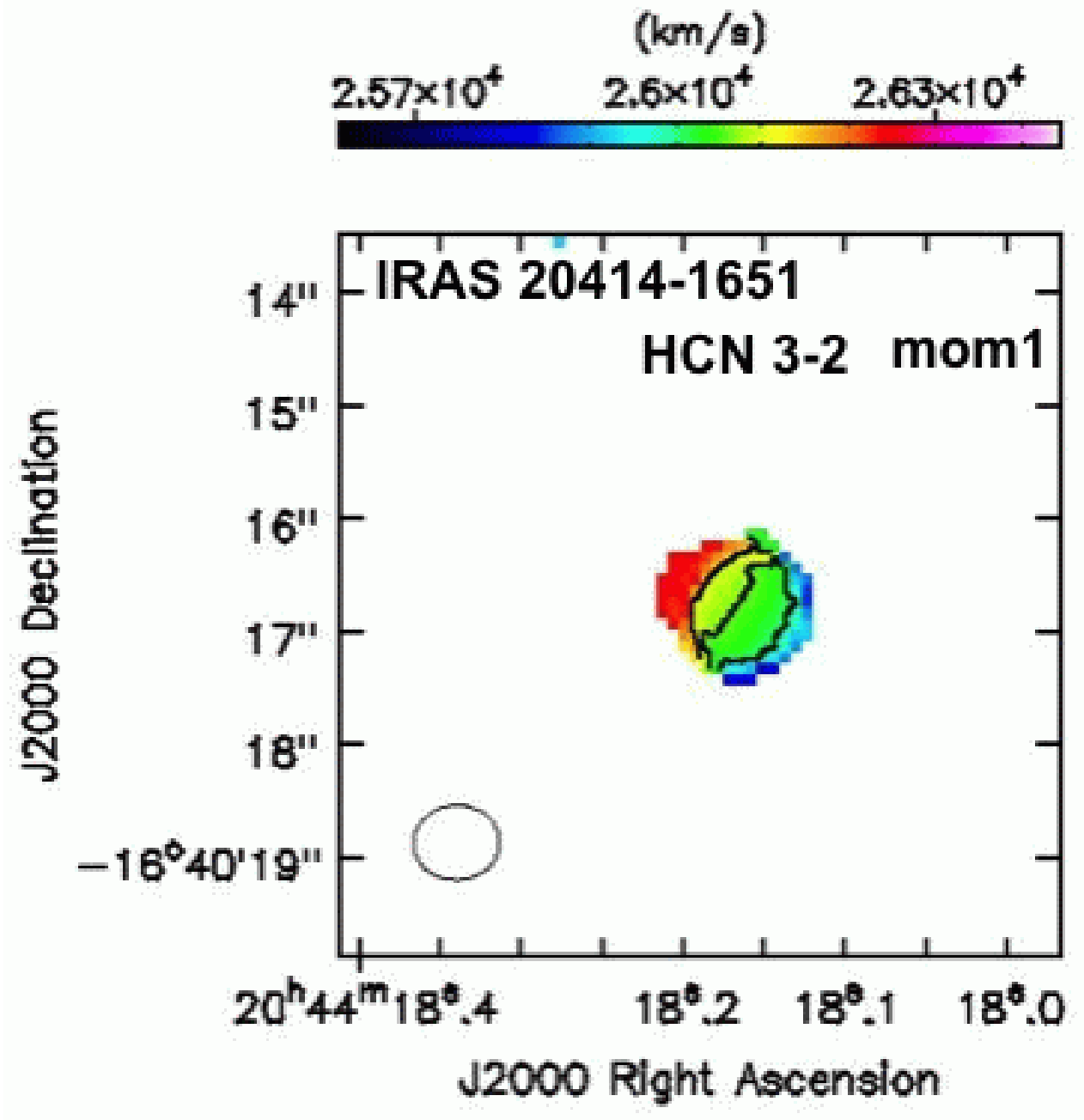} 
\includegraphics[angle=0,scale=.25]{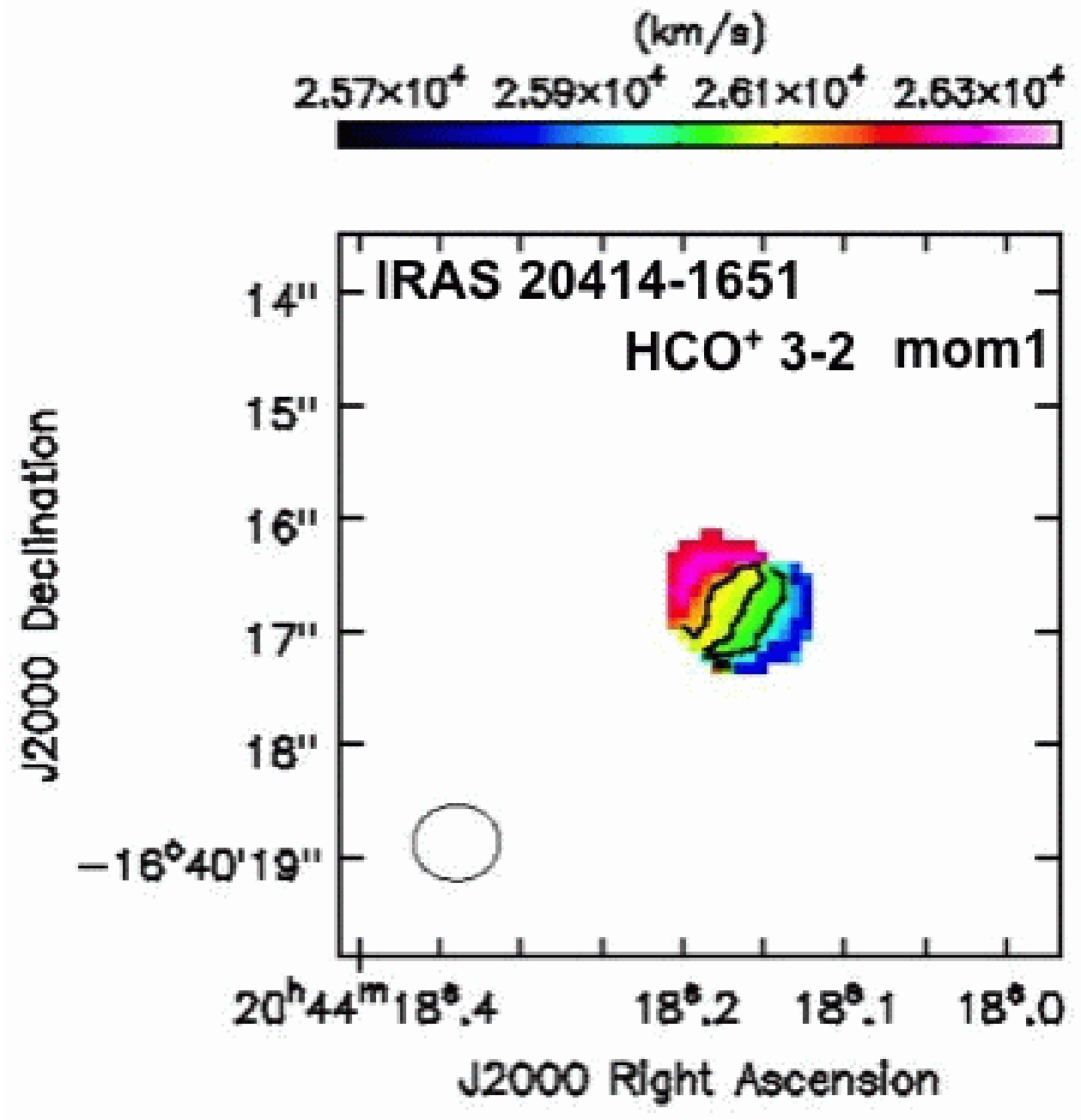} 
\includegraphics[angle=0,scale=.25]{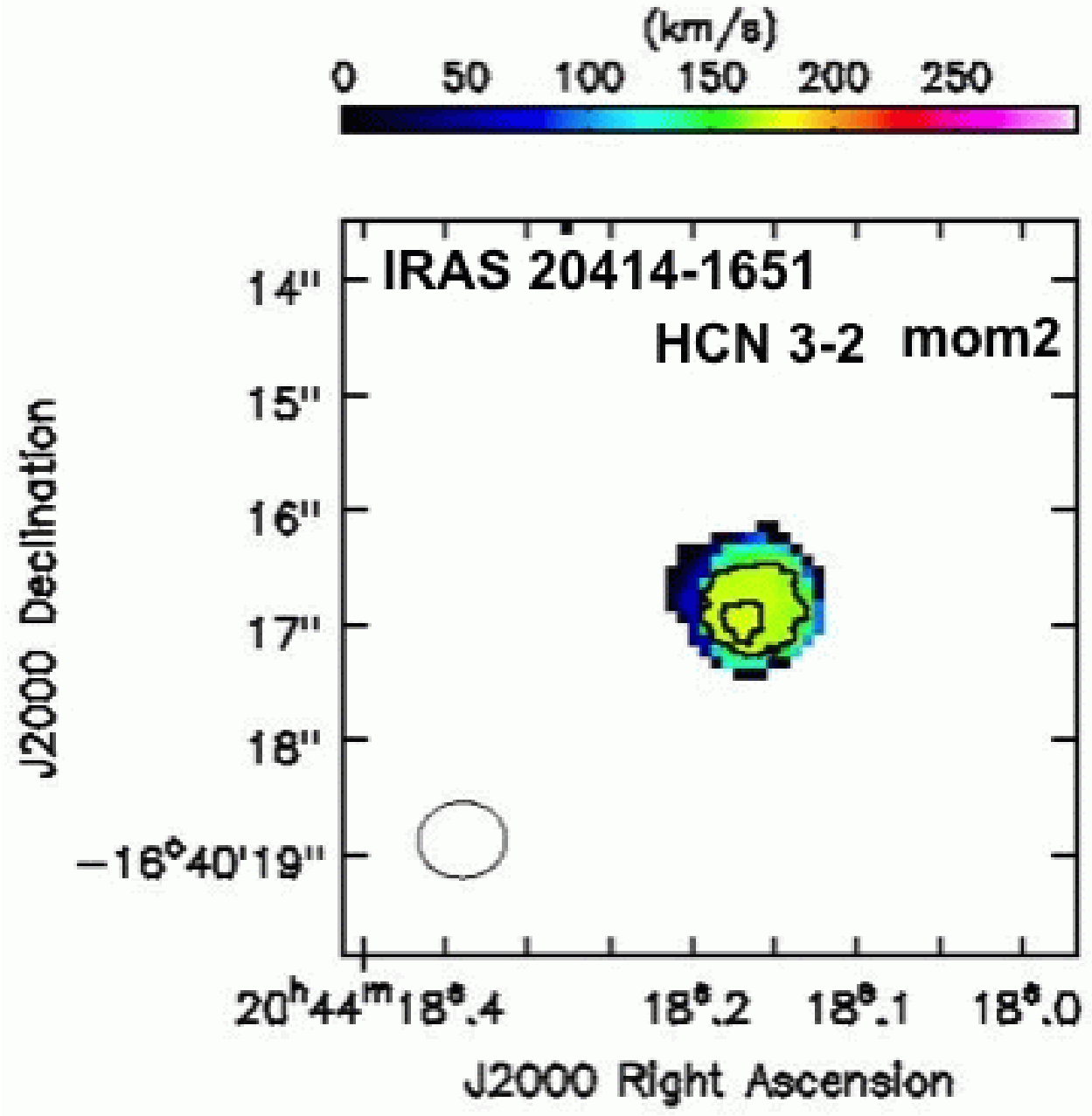} 
\includegraphics[angle=0,scale=.25]{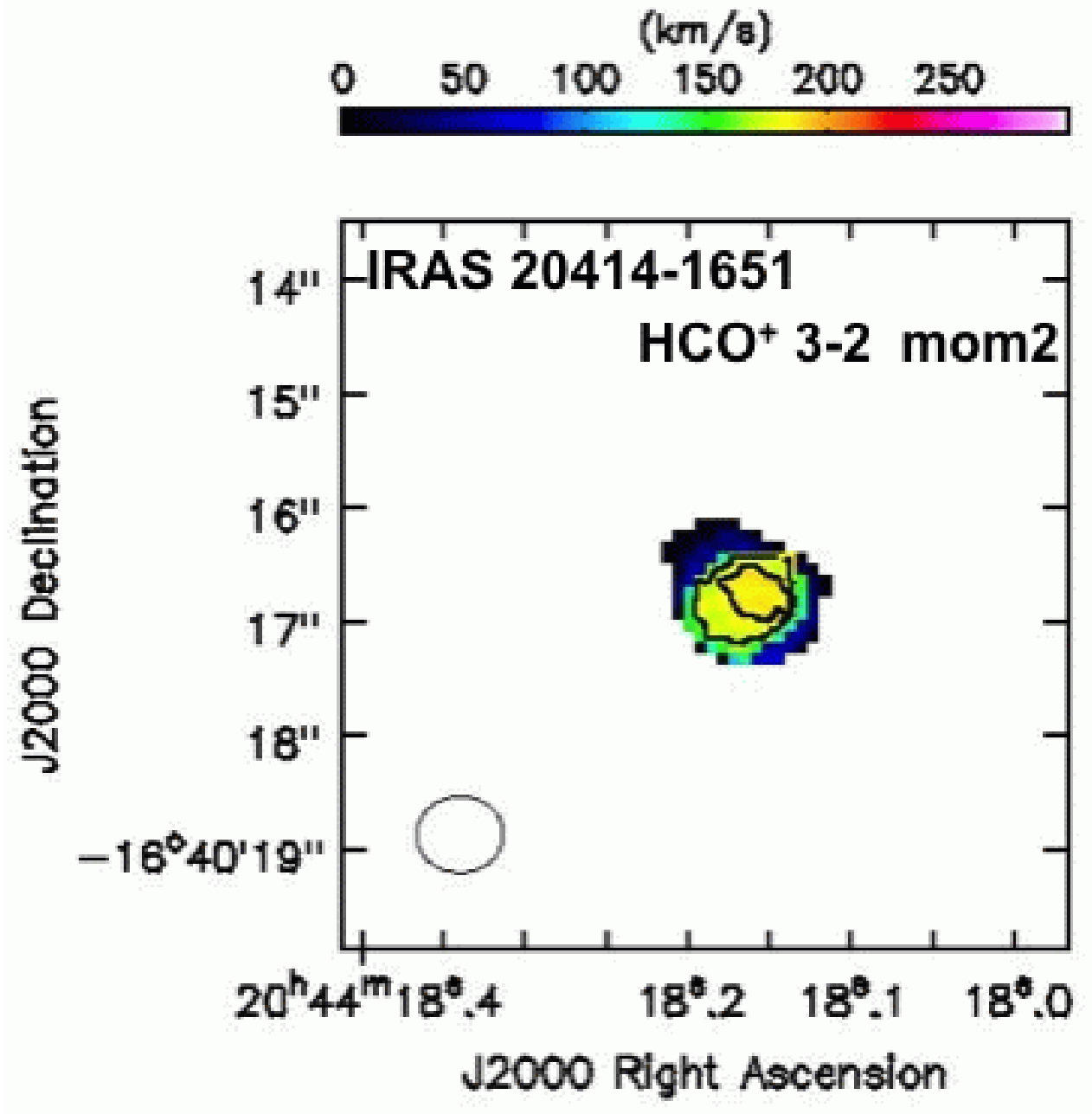} \\
\caption{
Intensity-weighted mean velocity (moment 1) and intensity-weighted
velocity dispersion (moment 2) maps of the HCN J=3--2 and HCO$^{+}$ J=3--2
emission lines.  
An appropriate cut-off was applied so that the resulting maps are not
dominated by noise. 
The left two panels are moment 1 maps. 
The contours represent 4850, 4925, 5000 km s$^{-1}$ for NGC 7469 HCN
J=3--2 and HCO$^{+}$ J=3--2, 
no contour for I Zw 1 HCN J=3--2, 
18335 km s$^{-1}$ for I Zw 1 HCO$^{+}$ J=3--2,
17500 km s$^{-1}$ for IRAS 08572$+$3915 HCN J=3--2, 
17480 km s$^{-1}$ for IRAS 08572$+$3915 HCO$^{+}$ J=3--2,
18300, 18500, 18700 km s$^{-1}$ for The Superantennae HCN J=3--2, 
18500, 18700 km s$^{-1}$ for The Superantennae HCO$^{+}$ J=3--2,  
21850 km s$^{-1}$ for IRAS 12112$+$0305 NE HCN J=3--2 and HCO$^{+}$ J=3--2,
23300 km s$^{-1}$ for IRAS 22491$-$1808 HCN J=3--2 and HCO$^{+}$ J=3--2, 
4750, 4850 km s$^{-1}$ for NGC 1614 HCN J=3--2 and HCO$^{+}$ J=3--2,
39960, 40000 km s$^{-1}$ for IRAS 12127$-$1412 HCN J=3--2 and HCO$^{+}$
J=3--2,   
16600 km s$^{-1}$ for IRAS 15250$+$3609 HCN J=3--2, 
16560 km s$^{-1}$ for IRAS 15250$+$3609 HCO$^{+}$ J=3--2, 
36460 km s$^{-1}$ for PKS 1345$+$12 HCN J=3--2 and HCO$^{+}$ J=3--2, 
22090 km s$^{-1}$ for IRAS 06035$-$7102 HCN J=3--2 and HCO$^{+}$ J=3--2,
40930 km s$^{-1}$ for IRAS 13509$+$0442 HCN J=3--2 and HCO$^{+}$
J=3--2, and 
26000, 26060, 26120 km s$^{-1}$ for IRAS 20414$-$1651 HCN J=3--2 
and HCO$^{+}$ J=3--2.  
The right two panels are moment 2 maps. 
The contours represent 
20, 45, 70 km s$^{-1}$ for NGC 7469 HCN J=3--2 and HCO$^{+}$ J=3--2, 
50, 70 km s$^{-1}$ for I Zw 1 HCN J=3--2, 
70 km s$^{-1}$ for I Zw 1 HCO$^{+}$ J=3--2, 
78, 96 km s$^{-1}$ for IRAS 08572$+$3915 HCN J=3--2, 
66, 81 km s$^{-1}$ for IRAS 08572$+$3915 HCO$^{+}$ J=3--2, 
150, 210 km s$^{-1}$ for The Superantennae HCN J=3--2 and HCO$^{+}$ J=3--2,
130 km s$^{-1}$ for IRAS 12112$+$0305 NE HCN J=3--2, 
120 km s$^{-1}$ for IRAS 12112$+$0305 NE HCO$^{+}$ J=3--2, 
120 km s$^{-1}$ for IRAS 22491$-$1808 HCN J=3--2, 
100 km s$^{-1}$ for IRAS 22491$-$1808 HCO$^{+}$ J=3--2, 
10 km s$^{-1}$ for NGC 1614 HCN J=3--2, 
30 km s$^{-1}$ for NGC 1614 HCO$^{+}$ J=3--2, 
110, 125 km s$^{-1}$ for IRAS 12127$-$1412 HCN J=3--2, 
102 km s$^{-1}$ for IRAS 12127$-$1412 HCO$^{+}$ J=3--2, 
90 km s$^{-1}$ for IRAS 15250$+$3609 HCN J=3--2, 
63 km s$^{-1}$ for IRAS 15250$+$3609 HCO$^{+}$ J=3--2, 
118 km s$^{-1}$ for PKS 1345$+$12 HCN J=3--2, 
120 km s$^{-1}$ for PKS 1345$+$12 HCO$^{+}$ J=3--2, 
104 km s$^{-1}$ for IRAS 06035$-$7102 HCN J=3--2, 
110 km s$^{-1}$ for IRAS 06035$-$7102 HCO$^{+}$ J=3--2, 
60 km s$^{-1}$ for IRAS 13509$+$0442 HCN J=3--2, 
62 km s$^{-1}$ for IRAS 13509$+$0442 HCO$^{+}$ J=3--2, 
164, 174 km s$^{-1}$ for IRAS 20414$-$1651 HCN J=3--2, 
and 
170, 184 km s$^{-1}$ for IRAS 20414$-$1651 HCO$^{+}$ J=3--2.
} 
\end{center}
\end{figure}

\begin{figure}
\begin{center}
\includegraphics[angle=0,scale=.9]{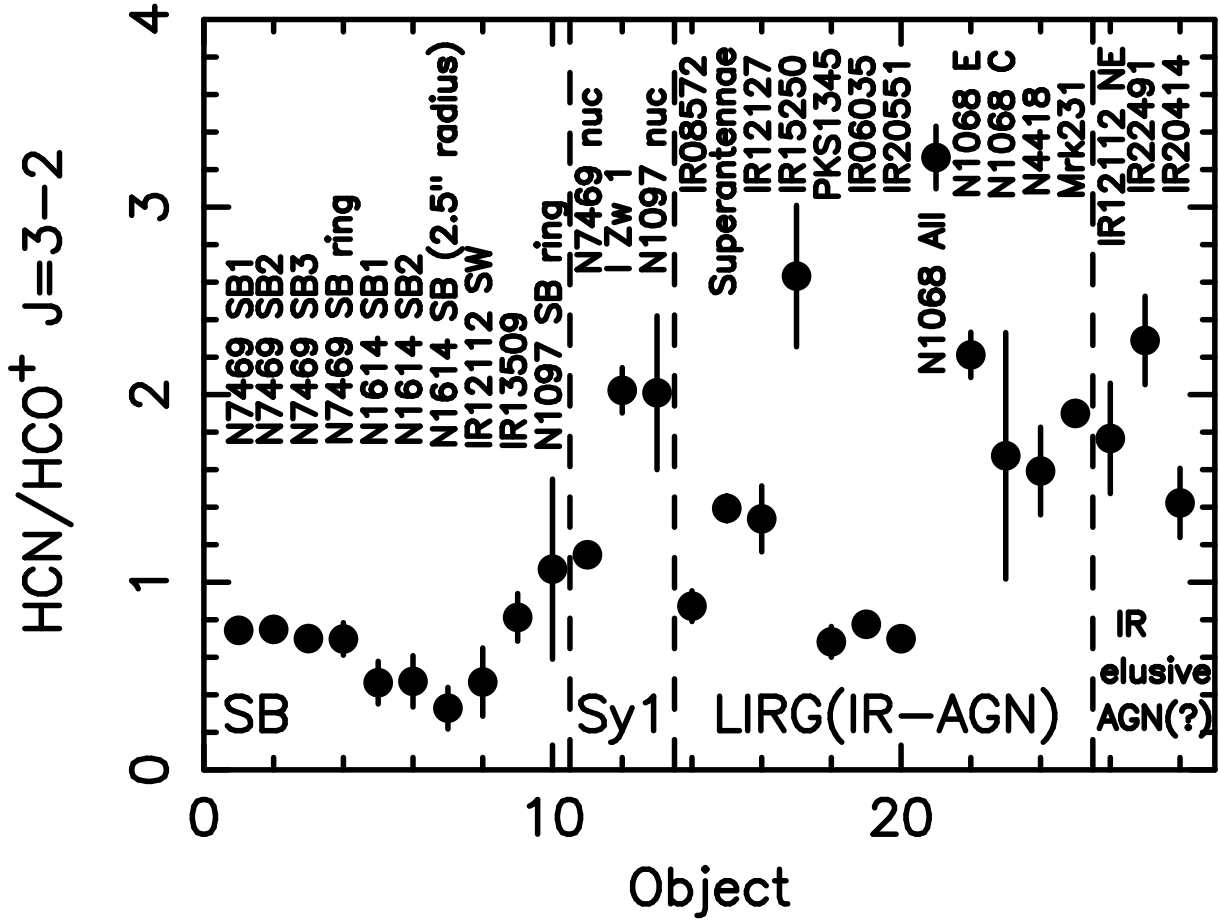} 
\end{center}
\caption{
The HCN J=3--2 to HCO$^{+}$ J=3--2 flux ratio of observed galaxies with
interferometry.  
Sources are categorized into starburst-dominated regions, denoted
as ``SB'' (left), optical Seyfert 1 galaxies, denoted as ``Sy1'' (second
left), and LIRGs which are infrared-diagnosed to contain luminous
obscured AGNs, denoted as ``LIRG(IR-AGN)'' (second right).
LIRGs which have no obvious infrared AGN indicators, but show signatures
of vibrationally excited HCN v$_{2}$=1f J=3--2 emission lines, 
are denoted as ``IR elusive AGN (?)'' (right).  
ALMA interferometric data for IRAS 20551$-$4250 \citep{ima16a} and 
NGC 1068 \citep{ima16b}, SMA interferometric data for NGC
1097 \citep{hsi12} and NGC 4418 \citep{sak10}, and PdBI 
interferometric data for Mrk 231 \citep{aal15a} are included, in
addition to the sources observed in this paper.  
For NGC 1068, data from the nuclear area-integrated emission within a
2$\farcs$4 radius circular region (denoted as ``N1068 All''),  
molecular gas emission peak at the eastern side of the AGN (denoted as
``N1068 E''), and the continuum peak as the putative location of an AGN
(denoted as ``N1068 C'') \citep{ima16b}, are shown.} 
\end{figure}

\begin{figure}
\begin{center}
\includegraphics[angle=0,scale=.41]{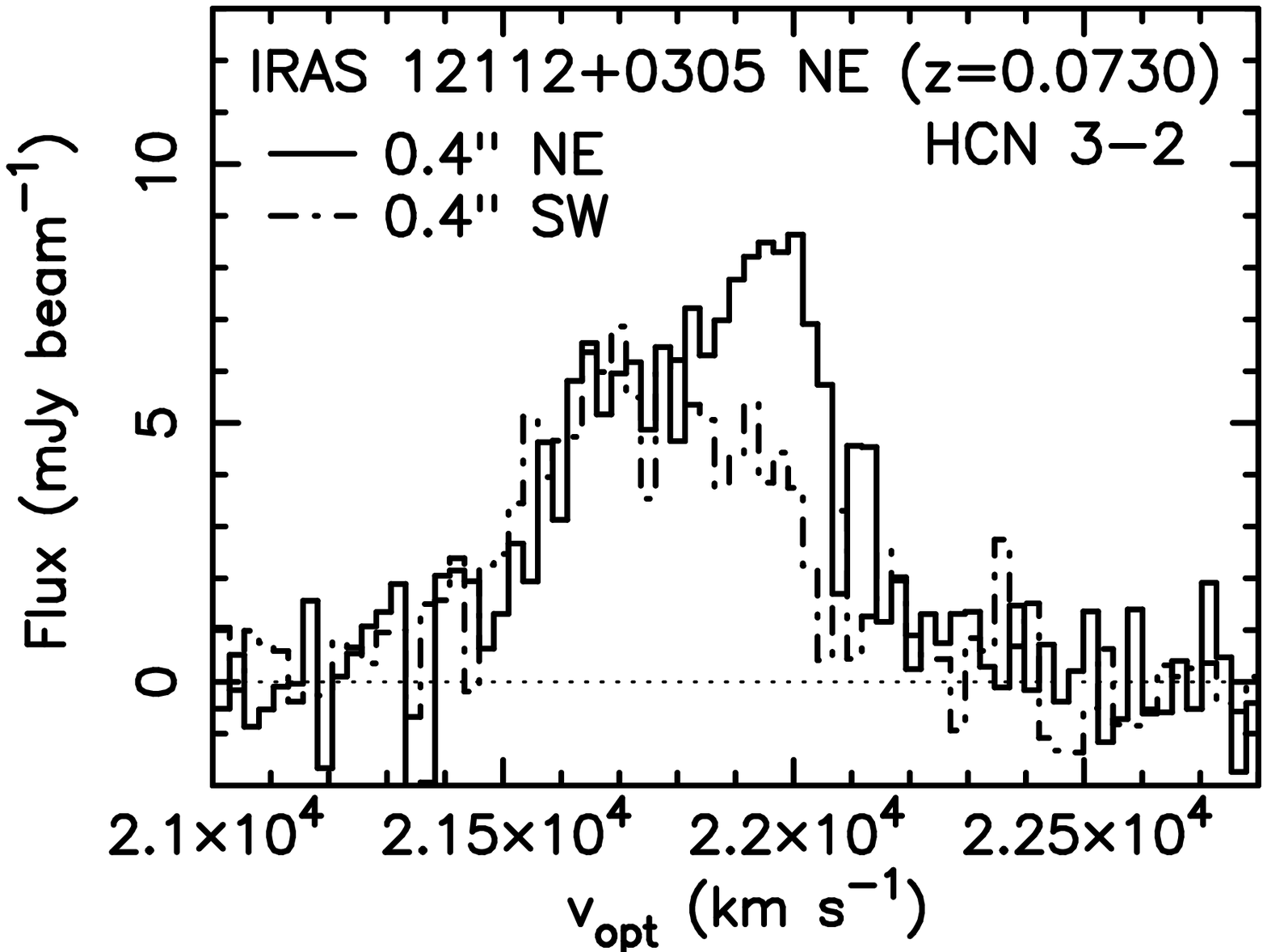} 
\includegraphics[angle=0,scale=.41]{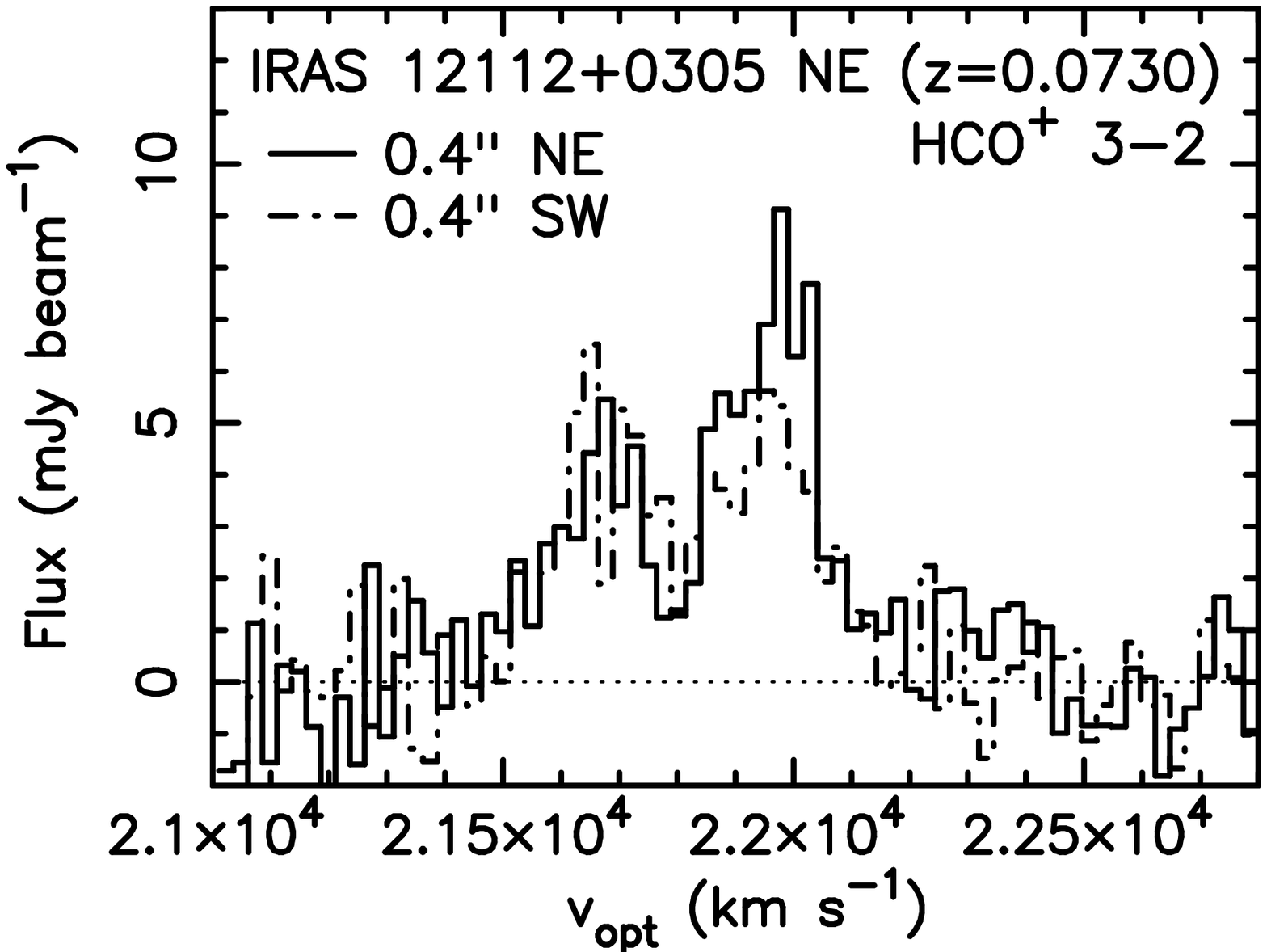} \\
\includegraphics[angle=0,scale=.41]{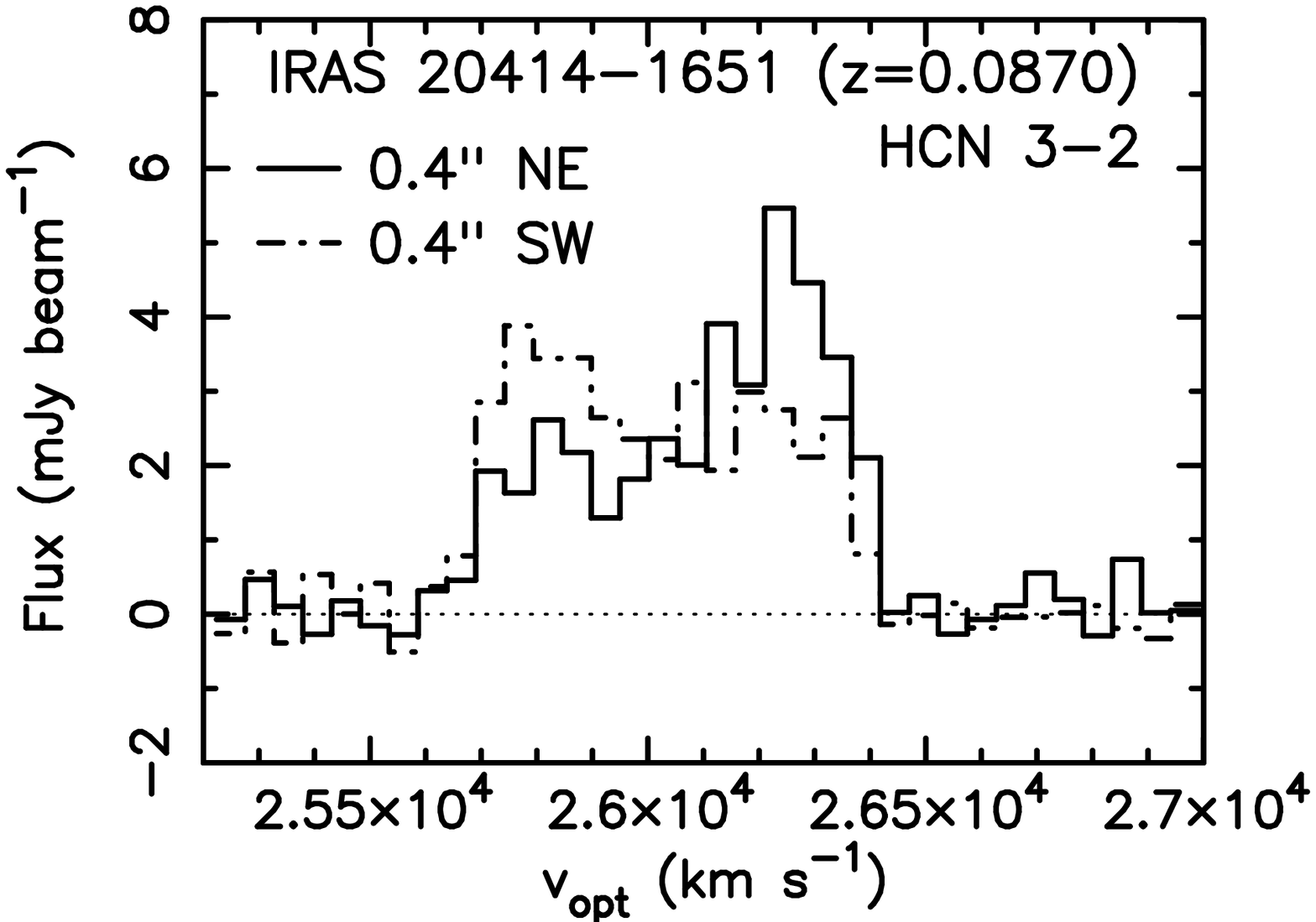} 
\includegraphics[angle=0,scale=.41]{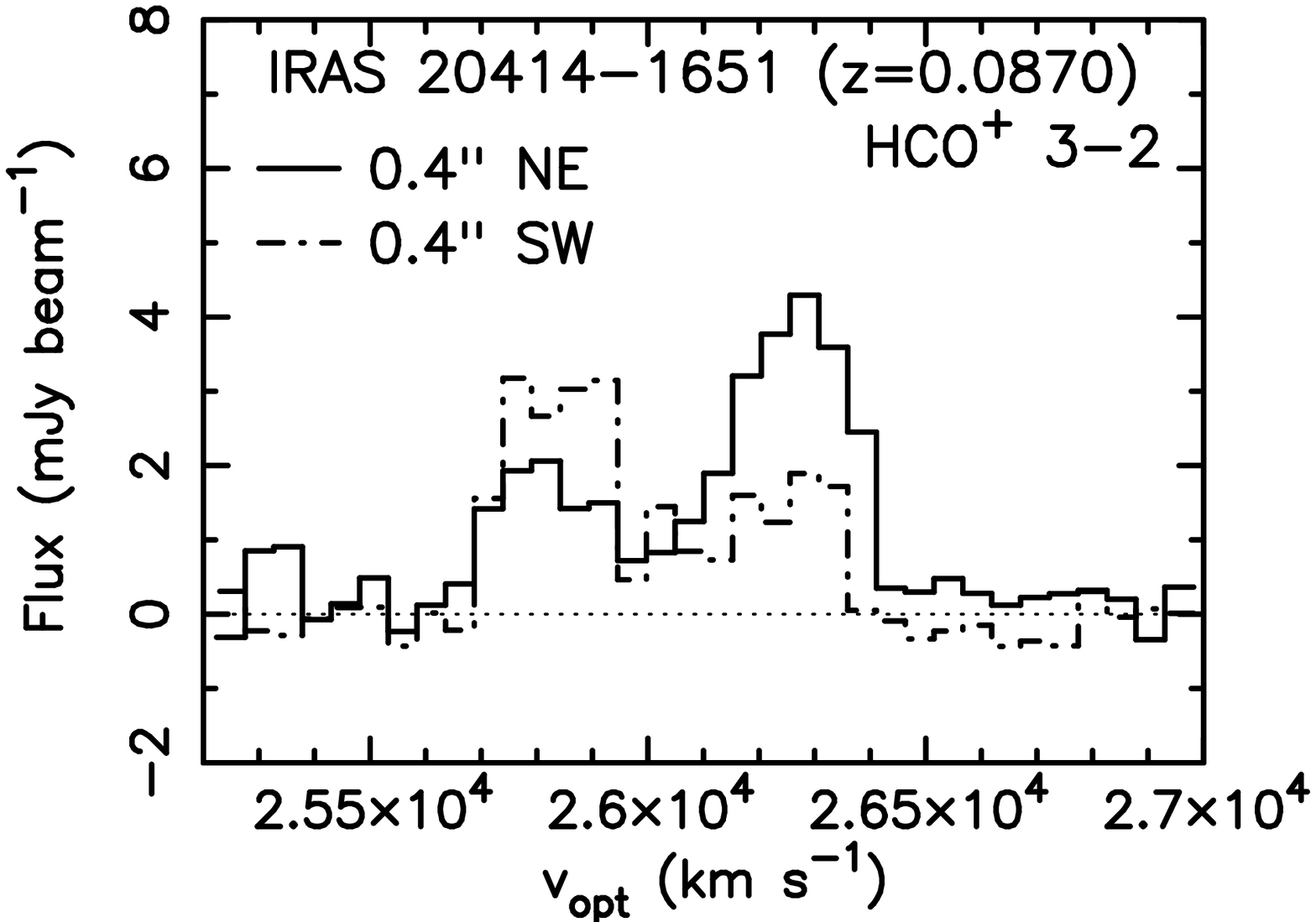} \\
\end{center}
\vspace{-0.6cm}
\caption{
Spectra within the beam size, at 3 pix (0$\farcs$3) east and 3 pix
(0$\farcs$3) north from the continuum peak position (denoted as 0.4'' NE
and displayed with the solid line), and at 3 pix (0$\farcs$3) west and 3
pix (0$\farcs$3) south from the continuum peak position (denoted as
0.4'' SW and displayed with the dash-dotted line) for IRAS 12112$+$0305
NE and IRAS 20414$-$1651. 
In the intensity-weighted mean velocity (moment 1) maps of both objects,
the north-eastern part is more redshifted than the south-western part.
}
\end{figure}

\begin{figure}
\begin{center}
\includegraphics[angle=0,scale=.41]{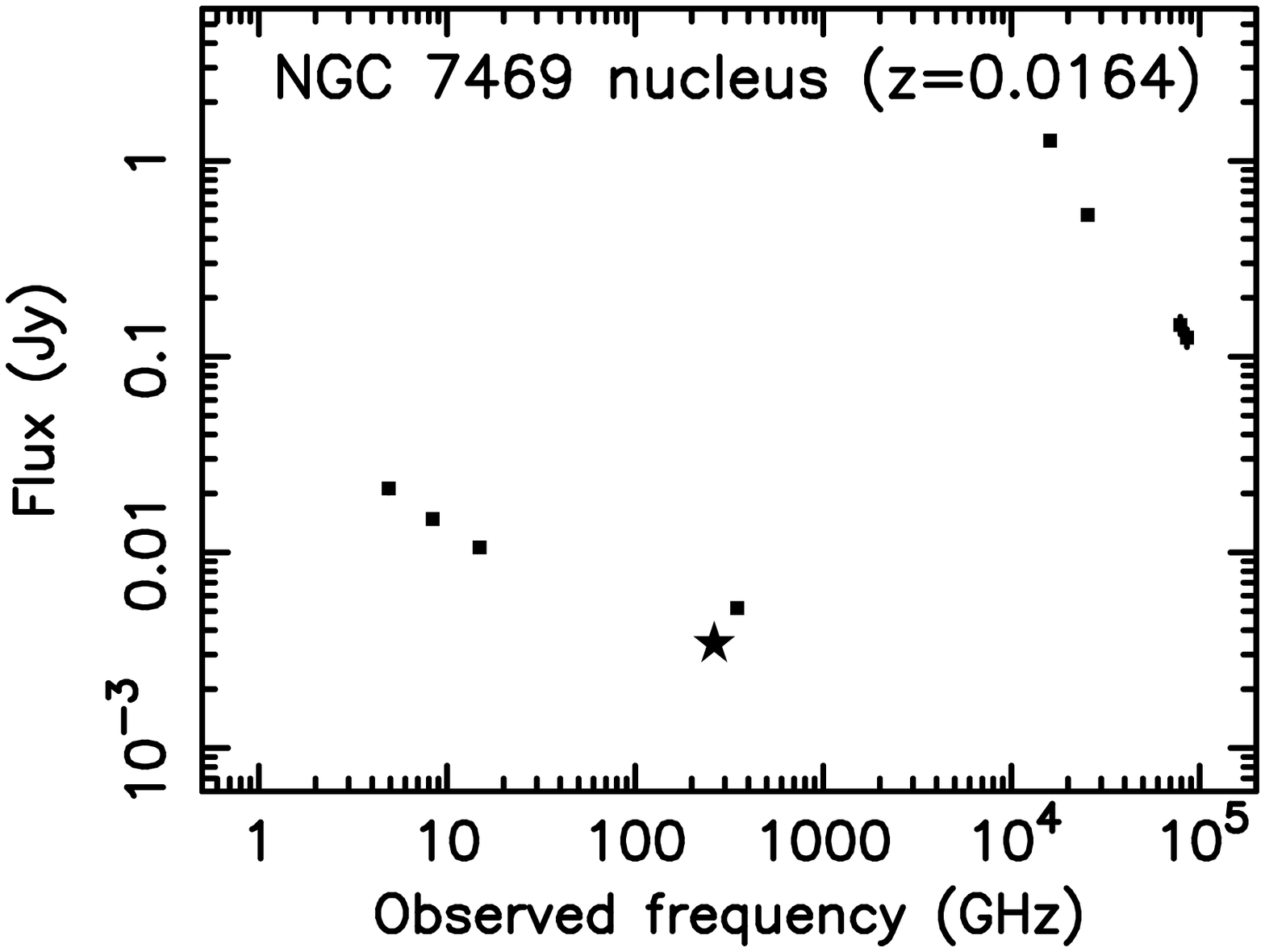} 
\includegraphics[angle=0,scale=.41]{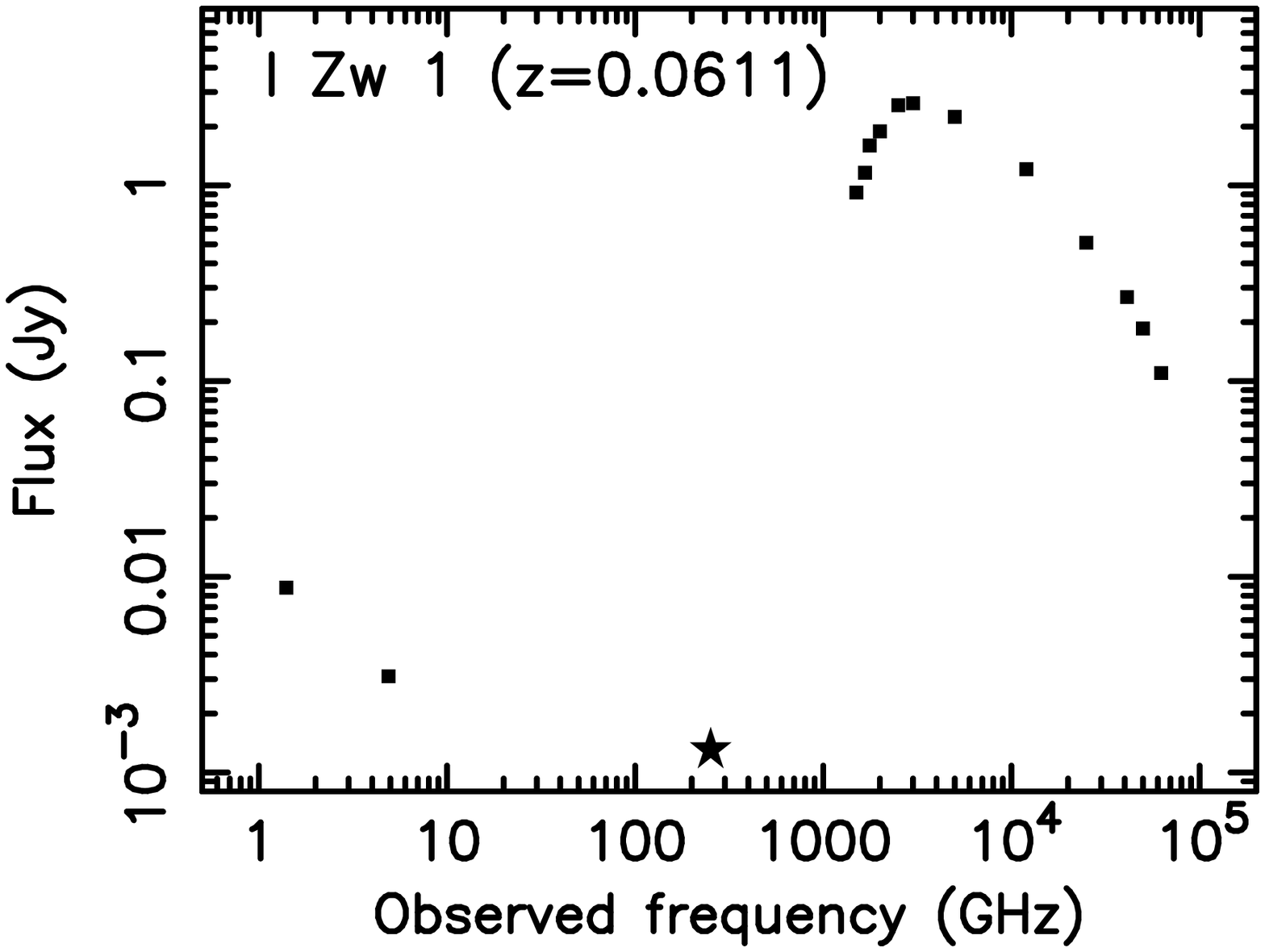} \\
\includegraphics[angle=0,scale=.41]{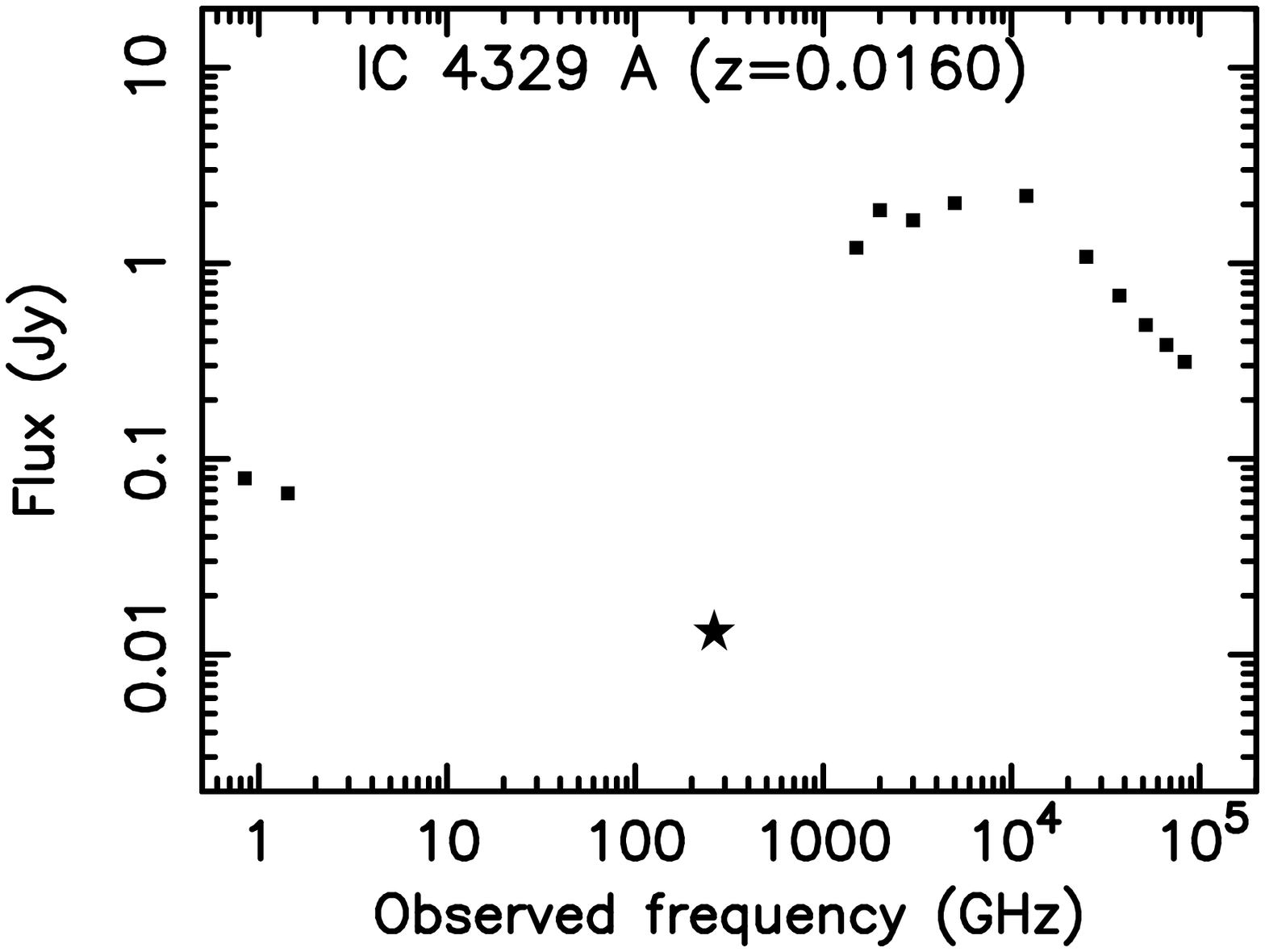} 
\end{center}
\caption{
Spectral energy distribution from 3 $\mu$m (10$^{5}$ GHz) to radio
($\sim$1 GHz). 
Filled stars are from our ALMA Cycle 1 observations and filled squares
are from the literature. 
For NGC 7469, following \citet{izu15}, we adopt flux measurement for the
nuclear component only \citep{mar98,reu10,izu15,wil91,ori10}, 
to remove the contamination from the spatially extended starburst ring.
For I Zw 1 and IC 4329 A, we use data from the NASA/IPAC Extragalactic
Database to derive the total flux at an individual frequency.} 
\end{figure}


\begin{thebibliography}{}
\bibitem[Aalto et al.(1995)]{aal95}
         Aalto, S., Booth, R. S., Black, J. H., \& Johansson,
         L. E. B. 1995, A\&A, 300, 369 
\bibitem[Aalto et al.(2015a)]{aal15a}
         Aalto, S., Garcia-Burillo, S., Muller, S., et al. 2015a, A\&A,
         574, A85
\bibitem[Aalto et al.(2015b)]{aal15b}
         Aalto, S., Costagliola, S., Martin F., et al. 2015b, A\&A, 584,
         A42
\bibitem[Alonso-Herrero et al.(2001)]{alo01}
         Alonso-Herrero, A., Engelbracht, C. W., Rieke, M. J., Rieke,
         G. H., \& Quillen, A. C. 2001, ApJ, 546, 952
\bibitem[Alonso-Herrero et al.(2011)]{alo11}
         Alonso-Herrero, A., Ramos Almeida, C., Mason, R., et al. 2011,
         ApJ, 736, Engelbracht, C. W., Rieke, M. J., Rieke,
\bibitem[Armus et al.(2007)]{arm07}
         Armus, L., Charmandaris, V., Bernard-Salas, J., et al. 2007,
         ApJ, 656, 148 
\bibitem[Asmus et al.(2014)]{asm14}
         Asmus, D., Honig, S. F., Gandhi, P., Smette, A., \& Duschl, W. J.
         2014, MNRAS, 439, 1648
\bibitem[Barvainis et al.(1996)]{bar96}
         Barvainis, R., Lonsdale, C., \& Antonucci, R. 1996, AJ, 111,
         1431 
\bibitem[Braito et al.(2003)]{bra03} 
         Braito, V., Franceschini, A., Della Ceca, R. et al. 2003, A\&A,
         398, 107 
\bibitem[Braito et al.(2009)]{bra09} 
         Braito, V., Reeves, J. N., Della Ceca, R. et al., 2009, A\&A,
         504, 53 
\bibitem[Brandl et al.(2006)]{bra06}
         Brandl, B. R., Bernard-Salas, J., Spoon, H. W. W. et al. 2006,
         ApJ, 653, 1129    
\bibitem[Brenneman et al.(2008)]{bre14}
         Brenneman, L. W., Madejski, G., Fuerst, F., et al. 2014, ApJ,
         788, 61
\bibitem[Brightman \& Nandra(2011)]{bri11}
         Brightman, M., \& Nandra, K. 2011, MNRAS, 414, 3084
\bibitem[Burtscher et al.(2013)]{bur13}
         Burtscher, L., Meisenheimer, K., Tristram, K. R. W. et
         al. 2013, A\&A, 558, 149
\bibitem[Carrol \& Goldsmith(1981)]{car81}
         Carrol, T. J., \& Goldsmith, P. F., 1981, ApJ, 245, 891  
\bibitem[Condon et al.(1991)]{con91}
         Condon, J. J., Anderson, M. L., \& Helou, G. 1991, ApJ, 376, 95
\bibitem[Condon et al.(1998)]{con98}
         Condon, J. J., Cotton, W. D., Greisen, E. W. et al. 1998, AJ,
         115, 1693
\bibitem[Costagliola et al.(2011)]{cos11}
         Costagliola, F., Aalto, S., Rodriguez, M. I., et al. 2011,
         A\&A, 528, A30 
\bibitem[Costagliola et al.(2015)]{cos15}
         Costagliola, F., Sakamoto, K.,  Muller, S., et al. 2015, 
         A\&A, 582, 91
\bibitem[Crawford et al.(1996)]{cra96}
         Crawford, T., Marr, J., Partridge, B., \& Strauss, M. A. 1996,
         ApJ, 460, 225
\bibitem[Dartois et al.(2007)]{dar07}
         Dartois, E., \& Munoz-Caro, G. M. 2007, A\&A, 476, 1235 
\bibitem[Davies et al.(1984)]{dav84}
         Davies, P. B., Hamilton, P. A., \& Rothwell, W. J. 1984,
         J. Chem. Phys., 81, 1598 
\bibitem[Davies et al.(2004)]{dav04}
         Davies, R. I., Tacconi, L. J., \& Genzel, R. 2004, ApJ, 602,
         148 
\bibitem[Diaz-Santos et al.(2008)]{dia08}
         Diaz-Santos, T., Alonso-Herrero, A., Colina, L., Packham, C., 
         Radomski, J. T., \& Telesco, C. M. 2008, ApJ, 685, 211   
\bibitem[Diaz-Santos et al.(2007)]{dia07}
         Diaz-Santos, T., Alonso-Herrero, A., Colina, L., Ryder, S. D.,
         \& Knapen, J. H. 2007, ApJ, 661, 149 
\bibitem[Downes et al.(1993)]{dow93}
         Downes, D., Solomon, P. M., \& Radford, S. J. E. 1993, ApJ,
         414, L13
\bibitem[Drake et al.(2004)]{dra04} 
         Drake, C. L., McGregor, P. J., Dopita, M. A., \& van Breugel,
         W. J. M. 2003, AJ, 126, 2237
\bibitem[Draine \& Lee(1984)]{dra84} 
         Draine, B. T., \& Lee, H. M. 1984, ApJ, 285, 89
\bibitem[Duc et al.(1997)]{duc97}
         Duc, P. -A., Mirabel, I. F., \& Maza, J. 1997, A\&AS, 124, 533  
\bibitem[Dudley \& Wynn-Williams(1997)]{dud97} 
         Dudley, C. C., \& Wynn-Williams, C. G. 1997, ApJ, 488, 720 
\bibitem[Esquej et al.(2014)]{esq14}
         Esquej, P., Alonso-Herrero, A., Gonzalez-Martin, O., et
         al. 2014, ApJ, 780, 86
\bibitem[Evans et al.(1999)]{eva99} 
         Evans, A. S., Kim, D. C., Mazzarella, J. M. et al. 1999, ApJ,
         521, 107L
\bibitem[Evans et al.(2002)]{eva02} 
         Evans, A. S., Mazzarella, J. M., Surace, J. A., \& Sanders,
         D. B. 2002, ApJ, 580, 749
\bibitem[Evans et al.(2006)]{eva06} 
         Evans, A. S., Solomon, P. M., Tacconi, L. J., Vavilkin, T., \& 
         Downes, D. 2006, AJ, 132, 2398
\bibitem[Farrah et al.(2009)]{far09}
         Farrah, D., Connolly, B., Connolly, N. et al. 2009, ApJ, 700,
         396 
\bibitem[Ferrarese \& Merritt(2000)]{fer00}
         Ferrarese, L., \& Merritt, D. 2000, ApJ, 539, L9 
\bibitem[Galliano et al.(2005)]{gal05}
         Galliano, E., Alloin, D., Pantin, E., Lagage, P. O., \& Marco,
         2005, A\&A, 438, 803
\bibitem[Garcia-Burillo et al.(2006)]{gar06}
         Garcia-Burillo, S., Gracia-Carpio, J., Guelin, M., et al. 2006,
         ApJ, 645, L17 
\bibitem[Garcia-Burillo et al.(2014)]{gar14}
         Garcia-Burillo, S., Combes, F., Usero, A., et al. 2014, A\&A,
         567, 125 
\bibitem[Genzel et al.(1995)]{gen95} 
         Genzel, R., Weitzel, L., Tacconi-Garman, L. E., et al. 1995,
         ApJ, 444, 129
\bibitem[Genzel et al.(1998)]{gen98} 
         Genzel, R., Lutz, D., Sturm, E. et al. 1998, ApJ, 498, 579 
\bibitem[Greve et al.(2009)]{gre09}
         Greve, T. R., Papadopoulos, P. P., Gao, Y., \& Radford,
         S. J. E., 2009, ApJ, 692, 1432
\bibitem[Gultekin et al.(2009)]{gul09} 
         Gultekin, K., Richstone, D. O., Gebhardt, K., et al. 2009, ApJ,
         698, 198 
\bibitem[Haan et al.(2011)]{haa11}
         Haan, S., Armus, L., Laine, S. et al. 2011, ApJS, 197, 27
\bibitem[Harada et al.(2010)]{har10}
         Harada, N., Herbst, E., \& Wakelam, V. 2010, ApJ, 721, 1570 
\bibitem[Harada et al.(2013)]{har13}
         Harada, N., Thompson, T. A., \& Herbst, V. 2013, ApJ, 765, 108
\bibitem[Henkel et al.(2014)]{hen14}
         Henkel, C., Asiri, H., Ao, Y., et al. 2014, A\&A, 565, A3
\bibitem[Henkel \& Mauersberger(1993)]{hen93a}
         Henkel, C., \& Mauersberger, R. 1993, A\&A, 274, 730
\bibitem[Henkel et al.(1993)]{hen93b}
         Henkel, C., Mauersberger, R., Wiklind, T., et al. 1993, A\&A,
         268, L17
\bibitem[Hildebrand(1983)]{hil83}
         Hildebrand, R. H. 1983, QJRAS, 24, 267
\bibitem[Honig et al.(2010)]{hon10}
         Honig, S. F., Kishimoto, M., Gandhi, P. et al. 2010, A\&A, 515,
         23
\bibitem[Hopkins et al.(2005)]{hop05}
         Hopkins, P. F., Hernquist, L., Cox, T. J., et al. 2005, ApJ,
         630, 705 
\bibitem[Hopkins et al.(2006)]{hop06}
         Hopkins, P. F., Hernquist, L., Cox, T. J., et al. 2006, ApJS,
         163, 1 
\bibitem[Hsieh et al.(2012)]{hsi12}
         Hsieh, P-Y., Ho, P. T. P., Kohno, K., Hwang, C-Y., Matsushita,
         S. 2012, ApJ, 747. 90
\bibitem[Huchra \& Burg(1992)]{huc92}
         Huchra, J., \& Burg, R. 1992, ApJ, 393, 90
\bibitem[Ichikawa et al.(2015)]{ich15}
         Ichikawa, K., Packham, C., Ramos Almeida, C., et al. 2015, ApJ,
         803, 57 
\bibitem[Imanishi(2002)]{ima02} 
         Imanishi, M. 2002, ApJ, 569, 44
\bibitem[Imanishi(2003)]{ima03}
         Imanishi, M. 2003, ApJ, 599, 918
\bibitem[Imanishi(2006)]{ima06c}
         Imanishi, M. 2006, AJ, 131, 2406
\bibitem[Imanishi \& Dudley(2000)]{imd00} 
         Imanishi, M., \& Dudley, C. C. 2000, ApJ, 545, 701 
\bibitem[Imanishi et al.(2006a)]{ima06a}
         Imanishi, M., Dudley, C. C., \& Maloney, P. R. 2006a, ApJ, 637, 
         114
\bibitem[Imanishi et al.(2007b)]{ima07b} 
         Imanishi, M., Dudley, C. C., Maiolino, R., Maloney, P. R., 
         Nakagawa, T., \& Risaliti, G. 2007b, ApJS, 171, 72
\bibitem[Imanishi et al.(2011a)]{ima11a} 
         Imanishi, M., Ichikawa, K., Takeuchi, T., et al. 2011a, PASJ,
         63, S447 
\bibitem[Imanishi et al.(2011b)]{ima11b} 
         Imanishi, M., Imase, K., Oi, N., \& Ichikawa, K. 2011b, AJ, 141,
         156
\bibitem[Imanishi et al.(2010)]{ima10}
         Imanishi, M., Nakagawa, T., Shirahata, M., Ohyama, Y., \&
         Onaka, T. 2010, ApJ, 721, 1233
\bibitem[Imanishi \& Nakanishi(2006)]{in06} 
         Imanishi, M., \& Nakanishi, K. 2006, PASJ, 58, 813
\bibitem[Imanishi \& Nakanishi(2013a)]{ima13a} 
         Imanishi, M., \& Nakanishi, K. 2013a, AJ, 146, 47
\bibitem[Imanishi \& Nakanishi(2013b)]{ima13b} 
         Imanishi, M., \& Nakanishi, K. 2013b, AJ, 146, 91
\bibitem[Imanishi \& Nakanishi(2014)]{ima14b} 
         Imanishi, M., \& Nakanishi, K. 2014, AJ, 148, 9
\bibitem[Imanishi et al.(2016c)]{ima16c}
         Imanishi, M., \& Nakanishi, K. et al. 2016c, in preparation
\bibitem[Imanishi et al.(2016b)]{ima16b}
         Imanishi, M., Nakanishi, K., \& Izumi, T. 2016b, ApJL, 822, L10 
\bibitem[Imanishi et al.(2016a)]{ima16a}
         Imanishi, M., Nakanishi, K., \& Izumi, T. 2016a, ApJ, 825, 44
\bibitem[Imanishi et al.(2004)]{ima04}
         Imanishi, M., Nakanishi, K., Kuno, N., \& Kohno, K. 2004, AJ,
         128, 2037 
\bibitem[Imanishi et al.(2006b)]{ima06b} 
         Imanishi, M., Nakanishi, K., \& Kohno, K. 2006b, AJ, 131, 2888
\bibitem[Imanishi et al.(2007a)]{ima07a} 
         Imanishi, M., Nakanishi, K., Tamura, Y., Oi, N., \& Kohno,
         K. 2007a, AJ, 134, 2366
\bibitem[Imanishi et al.(2009)]{ima09} 
         Imanishi, M., Nakanishi, K., Tamura, Y., \& Peng, C. -H. 2009,
         AJ, 137, 3581 
\bibitem[Imanishi \& Saito(2014)]{ima14a}
         Imanishi, M., \& Saito, Y. 2014, ApJ, 780, 106  
\bibitem[Imanishi \& Wada(2004)]{iw04} 
         Imanishi, M., \& Wada, K. 2004, ApJ, 617, 214
\bibitem[Iono et al.(2013)]{ion13} 
         Iono, D., Saito, T., Yun, M. S., et al. 2013, PASJ, 65, L7
\bibitem[Izumi et al.(2015)]{izu15}
         Izumi, T., Kohno, K., Aalto, S., et al. 2015, ApJ, 811, 39
\bibitem[Izumi et al.(2016)]{izu16}
         Izumi, T., Kohno, K., Aalto, S., et al. 2016, ApJ, 818, 42
\bibitem[Jia et al.(2012)]{jia12}
         Jia, J., Ptak, A., Heckman, T. M., Braito, V., \& Reeves,
         J. 2012, ApJ, 759, 41
\bibitem[Kawaguchi et al.(1985)]{kaw85}
         Kawaguchi, K., Yamada, C., Saito, S., \& Hirota, E., 1985,
         J. Chem. Phys, 82, 1750
\bibitem[Kewley et al.(2001)]{kew01}
         Kewley, L. J., Heisler, C. A., Dopita, M. A., \& Lumsden,
         S. 2001, ApJS, 132, 37 
\bibitem[Kim \& Sanders(1998)]{kim98}
         Kim, D. -C., \& Sanders, D. B., 1998, ApJS, 119, 41  
\bibitem[Kim et al.(1995)]{kim95}
         Kim, D. -C., Sanders, D. B., Veilleux, S., Mazzarella, J. M.,
         \& Soifer, B. T. 1995, 98, 129
\bibitem[Kim et al.(2002)]{kim02}
         Kim, D. -C., Veilleux, S., \& Sanders, D. B., 2002, ApJS, 143, 277
\bibitem[Kohno(2005)]{koh05}
         Kohno, K. 2005, in AIP Conf. Ser. 783, 
         The Evolution of Starbursts, ed. S. H\"uttemeister, E. Manthey,
         D. Bomans, \& K. Weis (New York: AIP), 203 (astro-ph/0508420)
\bibitem[Komatsu et al.(2009)]{kom09}
         Komatsu, E., Dunkley, J., Nolta, M. R., et al. 2009, ApJS, 180,
         330 
\bibitem[Krips et al.(2008)]{kri08}
         Krips, M., Neri, R., Garcia-Burillo, S., Martin, S., Combes,
         F., Gracia-Carpio, J., \& Eckart, A. 2008, ApJ, 677, 262 
\bibitem[Lintott \& Viti(2006)]{lin06}
         Lintott, C., \& Viti, S. 2006, ApJ, 646, L37
\bibitem[Ma et al.(1998)]{ma98}
         Ma, C., Arias, E. F., Eubanks, T. M. et al. 1998, AJ, 116, 516
\bibitem[Magorrian et al.(1998)]{mag98}
         Magorrian, J., Tremaine, S., Richstone, D., et al. 1998, ApJ, 
         115, 2285
\bibitem[Malkan et al.(1998)]{mal98}
         Malkan, M. A., Gorjian, V., \& Tam, R., 1998, ApJS, 117, 25
\bibitem[Marco \& Alloin(1998)]{mar98}
         Marco, O., \& Alloin, D. 1998, A\&A, 336, 823
\bibitem[Marshall et al.(2007)]{mar07}
         Marshall, J. A., Herter, T. L., Armus, L., et al. 2007, ApJ,
         670, 129
\bibitem[Martin et al.(2010)]{mar10}
         Martin, S., Aladro, R., Martin-Pintado, J., \& Mauersberger,
         R. 2010, A\&A, 522, A62
\bibitem[Martin et al.(2016)]{mar16}
         Martin, S., Aalto, S., Sakamoto, K., et al. 2016, A\&A, 590, A25
\bibitem[McConnell \& Ma(2013)]{mcc13}
         McConnell, N. J. \& Ma, C-P. 2013, ApJ, 764, 184
\bibitem[Meijerink \& Spaans(2005)]{mei05}
         Meijerink, R., \& Spaans, M. 2005, A\&A, 436, 397 
\bibitem[Meijerink et al.(2007)]{mei07}
         Meijerink, R., Spaans, M., \& Israel, F. P. 2007, A\&A, 461,
         793
\bibitem[Miles et al.(1996)]{mil96}
         Miles, J. W., Houck, J. R., Hayward, T. L., \& Ashby,
         M. L. N. 1996, ApJ, 465, 191  
\bibitem[Mirabel et al.(1991)]{mir91}
         Mirabel, I. F., Lutz, D., \& Maza, J. 1991, A\&A, 243, 367 
\bibitem[Moorwood(1986)]{moo86}
         Moorwood, A. F. M. 1986, A\&A, 166, 4
\bibitem[Nagy et al.(2015)]{nag15}
         Nagy, Z., van der Tak, F. F. S., Fuller, G. A., \& Plume,
         R. 2015, A\&A, 577, A127
\bibitem[Nardini et al.(2008)]{nar08}
         Nardini, E., Risaliti, G., Salvati, M., et al. 2008, MNRAS,
         385, L130
\bibitem[Nardini et al.(2009)]{nar09}
         Nardini, E., Risaliti, G., Salvati, M., et al. 2009, MNRAS,
         399, 1373 
\bibitem[Nardini et al.(2010)]{nar10}
         Nardini, E., Risaliti, G., Watabe, Y., Salvati, M., \& Sani,
         E. 2010, MNRAS, 405, 2505
\bibitem[Orienti \& Prieto(2010)]{ori10}
         Orienti, M., \& Prieto, M. A. 2010, MNRAS, 401, 2599 
\bibitem[Papadopoulos et al.(2007)]{pap07}
         Papadopoulos, P. P. et al. 2007, ApJ, 656, 792
\bibitem[Pereira-Santaella et al.(2012)]{per15}
         Pereira-Santaella, M., Colina, L., Alonso-Herrero, A. et
         al. 2015, MNRAS, 454, 3679
\bibitem[Pinoncelli et al.(2005)]{pin05} 
         Pinoncelli, E., Jimenez-Bailon, E., Guainazzi, M., et al. 2005,
         A\&A, 432, 15
\bibitem[Privon et al.(2015)]{pri15}
         Privon, G. C., Herrero-Illana, R., Evans, A. S., et al. 2015,
         ApJ, 814, 39
\bibitem[Ranalli et al.(2003)]{ran03} 
         Ranalli, P., Comastri, A., \& Setti, G. 2003, A\&A, 399, 39 
\bibitem[Rangwala et al.(2011)]{ran11}
         Rangwala, N., Maloney, P. R., Glenn, J., et al. 2011, ApJ, 743, 94 
\bibitem[Reunanen et al.(2007)]{reu07}
         Reunanen, J., Tacconi-Garman, L. E., \& Ivanov, V. D. 2007,
         MNRAS, 382, 951
\bibitem[Reunanen et al.(2010)]{reu10}
         Reunanen, J., Prieto, M. A., \& Siebenmorgen, R. 2010, MNRAS,
         402, 879 
\bibitem[Risaliti et al.(2003)]{ris03}
         Risaliti, G., Maiolino, R., Marconi, A. et al. 2003, ApJ, 595,
         L17 
\bibitem[Roche et al.(1991)]{roc91}
         Roche, P. F., Aitken, D. K., Smith, C. H., \& Ward, M. J., 
         1991, MNRAS, 248, 606
\bibitem[Rodriguez-Ardila \& Viegas(2003)]{rod03}
         Rodriguez-Ardila, A., \& Viegas, S. M. 2003, MNRAS, 340, L33
\bibitem[Roy et al.(1998)]{roy98}
         Roy, A. L., Norris, R. P., Kesteven, M. J., Troup, E. R., \&
         Raynolds, J. E. 1998, MNRAS, 301, 1019 
\bibitem[Rush et al.(1993)]{rus93}
         Rush, B., Malkan, M. A., \& Spinoglio, L. 1993, ApJS, 89, 1
\bibitem[Sakamoto et al.(2010)]{sak10}
         Sakamoto, K., Aalto, S., Evans, A. S., Wiedner, M., \& Wilner,
         D. 2010, ApJ, 725, L228
\bibitem[Sakamoto et al.(2009)]{sak09}
         Sakamoto, K., Aalto, S., Wilner, D. J., et al. 2009, ApJ, 700,
         L104
\bibitem[Sanders \& Mirabel(1996)]{sam96}
         Sanders, D. B., \& Mirabel, I. F. 1996, ARA\&A, 34, 749
\bibitem[Schmidt \& Green(1983)]{sch83}
         Schmidt, M., \& Green, R. F. 1983, ApJ, 269, 352
\bibitem[Scoville et al.(2000)]{sco00}
         Scoville, N. Z., Evans, A. S., Thompson, R., et al. 2000, AJ,
         119, 991 
\bibitem[Scoville et al.(2015)]{sco15}
         Scoville, N., Sheth, K., Walter, F., et al. 2015, ApJ, 800, 70
\bibitem[Shang et al.(2011)]{sha11}
         Shang, Z., Brotherton, M. S., Wills, B. J., et al, 2011, ApJ,
         196, 2  
\bibitem[Soifer et al.(2003)]{soi03}
         Soifer, B. T., Bock, J. J., Marsh, K., Neugebauer, G.,
         Matthews, K., Egami, E., \& Armus, L. 2003, AJ, 126, 143
\bibitem[Soifer et al.(2000)]{soi00}
         Soifer, B. T., Neugebauer, G., Matthews, K., et al. 2000, AJ,
         119, 509  
\bibitem[Soifer et al.(1987)]{soi87}
         Soifer, B. T., Sanders, D. B., Madore, B. F., et al. 1987, ApJ,
         320, 238
\bibitem[Soifer et al.(2001)]{soi01}
         Soifer, B. T., Neugebauer, G., Matthews, K. et al. 2001, AJ,
         122, 1213 
\bibitem[Solomon et al.(1987)]{sol87}
         Solomon, P. M., Rivolo, A. R., Barrett, J., et al. 1987, ApJ,
         319, 730 
\bibitem[Spoon et al.(2002)]{spo02}
         Spoon, H. W. W., Keane, J. V., Tielens, A. G. G. M., Lutz, D.,
         Moorwood, A. F. M., \& Laurent, O.  2002, A\&A, 385, 1022
\bibitem[Spoon et al.(2006)]{spo06}
         Spoon, H. W. W., Tielens, A. G. G. M., Armus, L., et al. 2006,
         ApJ, 638, 759
\bibitem[Stierwalt et al.(2013)]{sti13}
          Stierwalt, S., Armus, L., \& Surace, J. A., et al. 2013, ApJS,
          206, 1
\bibitem[Strauss et al.(1992)]{str92}
         Strauss, M. A., Huchra, J. P., Davis, M., et al. 1992, ApJS,
         83, 29
\bibitem[Teng et al.(2015)]{ten15}
         Teng, S. H., Rigby, J. R., Stern, D,. et al. 2015, ApJ, 814, 56
\bibitem[Teng et al.(2014)]{ten14}
         Teng, S. H., Brandt, W. N., Harrison, F. A. et al., 2014, ApJ,
         785, 19
\bibitem[Teng et al.(2009)]{ten09}
         Teng, S. H., Veilleux, S., Anabuki, N., et al. 2009, ApJ, 691, 261
\bibitem[White \& Rees(1978)]{whi78}
         White, S. D. M., \& Rees, M. J. 1978, MNRAS, 183, 341
\bibitem[Wilson et al.(1991)]{wil91}
         Wilson, A. S., Helfer, T. T., Haniff, C. A., \& Ward,
         M. J. 1991, ApJ, 381, 79 
\bibitem[Yamada et al.(2007)]{yam07}
         Yamada, M., Wada, K., \& Tomisaka, K. 2007, ApJ, 671, 73 
\bibitem[Vega et al.(2008)]{veg08}
         Vega, O., Clemens, M. S., Bressan, A., et al. 2008, A\&A, 484,
         631 
\bibitem[Veilleux et al.(1999)]{vei99} 
         Veilleux, S., Kim, D. -C., \& Sanders, D. B. 1999, ApJ, 
         522, 113
\bibitem[Veilleux et al.(1995)]{vei95} 
         Veilleux, S., Kim, D. -C., Sanders, D. B., Mazzarella, J. M., \&
         Soifer, B. T. 1995, ApJS, 98, 171 
\bibitem[Veilleux et al.(2013)]{vei13} 
         Veilleux, S., Melendez, M., Sturm, E., et al. 2013, ApJ, 776,
         27  
\bibitem[Veilleux et al.(1997)]{vei97} 
         Veilleux, S., Sanders, D. B., \& Kim, D. -C. 1997, ApJ, 484, 92  
\bibitem[Veilleux et al.(2009)]{vei09}
         Veilleux, S., Rupke, D. S. N., Kim, D.-C., et al. 2009, ApJS,
         182, 628
\bibitem[Voit(1992)]{voi92}
         Voit, G. M. 1992, MNRAS, 258, 841
\bibitem[Wu et al.(2009)]{wu09}
         Wu, Y., Charmandaris, V., Huang, J., Spinoglio, L., \&
         Tommasin, S. 2009, ApJ, 701, 658
\bibitem[Yuan et al.(2010)]{yua10} 
         Yuan, T. -T., Kewley, L. J., \& Sanders, D. B. 2010, ApJ, 709,
         884 
\bibitem[Ziurys \& Turner(1986)]{ziu86}
         Ziurys, L. M., \& Turner, B. E. 1986, ApJ, 300, L19 
\end{thebibliography}
\end{document}